\documentstyle[11pt,fleqn]{article}
\newtheorem{Theorem}{Theorem}
\newtheorem{Lemma}{Lemma}
\renewcommand{\theequation}{\mbox{\arabic{section}.\arabic{equation}}}
\begin{document}

\begin{titlepage}
\raggedleft{\sl UTF 325, \rm to appear in Physics Reports}

\bigskip
\begin{center} \Large\bf
Quantum Fields and Extended Objects \\
in Space-Times \\
with Constant Curvature Spatial Section
\end{center}

\bigskip
\begin{center}\Large\rm
Andrei A. Bytsenko
\footnote{e-mail: root@fmf.stu.spb.su  (subject: Prof. A.A. Bytsenko)}\\
\smallskip\large
Department of Theoretical Physics,
State Technical University \\
St. Petersburg 195251, Russia
\end{center}

\begin{center}\Large\rm
Guido Cognola\footnote{e-mail: cognola@science.unitn.it},
Luciano Vanzo\footnote{e-mail: vanzo@science.unitn.it}
and Sergio Zerbini\footnote{e-mail: zerbini@science.unitn.it} \\
\smallskip\large
Dipartimento di Fisica, Universit\`a di Trento \\
and Istituto Nazionale di Fisica Nucleare, \\
Gruppo Collegato di Trento, Italia
\end{center}
\rm\normalsize
\bigskip
\date{May 1995}

\begin{abstract}
The heat-kernel expansion and $\zeta$-regularization techniques for
quantum field theory and extended objects on curved space-times are
reviewed. In particular, ultrastatic  space-times with spatial section
consisting in manifold with constant curvature are discussed in
detail. Several mathematical results, relevant to physical
applications are presented, including exact solutions of the
heat-kernel equation, a simple exposition of hyperbolic geometry and
an elementary derivation of the Selberg trace formula.
With regards to the physical applications,
the vacuum energy for scalar fields, the
one-loop renormalization of a self-interacting scalar
field theory on a hyperbolic space-time,
with a discussion on the topological symmetry breaking,
the finite temperature effects and
the Bose-Einstein condensation, are considered.
Some attempts to generalize the results to extended objects are
also presented, including some remarks on path integral
quantization, asymptotic properties of extended objects and
a novel representation for the one-loop (super)string free energy.
\end{abstract}

\end{titlepage}

\normalsize\rm
\tableofcontents
\newpage
\setcounter{equation}{0}
\section{Introduction}

The problem of reconciling general relativity with quantum theory,
which has been deeply studied in the early twentieth century, has not
yet found a consistent and satisfactory solution. The theory of
quantum fields in curved space-time deals with quantum matter field in
an external gravitational field and may be considered as a preliminary
step toward the complete quantum theory of gravity (see for example
Refs.~\cite{dewi75-19-295,birr82b,full89b}).

Electromagnetic, weak and strong interactions may be unified in the so
called grand unified theory and this may be achieved in the well
established framework of relativistic quantum field theory. Roughly
speaking, this framework deals with point objects (particles) and
their related description in terms of local quantum fields. However,
it is well known that the gravitational interaction, which classically
is well described by general relativity, cannot be consistently
described within this framework (non perturbative renormalizability).
For this reason, the interest for extended objects has been grown.

The most simple extended object is the (super)-string \cite{gree87b}.
An interesting feature of string is that its energy-mass spectrum
contains a massless spin-two particle, which may be interpreted as a
graviton. Moreover, a fundamental length (Regge slope) appears in a
natural way and so the ultraviolet problems seem to be cured. When the
fundamental length goes to zero the usual field theory can be
recovered. The consequence of these facts leads to the hope that the
second quantized version of string theory will provide a consistent
picture of all elementary interactions at quantum level (see for
example Ref.~\cite{gree87b}).

A further insight for the construction of a consistent quantum theory
of all known elementary interactions has come with the proposal of the
modern Kaluza-Klein theories, which may provide a unified fundamental
theory in 4-dimensions by starting from an underlying theory in higher
dimensions. Some years ago these ideas were used in the framework of
field theory, including gravitation and extended objects, that is
(super)strings and (super)$p$-branes (generalization of strings),
which at the moment are receiving a great interest not only among
physicists of fundamental particles, but also among cosmologists.

In this report, we shall maily concentrate on space-times admitting a
constant curvature spatial section. The main motivation for studying
fields and (super)strings (at zero and finite temperature) on
ultrastatic space-times in which the spatial section is a manifold
with constant curvature, stems from the fact that, for a fixed value
of the cosmological time, such manifolds describe locally spatially
homogeneous isotropic universes.

More recently, there have been attempts to investigate also the role
of topology in the physics of quantum fields and  extended objects and
the phenomenon of topological mass generation and dynamical symmetry
breaking have been discovered. The non trivial topological space-times
which mainly have been considered till now, are the ones with the
torus and sphere geometries. In these cases, the spectrum of the
relevant operators is generally known. So far, this fact has been
fruitfully used in all the investigations. It has to be noted however,
that within the space-time with constant curvature spatial sections,
the compact hyperbolic manifolds have also to be considered. The
distinguish features of them are the huge number of possible
geometries and the highly non trivial topologies with respect to the
previous ones and for this reason they may provide many interesting
alternative solutions for the construction of models of our universe.
In these cases the spectrum of the relevant operators is, in general,
not explicitly known. However, there exists a powerful mathematical
tool, the Selberg trace formula \cite{selb56-20-47}, which permits to
evaluate some interesting physically global quantities, like the
vacuum energy or the one-loop effective action. Unfortunately, for a
compact hyperbolic manifold, it is difficult to get an explicit and
manageable form for local quantities like the propagators, which are
explicitly known for the torus, sphere and non compact hyperbolic
geometries (see for example
\cite{camp90-196-1}).

As we have already mentioned, in the last decades there has been a
great deal of investigations on the properties of interacting quantum
field theories in a curved space-time (see Ref.~\cite{gibb79b}).
Several techniques have been employed, among these we would like to
mention the background-field method \cite{dewi75-19-295} within
path-integral approach, which is very useful in dealing with the
one-loop approximation and which permits the one-loop effective action
to be evaluated. As a consequence, all physical interesting quantities
can be, in principle, evaluated. Throughout the paper, we mainly shall
make use of the path-integral quantization.

The (Euclidean) one-loop effective action, as derived from the
path-integral, is a ill defined quantity, being related to the
determinant of the fluctuation operator, which is a second order
elliptic non negative differential operator. Since it is not a bounded
operator, its real eigenvalues $\lambda_n$ diverge for large $n$.
Hence, the naive definition of the determinant leads to a formally
divergent quantity. It turns out that these divergences are controlled
by geometrical quantities, which are referred to as Seeley-DeWitt or
spectral coefficients. In order to control the divergences and extract
the finite part, many regularizations have been proposed and used. One
of the most promising, which works very well also in a curved
manifold, is the so called ``zeta-function regularization'' (see
\cite{hawk77-55-133}), which gives a rigorous definition of the
determinant of the fluctuation operator and therefore of the
regularized one-loop effective action. This latter has to be
renormalized, since it contains an arbitrary normalization parameter,
coming from the path integral measure. Here, other regularizations
will be analyzed and their close relation with the $\zeta$-function
one will be pointed out.

It should be stressed that, on a generic Riemannian manifold, the
$\zeta$-function related to the Laplace-Beltrami operator is generally
unknown, but nevertheless, some physical interesting quantities can be
related to its Mellin inverse transform, which has a computable
asymptotic expansion (heat kernel expansion). The coefficients of this
asymptotic expansion are the ones which control the ultraviolet
one-loop divergences (see for example \cite{dewi75-19-295,park79b}.
Thus, they are suitable only in  describing local properties
(ultraviolet behaviour), but some important physical issues, like
phase transitions in cosmological models, quantum anomalies and
symmetry breaking due to changes in topology, are insensitive to them.

The situation is really better on space-time with constant curvature
spatial sections, where the effective action can be explicitly
constructed, because the spectrum is explicitly known (zero and
positive curvature) or by making use of Selberg trace formula (compact
hyperbolic manifolds).

The zeta-function regularization technique is also  useful  in the
study of quantum fields or strings at finite temperature. It provides
complex integral representations (Mellin-Barnes) for one-loop
thermodynamic quantities, from which low and high temperature
expansions can be directly obtained.

With regards to the mathematical results we shall employ, for the
reader's convenience, part of the paper (mainly Secs.~\ref{S:PI} and
\ref{S:CCMan}) is devoted to the summary of some useful techniques,
with particular attention to the physical applications. We derive the
main features of $\zeta$-function and its relation with the heat
kernel expansion and report the results we need in several Appendices.
 In particular, for torus $T^N$, sphere $S^N$ and hyperbolic space
$H^N$, we give heat kernel and $\zeta$-function for the
Laplace-Beltrami operator in closed form. We briefly review the
Minakshisundaram-Pleijel-Schwinger-Seeley-DeWitt methods which permit
to obtain the heat parametrix for a differential operator on a compact
manifold without boundary (see, for example, the review article
\cite{park79b} and the monography \cite{birr82b}. With regard to this,
important results on a partly summed form of the heat-kernel expansion
have been presented in Refs.
\cite{park85-31-953,park85-31-2424,jack85-31-2439}.

After having discussed some properties of the hyperbolic space $H^N$,
the Harish-Chandra measure associated with scalar and spinor fields,
is derived solving the Laplace and Dirac equations on $H^N$ and
evaluating explicitly the corresponding density of states. We also
present a brief survey of the geometry and topology of compact
hyperbolic manifolds of the form $H^N/\Gamma$, $\Gamma$ being  group
of isometries of $H^N$ (see for example \cite{eliz94b}). An elementary
and self-contained discussion of Selberg trace formula for the
strictly hyperbolic manifold $H^N/{\Gamma}$ and the properties of the
Selberg $Z$-function and its logarithmic derivative are also
presented. Here, our aim is merely didactic and we refer to the
original mathematical literature for a more rigorous exposition. In
particular, for the two cases $H^2/\Gamma$ and $H^3/\Gamma$, which are
potentially interesting for a physical viewpoint, we also include
elliptic elements in the group $\Gamma$. In this way we shall deal
more properly with orbifold.

As for as the physical applications are concerned, making use of the
$\zeta$-function regularization, we derive the expressions of vacuum
and free energies (finite temperature effects within the canonical and
grandcanonical ensemble) for scalar and spinor fields, on arbitrary
$D$-dimensional ultrastatic Riemannian manifolds. The relation of
these results with the more general static case is pointed out, if
does not exist any Killing horizon in the manifold. Making use of the
Selberg trace formula, such expressions are specialized to compact
manifolds  of the form $\mbox{$I\!\!R$}\times H^2/\Gamma$ and
$\mbox{$I\!\!R$}\times H^3/\Gamma$ or, more generally, on Kaluza-Klein
space-times with the topology $\mbox{$I\!\!R$}^{D-p}\times H^p/\Gamma$.
It is found that, for trivial line bundles, one can obtain a negative
contribution to the vacuum energy. On such manifolds with compact
hyperbolic spatial section, we also compute the effective potential.
In the special but important case of $\mbox{$I\!\!R$}\times
H^3/\Gamma$, we carry on the one-loop renormalization program for the
$\lambda\phi^4$ self-interacting field theory. The role of topology is
analyzed in this context and the possibility of mass generation,
symmetry breaking or restoration is discussed.

The finite temperature effects are studied by employing a complex
integral representation  for the free energy, which will be referred
to as Mellin-Barnes representation. We show how it is particular
useful in deriving the high temperature expansion.

The inclusion in the previous theories of the chemical potential is
quite straightforward and this permits the grand canonical potential
and all related thermodynamic quantities to be computed. In
particular, we give high and low temperature expansions for the
thermodynamic potential  on $\mbox{$I\!\!R$}\times H^3$ and we discuss
in detail the problem of the Bose-Einstein condensation for a
relativistic ideal gas. The critical temperature is evaluated and its
dependence on curvature is exhibited.

Some attempts are made in order to generalize these results to the
extended objects. We discuss a semiclassical approximation to
path-integral quantization for the bosonic sector, when a closed
$p$-brane sweeps out on a $D=(p+1)$-dimensional compact hyperbolic
manifold. Within the same approximation, we present a general formula
of the static potential for $p$-branes compactified on Kaluza-Klein
space-times of the form $T^q\times\Pi_\alpha
S^{N_\alpha}\times\Pi_\beta(H^2/\Gamma_\beta)\times\mbox{$I\!\!R$}^{D-p}$.

It is shown that the quantum behaviour of the potential and its
extrema depend on the choice of twists.

Starting from the semiclassical quantization of $p$-branes (torus
compactification leads to discrete mass spectrum) and making use of
some rigorous results of number theory (the theorem of Meinardus) we
evaluate the  asymptotic form of the density of states. In particular,
the explicit form of the prefactors of total level degeneracy is
derived. The universal feature of the density, which grows slower than

the one for black holes, but faster than that for strings, leads to
the breakdown on the canonical description for $p>1$. With this
regard, we also comment on the possible connection between black holes
and $p$-branes.

As far as the finite temperature effects for extended objects are
concerned, due to the asymptotic behaviour of the state level density,
we have argued above that only the $p=1$ (string) seems to be
consistently tractable within the canonical ensemble. For this reason,
by using the analogous of the Mellin-Barnes representation for
one-loop thermodynamic quantities, we exhibit a novel representation
for the free energy of open and closed bosonic strings and open
superstrings, in terms of a Laurent series. It is shown that in this
formalism, the Hagedorn temperatures arise as the radius of
convergence of the corresponding Laurent series (convergence
condition). The high genus (interacting) string case is also discussed
and the independence of the Hagedorn temperature on the genus is
pointed out.

\paragraph{Notations.} \begin{description} \item{-} Throughout the
paper we shall use units in which the speed of light, Planck and
Boltzmann constants have the values $c=\hbar=k=1$. \item{-} $K_\nu(z)$
are the Mc Donald functions, $\Gamma(s)$ is the Euler gamma function,
$\zeta_{R,H}$ are the Riemann-Hurwitz zeta-functions, $Z_{ {\cal
R}}(z,\vec g,\vec h)$ is the Epstein $Z$-function, $Z(s)$ and $\Xi(s)$
are the Selberg functions. \item{-} $K_n(A)$ and $\zeta(s|A)$ are the
spectral coefficients and the $\zeta$-function respectively, related
to an elliptic operator $A$. \item{-} By $\cal M^N$ we shall indicate
a $N$ dimensional Riemannian manifold, with Euclidean metric $g_{ij}$,
$g$ being its determinant ($i,j=1,...,N$). $A$ or $L_N$ are  generic
differential operators and $\Delta_N$ is the Laplace-Beltrami operator.
\item{-} In the physical applications, we shall deal with space-times
$\cal M^D$ with metric $g_{\mu\nu}$, ($\mu,\nu=0,...,D-1$) and with
spatial section $\cal M^N$, $N=D-1$. \item{-} We shall mainly deal
with manifolds $\cal M^N$ with constant curvature $\kappa$.  Then we
shall set $\L_N=\Delta_N+\alpha^2+\kappa\varrho_N^2$, where
$\varrho_N=(N-1)/2$ and $\alpha$ is an arbitrary constant.
\end{description}

\newpage
\setcounter{equation}{0}
\section{Path integral and regularization techniques in curved space-times}
\label{S:PI}

Path-integral  quantization \cite{feyn65b} is a very powerful approach
to the quantization of a generic physical system. By means  of it, it
is possible to investigate, in an elegant and economic way, several
issues including for example gauge theory \cite{fadd67-25-30}, finite
temperature effects \cite{bern74-9-3312}, quantum anomalies
\cite{fuji80-44-1733}, effective action \cite{aber73-9-1} and quantum
gravity \cite{hawk78-18-1747,hawk79b}.

In the present section we discuss in some detail the quantization, via
path integral, of matter systems living on curved background
manifolds. With regard to this, the evaluation of the propagator and
heat-kernel in curved space-time by a path integral was presented in
Ref. \cite{beke81-23-2850}, where the Einstein universe (having
constant curvature spatial section) was solved exactly.

Since in the external field or in the Gaussian approximation of the
full theory, the one-loop effective action can be formally expressed
by means of determinants of elliptic differential operators, one needs
a regularization scheme in order to give a rigorous meaning to such
determinants. For this purpose, $\zeta$-function regularization
\cite{ray71-7-145,ray73-98-154,hawk77-55-133,eliz94b}, as we  have
already anticipated, is very convenient and it shall be extensively
used throughout the paper. Some properties and representations will be
derived in Sec.~\ref{S:ZF}. For the sake of completeness, other
(equivalent) regularizations of operator determinants will be
discussed in Sec.~\ref{S:RT}.

In this Section, the one-loop effective potential is introduced and,
for a $\lambda\phi^4$ theory, the renormalization group equations are
derived. The problems concerning the regularization of physical
quantities like the vacuum energy are analyzed in some detail. Zero
temperature as well as finite temperature effects are considered and
partition function, free energy and thermodynamic potential, for which
we give some useful representations, are investigated.

\subsection{The path integral} \label{S:ZT}

To start with,  let $\phi$ a matter quantum field living on a given
curved background $D$ dimensional manifold with metric $g_{\mu\nu}$.
Its physical properties are  conveniently described by means of the
path integral functional (in the following, an infinite normalization
constant we will be systematically neglected) \begin{equation}
Z[g]=\int d[\phi]\,e^{iS[\phi,g]}, \nonumber\end{equation} where
$S[\phi,g]$ is the classical action and the functional integral is
taken over all matter fields satisfying suitable boundary or
periodicity conditions. By the generic argument $g$ we leave
understood the functional dependence of the  external background field.

We recall that when the space-time is asymptotically flat and the
integral is over the fields infinitesimally closed to the classical
vacuum at infinity, $Z[g]$ is supposed to give the vacuum-to-vacuum
amplitude \cite{dewi79b}. Feymann boundary conditions are implicitly
assumed in the action functional. The related functional $W[g]=-i\ln
Z[g]$ plays also un important role, since it generates the effective
field equations, namely the classical ones plus quantum corrections.
For a non-asymptotically flat space-time, the physical meaning of the
functional $Z[g]$ as vacuum persistence amplitude is less clear, but
the functional $W[g]$ is still supposed to describe the effective
action. Of particular interest is the case when a complex metric
(manifold) can be found such that there exist both a Lorentzian and an
Euclidean real sections. In this case, one can relate the functional
$Z[g]$ to the Euclidean "partition function" \begin{eqnarray}
Z_E[g]=\int d[\phi]\,e^{-S_E[\phi,g]}, \nonumber\end{eqnarray} for
which the methods of the theory of the elliptic operators can be
applied \cite{hawk77-55-133}.  It should be noted that, in the general
case, this is not an unique procedure. However, for a static metric,
it is always possible to find the associated Euclidean section by
simply looking at the form of the metric tensor. Thus, in general, it
could not be possible to define consistently the partition function. A
sufficient condition is to restrict the analisys to the so called
"strongly elliptic metrics", namely complex metrics for which a
function $f$ exists such that $\,\mbox{Re}\,(fg_{\mu\nu})$ is positive
definite. In this case, the path integral is well defined
\cite{hawk77-55-133}.

Throughout the paper we shall dealing with matter field in a external
gravitational field or we shall make use of the one-loop approximation.
In the latter case, since the dominant contributions to $Z_E[g]$ will
come from fields which are near the classical solution $\phi_c$, which
extremizes the action, for our purposes, it will be sufficient to
consider classical actions which are quadratic in the quantum
fluctuations $\tilde\phi=\phi-\phi_c$. In fact, a functional Taylor
expansion of the action around $\phi_c$ gives \begin{equation}
S[\phi,g]=S_c[\phi_c,g]
+\left.\frac{\delta^2S[\phi,g]}{\delta\phi^2}\right|_{\phi=\phi_c}
\frac{\tilde\phi^2}{2} +\mbox{ higher order terms in }\tilde\phi,
\nonumber\end{equation} $S_c[\phi_c,g]$ being the classical action and
so the one-loop approximated theory is determined by the "zero
temperature partition function" \begin{equation}
Z[\phi_c,g]\sim\exp(-S_c[\phi_c,g])\int d[\tilde\phi]\,
\exp\left(-\frac{1}{2}\int\tilde\phi A\tilde\phi\,d^Dx\right).
\label{PI0} \end{equation} To ensure diffeomorphism invariance of the
functional measure, the functional integral dummy variables
$\tilde\phi$ are chosen to be scalar densities of weight $-1/2$
\cite{fuji80-44-1733} (see also Ref.~\cite{toms87-35-3796}). In
Eq.~(\ref{PI0}) the analogue of the Wick rotation of the time axis in
the complex plane has to be understood. In this manner, the metric
$g_{\mu\nu}$ becomes Euclidean and the small disturbance operator $A$
selfadjoint and non-negative. For non self-interacting fields
minimally coupled to gravity, $A$ is equal to the field operator, that
is the Klein-Gordon operator for scalar fields and the Dirac operator
for fermion fields.

With these assumptions, the partition function can be formally
computed in terms of the real eigenvalues $\lambda_n$ of the operator
$A$. This can be easily done by observing that the functional measure
$d[\tilde\phi]$ reads \begin{equation}
d[\tilde\phi]=\prod_n\frac{dc_n}{\sqrt{2\pi\ell}},
\nonumber\end{equation} $c_n$ being the expansion coefficients of
$\tilde\phi$ in terms of the eigenvectors of $A$ and $\ell$ an
arbitrary parameter necessary to adjust dimensions. It will be fixed
by renormalization. In this manner, for neutral scalar fields one has
\cite{hawk77-55-133} \begin{equation}
Z^{(1)}=\prod_n\frac{1}{\sqrt{2\pi\ell}} \int_{-\infty}^{\infty}dc_n
e^{-\frac{1}{2}\lambda_n c_n^2} =\left[\det(A\ell^2)\right]^{-1/2}
\:.\nonumber\end{equation} The functional $Z^{(1)}[\phi_c,g]$
determines the one-loop quantum corrections to the classical theory.
An expression very similar to the latter is valid for charged scalar
fields and also for anticommuting spinor fields. To consider all the
cases, we may write \begin{eqnarray}
Z^{(1)}=\left[\det(A\ell^2)\right]^{-\nu}\:,\qquad\qquad\qquad\qquad
\ln Z^{(1)}=-\nu\ln\det(A\ell^2)\:, \label{logPI} \end{eqnarray} where
$\nu=-1,1,1/2$ according to whether one is dealing with Dirac spinor,
charged or neutral scalar fields respectively. Eq.~(\ref{logPI}) is
valid also for a multiplet of matter fields. In such a case $A$ is a
matrix of differential operators.

\subsection{The $\zeta$-function regularization} \label{S:ZFR}

Looking at equations above, we see that in order to define the
(one-loop) quantum theory via path integral, one needs a suitable
regularization of determinant of (elliptic) differential operator,
which is formally infinity, since the naive definition of the product
over the eigenvalues gives rise to badly divergent quantity. This
stems from the fact that the number $ {\cal N}(\lambda)$ of
eingenvalues whose value is less than $\lambda$ admits an asymptotic
expansion of the form
\cite{horm68-121-193,gilk75-10-601,hawk78-18-1747,cogn89-223-416}
\begin{eqnarray}  {\cal N}(\lambda)\simeq \sum_{n<N}
\frac{K_n}{\Gamma{(\frac{N-n}2)+1}} \,\lambda^{\frac{N-n}2}
\,,\qquad\qquad\lambda\to\infty \,,\nonumber\end{eqnarray} where $K_n$
are the integrated spectral coefficients we are going to introduce in
the next subsection. As one can see, they control the degree of
divergence of the determinant. Throughout the paper, we shall mainly
make the choice of $\zeta$-function regularization. In Sec.~\ref{S:RT}
we shall also discuss a class of other possible and equivalent
regularizations. We recall that $\zeta(s|A)$ is defined in terms of
the complex power of the operator $A$ or, via a Mellin transform, by
the trace of the heat kernel related to $A$.

\subsubsection{Complex powers and heat kernel of elliptic operators}
\label{S:CPEO}

Here we just remind the definition of complex power of an elliptic
operator as given by Seeley in Ref.~\cite{seel67-10-172} and its link
with the heat kernel. We  briefly discuss some properties of it, which
we need in the following, referring the interested reader to the
literature for more details.

Let $A$ be an invertible elliptic differential (or
pseudo-differential) operator of positive order $p$, defined on a
N-dimensional compact manifold ${\cal M}^N$, with boundary or without
boundary. Then, for $\,\mbox{Re}\, s<0$, one can define a semigroup
$A^s$ by \begin{equation} A^s=-\frac{1}{2\pi
i}\int_{\Gamma}\frac{z^s}{A-z}dz\:, \nonumber\end{equation} and this
semigroup can be extended to a group containing $A$ and the identity.
The curve $\Gamma$ goes from $\infty$ to $\infty$ and around the
origin along a small circle, enclosing the spectrum of $A$. If $A^s$
is a pseudo-differential operator of order $p\cdot s$ the Seeley
method permits to build up a parametrix for its kernel $A(x,x';s)$,
which is a continuous function for $\,\mbox{Re}\,(p\cdot s)<-N$.

The Seeley method for the construction of the parametrix of $A^s$ and
as a consequence, a related asymptotic expansion for $\det A$, is very
general in principle, but quite complicated with regard to the
computational point of view.  Several alternative approaches to the
problem have been proposed, but mostly associated with a related
quantity. In fact the kernel of $A^s$ can be expressed, by means of a
Mellin transform, as a function of the heat kernel
$K_t(x,x'|A)=e^{-tA}(x,x')$. That is \begin{equation}
A^{-s}(x,x')=\frac{1}{\Gamma(s)} \int_0^\infty
t^{(s-1)}[K_t(x,x'|A)-P_0(x,x'|A)]\, dt\:, \label{MT} \end{equation}
$K_t(x,x'|A)$ being a bi-spinor or bi-tensor density of weight $-1/2$
and $P_0$ the projector onto zero modes. The presence of zero modes
does not create particular problems. Here we continue to suppose the
operator $A$ to be invertible, then zero modes are absent and the
projector $P_0$ can be ignored. When necessary zero modes shall be
carefully considered.

$K_t(x,x')$ satisfies the heat-type equation (the argument $A$ is
understood) \begin{equation} \partial_tK_t(x,x')+A\,K_t(x,x')=0 \:,
\qquad\qquad \lim\limits_{t\to 0_+}K_t(x,x')=[g(x)]^{1/4}
\delta(x,x')[g(x')]^{1/4} \:,\label{calore} \end{equation} the
solution of which admits an asymptotic expansion of the kind (see for
example Ref.~\cite{grei71-41-163}) \begin{equation}
K_t(x,x)\simeq\frac{\sqrt{g(x)}}{(4\pi)^{N/p}}
\sum_{0}^{\infty}{k_n(x)t^{(n-N)/p}} \:.\label{HKE} \end{equation}
This expansion is also valid when the manifold has a boundary. Thus
$K_t(x,x')$, together Eq.~(\ref{calore}), must satisfy suitable
boundary conditions.

A useful property of the heat kernel is factorization on a product
space, that is \begin{equation} K^{{\cal M}_1\times{\cal
M}_2}_t((x_1,x_2),(x'_1,x'_2)|A)= K^{{\cal M}_1}_t(x_1,x'_1|A_{1})
K^{{\cal M}_2}_t(x_2,x'_2|A_{2}) \:,\label{HKfact} \end{equation}
where the operators $A_{1}$ and $A_{2}$  are defined on ${\cal M}_1$
and ${\cal M}_2$ respectively and $A=A_{1}+A_{2}$. Eq.~(\ref{HKfact})
is equivalent to the fact that $\,\mbox{Tr}\,
e^{-t(A_{1}+A_{2})}=\,\mbox{Tr}\, e^{-tA_{1}} \,\mbox{Tr}\,
e^{-tA_{2}}$ when the operators $A_{1}$ and $A_{2}$ commute.

On a manifold without boundary, all $k_n(x)$ coefficients with odd $n$
are vanishing and so Eq.~(\ref{HKE}) assumes a simpler form. In the
present work we shall essentially deal with second order differential
operators ($p=2$) on manifolds without boundary ($k_{2n+1}=0$).
Putting $k_{2n}(x)=a_n(x)$ one gets \cite{mina49-1-242,dewi65b}
\begin{equation} K_t(x,x)\simeq\frac{\sqrt{g(x)}}{(4\pi t)^{N/2}}
\sum_{0}^{\infty}a_n(x)\;t^n \:,\label{HKEnB}\end{equation} the
expansion coefficients $a_n(x)$ being local invariant matrices
depending on curvature and torsion tensors and their covariant
derivatives. They all, but $a_0$ vanish in the flat-free case, while
$a_0$ is always equal to the identity matrix.

We have been mentioned (see also Ref. \cite{dewi75-19-295}) that these
coefficients control the one-loop divergences of the effective action
and of the related quantities, like the stress energy momentum tensor.
Hence, the explicit knowledge of the $a_n$ is important in the
physical applications. In general, a closed form for the heat kernel
is unknown also for simple operators like the Laplacian and for this
reason the related asymptotic expansion has been extensively
investigated . Many methods for computing the $K_n$ on smooth
Riemannian manifolds have been proposed by physicists and
mathematicians and some coefficients have been computed for Riemannian
manifolds with and without boundary as well as for the case of
Riemann-Cartan manifolds.

Here we would like just to mention the general procedure related to
Seeley method \cite{atiy73-19-279,roma79-41-190}, and the (well known
to physicists) Schwinger \cite{schw51-82-664} and DeWitt
\cite{dewi65b} technique, which work very well for manifolds without
boundary. When the manifold has a boundary the computation of the
spectral coefficients is a much more difficult task than the
boundaryless case \cite{mcke67-1-43,grei71-41-163}. The literature on
this subject is very vast and we refer the reader to
Ref.~\cite{bran90-15-245} (and references therein), where some
spectral coefficients in the case of manifolds with boundary have been
computed by a very powerful method on any smooth Riemannian manifold.
For a general derivation of heat coefficients on a Riemann-Cartan
manifold see Ref.~\cite{cogn88-214-70}. Extension of previous methods
to manifolds with boundary can be found in
Refs.~\cite{cogn90-241-381,mcav91-8-603} and references cited therein.

\subsubsection{The $\zeta$-function} \label{S:ZF}

Let $A$ be an invertible operator with the properties of
Sec.~\ref{S:CPEO}. The (generalized Riemann) $\zeta$-function related
to $A$, first investigated in Ref.~\cite{mina49-1-320}, can be defined
by \begin{equation} \zeta(s|A)=\,\mbox{Tr}\, A^{-s}=\int_{{\cal M}}
A^{-s}(x,x) d^Nx =\frac{1}{\Gamma(s)}\int_{0}^{\infty}
t^{s-1}\,\mbox{Tr}\, e^{-tA}\,dt \:,\label{ZFdef} \end{equation}
where, in writing the last term (the Mellin transform of the heat
kernel), we have made use of Eq.~(\ref{MT}) disregarding zero modes.
When $x\neq x'$, $A^{-s}(x,x')$ is an entire function of $s$, while
$A^{-s}(x,x)$ is a meromorphic function having simple poles on the
real axis. Positions and number of poles depend on the order of
operator and dimension of ${\cal M}$. Then we see that according to
definition (\ref{ZFdef}), also $\zeta(s|A)$ is a meromorphic function
with simple poles on the real axis. In particular, $\zeta(s|A)$ is
analytic at $s=0$. When the operator $A$ admits a complete set of
eigenvectors with eigenvalues $\lambda_n$, then Eq.~(\ref{ZFdef}) is
the analytic continuation of the  definition (valid for $\,\mbox{Re}\,
s$ sufficiently large) \begin{equation}
\zeta(s|A)=\sum_n\lambda_n^{-s}\,, \label{ZFdef2} \end{equation} from
which \begin{equation} \zeta'(s|A)=-\sum_n\lambda_n^{-s}\,\ln\lambda_n
\:.\label{ZFdef23} \end{equation} When $A$ is not an invertible
operator, the null eigenvalues must be omitted in Eq.~(\ref{ZFdef2}).

{}From definition (\ref{ZFdef2}), we easily obtain the scaling property
\begin{equation} \zeta(s|A\ell^2)=\ell^{-2s}\zeta(s|A)
\:,\label{ZFscaling} \end{equation} valid for any $\ell$, while from
Eqs.~(\ref{ZFdef23}) and (\ref{ZFscaling}), we see that a natural
definition of determinant of $A$, in the sense of analytic
continuation, is given by means of formula \cite{hawk77-55-133}
\begin{equation} \ln\det(A\ell^2)=-\zeta'(0|A)+\zeta(0|A)\ln\ell^2\:.
\nonumber\end{equation} The above equation shows that $\zeta(0|A)$
governs the scale dependence of the quantity $\ln\det(A\ell^2)$.

The residues of the poles of $\zeta$-function are directly related to
heat kernel coefficients. Integrating the heat expansion, we get a
parametrix for $K(t|A)=\,\mbox{Tr}\, e^{-tA}$ of the kind (here we
limit our analisys to a second order differential operator, namely
$p=2$) \begin{equation}
K(t|A)\simeq\sum_{n=0}^{\infty}K_n(A)t^{(n-N)/2} \:,\label{HK}
\end{equation} where $K_n(A)$ are the (integrated) spectral
coefficients, that is \begin{equation}
K_n(A)=\frac{1}{(4\pi)^{N/2}}\int_{\cal M}k_n(x|A) \sqrt{g(x)}\,dx
\:.\nonumber\end{equation} In order to derive the meromorphic
structure of $\zeta(s|A)$, the standard procedure is to split the
integration over $t$ of Eq.~(\ref{ZFdef}) in two parts, $(0,1)$ and
$(1,\infty)$. $K(t|A)$ is a smooth function of $t$ for $t\to\infty$
and so the integral from $1$ to $\infty$ in Eq.~(\ref{MT}) is an
entire function of $s$. The rest of the integral can be explicitly
computed using expansion (\ref{HK}), which is valid for small $t$.
Thus, for a second order differential operator, we get \begin{eqnarray}
\zeta(s|A)&\simeq&\frac{1}{\Gamma(s)}\left[ \sum_{n=0}^{\infty}K_n(A)
\int_0^1 t^{s-(N-n)/2-1}dt+\int_1^\infty t^{s-1}K(t|A)\,dt
\right]\nonumber\\ &=&\sum_{n=0}^{\infty}
\frac{K_n(A)}{\Gamma(s)(s-\frac{N-n}{2})} +\frac{\hat
G(s|A)}{\Gamma(s)} =\frac{K_N(A)}{\Gamma(s+1)}
+\frac{G(s|A)}{\Gamma(s)} \label{ZFpoles} \:,\end{eqnarray} where
\begin{eqnarray} \hat{G}(s|A)&=&\int_{1}^{\infty}t^{s-1}\,K(t|A)\,dt
\label{fsA}\end{eqnarray} is an entire function of $s$ while $G(s|A)$
has poles at all the points $s=(N-n)/2$, but $s=0$. In fact $G(0|A)$
is the principal part of $\Gamma(s)\zeta(s|A)$ at $s=0$. On the
contrary, due to the presence of $\Gamma(s)$ in Eq.~(\ref{ZFpoles}),
not all the points $s=(N-n)/2$ are simple poles of $\zeta(s|A)$.
Moreover, in the absence of boundaries, all $K_n$ with odd $n$
vanishes, so in this case, for even N we have N/2 poles situated at
the integers $n=1,2,...,N/2$ with residues $K_{N-2n}/\Gamma(n)$ while
for odd N we have infinite poles at all half integers $n\leq N/2$ with
residues $K_{N-2n}/\Gamma(n)$. In particular, $\zeta(s|A)$ is finite
for $s=0$ and one has the useful relations \begin{eqnarray}
\zeta(0|A)=K_N(A) \nonumber\:,\end{eqnarray} \begin{eqnarray}
\zeta'(0|A)=\gamma K_N(A)+G(0|A)=-\ln\det A
\:.\label{ze'0}\end{eqnarray}

\paragraph{Zero-modes.} As we have already mentioned before, zero
modes must be omitted in the definition of $\zeta$-function. Now we
briefly describe a possible way to proceed in order to define the
determinant of the operator $A$ when $ {\cal N}$ zero-modes are
present. First of all we introduce the positive operator
$A_\lambda=A+\lambda$ ($\lambda>0$) and define the regularized
determinant of $A$  by means of equation \begin{eqnarray} \ln\det
A&=&-\lim_{\lambda\to0} \left[\zeta'(0|A_\lambda)+ {\cal
N}\ln\lambda\right]\nonumber\\ &=&-\gamma K_N(A) -\lim_{\lambda\to
0}\left[ G(0|A_\lambda)+ {\cal N}\ln\lambda\right]
\:.\label{logdetZM}\end{eqnarray} In this way all zero-modes are
automatically subtracted. In writing the latter equation we have used
Eq.~(\ref{ze'0}) and the fact that the logarithmic divergence for
$\lambda\to0$ comes from $G(A_\lambda)$ (actually from $\hat
G(0|A_\lambda)$). In fact, it is easy to see looking at
Eq.~(\ref{fsA}) that $\hat G(0|A_\lambda)\sim- {\cal N}\ln\lambda$
plus a regular function in the limit $\lambda\to0$.

The next step is to observe that the function $\exp(-G(0|A_\lambda)$
has a zero of order $ {\cal N}$ in $\lambda=0$. Thus, deriving
Eq.~(\ref{logdetZM}) $ {\cal N}$ times with respect to $\lambda$ and
taking the limit $\lambda\to0$ we finally get \begin{eqnarray} \det A=
\frac{e^{-\gamma K_N(A)}}{ {\cal N}!}\lim_{\lambda\to0} \frac{d^{
{\cal N}}}{d\lambda^{ {\cal N}}}\,\exp[-G(0|A_\lambda)]
\:.\nonumber\end{eqnarray}

\paragraph{The factorization property.} We consider two positive
elliptic operators $A_1$ and $A_2$ acting on functions in $\cal
M^{N_1}$ and $\cal M^{N_2}$ ($N_1\leq N_2$). According to the
factorization property, Eq.~(\ref{HKfact}), the heat kernel related to
the operator $A=A_1+A_2$ is the product of the two kernels related to
the operators $A_1$ and $A_2$ or, what is the same, \begin{eqnarray}
K(t|A)=K(t|A_1)\,K(t|A_2) \:.\nonumber\end{eqnarray} Using such a
relation for the trace of the heat kernel now we obtain an interesting
representation for $\zeta(s|A)$ in terms of $\zeta(s|A_1)$ and
$\zeta(s|A_2)$. Of course, when $\cal M_1$ or/and $\cal M_2$ are
non-compact, we have to consider $\zeta$-function densities.

We start with Mellin representation of $\zeta$-function,
Eq.~(\ref{ZFdef}), that is \begin{eqnarray}
\zeta(s|A)=\frac1{\Gamma(s)}\int_0^\infty t^{s-1}
K(t|A_1)\,K(t|A_2)\,dt \:.\label{ze-A1A2}\end{eqnarray} The right hand
side of the latter equation can be transformed in an integral in the
complex plane using Mellin-Parseval identity, Eq.~(\ref{M-P}). To this
aim we choose $f(t)=K(t|A_1)t^{s/2}$, $g(t)=K(t|A_2)t^{s/2-1}$ and
rewrite Eq.~(\ref{ze-A1A2}) by means of Eq.~(\ref{M-P}). In this way
we obtain what we call the Mellin-Barnes representation for the
$\zeta$-function \begin{eqnarray} \zeta(s|A)=\frac1{2\pi i\Gamma(s)}
\int_{\,\mbox{Re}\, z=c}\Gamma(\frac s2+z)\zeta(\frac s2+z|A_1)
\Gamma(\frac s2-z)\zeta(\frac s2-z|A_2) \:,\nonumber\end{eqnarray}
where $c$ is a real number in the common strip in which $\hat f(z)$
and $\hat g(1-z)$ are analytical (see Appendix \ref{S:UR}). In our
case this corresponds to the condition
$-\frac{s-N_1}2<c<\frac{s-N_2}2$. Such a condition is  restrictive
but, as we shall see in the following, it can be relaxed in the
applications of physical interest.

The generalization of the latter formula to the case in which $A_1$
or/and $A_2$ have also zero-modes can be derived by the same
technique. It reads \begin{eqnarray} \zeta(s|A)&=& {\cal
N}_1\zeta(s|A_1)+ {\cal N}_2\zeta(s|A_2) \nonumber\\&& +\frac1{2\pi
i\Gamma(s)} \int_{\,\mbox{Re}\, z=c}\Gamma(\frac s2+z)\zeta(\frac
s2+z|A_1) \Gamma(\frac s2-z)\zeta(\frac s2-z|A_2)\,dz
\:,\label{ZFA1A2}\end{eqnarray} $ {\cal N}_1$ and $ {\cal N}_2$ being
the number of zero-modes of $A_1$ and $A_2$ respectively. As usual,
all zero-modes have to be omitted in the evaluation of
$\zeta$-functions.

\paragraph{Example: $S^1\times\cal M^N$.} \label{S:S1XMN}

It is of particular interest for physical applications (finite
temperature quantum field theories), the case $\cal M^D=S^1\times\cal
M^N$, i.e. the manifold is the product between a circle $S^1$ and a
(compact) $N$-dimensional manifold $\cal M^N$. Then
$A=-\partial_\tau^2+L_N$, where $\tau$ runs on $S^1$ and $L_N$ is an
operator in $\cal M^N$. If we assume that the field to satisfy
periodic conditions (only for illustrative purposes) on $S^1$, then
the eigenvalues of $A$ have the form \begin{eqnarray}
\lambda_{n,j}=\left(\frac{2\pi n}{\beta}\right)^2+\omega_j^2\:,
\qquad\qquad n=0,\pm1,\pm2,\dots\:, \label{eigenS1}\end{eqnarray}
$\beta$ (the inverse of the temperature $T$) being the circumference
of $S^1$ and $\omega_j^2$ the eigenvalues of $L_N$. In this case we
have \begin{eqnarray} \zeta(s|A)&=&\sum_{n=-\infty}^\infty
\zeta(s|L_N+[2\pi n/\beta]^2) \label{ZFb}\\ &=&
\frac{\beta\Gamma(s-\frac{1}{2})}{\sqrt{4\pi}\Gamma(s)}
\zeta(s-\frac{1}{2}|L_N)+\frac\beta{\pi}\sum_{n=1}^\infty\int_{-\infty}^\infty
e^{in\beta t}\,\zeta(s|L_N+t^2)\,dt \label{ZF-Poisson}\\ &=&
\frac{\beta\Gamma(s-\frac{1}{2})}{\sqrt{4\pi}\Gamma(s)}
\zeta(s-\frac{1}{2}|L_N)+\frac\beta{\sqrt{\pi}\Gamma(s)}
\sum_{n=1}^\infty\int_0^\infty t^{s-3/2}\,e^{-n^2\beta^2/4t}\,
K(t|L_N)\,dt \:.\label{ZF-Jacobi}\end{eqnarray} Eq.~(\ref{ZF-Poisson})
have been derived from Eq.~(\ref{ZFb}) using the Poisson summation
formula, Eq.~(\ref{PSF}), while Eq.~(\ref{ZF-Jacobi}) have been
directly obtained by using the Mellin representation (\ref{ZFdef}) for
$\zeta$-function in Eq.~(\ref{ZF-Poisson}).

Using Eq.~(\ref{ZFA1A2}) we could obtain a complex integral
representation of $\zeta$-function for this particular case, but we
prefer to derive it directly from  Eq.~(\ref{ZF-Jacobi}), since in
this case the restriction on $c$ is not so stringent. This
representation will be particularly useful for studying the high
temperature expansion. To this aim we choose $f(t)=t^{s-3/2}K(t|L_N)$,
$g(t)=e^{-n^2\beta^2/4t}$ and use Eq.~(\ref{ZF-Jacobi}) and
Mellin-Parseval identity, Eq.~(\ref{M-P}). As a result \begin{eqnarray}
\zeta(s|A)&=&\frac{\beta\Gamma(s-\frac{1}{2})}{\sqrt{4\pi}\Gamma(s)}
\zeta(s-\frac{1}{2}|L_N)+\frac{1}{\sqrt{\pi}\Gamma(s)} \frac{1}{2\pi
i}\times\nonumber\\&&\qquad \int_{\,\mbox{Re}\, z=c}\zeta_R(z)
\Gamma(\frac z2)\Gamma(\frac{z-1}{2}+s) \zeta(\frac{z-1}{2}+s|L_N)\;
\left(\frac\beta2\right)^{-(z-1)}\;dz
\:,\label{ZF-Barnes}\end{eqnarray} where $\zeta_R$ represents the
usual Riemann $\zeta_R$-function and
$c>N+1$.

{}From Eqs.~(\ref{ZFpoles}), (\ref{ZF-Poisson}), (\ref{ZF-Jacobi}) and
(\ref{ZF-Barnes}) we get three representations for the derivative of
$\zeta$, namely \begin{eqnarray} \zeta'(0|A) &=&
-\beta\zeta^{(r)}(-\frac{1}{2}|L_N)+\frac{\beta}\pi\sum_{n=1}^\infty\int_{-\infty}^\infty
e^{in\beta t}\,\zeta'(0|L_N+t^2)\,dt \label{ZF1-Poisson}\\
&=&-\beta\zeta^{(r)}(-\frac{1}{2}|L_N)+\frac\beta{\sqrt{\pi}}
\sum_{n=1}^\infty\int_0^\infty t^{-3/2}\,e^{-n^2\beta^2/4t}\,
K(t|L_N)\,dt \label{ZF1-Jacobi}\\ &=&
-\beta\zeta^{(r)}(-\frac{1}{2}|L_N)+\frac{\beta}{\pi
i}\int_{\,\mbox{Re}\, z=c}\zeta_R(z)
\Gamma(z-1)\zeta(\frac{z-1}{2}|L_N)\; \beta^{-z}\;dz
\:,\label{ZF1-Barnes}\end{eqnarray} where we have introduced the
definition \begin{eqnarray} \zeta^{(r)}(-\frac12|L_N)=
\,\mbox{PP}\,\zeta(-\frac12|L_N)
+(2-2\ln2)\,\mbox{Res}\,\zeta(-\frac12|L_N) \:,\nonumber\end{eqnarray}
$\,\mbox{PP}\,$ and $\,\mbox{Res}\,$ being respectively the finite
part and the residue of the function at the specified point. Note that
the residue of $\zeta(s|L_N)$ at $s=-1/2$ is equal to
$-K_{N+1}(L_N)/\sqrt{4\pi}$, $K_{N+1}(L_N)$ being the coefficient of
$\sqrt{t}$ in the asymptotic expansion of $K(t|L_N)$. Thus, when this
coefficient is vanishing, there is no pole at $s=-1/2$ and the
$\,\mbox{PP}\,$ prescription gives the value of $\zeta(s|L_N)$ at the
same point.

For use later, we define, for any constant $\ell$ \begin{eqnarray}
\zeta^{(r)}(s|L_N\ell^2)=
\,\mbox{PP}\,\zeta(s|L_N)+(2-2\ln2\ell)\,\mbox{Res}\,\zeta(s|L_N) \:.
\label{ZF-r} \end{eqnarray}

\paragraph{Example: $\mbox{$I\!\!R$}^p\times\cal M^N$.} In many
physical problems (for example in Kaluza-Klein theories), the manifold
$\cal M$ is the product of the flat $p$-dimensional manifold
$\mbox{$I\!\!R$}^p$ and a compact $N$-dimensional manifold $\cal M^N$
and the differential operator of interest takes the form
$A=-\Delta_p+L_N$, where $\Delta_p$ is the Laplace operator acting on
functions in $\mbox{$I\!\!R$}^p$ and $L_N$ is an operator acting on
functions in $\cal M^N$. In this case, using Eq.~(\ref{ze-A1A2}), we
obtain the useful relation \begin{eqnarray}
\frac{\zeta(s|A)}{\Omega_p}&=&\frac{1}{(4\pi)^\frac p2\Gamma(s)}
\int_{0}^{\infty} t^{s-\frac p2-1}K(t|L_N)\,dt\nonumber\\
&=&\frac{\Gamma(s-\frac p2)}{(4\pi)^\frac p2\Gamma(s)} \zeta(s-\frac
p2|L_N) \:,\label{ZFfact} \end{eqnarray} where $\Omega_p$ is a large
volume in $\mbox{$I\!\!R$}^p$. The left hand side of
Eq.~(\ref{ZFfact}) then represents a density in $\mbox{$I\!\!R$}^p$.

In the particular case in which $p=1$, we write down the derivative at
$s=0$ of the latter equation. Using Eq.~(\ref{ZFscaling}) we easily
obtain \begin{eqnarray} \frac{\ln\det A}{\Omega_1}
=-\frac{\zeta'(0|A)}{\Omega_1} =\zeta^{(r)}(-1/2|L_N)
\:.\label{ZF1-N+1}\end{eqnarray}

\subsection{Other regularization techniques} \label{S:RT}

Here we shall discuss a class of regularizations based on the
Schwinger representation (see for example  \cite{ball89-182-1}), which
contains $\zeta$-function regularization as particular case. In order
to work with dimensionless quantities, we put $B=A\ell^2$. The
regularized determinant of the operator $B$ may be defined by
\begin{equation} (\ln\det B)_\varepsilon = -\int_0^\infty
t^{-1}\varrho(\varepsilon,t) \,\mbox{Tr}\, e^{-tB}\,dt
\:,\label{logdetBep} \end{equation} where $\varrho(\varepsilon,t)$ is
a suitable regularization function of the dimensionless parameter $t$,
which has to satisfy the two requirements we are now going to discuss.
First, for fixed $t>0$, the limit as $\varepsilon$ goes to zero must
be equal to one. Second, for fixed and sufficiently large
$\varepsilon$, $\varrho(\varepsilon,t)$ has to regularize the
singularity at $t=0$ coming from the heat kernel expansion (\ref{HK})
related to $B$. The analytic continuation will be used to reach small
values of $\varepsilon$. As we shall see in the following, these
requirements do not uniquely determine the regularization function
$\varrho$.

Using Eq.~(\ref{HK}) in Eq.~(\ref{logdetBep}) one can easily see that
the number of divergent terms for $\varepsilon\to 0$ is equal to $Q+1$,
$Q$ being the integer part of $D/2$. They are proportional to the
spectral coefficients $K_0,\dots K_Q$ (which contain the full
dependence on the geometry), the prefactors depending on the
regularization function  \cite{ball89-182-1}. In Appendix~\ref{S:RF},
this general result is explicitly verified in several examples.

\paragraph{Regularizations in $\mbox{$I\!\!R$}^D$.} In the rest of
this Section, in the particular but important case  $A=-\Delta_D +M^2$
in $\mbox{$I\!\!R$}^D$, we would like to show that the finite part of
$(\ln\det B)_\varepsilon$ (that is the effective potential, see
following Sec.~\ref{S:OLEA}) is uniquely determined, modulo a constant
which can be absorbed by the arbitrary scale parameter $\ell$ and can
be evaluated without making use of the explicit knowledge of the
regularization function. As usual, in $\mbox{$I\!\!R$}^D$ we consider
a large region $ {\cal R}$ of volume $\Omega_{\cal R}$ and the limit
$\Omega_{\cal R}\to\infty$ shall be taken at the end of calculations.
It has to be stressed that, apart from a topological contribution,
this calculation is already enough to give the one loop-effective
potential in a constant curvature space-time, i.e. the identity
contribution.

{}From the known results on $\mbox{$I\!\!R$}^D$ (see Eq.~(\ref{HKRN}))
we have \begin{equation} \frac{\mbox{Tr}\, e^{-tB}}{\Omega_{\cal R}}
=\frac{e^{-t\ell^2M^2}}{(4\pi t\ell^2)^{D/2}}\,,
\nonumber\end{equation} where $M^2$ is a positive constant. As a
consequence, from  Eq.~(\ref{logdetBep}), we obtain the regularized
expression \begin{equation} \frac{(\ln\det
B)_\varepsilon}{2\Omega_{\cal R}}
=-\frac{1}{2}\left(\frac{1}{4\pi\ell^2}\right)^{D/2} \int_0^\infty
t^{-(1+D/2)}\varrho(\varepsilon,t) e^{-t\ell^2M^2}\,dt \equiv
f_D(\varepsilon,M)\,.
\nonumber\end{equation}

Now it is convenient to distinguish between even ($D=2Q$) and odd
($D=2Q+1$) dimensions. In order to derive a more explicit form of the
divergent terms for $\varepsilon\to 0$, let us consider the
$Q^{th}$-derivative of $f(\varepsilon,M)$ with respect to $M^2$. One
gets \begin{equation} \frac{d^Qf_D(\varepsilon,M)}{dM^{2Q}}
=\frac{(-1)^Q}{2}
\left(\frac{1}{4\pi\ell^2}\right)^{D/2}B_{e,o}(\varepsilon,M)\,,
\label{dQV} \end{equation} where \begin{equation}
B_e(\varepsilon,M)=-\int_0^\infty t^{-1}\varrho(\varepsilon,t)
e^{-tM^2}dt\, = \ln M^2+b+c_Q(\varepsilon)+O(\varepsilon)\,,
\label{Be}\end{equation}

\begin{equation} B_o(\varepsilon,M)=-\int_0^\infty
t^{-3/2}\varrho(\varepsilon,t) e^{-tM^2}dt\, =
2M\sqrt{\pi}+b+c_Q(\varepsilon)+O(\varepsilon)\,. \label{Bo}
\end{equation} In the derivation of the above equations we have made
use of the properties of the regularization functions. Furthermore, in
Eqs.~(\ref{Be}) and (\ref{Bo}) $b$ is a constant and
$c_Q(\varepsilon)$ is a function of $\varepsilon$, but not of $M$. We
have used the same symbols for even and odd dimensions, but of course
they represent different quantities in the two cases.

Making use of Eqs.~(\ref{Be}) and (\ref{Bo}) in Eq.~(\ref{dQV}) a
simple integration gives \begin{equation}
f_{2Q}(\varepsilon,M)=\frac{(-1)^Q}{2Q!(4\pi)^Q}
\left[\ln(\ell^2M^2)-C_Q+b\right]M^D
+\ell^{-D}\sum_{n=0}^Qc_n(\varepsilon)(\ell M)^{2n} + O(\varepsilon)\,,
\nonumber\end{equation} \begin{equation} f_{2Q+1}(\varepsilon,M)
=\frac{(-1)^Q\Gamma(-Q-1/2)}{2(4\pi)^{Q+1/2}}M^D
+\ell^{-D}\sum_{n=0}^Qc_n(\varepsilon)(\ell M)^{2n} + O(\varepsilon)\,,
\nonumber\end{equation} where we have set
$C_Q=\sum_{n=1}^{Q}\frac{1}{n}$.

The dimensionless integration constants $c_n(\varepsilon)$, which are
divergent for $\varepsilon\to0$, define the counterterms which must be
introduced in order to remove the divergences. For the physical
interesting case $D=4$, one gets \begin{equation} \frac{(\ln\det
B)_\varepsilon}{2\Omega_{\cal R}} =\frac{M^4}{64\pi^2}
\left[\ln(\ell^2M^2)-\frac{3}{2}+b\right]
+c_2(\varepsilon)f^4+c_1(\varepsilon)M^2\ell^{-2}
+c_0(\varepsilon)\ell^{-4}\,, \nonumber\end{equation} in agreement
with the well known result obtained in
Refs.~\cite{cole73-7-1888,ilio75-47-165}, where some specific
regularizations were used for $D=4$.

Some remarks are in order. The constants $b$ and $c_n(\varepsilon)$
depend on the  choice  of the regularization function and on the
Seeley-Dewitt coefficient $a_n(x,B)$, but $b$ can be absorbed by the
arbitrary scale parameter $\ell$. As a result, the finite part of
$(\ln\det B)_\varepsilon$ (the effective potential) does not depend on
the regularization, as expected. In Appendix~\ref{S:RF} several
examples of admissible regularizations of the kind we have discussed
are reported as an illustration of the above general result. In
particular we show that $\zeta$-function, as well as other known
regularizations, can be derived from Eq.~(\ref{logdetBep}) with a
suitable choice of $\varrho(\varepsilon)$.

In the rest of the paper, we shall make use of the $\zeta$-function
regularization.

\subsection{The one-loop effective action and the renormalization
group equations} \label{S:OLEA}

In this subsection we assume $A$ to be an elliptic second order
differential operator acting on fields in $\cal M^D$ and depending on
the classical solution $\phi_c$. It will be of the form
$A=-\Delta_D+V''(\phi_c,g)$ (the prime indicates the derivative with
respect to $\phi$), $V(\phi,g)$ being a scalar function describing the
self-interaction of the field, the coupling with gravity and
containing furthermore local expressions of dimension $D$ involving
curvature tensors and non-quadratic terms in the field. These latter
terms have, in general, to be included in order to ensure the
renormalizability of the theory \cite{utiy62-3-608}. When $V(\phi,g)$
is almost quadratic in the matter fields, then $A$ is just the
classical field operator.

The one loop effective action takes the form  \cite{aber73-9-1}
\begin{eqnarray} \Gamma[\phi_c,g]
&=&S_c[\phi_c,g]+\frac{1}{2}\ln\det(A\ell^2)\nonumber\\
&=&\int\left[V_{eff}(\phi_c,g)+\frac{1}{2}Z(\phi_c,g)
g^{ij}\partial_i\phi_c\partial_j\phi_c +\cdots\right]\sqrt{g}\,d^Dx
\:.\label{OLEA} \end{eqnarray} Eq.~(\ref{OLEA}) defines the one-loop
effective potential $V_{eff}(\phi_c)$. It is given by \begin{eqnarray}
V_{eff}(\phi_c,g)=V_c(\phi_c,g)+V^{(1)}(\phi_c,g)
=V(\phi_c,g)+\frac{\ln\det(A\ell^2)}{2\Omega(\cal M)}\,, \label{OLEP}
\end{eqnarray} $\Omega(\cal M)$ being the volume of $\cal M$. The
quantum corrections $V^{(1)}(\phi_c,g)$ to the classical potential
$V(\phi_c,g)$ are of order $\hbar$ and are formally divergent.

\paragraph{Renormalization group equations.} For the sake of
completeness, now we derive the renormalization group equations for
the $\lambda\phi^4$ theory  \cite{nels82-25-1019} on a 4-dimensional
compact smooth manifold without boundary, the extension to manifolds
with boundary being quite straightforward. The derivation which we
present is valid only at one-loop level. For a more general discussion
see for example Ref.~\cite{buch89-12-1,buch92b} and references cited
therein.

We regularize the one-loop effective action (\ref{OLEA}) by means of
$\zeta$-function. In this way we have \begin{equation}
\Gamma[\phi_c,g]=S_{c}[\phi_c,g]-\frac{1}{2}\zeta'(0|A\ell^2)
\:.\nonumber\end{equation}

The more general classical action has the form \begin{eqnarray}
S_c(\eta)&=&\int\left[\Lambda-\frac{\phi_c\Delta\phi_c}{2}
+\frac{\lambda\phi_c^4}{24}+\frac{m^2\phi_c^2}{2} +\frac{\xi
R\phi_c^2}{2}\right]\sqrt{g}d^4x\nonumber\\
&&\qquad\qquad\qquad\qquad+\int\left[\varepsilon_1R+\frac{\varepsilon_2R^2}{2}
+\varepsilon_3W\right]\sqrt{g}d^4x\,, \nonumber\end{eqnarray}
$W=C^{ijrs}C_{ijrs}$ being the square of the Weyl tensor and
$\eta=\eta_q\equiv(\Lambda,\lambda,m^2,\xi,\varepsilon_1,\varepsilon_2,\varepsilon_3)$
($q=0,\dots,6$) the collection of all coupling constants. We disregard
the Gauss-Bonnet invariant since its integral over the manifold is
proportional to the Euler-Poincar\'e characteristic, namely, it is
both a topological invariant as well as scale independent. As a
consequence it does not affect the scale dependence of the effective
action and we can dispense with it.

Now we consider a conformal transformation $\tilde
g_{\mu\nu}=\exp(2\sigma)g_{\mu\nu}$ with $\sigma$ a constant (scaling).
By the conformal transformation properties of the fields, one can
easily check that $\tilde S_{c}(\eta)=S_{c}(\tilde\eta)$, where
$\tilde\eta$ are all equal to $\eta$ except  $\Lambda$, $m^2$, and
$\varepsilon_1$. For these, we have \begin{equation}
\tilde\Lambda=\Lambda e^{4\sigma}\:,\qquad\qquad \tilde m^2=m^2
e^{2\sigma}\:,\qquad\qquad
\tilde\varepsilon_1=\varepsilon_1\,e^{2\sigma}\:.
\nonumber\end{equation} In the same way, for the eigenvalues
$\tilde\alpha_n(\eta)$ of the small disturbance operator $\tilde
A(\eta)$ we have
$\tilde\alpha_n(\eta)=e^{-2\sigma}\alpha_n(\tilde\eta)$. From this
transformation rule for the eigenvalues, we immediately get the
transformations for $\zeta(s|A\ell^2)$ and $\zeta'(s|A\ell^2)$. They
read \begin{eqnarray} \zeta(s|\tilde A\ell^2)
=e^{2s\sigma}\zeta(s|A(\tilde\eta)\ell^2)\,, \nonumber\end{eqnarray}
\begin{eqnarray} \zeta'(s|\tilde A\ell^2) =e^{2s\sigma}\left[
\zeta'(s|A(\tilde\eta)\ell^2)
+2\sigma\zeta(s|A(\tilde\eta\ell^2)\right] \nonumber\end{eqnarray} and
finally \begin{equation} \tilde\Gamma(\eta)=\Gamma(\tilde\eta)
-\sigma\zeta(0|A(\tilde\eta))\,. \nonumber\end{equation}

We have seen in Sec.~\ref{S:ZF} that in $N$ dimensions $\zeta(0|A)$ is
related to the spectral coefficient $k_{N}(x|A)$ by means of the
equation \begin{equation} \zeta(0|A)=\frac{1}{(4\pi)^{N/2}}\int
k_{N}(x|A)\sqrt{g}d^Nx \:.\nonumber\end{equation} What is relevant for
our case is $k_4(x)=a_2(x|A(\tilde\eta))$ which is well known to be
(see Eq.~(\ref{a2})) \begin{eqnarray}
a_2(x|A(\tilde\eta))&=&-\frac{\lambda\Delta\phi_c^2}{12}
+\frac{\lambda\tilde m^2\phi_c^2}{2}
+\frac{\lambda(\xi-1/6)R\phi_c^2}{2}+\frac{\lambda^2\phi_c^4}{8}
+\frac{\tilde m^4}{2}\nonumber\\ &&+\tilde
m^2(\xi-\frac{1}{6})R+\frac{(\xi-1/6)^2R^2}{2}
+\frac{W}{120}-\frac{G}{360}+\frac{(5-\xi)\Delta R}{6}\,,
\nonumber\end{eqnarray} where $G=R^{ijrs}R_{ijrs} -4R^{ij}R_{ij}+R^2$
is the Gauss-Bonnet invariant. Hence, integrating $a_2$ over the
manifold and disregarding total divergences, one finally gets
\begin{equation} \tilde\Gamma(\eta)=S_{c}(\tilde\eta(\sigma))
-\frac{1}{2}\zeta'(0|\tilde A) =\Gamma(\tilde\eta(\sigma))+O(\hbar^2)
\nonumber\end{equation} The new parameters $\eta(\sigma)$ are related
to the old ones $\eta=\eta(0)$ by \begin{eqnarray}
\Lambda(\sigma)-\Lambda=\frac{m^4}{2}\hat\sigma\:,
\qquad\qquad\lambda(\sigma)-\lambda=3\lambda^2\hat\sigma\:,
\nonumber\end{eqnarray} \begin{eqnarray} m^2(\sigma)-m^2=\lambda
m^2\hat\sigma\:,
\qquad\qquad\xi(\sigma)-\xi=(\xi-\frac{1}{6})\lambda\hat\sigma\:,
\label{IRGE}\end{eqnarray} \begin{eqnarray}
\varepsilon_1(\sigma)-\varepsilon_1=(\xi-\frac{1}{6})m^2\hat\sigma\:,
\qquad\qquad\varepsilon_2(\sigma)-\varepsilon_2=(\xi-\frac{1}{6})^2\hat\sigma\:,
\nonumber\end{eqnarray} \begin{eqnarray}
\varepsilon_3(\sigma)-\varepsilon_3=\frac{1}{30}\hat\sigma\:.
\nonumber\end{eqnarray} Here we have set $\hat\sigma=-\sigma/16\pi^2$.
With the substitution $\sigma\to-2\tau$, the latter equations have
been given in Refs.~\cite{ocon83-130-31,hu84-30-743}. These  formulae
tell us that all parameters (coupling constants) develop, as a result
of quantum effects, a scale dependence, even if classically they are
dimensionless parameters. This means that in the quantum case, we have
to define the coupling constants at some particular scale.
Differentiating Eq.~(\ref{IRGE}) with respect to $\sigma$, one
immediately gets the renormalization group equations
\cite{ocon83-130-31,hu84-30-743}.

\subsection{Static and ultrastatic space-times}

We have seen that the renormalization procedure, in general requires a
bare cosmological constant in the gravitational action. Thus one has
to consider solutions of Einstein equations with a non vanishing
cosmological constant. Among the globally static solutions of Einstein
equations in vacuum, besides the Minkowski space-time with $\Lambda=0$,
de Sitter $(\Lambda>0)$ and anti-de Sitter $(\Lambda<0)$ space-times
are particularly interesting. The first one is simply connected, while
the second one has a simple connected universal covering (see for
example Ref.~\cite{hawk73b}). These manifolds are also maximally
symmetric and consequently they have constant curvature. One may also
consider other static space-times by quotienting with a discrete group
of isometries, but then one can encounter pathologies, like the
existence of closed time-like geodesics. It is an interesting fact
that the Euclidean sections corresponding to Minkowski, de Sitter and
anti-de Sitter space-time are the three constant curvature space forms
$\mbox{$I\!\!R$}^4$, $S^4$ and $H^4$ respectively. As a consequence,
the Euclidean field theory on these manifolds can have some relevance.
However, in this subsection,  we shall deal with an arbitrary and
globally static space-time and we shall briefly discuss how it is
possible, making use of conformal transformation techniques, to
restrict ourselves to ultrastatic space-times.

By definition, an ultrastatic space-time admits a globally defined
coordinate system in which the components of the metric tensor are
time independent and the condition $g_{00}=1$ and $g_{0i}=0$ hold
true. This means that the metric admits a global time-like orthogonal
Killing vector field. In all physical applications of the present
paper, we shall deal with an ultrastatic space-time $\cal
M^D=\mbox{$I\!\!R$}\times\cal M^N$ ($D=N+1$). This is not a true
restriction, since any static metric may be transformed in an
ultrastatic one (optical metric) by means of a conformal
transformation \cite{gibb78-358-467}. This fact permits us to compute
all physical quantities in an ultrastatic manifold and, at the end of
calculations, transform  back them to a static one, with an arbitrary
$g_{00}$. This has been done, for example, in
Refs.~\cite{dowk78-11-895,page82-25-1499,brow85-31-2514,brow86-33-2840,gusy87-46-1097,dowk88-38-3327,dowk89-327-267}
and in Ref.~\cite{kirs91-8-2239}, where the existence of boundaries
has also been taken into account. In particular, the approximation for
the heat-kernel on a static Einstein space-time was first introduced
in Ref. \cite{page82-25-1499}, where  the approximate propagator has
been obtained in closed form by making use of the results contained in
Ref. \cite{beke81-23-2850}. Here we would like to review the
techniques of those papers.

To start with we consider a scalar field $\phi$ on a
$D=N+1$-dimensional static space-time defined by the metric
\begin{eqnarray}
ds^2=g_{00}(\vec{x})(dx^0)^2+g_{ij}(\vec{x})dx^idx^j\:, \qquad\qquad
\vec{x}={x^j}\:,\qquad\qquad i,j=1,...,N\:, \nonumber\end{eqnarray}

The one-loop partition function is given by \begin{equation}
Z[g]=\exp(-S_c[\phi_c,g])\int d[\tilde\phi]\,
\exp\left(-\frac12\int\tilde\phi A\tilde\phi\,d^Dx\right)
\:,\nonumber\end{equation} where the operator $A$ has the form
\begin{eqnarray} A=-g^{00}(\partial_\tau-\mu)^2-\Delta_N+m^2+\xi R
\:,\nonumber\end{eqnarray} $\Delta_N$ being the Laplace-Beltrami
operator on the $N$ dimensional hypersurface $x^0=\tau=$~const, $m$
and $\xi$ arbitrary parameters and $R$ the scalar curvature of the
manifold. For the sake of completeness, we consider the combination
$\partial_\tau-\mu$, which is relevant in the finite temperature
theories with chemical potential.

The ultrastatic metric $g'_{\mu\nu}$ can be related to the static one
by the conformal transformation \begin{eqnarray}
g'_{\mu\nu}(\vec{x})=e^{2\sigma(\vec{x})}g_{\mu\nu}(\vec{x})
\:,\nonumber\end{eqnarray} $\sigma(\vec{x})$ being a scalar function.
We have to choose $\sigma(\vec{x})=-\frac{1}{2}\ln g_{00}$. In this
manner, $g'_{00}=1$ and $g'_{ij}=g_{ij}/g_{00}$ (optic metric).

Recalling that by a conformal transformation \begin{eqnarray}
R'&=&e^{-2q\sigma}\left[ R-2(D-1)\Delta_D\sigma -(D-1)(D-2)g^{\mu\nu}
\partial_\mu\sigma\partial_\nu\sigma\right]\:, \nonumber\end{eqnarray}
\begin{eqnarray} \tilde\phi'&=&e^{q\sigma}\tilde\phi
\:,\nonumber\end{eqnarray} \begin{eqnarray}
\Delta'_D\tilde\phi'&=&e^{-q\sigma}\left[\Delta_D
-\frac{D-2}{2}\Delta_D\sigma -\frac{(D-2)^2}{4}g^{\mu\nu}
\partial_\mu\sigma\partial_\nu\sigma \right]\tilde\phi\nonumber\\
&=&e^{-\sigma}\left[\Delta_D+\xi_D(e^{2\sigma}R'-R)
\right]\tilde\phi\nonumber \:,\nonumber\end{eqnarray} where
$\xi_D=(D-2)/4(D-1)$ is the conformal factor, $R'$ and $\Delta'_D$ the
scalar curvature and Laplace operator in the metric $g'$, one obtains
\begin{eqnarray}
A\tilde\phi=e^\sigma\left\{-g'^{00}(\partial_\tau-\mu)^2-\Delta'_N
+\xi_DR'+e^{-2\sigma}\left[ m^2+(\xi-\xi_D)R\right]\right\}\tilde\phi'
\:.\nonumber\end{eqnarray} From the latter equation we have
$\tilde\phi A\tilde\phi=\tilde\phi'A'\tilde\phi'$, where, by definition
\begin{eqnarray} A'=e^{-\sigma}A
e^{-\sigma}=-g'^{00}(\partial_\tau-\mu)^2-\Delta'_N
+\xi_DR'+e^{-2\sigma}\left[ m^2+(\xi-\xi_D)R\right]
\nonumber\:.\end{eqnarray} This means that the action $S'=S$ by
definition. Note that classical conformal invariance requires the
action to be invariant in form, that is $S'=S[\phi,g]$, as to say
$A'=A$. As is well known, this happens only for conformally coupled
massless fields ($\xi=\xi_D$).

For the one-loop partition function we have \begin{eqnarray}
Z'=J[g,g']\,Z \:,\nonumber\end{eqnarray} where $J[g,g']$ is the
Jacobian of the conformal transformation. Such a Jacobian can be
computed for any infinitesimal conformal transformation (see for
example \cite{gusy87-46-1097} and references cited therin). To this
aim it is convenient to consider a family of continuous conformal
transformations \begin{eqnarray} g^q_{\mu\nu}=e^{2q\sigma}g_{\mu\nu}
\nonumber\end{eqnarray} in such a way that the metric is $g_{\mu\nu}$
or $g'_{\mu\nu}$ according to whether $q=0$ or $q=1$ respectively. In
this manner one has \begin{eqnarray} \ln J[g_q,g_{q+\delta q}]
=\ln\frac{Z_{q+\delta q}}{Z_q} =\frac{\delta q}{(4\pi)^{D/2}} \int
k_D(x|A_q)\sigma(x)\sqrt{g^q}d^Dx \label{deJ} \:,\end{eqnarray} where
$k_D(x|A_q)$, is the Seeley-DeWitt coefficient, which in the case of
conformal invariant theories, is proportional to the trace anomaly.

The Jacobian for a finite transformation can be obtained from
Eq.~(\ref{deJ}) by an elementary integration in $q$
\cite{gusy87-46-1097}. In particular we have \begin{eqnarray} \ln
J[g,g']=\frac{1}{(4\pi)^{D/2}} \int_0^1dq\int
k_D(x|A_q)\sigma(x)\sqrt{g^q}\,d^Dx \:,\label{lnJ}\end{eqnarray} and
finally \begin{eqnarray} \ln Z=\ln Z'-\ln J[g,g']
\:.\label{lnZ-Zbar}\end{eqnarray} Then we see that in principle the
knowledge of the partition function $Z'$ in the ultrastatic manifold
and the heat coefficient $k_D(A_q)$ are sufficient in order to get the
partition function in the static manifold. We know that the heat
coefficients depend on invariant quantities build up with curvature
(field strength) and their derivatives. As a consequence, they do not
depend on the parameter $\mu$, since we may regard such a parameter as
the temporal component of a (pure gauge) abelian potential. Thus we
can simply put $\mu=0$ in the computation of the Jacobian determinant
$J[g,g']$ using Eq.~(\ref{lnJ}).

\subsection{Finite temperature effects} \label{S:FT}

A reason to consider finite temperature quantum field theories is
mainly based on the recent developments of cosmological models.
According to the standard Big-Bang cosmology and the more recent
inflationary models, the very early universe has passed through a
phase of thermal equilibrium at very high temperature and density,
where the symmetry was restored but with a large cosmological
constant. As the universe has become cool, it has gone through
several phase transitions (see for example Ref.~\cite{bran85-57-1}).
Although the usual thermodynamical concepts may be inappropriate in
the presence of very strong gravitational interactions, the need for
considering finite temperature field theory in a curved background has
been arisen. Strictly speaking, it has been shown that thermal
equilibrium can be maintained for conformally invariant field theories
in conformally flat expanding space-times \cite{hu82-108-19}.
Otherwise the expansion must be nearly adiabatic
\cite{park74-9-341,park74-10-3905,hu81-103-331,hu83-123-189,ande87-36-2963}.
On the other hand, particle production in a hot Friedman universe is
exponentially suppressed due to the increase of the mass by thermal
effects at temperatures smaller than the Planck one
\cite{lind83-123-185} and the expansion should be nearly adiabatic.
Phase transitions in de Sitter space-time have been  first considered
in Ref.~\cite{alle83-226-228} with the important result that critical
behaviour strongly depends on curvature. Very little is known about
them in anti-de Sitter space-time, mainly because de Sitter is more
relevant than anti-de Sitter in inflationary scenarios.

Now we would like to present a survey of finite temperature quantum
field theory and discuss some useful representations of
thermodynamical quantities.

\subsubsection{The free energy}

To begin with, let us consider a (scalar) field  in thermal
equilibrium at finite temperature $T=1/\beta$. It is well known that
the corresponding partition function $Z_\beta$ may be obtained, within
the path integral approach, simply by Wick rotation $\tau=ix^0$ and
imposing a $\beta$ periodicity in $\tau$ for the field
$\phi(\tau,x^i)$ ($i=1,...,N$, $N=D-1$)
\cite{bern74-9-3312,dola74-9-3320,wein74-9-3357,kapu89b}. In this way
the one loop approximation reads \begin{equation}
Z_\beta[\phi_c,g]=e^{-S_c[\phi_c,g]}
\int_{\tilde\phi(\tau,x^i)=\tilde\phi(\tau+\beta,x^i)}
d[\tilde\phi]\,\exp\left(-\int_0^\beta d\tau\int\tilde\phi
A\tilde\phi\,d^Nx\right) \:.\nonumber\end{equation}

If the space-time is ultrastatic then we are dealing with the manifold
$S^1\times\cal M^N$. The relevant differential operator is
$A=L_D=-\partial_\tau^2+L_N$ and its eigenvalues are given by
Eq.~(\ref{eigenS1}). Then, using Eqs.~(\ref{logPI}) and
(\ref{ZF1-Poisson})-(\ref{ZF1-Barnes}) we obtain
\cite{alle86-33-3640,byts92-7-397} \begin{eqnarray} \ln
Z^{(1)}_\beta+\nu\beta\zeta^{(r)}(-1/2|L_N\ell^2)
&=&\frac{\nu\beta}\pi\sum_{n=1}^\infty\lim_{s\to0}\int_{-\infty}^\infty
e^{in\beta t}\,\zeta(s|L_N+t^2)\,dt \label{logPF-Poisson} \\&=&
\frac{\nu\beta}{\sqrt{\pi}} \sum_{n=1}^\infty\int_0^\infty
t^{-3/2}\,e^{-n^2\beta^2/4t}\, K(t|L_N)\,dt \label{logPF-Jacobi} \\&=&
\frac{\nu\beta}{\pi i}\int_{\,\mbox{Re}\, z=c}\zeta_R(z)
\Gamma(z-1)\zeta(\frac{z-1}{2}|L_N)\; \beta^{-z}\;dz
\:,\label{logPF-Barnes} \end{eqnarray} where $\nu=1$ ($\nu=1/2$) for
charged (neutral) scalar fields.
Eqs.~(\ref{logPF-Poisson})-(\ref{logPF-Barnes}) not only define the
finite temperature properties of quantum fields but, as we shall see
in Sec.~\ref{S:RVE}, they will be our starting point for the
computation of the regularized vacuum energy.

The free energy is related to the canonical partition function by
means of equation \begin{eqnarray} F(\beta)=-\frac{1}{\beta}\ln
Z_\beta=F_0+F_\beta \:,\nonumber\end{eqnarray} where $F_\beta$
represents the temperature dependent part (statistical sum) and so
Eqs.~(\ref{logPF-Poisson})-(\ref{logPF-Barnes}) give different
representations of $F_\beta$.

\subsubsection{The thermodynamic potential}

The generalization to the more general case of a charged (scalar) field
with a non vanishing chemical potential $\mu$, in thermal equilibrium
by some unspecified process at finite temperature $T=1/\beta$ is quite
immediate. The grand canonical partition function has the path
integral representation
\cite{free77-16-1130,kapu81-24-426,acto85-157-53,acto86-256-689}
\begin{eqnarray}
Z_{\beta,\mu}&=&e^{-S_c}\int_{\phi(\tau,x^i)=\phi(\tau+\beta,x^i)}
d[\bar\phi]d[\phi]\, \exp\left({-\int_0^\beta d\tau\int\bar\phi
A(\mu)\phi\,d^Nx}\right) \nonumber\\ &=&\exp[-\beta\Omega(\beta,\mu)]
\:,\nonumber\end{eqnarray} where now, the operator $A$ depends on the
chemical potential $\mu$, i.e. $A(\mu)=-(\partial_\tau-\mu)^2+L_N$ and
the latter equation defines the thermodynamic potential
$\Omega(\beta,\mu)$. The operator  $A(\mu)$  is still elliptic but not
hermitian, in fact it is normal and its eigenvalues are complex and
read \begin{equation} \lambda_{n,j}=\left(\frac{2\pi
n}{\beta}+i\mu\right)^2 +\omega_j^2 \qquad\qquad n=0,\pm 1,\pm 2,\dots
\:,\nonumber\end{equation} $\omega_j^2$ being the eigenvalues of
$L_N$. Nevertheless, one can still define the related $\zeta$-function.
One formally has \begin{eqnarray}
\Omega(\beta,\mu)&=&\frac{1}{\beta}S_c[\phi_c,g]
+\frac{1}{\beta}\ln\det[A(\mu)\ell^2] \:,\nonumber\end{eqnarray}
\begin{eqnarray} \ln\det A&=&\sum_{n,j} \ln{\left[\omega_j^2+(2\pi
n/\beta+i\mu)^2\right] }=\sum_{n,j}\int
\frac{d\omega_j^2}{\omega_j^2+(2\pi n/\beta+i\mu)^2}\nonumber\\
&=&\frac\beta2\sum_j\int\left[ \coth\frac{\beta(\omega_j+\mu)}2
+\coth\frac{\beta(\omega_j-\mu)}2\right] \frac{d\omega_j}{\omega_j}
\nonumber\end{eqnarray} and integrating on $\omega_j$ and summing over
all $j$ the  well known result \begin{eqnarray}
\Omega(\beta,\mu)&=&\sum_j\omega_j+\frac{1}{\beta}\sum_j
\ln\left(1-e^{-\beta(\omega_j+\mu)}\right)
+\frac{1}{\beta}\sum_j\ln\left(1-e^{-\beta(\omega_j -\mu)}\right)
\nonumber\\&=&
\sum_j\omega_j+\frac{1}{\beta}\,\mbox{Tr}\,\ln\left(1-e^{-\beta
Q^+}\right) +\frac{1}{\beta}\,\mbox{Tr}\,\ln\left(1-e^{-\beta
Q^-}\right) \label{Ombemu} \end{eqnarray} follows. Above, we have
introduced the two pseudo-differential operators
$Q^{\pm}=L_N^{1/2}\pm\mu$, whose  eigenvalues are $\omega_j\pm\mu$. We
see that $\Omega(\beta,\mu)$ is the sum of vacuum energy
(zero-temperature contribution, formally divergent) and two
finite-temperature contributions, one for particles and one for
anti-particles. We also see from Eq.~(\ref{Ombemu}), that the
temperature contribution, as a function of the complex parameter
$\mu$, has branch points when $\mu^2$ is equal to an eigenvalue
$\omega_j^2$. Then $\Omega_\beta(\beta,\mu)$ is analytic in the $\mu$
complex plain with a cut from $\omega_0$ to $\infty$ and from
$-\omega_0$ to $-\infty$. Thus, the physical values of $\mu$ are given
by $|\mu|\leq\omega_0$.

If one makes use of the $\zeta$-function regularization, one obtains
\begin{eqnarray} \Omega(\beta,\mu)=\frac{1}{\beta}S_c[\phi_c,g]
-\frac{1}{\beta}\zeta'(0|A(\mu)\ell^2) \:.\label{Omreg} \end{eqnarray}
Using Eq.~(\ref{PSF}) and Mellin representation of $\zeta$-function,
Eq.~(\ref{ZFdef}), after some  manipulations similar to the ones of
Sec.~\ref{S:ZF}, one gets the three useful representations for the
temperature dependent part $\Omega_\beta(\beta,\mu)$ of the
thermodynamic potential (\ref{Omreg})
\cite{byts92-7-397,cogn92-7-3677} \begin{eqnarray}
\Omega_\beta(\beta,\mu)&=&\Omega(\beta,\mu)-\Omega_0\nonumber\\
&=&-\frac{1}\pi\sum_{n=1}^\infty\int_{-\infty}^\infty e^{in\beta
t}\,\zeta'(0|L_N+[t+i\mu]^2)\,dt \label{Om-Poisson}
\\&=&-\frac{1}{\sqrt{\pi}} \sum_{n=1}^\infty\cosh
n\beta\mu\int_0^{\infty}t^{-3/2} e^{-n^2\beta^2/4t}\,K(t|L_N)\,dt
\label{Om-Jacobi}\\ &=&-\frac{1}{\pi i}\sum_{n=0}^{\infty}
\frac{\mu^{2n}}{(2n)!} \int_{\,\mbox{Re}\, s=c}\zeta_R(s)\Gamma(s+2n-1)
\zeta(\frac{s+2n-1}{2}|L_N)\;\beta^{-s}\;ds \:.\label{Om-Barnes}
\end{eqnarray} Here
$\Omega_0=\frac{1}{\beta}S_c[\phi_c,g]+\zeta^{(r)}(-1/2|L_N\ell^2)$ is
the zero-temperature contribution (classical and vacuum energy).
Eqs.~(\ref{Om-Poisson})-(\ref{Om-Barnes}) are valid for any
$|\mu|<\omega_0$, $\omega_0$ being the smallest eigenvalue of $L_N$.
When $|\mu|=\omega_0$ a more careful treatment is needed. The physical
reason for this behaviour is the occurrence of Bose-Einstein
condensation. Of course, in the limit $\mu\to0$,
Eqs.~(\ref{Om-Poisson})-(\ref{Om-Barnes}) reduces to
Eqs.~(\ref{logPF-Poisson})-(\ref{logPF-Barnes}) which has been
discussed in the previous section.

For the sake of completeness, we  report another integral
representation of the thermodynamic potential, which can be obtained
in a similarly to the one used for deriving Eq.~(\ref{Om-Barnes})
\cite{cogn92-7-3677}. To this aim, we observe that the poles of
$\zeta(s|Q^{\pm})$ (related to the two pseudo-differential operators
$Q^{\pm}$) are given by Seeley theorem \cite{seel67-10-172} at the
points $s=N-k$ ($k=0,1,\dots$) with residues which are polynomials in
$\mu$. The formula which generalizes the result of
Ref.~\cite{habe82-23-1852} to compact manifolds then reads
\begin{equation} \Omega_\beta(\beta,\mu) =-\frac{1}{2\pi
i}\int_{\,\mbox{Re}\, s=c}\zeta_{R}(s)\Gamma(s-1)
\left[\zeta(s-1|Q^{+})+\zeta(s-1|Q^{-})\right]\beta^{-s}\,ds
\:.\label{Om-B2} \end{equation} This should be used to discuss
$\Omega_\beta(\beta,\mu)$ as a function of complex $\mu$.

\subsection{The regularization of vacuum energy} \label{S:RVE}

The vacuum energy density (the first term on the right hand side in
Eq.~(\ref{Ombemu})) is the only source of divergences in the
thermodynamic potential. Therefore we shall briefly discuss how to
give it a mathematical meaning. We present a general formula for the
vacuum energy of a scalar field defined on an ultrastatic space-time
with compact spatial section consisting in general in a manifold with
boundary.

We start with some general considerations. In the last decade there
has been un increasing interest in investigating vacuum effects or
zero point fluctuations in the presence of boundaries
\cite{plun86-134-87} or in space-time with non trivial topology. The
introduction of boundaries and related boundary conditions on quantum
fields may be considered as an excellent idealization of complicated
matter configurations. This is certainly true for the original Casimir
configuration, namely two neutral parallel conducting plates placed at
a distance L. In fact the interpretation due to Casimir of the
attractive force present in such a configuration (experimentally
verified) seems to support such a point of view. In this case, the
boundary conditions associated with the electromagnetic field are the
perfect conductor boundary conditions and the related Casimir energy
is negative. Casimir conjectured the same thing to be hold for a
spherical shell. In this way the fine structure constant could have
been determined.

The computation for the spherical shell was performed in
Refs.~\cite{boye68-174-1764,bali78-112-165,milt80-22-1441} and the
result was a positive vacuum energy, proving that the Casimir
conjecture was wrong (see however Ref.~\cite{kuni81-177-528}, where
interpreting the Casimir effect as a screening effect, a negative
energy is obtained for a scalar field and also
Ref.~\cite{cett93-108-447} for a recent treatment). Other computations
associated with different geometries of the boundary seems to support
the idea that the sign of Casimir energy depends on the cavity in a
non trivial way.

On general ground, as it has been stressed by DeWitt
\cite{dewi75-19-295,kay79-20-3052}, the introduction of boundaries or
the identification of surfaces modifies the global topology of a local
space-time. As a consequence one is effectively dealing with  quantum
field theory  on a curved manifold.

We would like to mention that Casimir energy plays an important role
in hadronic physics where, due to the confinement of quark and gluon
fields, vacuum effects cannot be neglected as soon as one is working
in the framework of bag models  \cite{veps90-187-109}. In such
situation, one is forced to consider a compact cavity. For
computational reasons, only the case of a spherical cavity has been
investigated in some detail (see for example
Ref.~\cite{bend76-14-2622}).

For the sake of simplicity, here we shall mainly deal with a scalar
field defined on a $D$ dimensional space-time ${\cal M}^D$ with a
$N=(D-1)$-dimensional smooth boundary. The metric is supposed to be
ultrastatic, then the results of Sec.~\ref{S:FT} can be applied.
Vector and spinor fields may be treated along the same line, paying
attention however to the related boundary conditions.

Given an arbitrarily shaped cavity, one may consider an internal and an
external problem. For the sake of simplicity, we shall assume that the
internal problem is associated with a simply connected manifold. The
external problem is usually associated with a double connected
manifold. The reason stems from the necessity to introduce a very
"large" manifold in order to always deal with a compact manifold. This
large manifold may be considered as a reference manifold, namely a
volume cut-off of Minkowski space-time. Its boundary can be pushed to
infinity at the end of the computation \cite{bali78-112-165}. As far
as the boundary conditions associated with the external problem, we
shall mainly leave understood the use of the same internal boundary
conditions, paying attention however to the proposal of
Ref.~\cite{syma81-190-1}, in which the external boundary condition is
different from the internal one.

To start with, we shall consider a compact arbitrarily shaped manifold
with boundary, which may be multi-connected. The main issue related to
vacuum energy $E_v$ is the necessity of a regularization scheme. In
fact, if one naively makes use of canonical quantization of a scalar
field, the result is (see for example Ref.~\cite{full73-7-2850})
\begin{equation} E_v=\nu\sum_j\omega_j \:,\label{VE} \end{equation}
$\omega_n^2$ being the eingenvalues of $L_N$ defined on ${\cal M}^N$,
the eigenfunctions of which are the mode functions satisfying suitable
conditions on the boundary of ${\cal M}^N$ and, as in Sec.~\ref{S:FT},
$\nu=1$ ($\nu=1/2$) for charged (neutral) scalar fields. As we have
seen in Sec.~\ref{S:FT}, also formal manipulations lead to such an
expression like Eq.~(\ref{VE}). Of course it is an ill defined object
and one needs some prescriptions in order to extract a finite
observable quantity.

In order to regularize Eq.~(\ref{VE}) by the use of $\zeta$-function,
one may try to write \begin{equation} E_v=\nu\lim_{s\to-\frac{1}{2}}
\zeta(s|L_N) \:.\label{VEreg} \end{equation} Indeed, in the case of
hyper-rectangular cavities and massless free fields, this expression
gives a result in agreement with other regularizations
\cite{ambj83-147-1}. However, Eq.~(\ref{VEreg}) does not work when one
is dealing with an arbitrarily shaped cavity. This can be easily
understood looking at Eq.~(\ref{ZFpoles}). From such an equation we
see in fact that for $s\to-1/2$, $\zeta(s|L_N)$ has in general a
simple pole, the residue of which is given by $-K_D(L_N)/\sqrt{4\pi}$.
As a consequence the regularization given by Eq.~(\ref{VEreg}) is
meaningless, unless $K_D(L_N)$ vanishes. This is just the case of
massless free fields in hyper-rectangular cavities.

A possible definition of vacuum energy in terms of $\zeta$-function,
which is valid for a manifold with an arbitrary smooth boundary, can
be obtained starting from the finite temperature partition function
$Z_\beta$. According to Ref.~\cite{cogn92-33-222}, we define
\begin{equation} E_v=-\lim_{\beta\to\infty}\partial_\beta\ln Z_\beta
\:.\nonumber\end{equation} Using for $\ln Z_\beta$ the regularized
expression (\ref{logPF-Jacobi}), one immediately gets (here
$S_c[\phi_c,g]=0$ because we are considering free fields)
\begin{equation} E_v=\nu\zeta^{(r)}(-1/2|L_N\ell^2) \:.\label{VEZreg}
\end{equation} With regard to this result, we would like to observe
that a similar prescription has been appeared in many places in the
literature (see for example \cite{dowk76-13-3224,dowk78-11-895}). Here
we would like to stress the proposal contained in
Ref.~\cite{blau88-310-163}, which is very close to our approach. In
other words, the path-integral $\zeta$-function derivation presented
here, leads directly to the "principal part prescription" of
Ref.~\cite{blau88-310-163}. It is interesting to observe that the term
in the vacuum energy, which depends on $\ell$ (see Eq.~(\ref{ZF-r})),
is proportional to the integral of the conformal anomaly
\cite{birr82b}.

The total contribution (interior plus exterior minus the reference one)
to vacuum energy is relevant from the physical point of view, for
example in the classic electromagnetic Casimir effect associated to a
conducting cavity or in some extensions of the bag model. If one is
dealing with a free massless field on a $3+1$ dimensional manifold, as
in the electromagnetic case, it turns out that the total Casimir
energy seems to be free from the ambiguity associated with $\ell$
\cite{bali78-112-165}. As we shall see in Sec.~\ref{S:OLEP}, the
ambiguity due to the free parameter $\ell$ can be removed by
renormalization when the theory has a natural scale. This is true for
massive or self-interacting fields, but also for massless fields on
curved manifolds.

\newpage
\setcounter{equation}{0}
\section{Constant curvature manifolds}
\label{S:CCMan}

The present section is devoted to the derivation of heat kernel and
$\zeta$-function related to the Laplace-Beltrami operator acting on
fields living on compact manifolds with constant curvature. In the
spirit of the report, we describe general aspects, relegating specific
results to the appendices.

For a detailed discussion of $N$ dimensional torus and sphere, we
refer the reader to the vast literature on the subject (for a recent
review see for example Ref.~\cite{camp90-196-1} and references
therein). Here we shall very briefly describe some useful techniques
and we shall give some known, but also less known representations for
$\zeta$-function, which shall be used in the physical applications. On
the contrary, compact hyperbolic manifolds shall be analyzed in some
detail, because we suppose the reader to be not familiar with the
hyperbolic geometry, the related isometry groups and the Selberg trace
formula (see, for example, Refs. \cite{bala86-143-109,gutz90b,eliz94b}
and references therein).

\subsection{The heat kernel and $\zeta$-function on the torus}
\label{S:ZFTN}

Here we perform the analytic continuation for the $\zeta$-function of
the operator $\L_N=-\Delta_{T^N}+\alpha^2$ acting on twisted scalar
fields on the torus $T^N$. $T^N$ is the direct product of $N$ circles
$S^1$ with radii $r_i$. In order to take into account of twists, we
introduce a $N$-vector $\vec q$ with components $0$ or $1/2$, in such
a manner that the eigenvalues of the Laplacian have the form $(\vec
k+\vec q)\cdot {\cal R}_N^{-1}(\vec k+\vec q)$, $ {\cal R}_N$ being
the diagonal matrix $diag( {\cal R}_N)=(r_1^2,\dots,r_N^2)$.

The trace of the heat kernel $K(t|\L_N)=\,\mbox{Tr}\,\exp(-t\L_N)$ on
the torus can be directly derived by the factorization property
(\ref{HKfact}) knowing the kernel on $S^1$. So we have \begin{eqnarray}
K(t|\L_N)=e^{-t\alpha^2}\sum_{\vec k} e^{-t(\vec k+\vec q)\cdot {\cal
R}_N^{-1}(\vec k+\vec q)} =\frac{\Omega_Ne^{-t\alpha^2}}{(4\pi
t)^{\frac N2}} \sum_{\vec k}e^{-2\pi i\vec k\cdot\vec q}
e^{-\pi^2(\vec k\cdot {\cal R}_N\vec k)/t}
\:,\label{KtTN}\end{eqnarray} where $\vec k\in\mbox{$Z\!\!\!Z$}^N$ and
$\Omega_N=(2\pi)^N\sqrt{\det {\cal R}_N}$ is the hypersurface of the
torus. The latter equation has been derived by using Eq.~(\ref{HKTN})
in Appendix~\ref{S:HKES}.

As it is well known, the $\zeta$-function on the torus is given in
terms of Epstein $Z$-function (see Appendix \ref{S:UR} for definition
and properties). In fact, one directly obtains \begin{eqnarray}
\zeta(s|\L_N)=Z_{ {\cal R}^{-1}_N}(\frac{2s}N;\vec q,0)
\:.\label{zeTN-Epstein}\end{eqnarray} Other useful representations can
be obtained by making use of Mellin representation (\ref{ZFdef}). They
read \begin{eqnarray}
\zeta(s|\L_N)-\frac{\Omega_N\Gamma(s-N/2)\alpha^{N-2s}} {(4\pi)^{\frac
N2}\Gamma(s)} =\frac{2\Omega_N\alpha^{N-2s}}{(4\pi)^{\frac
N2}\Gamma(s)} \sum_{\vec k\neq0} \frac{e^{-2\pi i\vec k\cdot\vec q}
K_{N/2-s}(2\pi\alpha[\vec k\cdot {\cal R}_N\vec k]^{\frac12})}
{\left(\pi\alpha[\vec k\cdot {\cal R}_N\vec k]^{\frac12}\right)^{\frac
N2-s}} \label{ZFTN-K} \\=\frac{\Omega_N\alpha^{N-2s}}
{(4\pi)^{\frac{N-1}2}\Gamma(s)\Gamma(\frac{N+1}2-s)} \sum_{\vec k\neq0}
e^{-2\pi i\vec k\cdot\vec q} \int_{1}^{\infty}(u^2-1)^{\frac{N-1}2-s}
e^{-2\pi\alpha u[\vec k\cdot {\cal R}_N\vec k]^{\frac12}}\,du
\:.\label{ZFTN}\end{eqnarray} The representation of $\zeta$ given by
Eq.~(\ref{ZFTN-K}) is valid for any $s$, since the Mc Donald functions
$K_\nu(z)$ are exponentially vanishing for $z\to\infty$ and so the
series is convergent (here $\alpha>0$). On the contrary, the
representation (\ref{ZFTN}) is only valid for $\,\mbox{Re}\, s<N/2$,
but this is what we need in the evaluation of physical quantities. It
has to be remarked that the right hand sides of the above formula, in
the limit $\alpha\to0$, give exactly the $\zeta$-function in the
massless case.

\paragraph{The $\zeta$-function on $T^Q\times\cal M^N$}

Here we just write down the Mellin-Barnes representation of
$\zeta$-function, Eq.~(\ref{ZFA1A2}), for this particular case, $\cal
M^N$ being an arbitrary compact manifold. The operator is assumed to
be of the form $A=-\Delta_Q+L_N$, with $\Delta_Q$ the Laplace operator
in $T^Q$ and $L_N$ a differential operator in $M^N$. For the
$\zeta$-function related to $\Delta_Q$ we use the representation in
terms of Epstein $Z$-function (see Eq.~(\ref{zeTN-Epstein})). Then,
supposing $L_N$ to be an invertible operator, from Eq.~(\ref{ZFA1A2})
we obtain \begin{eqnarray} \zeta(s|A)=\frac1{2\pi i\Gamma(s)}
\int_{\,\mbox{Re}\, z=c}\Gamma(\frac s2+z) \zeta(\frac
s2+z|L_N)\Gamma(\frac s2-z) Z_{ {\cal R}_Q^{-1}}(\frac{s-2z}Q|\vec
q,0)\,dz \:.\label{ZFTQMN}\end{eqnarray} Here we have implicitly
assumed also $\Delta_Q$ to be an invertible operator. This happens
when there is at least one twist. If $\vec q=0$ (no twists), there is
a zero-mode and to take account of it on the right hand side of the
latter equation we must add $\zeta(s|L_N)$. In Eq.~(\ref{ZFTQMN}), $c$
must satisfy the condition $-\frac{s-Q}2<c<\frac{s-N}2$, which is
quite restrictive for physical aims. Such a restriction can be relaxed
if we compute $\zeta$-function by a technique similar to the one used
in Sec.~\ref{S:S1XMN}. In this way we have \begin{eqnarray}
\zeta(s|A)&=&\frac{\Omega_Q}
{(4\pi)^{Q/2}\Gamma(s)}\left[\phantom{\int} \Gamma(s-\frac
Q2)\zeta(s-\frac Q2|L_N)\right.\nonumber\\&&\left. +\frac1{2\pi i}
\int_{\,\mbox{Re}\, z=c}\pi^{-\frac z2}\Gamma(s+\frac{z-Q}2)
\zeta(s+\frac{z-Q}2|L_N)\Gamma(\frac z2) Z_{ {\cal R}_Q}(\frac
zQ|0,-\vec q)\,dz\right] \:,\nonumber\end{eqnarray} which reduces to
Eq.~(\ref{ZF-Barnes}) for $Q=1$ and $q=0$. Now $c>N+Q$. Similar
expressions can be also obtained for manifolds with other kinds of
sections, for example spheres or hyperbolic manifolds.

\subsection{Representations and recurrence relations for
$\zeta$-function on the sphere} \label{S:ZFRR}

Here we shall review some techniques used in the literature, which
provide useful analytical extensions for $\zeta$-function of Laplacian
on spheres (compact rank-one symmetric spaces). We shall explain the
known method based on binomial expansion, which has been recently used
in Ref.~\cite{chan93-407-432} in order to evaluate $\zeta$-function on
orbifold-factored spheres $S^N/\Gamma$, $\Gamma$ being a group of
isometry. We shall closely follow Ref.~\cite{camp90-196-1}. We would
like to mention the pionering work by Minakshisundaram, who first
introduced these kind of generalized zeta functions
\cite{mina49-13-41,mina52-4-26}. His results are very close to the
ones we shall obtain for the hyperbolic case. Recently, these results
have been derived by a different technique also in
Ref.~\cite{camp93-47-3339}. Finally we shall also give an integral
representation in the complex plane for the trace of an arbitrary
function of the Laplacian acting on scalar fields in $S^N$ and we
shall derive some recurrence relations for the trace of the heat
kernel and for the $\zeta$-function density. Here we normalize the
constant curvature to $\kappa=1$.

To start with, we recall that for compact rank-one symmetric spaces
the spectrum and its degeneration are at disposal (see for example
Refs.~\cite{cahn76-51-1,camp90-196-1}). In particular, for the
eigenvalues $\lambda_n$ and their degeneration $d_n^N$ of the Laplace
operator $-\Delta_N$ on $S^N$ we have \begin{eqnarray}
\lambda_n=n(n+2\varrho_N)\:,\qquad\qquad
d_n^N=\frac{2(n+\varrho_N)\Gamma(n+2\varrho_N)}
{\Gamma(2\varrho_N+1)\Gamma(n+1)} \:,\qquad\qquad n\geq0
\:,\label{la-dN}\end{eqnarray} with $\varrho_N=(N-1)/2$.
Eqs.~(\ref{la-dN}) are valid for any $N$, but $d_0^1=1$. One can see
that the degeneration of eigenvalues for odd $N$ (respectively even
$N$) is an even (respectively odd) polynomial of $N-1$ degree in
$n+\varrho_N$. In fact we have \begin{eqnarray} d_0^1=1\:,\qquad\qquad
d_n^1=2\:,\qquad\qquad d_n^2=2(n+\varrho_2)\:, \nonumber\end{eqnarray}
\begin{eqnarray} d_n^N&=&\frac{2}{(N-1)!}
\prod_{k=0}^{\frac{N-3}2}[(n+\varrho_N)^2-k^2]
\:,\qquad\qquad\mbox{for odd }N\geq3\:,\\
d_n^N&=&\frac{2(n+\varrho_N)}{(N-1)!}
\prod_{k=0}^{\frac{N-4}2}[(n+\varrho_N)^2-(k+\frac12)^2]
\:,\qquad\qquad\mbox{for even }N\geq4 \:.\nonumber\end{eqnarray} Then
we can define the coefficients $a_k^N$ by means of equation
\begin{eqnarray} d_n^N=\Omega_Nc_n^N
=\Omega_N\sum_{k=1}^{N-1}a_k^N(n+\varrho_N)^{k}
\:,\label{akN}\end{eqnarray} valid for any $N\geq2$. Here
$\Omega_{N-1}=2\pi^{\frac{N}2}/\Gamma(\frac{N}2)$ is the volume
(hypersurface) of $S^{N-1}$, which we shall explicitly write in all
formulae in order to point out the strictly similarity with the
compact hyperbolic case, which shall be extensively treated in
Sec.~\ref{S:HKZFHM}. The case $N=1$ is quite trivial and will be
treated separately. It can be also considered as a particular case of
previous section. To consider both even and odd cases, we have written
a general polynomial, but of course only even or odd coefficients
$a_k^N$ are non vanishing, according to whether $N$ is odd or even.

By definition, heat kernel and $\zeta$-function are given by
\begin{eqnarray} K(t|\L_N)=\Omega_Ne^{-t\alpha^2}\sum_{n=0}^{\infty}
c_n^N\,e^{-t(n+\varrho_N)^2} \:,\label{KtSN}\end{eqnarray}
\begin{eqnarray} \zeta(s|\L_N)=\Omega_N\sum_{n=0}^{\infty} c_n^N
[(n+\varrho_N)^2+\alpha^2]^{-s} \:.\label{ZFSN}\end{eqnarray} For more
generality, we have considered the massive operator
$\L_N=-\Delta_{S^N}+\alpha^2+\kappa\varrho_N^2$, $\alpha$ being an
arbitrary constant. The choices $\alpha^2=m^2-\varrho_N^2$ corresponds
to minimal coupling while the choice  $\alpha^2=0$ correspond to
conformal coupling (we think of $S^N$ as the spatial section of an
ultrastatic $N+1$ dimensional manifold). As usual zero modes must be
omitted in Eq.~(\ref{ZFSN}). Because of homogeneity of $S^N$,
$K_t^{S^N}(x,x)$ does not depend on $x$ and so, a part the volume,
Eq.~(\ref{KtSN}) determines also the heat kernel in the coincidence
limit.

\subsubsection{The series representation}

Using Eq.~(\ref{akN}) in Eq.~(\ref{ZFSN}) and making a binomial
expansion we obtain a representation of $\zeta(s|\L_N)$ as an infinite
sum of Riemann-Hurwitz $\zeta_H$-functions (See Appendix \ref{S:UR},
Eq.~(\ref{RHzf})). For any $N\geq2$ and $|\alpha^2|<\varrho_N^2$
(remember that the curvature is normalized to $\kappa=1$) it reads
\begin{eqnarray} \zeta(s|\L_N)=\Omega_N\sum_{n=0}^\infty
\sum_{k=1}^{N-1}a_k^N
\frac{(-1)^n\Gamma(s+n)\alpha^{2n}}{\Gamma(n+1)\Gamma(s)}
\zeta_H(2s+2n-k;\varrho_N) \:.\label{ZF-Hurwitz}\end{eqnarray} By the
same method, for $N=1$ we obtain \begin{eqnarray}
\zeta(s|\L_1)&=&\frac1{\alpha^{2s}} +2\sum_{n=1}^\infty
\frac{(-1)^n\Gamma(s+n)\alpha^{2n}}{\Gamma(n+1)\Gamma(s)}
\zeta_R(2s+2n) \:.\nonumber\end{eqnarray} Eq.~(\ref{ZF-Hurwitz})
notably simplify in the case of conformal coupling ($\alpha=0$). In
fact we have \begin{eqnarray}
\zeta(s|\L_N)=\Omega_N\sum_{k=1}^{N-1}a_k^N \zeta_H(2s-k;\varrho_N)
\:.\nonumber\end{eqnarray} For the massless minimal coupling
($\alpha^2=-\varrho_N^2$), one has to pay attention to the zero mode.
After the subtraction of it one obtains \begin{eqnarray}
\zeta(s|\L_N)=\Omega_N\sum_{n=0}^\infty \sum_{k=1}^{N-1}a_k^N
\frac{\Gamma(s+n)\varrho_N^{2n}}{\Gamma(s)\Gamma(n+1)}
\zeta_H(2s+2n-k;\varrho_N+1) \:.\nonumber\end{eqnarray} These
equations are valid for $N\geq2$. For $N=1$ we have the simpler result
$\zeta(s|\L_1)=2\zeta_R(2s)$.

\subsubsection{Recursive representation}

Now we are going to derive a representation of $\zeta(s|\L_N)$ in
terms od $\zeta(s|\L_1)$ or $\zeta(s|\L_2)$ according to whether $N$
is odd or even. Similar results can be found in
Ref.~\cite{camp90-196-1}. We first consider the odd dimensional case
with $N\geq3$. Using again Eq.~(\ref{ZFSN}) and observing that
$\varrho_N$ is an integer we easily get \begin{eqnarray}
\zeta(s|\L_N)&+&\Omega_N\sum_{n=1}^{\varrho_N-1}
\sum_{k=1}^{\varrho_N}a_{2k}^Nn^{2k} (n^2+\alpha^2)^{-s} \nonumber\\
&=&\left.\Omega_N\sum_{n=1}^{\infty}
\sum_{k=1}^{\varrho_N}\frac{(-1)^ka_{2k}^N}{\Gamma(s)}
\frac{d^k}{d\lambda^k}\int_{0}^{\infty} t^{s-k-1}e^{-\lambda
t(n^2+\alpha^2/\lambda)}\,dt\,\right|_{\lambda=1} \nonumber\\
&=&\Omega_N\sum_{k=1}^{\varrho_N}
\frac{a_{2k}^N\Gamma(s-k)\alpha^{-2s}}{2\Gamma(s)}
\left(\alpha^4\frac{d\phantom{\alpha^2}}{d\alpha^2}\right)^k
\left[\alpha^{2(s-k)}\zeta(s-k|\L_1)\right]
\:.\label{ZFSN-odd}\end{eqnarray}

A similar equation can be obtained also in the even dimensional case
with $N\geq4$. By taking into account that now $\varrho_N$ is a
half-integer one obtains \begin{eqnarray}
\zeta(s|\L_N)&+&\Omega_N\sum_{n=0}^{[\varrho_N]-1}
\sum_{k=0}^{[\varrho_N]}a_{2k+1}^N(n+\frac12)^{2k+1}
[(n+\frac12)^2+\alpha^2)]^{-s} \nonumber\\
&=&\left.\Omega_N\sum_{n=1}^{\infty}
\sum_{k=0}^{[\varrho_N]}\frac{(-1)^ka_{2k+1}^N(n+1/2)}{\Gamma(s)}
\frac{d^k}{d\lambda^k}\int_{0}^{\infty} t^{s-k-1}e^{-\lambda
t[(n+1/2)^2+\alpha^2/\lambda]}\,dt\,\right|_{\lambda=1} \nonumber\\
&=&\Omega_N\sum_{k=0}^{[\varrho_N]}
\frac{a_{2k+1}^N\Gamma(s-k)\alpha^{-2s}}{2\Gamma(s)}
\left(\alpha^4\frac{d\phantom{\alpha^2}}{d\alpha^2}\right)^k
\left[\alpha^{2(s-k)}\zeta(s-k|\L_2)\right]
\:,\label{ZFSN-even}\end{eqnarray} where $[\varrho_N]$ represents the
integer part of $\varrho_N$. Eqs.~(\ref{ZFSN-odd}) and
(\ref{ZFSN-even}) are valid in the absence of zero-modes. Zero-modes
must be not consider in the definition of $\zeta$-function and this is
equivalent to disregard (possible) singularities in the latter
equations.

We know that the knowledge of $\zeta$-function on $S^1$ and $S^2$ is
sufficient in order to get the $\zeta$-function on any sphere. The
results for $S^1$, $S^2$ and $S^3$ are derived in Appendix
\ref{S:ZF-ExComp}.

\subsubsection{A complex integral representation}

To finish the section, we exhibit an integral representation and some
recurrence relations for $\zeta$-function, which look like to the ones
we shall obtain for the hyperbolic case in Sec.~\ref{S:HKZFHM}.

First of all, we observe that \begin{eqnarray}
c_n^{N+2}=\frac{(n+1)(n+N)}{2\pi N}\,c_{n+1}^N
\:,\nonumber\end{eqnarray} from which we easily get the recurrence
relations \begin{eqnarray} \frac{K(t|\L_{N+2})}{\Omega_{N+2}}=
-\frac1{2\pi N\Omega_N}
\left[\partial_t+\alpha^2+\kappa\varrho_N^2\right]\,K(t|\L_N)
\:,\label{RRK-SN}\end{eqnarray} \begin{eqnarray}
\frac{\zeta(s|\L_{N+2})}{\Omega_{N+2}}= -\frac{1}{2\pi
N\Omega_N}\left[(\alpha^2+\kappa\varrho_N^2)\zeta(s|\L_N)
-\zeta(s-1|\L_N)\right] \:.\label{ZFRR}\end{eqnarray} It is easy to
see that Eq.~(\ref{ZFRR}) is also valid if we substitute the
corresponding regularized quantities according to Eq.~(\ref{ZF-r}). As
we shall see if Sec.~\ref{S:HKZFHM} with small changes,
Eqs.~(\ref{RRK-SN}) and (\ref{ZFRR}) are also valid on $H^N$. Knowing
the $\zeta$-function on $S^1$ and $S^2$ and using recurrence formulae
above, we obtain the $\zeta$-function on the sphere in any dimension.

In order to get the integral representation in the complex plane, we
consider an analytic function $h(z^2)$ such that $\sum
h(\lambda_n+\varrho_N^2)$ exists, the sum being extended to all
eigenvalues (counted with their multiplicity) of $-\Delta_N$ on the
sphere $S^N$. Then one can easily check that for any $N>1$ the
following complex integral representation holds: \begin{eqnarray}
\sum_{n}h(\lambda_n+\varrho_N^2)=\frac{\Omega_N}{2\pi i}\int_{\Gamma}
h(z^2)\Phi_N^S(z)\,dz \:,\label{TFSN}\end{eqnarray} where $\Gamma$ is
an open path in the complex plane going (clockwise) from $\infty$ to
$\infty$ around the positive real axix enclosing the point
$z=\varrho_N$ and \begin{eqnarray}
\Phi_N^S(z)=\frac{2z\Gamma(\varrho_N+z)\Gamma(\varrho_N-z)}
{(4\pi)^{N/2}\Gamma(N/2)} \cos\pi(\varrho_N-z)
\:,\nonumber\end{eqnarray} which satisfy the recurrence formula
\begin{eqnarray} \Phi_{N+2}^S(z)=\frac{\varrho_N^2-z^2}{2\pi
N}\Phi_N^S(z) \:.\nonumber\end{eqnarray} We have \begin{eqnarray}
\Phi_2^S(z)=\frac{z\tan\pi z}2\,,\qquad\qquad
\Phi_3^S(z)=-\frac{z^2\cot\pi z}{2\pi}\, \:.\nonumber\end{eqnarray}

Using Eq.~(\ref{TFSN}) we obtain for example \begin{eqnarray}
K(t|\L_2)=\frac{\Omega_2e^{-t\alpha^2}}{8it}
\int_{\Gamma}\frac{e^{-tz^2}}{\cos^2\pi z}\,dz
\:,\label{Kts2}\end{eqnarray} \begin{eqnarray}
\zeta(s|\L_2)=\frac{\Omega_2}{8i(s-1)}
\int_{\Gamma}\frac{(z^2+\alpha^2)^{-(s-1)}}{\cos^2\pi z}\,dz
\:.\label{ZFs2}\end{eqnarray} The latter equation is valid for any $s$
if we choose a suitable path, for example $z=a+re^{\pm i\pi/4}$, $a$
being an arbitrary real number satisfying $|a|<1/2$, $|\alpha|<a<1/2$,
$1/2<a<3/2$ according to whether $\alpha^2>0$, $0\geq\alpha^2>-1/4$,
$\alpha^2=-1/4$.

\subsection{Hyperbolic manifolds} \label{H:HM} The geometry of
$N$-dimensional torus and sphere are quite known. On the contrary, the
hyperbolic one is less familiar. For this reason, in the following two
subsections, we shall present an elementary and self-contained survey
of some issues on hyperbolic geometry. For further details, we refer
to the cited references.

By definition, hyperbolic manifolds are the Riemannian space forms
with constant negative curvature. Actually, they can be defined for
any metric signature in which case they are called pseudo-Riemannian
hyperbolic space forms. The Riemann tensor of such spaces is locally
characterized by the condition
$R_{ijkl}=\kappa(g_{ik}g_{jl}-g_{il}g_{jk})$ with $\kappa$ a constant.
We denote the signature of a non degenerate metric $g_{ij}$ by
$(n,N-n)$, where $n$ is the number of negative eigenvalues of $g_{ij}$
($i,j=0,...,N-1$). The unique simply connected and flat
pseudo-Riemannian manifold of signature $(n,N-n)$ is just
$\mbox{$I\!\!R$}^N$ endowed with the standard diagonal metric with the
given signature. The following theorem is quoted from \cite{wolf77b}:
\begin{Theorem} Let $\kappa$ be non zero and $a>0$ defined by
$\kappa=ea^{-2}$, $e=\pm1$ and \begin{eqnarray}
\Sigma_n^N=\left\{x\in\mbox{$I\!\!R$}^N:g_{ij}x^ix^j=ea^2\right\},
\qquad\qquad N\geq3\nonumber \:,\end{eqnarray} where $g_{ij}$ has
signature $(n,N-n)$ and $x^i=(x^0,...,x^{N-1})$ are rectilinear
coordinates on $\mbox{$I\!\!R$}^N$. Then every $\Sigma_n^N$, endowed
with the induced metric, is a complete pseudo-Riemannian manifold of
constant curvature $\kappa$ and signature $(n,N-n-1)$ if $e=1$ or
$(n-1,N-n)$ if $e=-1$. The geodesics of $\Sigma_n^N$ are the
intersections $\Pi\cap \Sigma_n^N$, $\Pi$ being a plane through the
origin in $\mbox{$I\!\!R$}^N$. The group of all isometries of
$\Sigma_n^N$ is the pseudo-orthogonal group $O(n,N-n)$ of the metric
$g_{ij}$. \end{Theorem} The class of spaces described by the theorem
are the fundamental models for constant curvature manifolds. The
models with $\kappa>0$ are the pseudo-spherical space forms, denoted
by $S_n^N$, the models with $\kappa<0$ are called pseudo-hyperbolic
space forms, denoted by $H_n^N$. It may be noted that
$\Sigma_N^N=\emptyset$ if $e=1$ and likewise $\Sigma_0^N=\emptyset$ if
$e=-1$. Thus we define $S_n^N=\Sigma_{n}^{N+1}$ and
$H_n^N=\Sigma_{n+1}^{N+1}$, so for both of them the signature is
$(n,N-n)$, the dimension is $N$ and no one is the empty set. From the
definition it follows that $H_n^N$ is simply connected for $n\neq0,1$.
On the other hand, $H_1^N$ is connected with infinite cyclic
fundamental group. From the theorem we see it has Lorentz signature
$(1,N-1)$ and contains closed time-like geodesics. In fact, $H_{1}^N$
is what in general relativity is called the $N$-dimensional anti-de
Sitter space-time (similarly, the space $S_1^N$ is de Sitter
space-time) \cite{hawk73b}. Finally we have $H_{0}^N$ which has two
simply connected, isometric components. We denote the component with
$x_0>a$ by $H^N$. This is the manifold we are mainly interested in. It
can be characterized as the unique simply connected Riemannian
manifold with constant negative curvature. Its isometry group is the
orthochronus Lorentz group $O^+(1,N)$ of matrices $\Lambda_{ab}$ such
that $\Lambda_{00}>0$, in order to preserve the condition $x_0>a$. An
interesting feature of this space is that it represents the Euclidean
section appropriate for anti-de Sitter space-time
\cite{dowk76-13-224}. Accordingly, it can be shown that Euclidean
quantum field theory on $H^N$ is the Wick rotation of a field theory
on the anti-de Sitter, although on this space-time there are other
field representations which cannot be obtained in this way.

Next, we describe all the connected spaces with constant negative
curvature and arbitrary signature. They are obtained from the spaces
$H_n^N$ by the operation of taking the quotient with respect to the
action of a group  of isometries. But of course the group action has
to be restricted in some way. For example, fixed points are going to
produce singularities in the quotient manifold, as can be seen with
the following example. Let $\Gamma$ be the cyclic group generated by a
rotation around the origin in $\mbox{$I\!\!R$}^2$ with angle
$2\pi/\alpha$, i.e. the set of all rotations whose angle is an integer
multiple of $2\pi/\alpha$. The quotient manifold (i.e. the set of
orbits) $\mbox{$I\!\!R$}^2/\Gamma$ is then a cone with angle
$2\pi/\alpha$ at the vertex, which is also the fixed point of
$\Gamma$. Note that the cone is a metric space but the infinitesimal
metric is singular at the vertex.

When a group $\Gamma$ acts on a space without fixed points, the group
is said to act freely. When every point has a neighborhood $V$ such
that $\{\gamma\in\Gamma:\gamma(V)\cap V\neq\emptyset\}$ is finite, the
group is said to act properly discontinuously (see, for example,
Ref.~\cite{bear83b}). These are the restrictions on the group action
which permit to avoid conical singularities and other pathologies in
the quotient manifold.

Let us denote by $\tilde{H}_{n}^N$ the universal covering space of
$H_{n}^{N}$. Then $\tilde{H}^N=H^N$ and $\tilde{H}_n^N=H_n^N$ for
$n\neq0,1$, since these spaces are already simply connected. Now we
have \cite{wolf77b}

\begin{Theorem} Let $M_n^N$ be a complete, connected pseudo-Riemannian
manifold of constant negative curvature $\kappa$, with signature
$(n,N-n)$ and dimension $N\geq2$. Then $M_n^N$ is isometric to a
quotient $\tilde{H}_n^N/\Gamma$, where $\Gamma$ is a group of
isometries acting freely and properly discontinuously on
$\tilde{H}_n^N$. \end{Theorem} The Riemannian case is due to Killing
and Hopf. In this case $\Gamma$ is a discrete subgroup of $O^+(1,N)$,
namely the elements of $\Gamma$ can be parametrized by a discrete
label (more exactly, the relative topology of $\Gamma$ in $O^+(1,N)$
is the discrete topology). Hence the "Clifford-Klein space form
problem", to classify all manifolds with constant negative curvature,
is reduced to the problem of finding all the discrete subgroups of
$O^+(1,N)$ acting freely and properly discontinuously on $H^N$.

The model spaces just defined have the further property to be
homogeneous. By definition, this means that for every pair of points
$x,y$ there is an isometry which moves $x$ into $y$. As a consequence
they are complete, i.e. there are no geodesics ending or beginning at
any point. In this connection, a remarkable fact concerning the
hyperbolic space $H^N$ is that it is the only connected and homogeneous
Riemannian manifold with constant negative curvature \cite{wolf77b}.
It follows that every quotient $H^N/\Gamma$, with $\Gamma\neq e$,
cannot be a homogeneous space and thus it has less symmetry than the
covering $H^N$. This fact restrict somewhat their use in cosmological
models \cite{elli71-2-7}.

Up to now we have described the global properties of hyperbolic
manifolds. In order to describe explicitly the metric tensor of $H^N$
we introduce some known model of this space.

\paragraph{The hyperboloid model.} Let $x_0,...,x_N$ on
$\mbox{$I\!\!R$}^{N+1}$  be the standard coordinates with metric
$ds^2=-dx_{0}^{2}+\sum_{i=1}^N\,dx_i^2$ such that the space $H^N$ is
the upper sheet of the $N$-dimensional hyperboloid
$x_{0}^{2}-\sum_{i=1}^N\,x_i^2=a^2$, with the induced metric. Then
$\sigma\in[0,\infty)$ and $\vec{n}\in S^{N-1}$, such that
\begin{eqnarray} x_0=a\cosh\sigma,
\qquad\qquad\vec{x}=a\vec{n}\sinh\sigma, \nonumber\end{eqnarray} will
define a global system of coordinates for this model of $H^N$. A
direct calculation leads to the metric tensor and volume element in
the form \begin{eqnarray} ds^2=a^2[d\sigma^2+\sinh^2\sigma
d\ell_{N-1}^{2}] \:,\label{mhyp} \end{eqnarray} \begin{eqnarray}
dV=a^{N-1}(\sinh\sigma)^{N-1}d\sigma d\Omega_{N-1}
\:,\nonumber\end{eqnarray} where $d\ell_{N-1}$ and $d\Omega_{N-1}$ are
the line and volume element of $S^{N-1}$ respectively. This metric is
of some interest, since the coordinate $\sigma$ measures the geodesic
distance of any point from the bottom of the hyperboloid, the point
$x_0=a$. It is also obvious that the group which fixes this point is
the group $O(N)$ of rotations around the $x_0$-axis. Thus we can
regard $H^N$ as the coset space $O^+(1,N)/O(N)$.

\paragraph{The cylinder model.} Assuming $\rho\in[0,\pi/2)$ and
$\tau\in(-\infty,+\infty)$, the global metric tensor in this model is
\begin{eqnarray} ds^2=a^2(\cos\rho)^{-2}[d\tau^2+d\rho^2 +\sin^2\rho
d\ell_{N-2}^{2}] \:,\label{mstrip} \end{eqnarray} where $d\ell_{N-2}$
is the line element of $S^{N-2}$. This metric is conformal to half the
cylinder $\mbox{$I\!\!R$}\times S^{N-1}$, with the product metric. In
2-dimensions, this is the strip $-\pi/2<\rho <\pi/2$, $-\infty
<\tau<\infty$ with metric \cite{bala86-143-109} \begin{eqnarray}
ds^2=a^2(\cos\rho)^{-2} [d\tau^2+d\rho^2] \:,\nonumber\end{eqnarray}
which still provides another model for $H^2$. The coordinate $\tau$
can be directly interpreted as imaginary time, because under Wick
rotation $\tau\rightarrow -it$, the metric (\ref{mstrip}) goes over to
the anti-de Sitter metric, which is in fact conformal to a region of
the Einstein static universe.

\paragraph{The ball model.} We shall seldom make use of this model but
we give it for completeness. We denote by $B^N$ the ball in
$\mbox{$I\!\!R$}^N$ of radius $2a$. We set \begin{eqnarray}
u=\frac{2a\sinh \sigma}{\cosh \sigma+1} \:,\qquad\qquad
u^2<4a^2\:.\nonumber\end{eqnarray} In terms of $u$, the metric tensor
Eq.~(\ref{mhyp}), takes the conformally flat form \begin{eqnarray}
ds^2=\left(1-\frac{u^2}{4a^2}\right)^{-2} [du^2+u^2d\ell^{2}_{N-1}]
\:.\nonumber\end{eqnarray} The discovery of this metric and the
observation that it has constant curvature, was one of Riemann great
contributions. The geodesics are the circles or lines that meet the
boundary of the disc orthogonally.

\paragraph{The Poincar\'e half-space model.} There is a conformal map
from the ball $B^N$ of radius $2a$ in $\mbox{$I\!\!R$}^N$ onto the
half space $x_N>0$ \cite{bear83b}. For convenience, we rename $x_N=r$.
Poincar\'e used this map to transform the ball metric into the
half-space metric \begin{eqnarray}
ds^2=\frac{a^2}{r^2}(dx_{1}^{2}+...+dx_{N-1}^{2}+dr^2)
\:.\nonumber\end{eqnarray} The distance $d(X,Y)$ between two points
$X=(x^1,...,x^{N-1},r)$ and $Y=(y^1,...,y^{N-1},s)$ in this model is
implicitly given by \begin{eqnarray}
\cosh\left[\frac{d(X,Y)}{a}\right]-1=\frac{|X-Y|^2}{2rs}
\:,\label{dist} \end{eqnarray} where $|X-Y|$ is the Euclidean
distance. The formula is easily found, first for $X=(0,...,0,r)$ and
$Y=(0,...,0,s)$ and then for any pair of points, using the
transitivity of isometries, since both sides of Eq.~(\ref{dist}) are
invariants. The geodesics of this metric are the circles or straight
lines that meet the boundary $r=0$ orthogonally. It is geometrically
obvious that for every pair of points there is a unique minimizing
geodesic connecting them.

This latter is the model we  shall employ in discussing the Selberg
trace formula. Of course, the curvature constant $\kappa=-1/a^2$ in
all the four models.

\subsubsection{The Laplace operator and the density of states}
\label{S:LDO}

In the present subsection, we normalize the curvature to $\kappa=-1$
for the sake of simplicity. The first step towards the  Selberg trace
formula is to compute the spectral decomposition of the Laplace
operator on $H^N$. This is equivalent to find the density of states,
since we expect this operator to have a continuous spectrum. This
density gives the simplest kind of trace formula, like \begin{eqnarray}
\,\mbox{Tr}\,(h(\Delta))=\int_{S}h(r)\Phi_N(r)dr
\:,\nonumber\end{eqnarray} where the density $\Phi_N(r)$ is defined
over the spectrum $S$ of $\Delta$ and for functions $h(r)$ for which
the trace and the integral exist. Here it is convenient to use the
hyperboloid model of $H^N$. A simple computation with the metric
(\ref{mhyp}) gives the Laplace operator \begin{equation}
\Delta=\frac{\partial^{2}}{\partial \sigma^{2}}+(N-1)
\coth\sigma\frac{\partial}{\partial\sigma}+(\sinh\sigma)^{-2}\Delta_{S^{N-1}}
\:,\nonumber\end{equation} where the last term denotes the Laplacian
on the unit sphere $S^{N-1}$. The eigenvalues equation is
$-\Delta\phi=\lambda\phi$. Note that if we replace $\lambda$ with
$\lambda+m^2$ we can also include the massive operator in our
calculations. Clearly the eigenfunctions have the form
$\phi=f_{\lambda}(\sigma)Y_{lm}$ where $Y_{lm}$ are the spherical
harmonics on $S^{N-1}$ and for simplicity we leave understood the
dependence of $f_{\lambda}$ on $l$, $l=0,1,2...$ and
$m=(m_1,...,m_{N-2})$ being the set of angular momentum quantum
numbers. For these harmonics we have \begin{eqnarray}
\Delta_{S^{N-1}}Y_{lm}=-l(l+N-2)Y_{lm} \:.\nonumber\end{eqnarray}
Thus, the radial wave functions will satisfy the ordinary differential
equation \begin{equation} f_{\lambda}^{''}+(N-1)\coth\sigma
f_{\lambda}^{\prime}
+\left[\lambda-\frac{l(l+N-2)^{2}}{\sinh^{2}\sigma}\right]f_{\lambda}=0
\:.\nonumber\end{equation} If the zero angular momentum wave functions
are normalized so that $f_{\lambda}(0)=1$, their density is known to
mathematicians as the Harish-Chandra or Plancherel measure.

We continue by transforming the radial wave equation into a known form.
On setting $\rho_N=(N-1)/2$, $\mu_{l}=1-l-N/2$,
$r=\sqrt{(\lambda-\rho_{N}^{2})}$, defining $v_{\lambda}(\sigma)$ by
\begin{equation}
f_{\lambda}(\sigma)=(\sinh\sigma)^{1-N/2}v_{\lambda}(\sigma)
\:,\nonumber\end{equation} and performing the change of variable
$x=\cosh\sigma$, we get  the differential equation of associated
Legendre functions \begin{equation}
\frac{d}{dx}\left[(1-x^{2})\frac{dv_{\lambda}}{dx}\right]
+\left[\nu(\nu+1)-\frac{\mu^{2}_{l}}{1-x^{2}}\right]v_{\lambda}=0
\:,\label{legen} \end{equation} with parameters $\nu=-1/2\pm ir$. The
invariant measure defining the scalar product between eigenfunctions is
\begin{equation}
(f_{\lambda},f_{\lambda'})=\Omega_{N-1}\int_{0}^{\infty}
f_{\lambda}^{*}f_{\lambda'}(\sinh\sigma)^{N-1}d\sigma
=\Omega_{N-1}\int_{1}^{\infty}v_{\lambda}^{*}v_{\lambda'}dx
\:,\nonumber\end{equation} where $\Omega_{N-1}$ is the volume of the
$N-1$ dimensional sphere.

The only bounded solutions of Eq.~(\ref{legen}) are the associated
Legendre functions of the first kind and denoted by $P_{\nu}^{\mu}(x)$
\cite{grad80b}. Hence, the radial wave functions are \begin{equation}
f_{\lambda}(\sigma)=\Gamma\left(\frac{N}{2}\right)
\left(\frac{\sinh\sigma}2\right)^{1-N/2} P_{-1/2+ir}^{\mu}(\cosh\sigma)
\:,\nonumber\end{equation} where the multiplicative constant has been
chosen in order to satisfy the normalization condition
$f_{\lambda}(0)=1$.

The density of states is determined from the knowledge of the radial
wave functions as follows. One chooses a variable $r$ parametrizing
the continuum spectrum so that $\lambda=\lambda(r)$ and computes the
scalar product \begin{equation}
(\phi_{\lambda},\phi_{\lambda^{\prime}})
=\frac1{\Phi_N(r)}\delta(r-r^{\prime}) \label{philala'}\,,
\end{equation} which defines the density of states $\Phi_N(r)$. Let us
choose $r=(\lambda-\rho_{N}^{2})^{1/2}$ as a label for the continuum
states. The Legendre functions are given in terms of hypergeometric
functions. For $\,\mbox{Re}\, z>1$ and $\mid 1-z\mid<2$ they are
\begin{equation} P_{\nu}^{\mu}(z)=\frac{1}{\Gamma(1-\mu)}
\left(\frac{z+1}{z-1}\right)^{\mu/2}F(-\nu,\nu+1;1-\mu;(1-z)/2)
\:.\nonumber\end{equation} These are analytic throughout the complex
plane, with a cut along the real axis from $-\infty$ up to 1. The
asymptotic behaviour for $\mid z\mid \gg 1$ is \begin{equation}
P_{\nu}^{\mu}(z)\approx
\frac{2^{\nu}\Gamma(\nu+1/2)}{\pi^{1/2}\Gamma(\nu-\mu+1)}
z^{\nu}+\frac{\Gamma(-\nu-1/2)}{2^{\nu+1}\pi^{1/2}
\Gamma(-\nu-\mu)}z^{-\nu-1} \:,\nonumber\end{equation} from which we
obtain the asymptotic behaviour of the eigenfunctions \begin{eqnarray}
f_{\lambda}(\sigma)\simeq\frac{2^N\Gamma(N/2)\Gamma(ir)}{4\pi^{1/2}\Gamma(\rho_N+ir)}
e^{-\rho_N\sigma+ir\sigma}+c.c. \:.\nonumber\end{eqnarray} As a
result, the radial functions will remain bounded at infinity provided
the parameter $r$ were real, which is equivalent to the condition
$\lambda\geq\rho_{N}^{2}$. Thus, the spectrum of the Laplacian has a
gap which is determined by the curvature and depends on $N$, although
the excitations of the corresponding wave operator on
$\mbox{$I\!\!R$}\times H^N$ still propagate on the light cone. Due to
this gap, the Green functions of $\Delta$ on $H^N$ are exponentially
decreasing at infinity. As a consequence, negative constant curvature
provides a natural infrared cut-off \cite{call90-340-366} for
interacting boson field theories.

Now we come to compute the scalar product between two radial
eigenfunctions. Using the fact that the product of two eigenfunctions
is the derivative of the their Wronskian $W[\cdot,\cdot]$, we get
\begin{equation} (f_{\lambda},f_{\lambda^{\prime}})=
\frac{2^{N-1}\pi^{\frac N2}\Gamma(\frac N2)}
{\nu^{*}(\nu^{*}+1)-\nu'(\nu'+1)} \lim_{x\rightarrow \infty}(1-x^{2})
W[P_{\nu^{*}}^{\mu}(x),P_{\nu^{\prime}}^{\mu}(x)]
\:,\label{flala'}\end{equation} where the limit is taken in the sense
of distributions. Now we can use the asymptotic form of the Legendre
functions to compute the limit. In this way Eq.~(\ref{flala'}) reduces
to Eq.~(\ref{philala'}) with the density of states given by
\begin{equation} \Phi_N(r)=\frac{2}{(4\pi)^{N/2}\Gamma(N/2)}\frac{\mid
\Gamma(ir+l+\rho_N)\mid^{2}}{\mid \Gamma(ir)\mid^{2}} \:.\label{harish}
\end{equation} When $l=0$ we have the density of zero angular momentum
radial functions and Eq.~(\ref{harish}) reduces to the Harish-Chandra
measure.

We stress that $dn=\Phi_N(r)dr$ is the number of states per unit
volume in the range $dr$. Hence the scalar $\zeta$-function per unit
volume reads \begin{eqnarray} \tilde\zeta(s|L_N)=a^{2s-N}
\int_0^{\infty}[r^2+(a\alpha)^2]^{-s}\Phi_N(r)dr
\:.\nonumber\end{eqnarray} It is independent from $X$ because $H^N$ is
homogeneous. As in Sec.~\ref{S:ZFRR},
$L_N=-\Delta_N+\alpha^2+\kappa\varrho_N^2$, but now $\alpha\geq0$ and
$\kappa<0$ ($a^2=|\kappa|^{-1})$. The particular cases $N=3$ and $N=4$
deserve some attention. For these two cases, we have respectively
\begin{eqnarray} \tilde\zeta(s|L_3)=\frac{\alpha^{3-2s}}{(4\pi)^{3/2}}
\frac{\Gamma(s-3/2)}{\Gamma(s)} \:,\nonumber\end{eqnarray}
\begin{eqnarray} \tilde\zeta(s|L_4)&=&\frac1{16\pi^2}
\left[\frac{\alpha^{4-2s}}{(s-1)(s-2)}
+\frac{\alpha^{2-2s}}{4a^2(s-1)}\right]\nonumber\\
&&\qquad\qquad-\frac{a^{2s-4}}{4\pi^2}
\int_0^\infty[r^2+(a\alpha)^2]^{-s} \frac{r(r^2+\frac{1}{4})}{e^{2\pi
r}+1}dr \:,\nonumber\end{eqnarray} where the meromorphic structure can
be immediately read off. The quantity $\zeta'(0|L_4)$ gives the
one-loop functional determinant for a scalar field on anti-de Sitter
space-time, while $\zeta(s|L_4)$ itself is related to the free energy
density on such a space-time.

\subsubsection{The Dirac operator and the density of states}
\label{S:DDO} Here we consider the Dirac-like equations on $H^N$. Let
$N$ be even. The $N$-dimensional Clifford algebra has only one
complex, irreducible representation in the $2^{N/2}$-dimensional
spinor space. These spinors are reducible with respect to the even
subalgebra (generated by products of an even number of Dirac matrices)
and split in a pair of $2^{N/2-1}$-component irreducible Weyl spinors.

The Dirac eigenvalue equation is \begin{eqnarray}
i\not\!\nabla\psi-m\gamma_0\psi=-\omega\psi \nonumber\:,\end{eqnarray}
where $m$ is a mass parameter and $i\not\!\nabla$ is the Dirac
operator on $H^N$. It can be regarded as the full time-dependent Dirac
equation on $\mbox{$I\!\!R$}\times H^N$ restricted on time-harmonic
fields. When we use the hyperboloid model of $H^{N}$, the submanifolds
with $r$ a constant are a family of spheres $S^{N-1}$ covering the
space. The covariant derivative of a spinor field on $H^{N}$ can be
decomposed into a radial part plus the covariant derivative along the
unit $(N-1)$-sphere. Making this decomposition in a Dirac-like
representation of gamma matrices, the equation takes the form of a
coupled system \begin{equation}
i\gamma_{1}\left(\partial_{\sigma}+\rho_N
\coth\sigma\right)\psi_{1}+\frac{1}{\sinh\sigma}i\not\!\nabla_s\psi_{1}
=-(\omega+m)\psi_{2} \:,\label{dirac1} \end{equation} \begin{equation}
i\gamma_{1}\left(\partial_{\sigma}+\rho_N
\coth\sigma\right)\psi_{2}+\frac{1}{\sinh\sigma}i\not\!\nabla_s\psi_{2}
=-(\omega-m)\psi_{1} \:,\label{dirac2} \end{equation} where
$\psi_{1,2}$ are the $2^{N/2-1}$-components Weyl spinors into which
the original representation decomposes and $i\not\!\nabla_s$ is the
Dirac operator on $S^{N-1}$. The spinors $\psi_{1,2}$ transform
irreducibly under $SO(N)$ so that we can put
$\psi_{1,2}=f_{1,2}(\sigma)\chi_{1,2}$, where $\chi_{1,2}$ are spinors
on $S^{N-1}$. Separation of variables is achieved on requiring
$i\not\!\nabla_s\chi_{1}=i\lambda\chi_{2}$ and
$\gamma_{1}\chi_{1}=\chi_{2}$. Since $\gamma_{1}\gamma_{1}=1$ and
$\{\gamma_{1},i\not\!\nabla_s\}=0$ we also have
$i\not\!\nabla_s\chi_{2}=-i\lambda\chi_{1}$ and
$\gamma_{1}\chi_{2}=\chi_{1}$.

The eigenvalues of the Dirac operator on $S^{N-1}$ are known to be
$\lambda=\pm(l+\rho_N)$, $l=0,1,2,...$ \cite{cand84-237-397}. Hence,
we obtain two radial equations which are equivalent to the
$2^{nd}$-order system \begin{eqnarray}
\frac{d}{dz}\left[(z^{2}-1)\frac{d\phi_1}{dz}\right]+\left[r^{2}
+\rho_{N}^{2}-\frac{N}{2}-\frac{(\lambda^{2}-\rho_{N}^{2}+\rho_N)
+\lambda z+(\rho_N-\frac{1}{2})^2z^{2}}{z^{2}-1}\right]\phi_1=0
\:,\label{sys1}\end{eqnarray} \begin{eqnarray}
\frac{d}{dz}\left[(z^{2}-1)\frac{d\phi_2}{dz}\right]+\left[r^{2}
+\rho_{N}^{2}-\frac{N}{2}-\frac{(\lambda^{2}-\rho_{N}^{2}+\rho_N)
-\lambda z+(\rho_N-\frac{1}{2})^2z^{2}}{z^{2}-1}\right]\phi_2=0
\:,\label{sys2} \end{eqnarray} where we have defined $z=\cosh\sigma$,
$\phi_{1,2}=(\sinh\sigma)^{N/2-1}f_{1,2}(\sigma)$ and
$r^2=\omega^2-m^2$.

For odd $N$, there are two $2^{(N-1)/2}$-components irreducible
spinors. Since we are regarding the Dirac equation on $H^N$ as the
restriction of the Dirac equation on $\mbox{$I\!\!R$}\times H^N$ on
time-harmonic fields, we must take into account both spinors. We can
say that they are the Weyl spinors of the unphysical space-time
$\mbox{$I\!\!R$}\times H^N$. Both belong to the same irreducible
representation of $SO(N)$, thus we can separate variables as above and
we obtain exactly the same equations as (\ref{dirac1}), (\ref{dirac2})
or (\ref{sys1}) and (\ref{sys2}). Of course, there is nothing wrong to
consider the Dirac equation directly on $H^N$ without thinking about
it as the spatial section of an unphysical space-time. However, for
$N=3$ we do have a physical space-time. Alternatively, one may
consider the spinor Laplacian as the relevant operator. This point of
view is taken in Ref.~\cite{camp92-148-283}. We can set $m=0$, then
the equations decouples into pairs of equivalent Dirac equations on
$H^N$ and we should recover the results of Ref.~\cite{camp92-148-283}.

The solutions of Eqs.~(\ref{sys1}) and (\ref{sys2}) are given in terms
of hypergeometric functions as follows. Let $f_{1,2}^{\pm}(r)$ the
solutions with $\lambda =\pm(l+\rho_N)$ and set $\alpha=l+N/2+ir$. We
find \begin{eqnarray}
f_{1}^{+}(\sigma)&=&A(1+z)^{l/2}(z-1)^{(l+1)/2}F\left(\alpha,\alpha^*;
l+\rho_N+\frac{3}{2};\frac{1-z}{2}\right)\:,\nonumber\\
f_{2}^{+}(\sigma)&=&B(1+z)^{(l+1)/2}(z-1)^{l/2}F\left(\alpha,\alpha^*;l+\rho_N+
\frac{1}{2};\frac{1-z}{2}\right)\:,\nonumber\\
f^{-}_{1}(\sigma)&=&C(1+z)^{(l+1)/2}(z-1)^{l/2}F\left(\alpha,\alpha^*;l+\rho_N+
\frac{1}{2};\frac{1-z}{2}\right)\:,\nonumber\\
f^{-}_{2}(\sigma)&=&D(1+z)^{l/2}(z-1)^{(l+1)/2}F\left(\alpha,\alpha^*;
l+\rho_N+\frac{3}{2};\frac{1-z}{2}\right) \:.\nonumber\end{eqnarray}
The ratios $A/B=i(\omega+m)(l+N/2)^{-1}$ and
$D/C=i(\omega-m)(l+N/2)^{-1}$ are determined by demanding that the
solutions of the second order system satisfy the first order Dirac
equation. Next we choose $B=(4\omega)^{-1/2}(\omega-m)^{1/2}$ and
$C=(4\omega)^{-1/2}(\omega+m)^{1/2}$. Then the asymptotic behaviour
for $r\rightarrow\infty$ of the radial solutions takes the symmetric
form \begin{eqnarray} f_{1}^{\pm}(\sigma)&\simeq
&\left(\frac{\omega+m}{2\omega}\right)^{1/2} {\cal C}(r)
e^{-\rho_N\sigma+ir\sigma}+c.c.\:,\nonumber\\
f_{2}^{\pm}(\sigma)&\simeq
&\left(\frac{\omega-m}{2\omega}\right)^{1/2} {\cal C}(r)
e^{-\rho_N\sigma+ir\sigma}+c.c. \:,\nonumber\end{eqnarray} where we
introduced the meromorphic function \begin{eqnarray}  {\cal
C}(r)=\frac{2^{N-1+l-2ir}\Gamma(l+N/2)\Gamma(2ir)}{\Gamma(N/2+l+ir)
\Gamma(ir)} \:.\label{meas} \end{eqnarray} One can see that the
solutions remain bounded at infinity if $r$ is real. Hence, the
spectrum of the Dirac operator on $H^N$ is $|\omega|\geq m$. For $m=0$
it extends over the entire real axis. Thus, we reach the conclusion
that unlike the scalar case, there is no gap for fermions on $H^N$.
Nevertheless, the solutions are exponentially vanishing at infinity.

The function (\ref{meas}) for $l=0$ determines the Plancherel measure
or the density of states with zero angular momentum, which has to be
used in the evaluation of the spinor $\zeta$-function. We define the
density $\mu_N(r)$ so that the spinor $\zeta$-function per unit volume
is given by \begin{eqnarray} \tilde\zeta(s|D_N)
=:\,\mbox{tr}\,(-\not\!\nabla^2+m^2)^{-s}(X,X)=2^{[\frac{N}{2}]}\int_0^\infty
(r^2+m^2)^{-s}\mu_N(r)dr\,. \nonumber\end{eqnarray} Here $D_N$
represents the massive Dirac operator on $H^N$, $[N/2]$ denotes the
integer part of $N/2$ and the trace is over the spinor indices. Again,
the $X$-independence comes from the homogeneity of $H^N$.

Assuming that the spherical spinors are already normalized and using
$r$ as a label for the continuous spectrum, we obtain the density
\begin{eqnarray} \mu_N(r)=\frac{\Gamma(N/2)2^{N-3}}{\pi^{N/2+1}}|
{\cal C}(r)|^{-2} \:,\nonumber\end{eqnarray} which, apart from a
slightly different definition of the measure,  agrees with results
given in Ref.~\cite{camp92-148-283}, where it has been also noted that
in the even case, the poles and residues of the density determine the
spectrum and degeneracies of the spinor Laplacian on the sphere $S^N$.
For odd $N$ the density is analytic.

In 3-dimensions the $\zeta$-function density becomes \begin{eqnarray}
\tilde\zeta(s|D_3)= \frac{2m^{3-2s}}{(4\pi)^{3/2}}
\frac{\Gamma(s-3/2)}{\Gamma(s)} \left(1+\frac{2s-3}{(2am)^2}\right)
\:.\nonumber\end{eqnarray} Note that the zero mass limit is not
uniform with respect to $s$, unlike the scalar case. The reason is the
absence of the gap in the spectrum for fermions. In 4-dimensions it
takes the form \begin{eqnarray} \tilde\zeta(s|D_4)
&=&\frac1{4\pi^2}\left[\frac{m^{4-2s}}{(s-1)(s-2)}
+\frac{m^{2-2s}}{a^2(s-1)}\right]\nonumber\\
&&\qquad\qquad+\frac{a^{2s-4}}{\pi^2}
\int_0^\infty[r^2+(am)^2]^{-s}\frac{r(r^2+1)} {e^{2\pi r}-1} \,dr
\:.\nonumber\end{eqnarray} Again, this function determines the
one-loop spinor determinant on anti-de Sitter space-time
\cite{camp92-45-3591} as well as the fermion free energy density.

\subsection{Compact hyperbolic manifolds} \label{S:CHM}

We have seen that every complete, connected hyperbolic Riemannian
manifold $ {\cal M}$ is a quotient of $H^N$ by a discontinuous group
$\Gamma$ of isometries. There is a huge number of such manifolds. In
the case $N=3$, their relevance as spatial sections for
Robertson-Walker-Freedman cosmologies has been discussed in
Ref.~\cite{elli71-2-7}. One interesting feature of these spaces lies
in the mismatch between non trivial topology and negative curvature,
which permits a variety of situations. In particular, they can
describe a finite universe that expands forever, contrary to the
belief that a locally hyperbolic universe must be infinite. To have
some ideas of the spaces $H^N/\Gamma$, we start with an elementary
classification of the isometries of $H^N$ since this is also the
second important step towards the Selberg trace formula.

We write a point in $H^N$ as $X=(x,r)$, where $x=(x_1,...,x_{N-1})$
belongs to the boundary $\partial H^N$ (the plane $r=0$) and $r>0$,
namely we are considering the half-space model of $H^N$. It is
understood that $\{\infty \}$ is considered to be a point of $\partial
H^N$, that is $\partial H^N=\mbox{$I\!\!R$}^{N-1}\cup \{\infty \}$.
This is done in order to make the inversion well defined at the origin.

Let $G_N$ be the group of motions of $\mbox{$I\!\!R$}^N\cup\{\infty\}$
which is generated by the following transformations:
\begin{description} \item{(i)} translations: $(x,r)\rightarrow
(x+y,r)$,            $\infty\rightarrow\infty$, \item{(ii)} rotations:
$(x,r)\rightarrow (\Lambda x,r)$, $\infty\rightarrow\infty$
($\Lambda\in O(N-1)$), \item{(iii)} dilatations: $X\rightarrow \lambda
X $, $\infty\rightarrow\infty$ ($\lambda>0$), \item{(iv)} inversions:
$X\rightarrow a+(X-a)/|X-a|^2$, $a\rightarrow \infty$,
$\infty\rightarrow a$ ($a\in\partial H^N$). \end{description} Three
elementary facts about $G_N$ are noteworthy \cite{bear83b,mask88b}:
\begin{description} \item{(a)} $G_N$ preserves both $H^N$ and
$\partial H^N$. As a group of motions of $\partial H^N$ (obtained
setting $r=0$ in (i) through (iv)) it is the conformal group of the
flat metric. The dimension is $N(N+1)/2$. \item{(b)} Regarding
$\partial H^N$ as the plane $r=0$ in
$\mbox{$I\!\!R$}^N\cup\{\infty\}$, there is a natural inclusion
$G_N\subset G_{N+1}$, the full conformal group of
$\mbox{$I\!\!R$}^N\cup\{\infty\}$. \item{(c)} $G_N$ is the group of
all isometries of $H^N$. \end{description} The first half of statement
(a) is trivial as well as the statement (b). It is a simple exercise
to show that elements of $G_N$ act as isometries of the Poincar\'e
metric. Since $N(N+1)/2$ is the maximal dimension permitted by a group
of isometries, the statement (c) follows. As a consequence, $G_N$ is
isomorphic to the Lorentz group $O^+(1,N)$. By inspection of the
elements of $G_N$, it is clear that every $\gamma\in G_N$ has at least
one fixed point in $H^N\cup\partial H^N$. We say that $\gamma$ is
elliptic if $\gamma$ has at least one fixed point in $H^N$ (rotations
and inversions are example of elliptic elements). If there is only one
fixed point and it lies in $\partial H^N$, then $\gamma$ is parabolic
(translations are examples of parabolic elements). Otherwise $\gamma$
is loxodromic (dilatations are examples of loxodromic elements). If
there are three or more fixed points in $\partial H^N$, then $\gamma$
must be elliptic \cite{mask88b} (i.e. it fixes some point in $H^N$).
Thus every loxodromic element has exactly two fixed points in
$\partial H^N$.

Since fixed points produce metric singularities in the quotient
manifold, we have that $H^N/\Gamma$ is a singularity-free Riemannian
space only if $\Gamma$ does not contain elliptic elements. Now we can
use the explicit form of the generators to find a canonical form for
the various isometries. We recall that two elements $\gamma$ and
$\gamma^{\prime}$ in a group are conjugate if
$\gamma=g\gamma^{\prime}g^{-1}$ for some $g$. Conjugation is an
equivalence relation so that the group is a disjoint union of
conjugacy classes.

First let $\gamma$ be elliptic and $X$ a fixed point. We can move $X$
to the point $(0,1)$ with an isometry $g$. Thus $g\gamma g^{-1}$ fixes
$(0,1)$. Since translations and dilatations do not fix $(0,1)$,
$\gamma$ is conjugate to a product of inversions and a rotation around
the vertical line $\{(0,r):r>0\}$. For future references, we call this
line $H^1$. A better description of elliptic elements may be the
following: we regard $\gamma$ as acting in $B^N$ with fixed point the
origin. Since the ball metric is radial, it is clear that $\gamma\in
O(N)$ is a rotation or a reflection around the origin. But there is a
conformal transformation $q\in G_{N+1}$ which sends $H^N$ onto $B^N$
and the point $(0,1)$ into the origin. Thus, every elliptic element is
conjugate in $G_{N+1}$ to a rotation or reflection around the origin.
This fact has an interesting consequence in $H^3$. A rotation in
3-dimensions always has an eigenvalue equal to 1 or $-1$, i.e. an axis
of rotation. This will be a radial geodesic in the ball $B^3$. Going
back to $H^3$, we conclude that every elliptic element is a rotation
around a unique invariant axis, namely around a circle or line that
meets $\partial H^N$ orthogonally.

Now we suppose $\gamma$ is a loxodromic element with fixed points $x$,
$y$ and $\ell$ is the geodesic in $H^N$ joining $x$ to $y$. This is
called the axis of $\gamma$. Clearly $\gamma(\ell)=\ell$, since
$\gamma$ maps geodesics into geodesics and there is only one geodesic
joining $x$ to $y$. We can map the axis to the vertical line $H^1$
with an isometry $g$. Thus $g\gamma g^{-1}$ leaves $H^1$ invariant
with fixed points $0$ and $\infty$. Translations and inversions do not
have this property, thus,  $\gamma$ must be conjugate to a rotation
around $H^1$ followed by a dilatation. If the rotation is trivial,
i.e. if $\gamma$ is conjugate to a dilatation, we say that $\gamma$ is
a hyperbolic isometry. In the Lorentz group of the hyperboloid model,
such a $\gamma$ is conjugate to a Lorentz boost.

Finally, if $\gamma$ is parabolic, we can move the fixed point at
$\infty$ with an isometry $g$. Hence, $g\gamma g^{-1}$ is an isometry
fixing only $\infty$. Since the inversion and dilatations do not have
this property, $\gamma$ must be conjugate to a rotation followed by a
translation, i.e. $g\gamma g^{-1}(X)= \Lambda(X)+a$. Actually, one can
choose the point $a$ and the matrix $\Lambda$ such that
$\Lambda(a)=a$. This is the normal form of a parabolic element.

Clearly, conjugation preserves the isometry types so that the conjugacy
classes are  classified as elliptic, loxodromic and parabolic classes.

Now we suppose that $ {\cal M}=H^N/\Gamma$ is a compact manifold. The
action of $\Gamma$ must be free, otherwise $ {\cal M}$ is not a
manifold, in general. Then $\Gamma$ is torsion-free, which means it
does not contain elements of finite order \cite{bore63-2-111}.
Moreover $\Gamma$ is discontinuous and thus it must be a discrete
subgroup of $G_N$ \cite{mask88b}.

What are the consequences of these facts? The most important for us is
that every element of $\Gamma$ must be a loxodromic isometry. Thus
$\Gamma$ will not contain any parabolic element. Such a group is
called co-compact. Very briefly, the argument runs as follows.
Consider the distance function $d(X,\gamma X)$ on $H^N$ where
$\gamma\in\Gamma$. Any arc joining $X$ and $\gamma X$ projects to a
closed homotopically non trivial loop on $ {\cal M}$. For, if it were
possible to shrink the loop continuously to a point, it would be
possible to move $\gamma$ continuously into the identity without exit
$\Gamma$. This cannot be done because $\Gamma$ is discrete. Thus the
function is strictly positive and there must be an $X_0$ at which
$d(X_0,\gamma X_0)$ is minimum. The reason is that the distance
function is continuous and $ {\cal M}$ is compact. It follows that the
minimal geodesic $J_{\gamma}$ joining $X_0$ and $\gamma X_0$ is
invariant under $\gamma$, since any other geodesic will have greater
length. $J_{\gamma}$ will intersect $\partial H^N$ in two points which
are fixed points of $\gamma$ and so $\gamma$ will be a loxodromic
isometry. Recalling the normal form of such an isometry as given
above, we see that if $ {\cal M}$ is compact, then every
$\gamma\in\Gamma$ is conjugate in $G_N$ to a transformation of the
form $D\Lambda$, where $D: (x,r)\rightarrow (\lambda x,\lambda r)$ is
a dilatation and $\Lambda$ is a rotation acting on $x$. The invariant
geodesic $J_{\gamma}$ is called the axis of $\gamma$. Moreover, if
$\gamma_1$ and $\gamma_2$ commute, they share a common axis. A second
fact is that any non trivial abelian subgroup of $\Gamma$ must be
infinite cyclic. In particular, the centralizer $ {\cal C}_{\gamma}$
of any non trivial element $\gamma$ must be infinite cyclic. $ {\cal
C}_{\gamma}$ is the set of elements $g\in\Gamma$ which commute with
$\gamma$. The third fact is the existence of a convex polyhedron $
{\cal F}$ in $H^N$ such that \cite{mask88b} (a) the hyperbolic volume
of $ {\cal F}$ is finite, (b) $\gamma( {\cal F})\cap {\cal
F}=\emptyset$ for all non trivial $\gamma\in\Gamma$, (c) the
translates of $ {\cal F}$ tessellate $H^N$, that is
$\bigcup_{\gamma\in\Gamma}\gamma( {\cal F})=H^N$, (d) for every face
$s$ of $ {\cal F}$ there is a face $s^{\prime}$ and an element
$\gamma_s\in\Gamma$ with $\gamma_s(s)=s^{\prime}$ and
$\gamma_s(s^{\prime})=s$. That is the faces of $ {\cal F}$ are paired
by elements of $\Gamma$. Note that the faces of $ {\cal F}$ are
portions of $(N-1)$-dimensional spheres or planes orthogonal to
$\partial H^N$. Such a polyhedron is called a fundamental domain for
$\Gamma$. Conversely, the Poincar\'e polyhedron theorem roughly states
that given a polyhedron with the above properties (plus additional
technical assumptions), the group generated by the face pairing
transformations is discrete with fundamental domain being the given
polyhedron \cite{mask88b}. As a result, these manifolds can be
obtained by identifying the faces of a finite volume polyhedron in a
suitable way.

\subsubsection{The Selberg trace formula for scalar fields}
\label{S:STFN}

We begin by stating the trace formula and then we try to illustrate
it, first restricting ourselves to a strictly hyperbolic group
$\Gamma$ such that $H^N/\Gamma$ is compact with fundamental domain $
{\cal F}_N$ and  the volume $\Omega( {\cal F}_N)$ of $ {\cal M}$ is
finite.

Let $h(r)$ be an even and holomorphic function in a strip of width
larger than $N-1$ about the real axis, such that
$h(r)=O(r^{-(N+\varepsilon)})$ uniformly in the strip as $r\to\infty$.
Let $\chi(\gamma): \Gamma\rightarrow S^1$ be a character of $\Gamma$.
Let us denote by $A_{\Gamma,\chi}$ the non negative, unbounded,
self-adjoint extension of the Laplace operator $\Delta_{\Gamma}$
acting in the Hilbert space ${ {\cal H}}(\Gamma,\chi)$ of square
integrable automorphic functions on a fundamental domain ${ {\cal
F}_N}$, relative to the invariant Riemannian measure. A scalar
function $\phi$ is called automorphic (relative to $\Gamma$ and $\chi$)
if for any $\gamma\in\Gamma$ and $x\in H^N$ the condition $\phi(\gamma
x)=\chi(\gamma)\phi(x)$ holds. We will refer automorphic functions
with $\chi\neq1$ as twisted scalar fields. For $\Gamma$ co-compact,
the operator $A(\Gamma,\chi)$ has pure discrete spectrum with isolated
eigenvalues $\lambda_j$, $j=0,1,2,..\infty$ of finite multiplicity.
With these assumptions, the Selberg trace formula holds (see
Ref~.\cite{chav84b}) \begin{equation}
\sum_{j=0}^{\infty}h(r_j)=\Omega({ {\cal F}_N}) \int_0^\infty
h(r)\Phi_N(r)\,dr +\sum_{\{\gamma\}}\sum_{n=1}^{\infty}
\frac{\chi^n(\gamma)l_{\gamma}} {S_N(n;l_{\gamma})}\hat h(nl_{\gamma})
\:.\label{STFN} \end{equation} Above, each number $r_j$ is that root
of $r_j^2=(\lambda_j-\rho_N^2)$ having positive imaginary part and the
sum over $j$ includes the eigenvalues multiplicity. The symbols
$S_N(n;l_{\gamma})$, $\{\gamma\}$, $l_{\gamma}$ will be defined soon,
while the function $\hat h(p)$ is the Fourier transform of $h(r)$
defined by \begin{eqnarray} \hat h(p)=\frac{1}{2\pi}
\int_{-\infty}^{\infty} e^{ipr}h(r)\;dr \:\nonumber\end{eqnarray} and
$\Phi_N(r)$ is the density of states we computed in the previous
section. In particular we have \begin{equation}
\Phi_2(r)=\frac{r}{2\pi}\tanh\pi r\:,
\qquad\qquad\Phi_3(r)=\frac{r^2}{2\pi^2} \label{phi23} \end{equation}
and $\Phi_N$ satisfies the recurrence relation \begin{equation}
\Phi_{N+2}(r)=\frac{\rho_N^2+r^2}{2\pi N}\Phi_N(r) \:,\label{RRPhiN}
\end{equation} which permits to obtain any $\Phi_N$ starting from
$\Phi_2$ or $\Phi_3$ according to whether $N$ is even or odd.

The left hand side of Eq.~(\ref{STFN}) is a result of functional
calculus. To see this, let we define $\tilde h(\lambda)=h(r)$ for
$\lambda=r^2+\rho_N^2$. Then the left hand side is the operator trace
of $\tilde h(A_{\Gamma,\chi})$, because the Laplacian is a
self-adjoint operator, so the trace must be given by the sum of
$\tilde h(\lambda)$ over the eigenvalues. Thus the left member of the
trace formula is determined by the spectrum of the Laplace operator.

The twist condition is the analog on $H^N/\Gamma$ of the "periodicity
up to a phase" for fields on the torus $\mbox{$I\!\!R$}^N/T$, where
$T$ is a discrete group of translations in $\mbox{$I\!\!R$}^N$. Hence,
the only things that need an explanation are the integral and the
double series on the right hand side of Eq.~(\ref{STFN}). We give two
descriptions and a partial interpretation of them.

\paragraph{The group-theoretic description.} We recall that every
element $\gamma'\in\Gamma$ is a loxodromic isometry and its
centralizer is infinite cyclic. Thus every element commuting with
$\gamma^{\prime}$, and $\gamma^{\prime}$ itself, will be the power of
a generator $\gamma$, namely $\gamma^{\prime}=\gamma^n$, for some $n$.
This element $\gamma$ is called primitive and $\{\gamma\}$ denotes the
primitive conjugacy class determined by $\gamma$, i.e. the set of all
elements in $\Gamma$ of the form $\gamma''=g\gamma g^{-1}$. Let us
denote the trivial class by $\{e\}$ and non trivial ones by
$\{\gamma\}$. It is simple to see that $\gamma^n$ is not conjugate to
$\gamma$ if $n>1$. This means that for any non trivial primitive class
, we have infinitely many other distinct classes determined by the
positive powers of $\gamma$. Now, the double series in
Eq.~(\ref{STFN}) is the sum over all the powers of any selected non
trivial primitive class followed by the sum over all the non trivial
primitive classes, in order to include all possible elements of
$\Gamma$. We have also learned that every $\gamma$ is conjugate to the
product of a dilatation $D_{\gamma}:
(x,r)\rightarrow(N_{\gamma}x,N_{\gamma}r)$ with $N_{\gamma}>1$ and a
rotation  $\Lambda_{\gamma}\in O(N-1)$ acting on $x$. The dilatation
factor $N_{\gamma}$ is called the norm of $\gamma$ and $l_{\gamma}=\ln
N_{\gamma}$ by definition. The factor $S_N(n;l_{\gamma})$ is defined
in terms of $D_{\gamma}$, $\Lambda_{\gamma}$ and $l_{\gamma}$ by
\begin{eqnarray}
S_N(n;l_{\gamma})=|\det(I-(N_{\gamma}\Lambda_{\gamma})^n)|e^{-n\rho_Nl_{\gamma}}
\label{factor} \:,\end{eqnarray} where the matrix
$I-(N_{\gamma}\Lambda_{\gamma})^n$ has dimension $(N-1)\times (N-1)$.
The integral is the contribution to the trace of the trivial class,
i.e. the identity element of the group and it will soon be computed.

\paragraph{The geometric description.} This follows if we remind the
definition of the axis of a $\gamma\in\Gamma$: it is the unique
invariant geodesic in $H^N$ joining $x$ to $\gamma x$. When we
identify $x$ with $\gamma x$ in the quotient manifold $ {\cal M}$, the
axis clearly projects to a closed geodesic in $ {\cal M}$. A simple
calculation with the Poincar\'e metric shows that the length of the
portion of the axis between $x$ and $\gamma x$ is just $l_{\gamma}$,
the logarithm of the dilatation factor $N_{\gamma}$. Thus, there is a
closed geodesic of length $l_{\gamma}$ associated with such a
$\gamma$. If $\gamma$ is primitive the geodesic will go once from $x$
to $\gamma x$ in $ {\cal M}$. If $\gamma=\delta^n$ with $\delta$
primitive, then $l_{\gamma}=\ln N_{\delta}^n=n\ln
N_{\delta}=nl_{\delta}$. Hence, the geodesic will go $n$ times from
$x$ to $\gamma x$ and $n$ is the winding number. It is evident that
the closed geodesic depends only on the conjugacy class $\{\gamma\}$.
If the class is primitive, the associated geodesic is likewise called
primitive. The converse is also true, since every geodesic loop in $
{\cal M}$, say $J$, must be the projection of a geodesic in $H^N$
joining $x$ and $\gamma x$ for some $\gamma\in\Gamma$ and $x\in H^N$
(because $J$ is closed). If $\gamma^{\prime}=g\gamma g^{-1}$ then the
axis of $\gamma^{\prime}$ projects again to $J$ (because $x$, $gx$,
$\gamma x$ and $g\gamma x$ all belong to the same orbit). As a result,
to every geodesic loop there corresponds a conjugacy class in
$\Gamma$. Thus, we can think of each $\{\gamma\}$ as a closed geodesic
and the double sum in Eq.~(\ref{STFN}) is geometrically the same as
the sum over all the primitive geodesics in $ {\cal M}$ (those passing
only once through any given point) together with the sum over the
winding numbers. Furthermore, the integral is the "direct path"
contribution to the trace. Note that it is determined by the volume of
the manifold and corresponds to the Weyl asymptotic leading term of
the spectral density. We see that we can regard Eq.~(\ref{STFN}) as a
generalization to $H^N/\Gamma$ of the familiar method of images on the
torus.

\paragraph{The path integral interpretation.} This stems from the use
of the trace formula to compute $\,\mbox{Tr}\,\exp(t\Delta_{\Gamma})$
and from its path integral representation, valid also on non simply
connected spaces \cite{laid71-3-1375}. The result for $\hat
h(nl_{\gamma})$ is essentially $\exp(-n^2l_{\gamma}^{2}/4t)$. This is
the exponential of the classical action for a geodesic of length
$nl_{\gamma}$, the factor $S_N(n;l_{\gamma})$ resembles the Van
Vleck-Morette determinant and the integral involving $\Phi_N$ the
direct path contribution. The character $\chi(\gamma)$ also must be
present in general, but the path integral may select one or more among
them. This happens, for instance, in the path integral treatment of
identical particles, where only completely symmetric or anti-symmetric
wave functions are allowed. It would be certainly interesting to
establish this connection rigorously.

Finally, let us see how the trace formula itself comes out. Let us
define the invariant function $u(X,Y)=[\cosh
d(X,Y)-1]/2=\sinh^2[d(X,Y)/2]$ (see Eq.~(\ref{dist})) and let $K$ be
an invariant integral operator, its kernel being given by $k(X,Y)$.
The invariance of the operator is equivalent to the fulfilment of the
condition $k(X,Y)=k(u(X,Y))$, i.e. the kernel is a function of the
geodesic distance. Now to any function $k(u)$ on $[0,\infty)$ with
compact support, we can associate the integral operator
$K_{\Gamma,\chi}$ in the space ${ {\cal H}}(\Gamma,\chi)$ defined by
the kernel \begin{eqnarray}
K_{\Gamma,\chi}(X,Y)=\sum_{\gamma\in\Gamma}\chi(\gamma)k(u(X,\gamma Y))
\nonumber\end{eqnarray} (the sum if finite because $k(u)$ has compact
support and $\Gamma$ is discrete). It can be proved that if
$K_{\Gamma,\chi}$ is trace-class, then a function $h(r)$ exists such
that \begin{eqnarray} \sum_jh(r_j)&=&\int_{ {\cal
F}_N}K_{\Gamma,\chi}(X,X)dX\nonumber\\ &=&\sum_{\gamma\in\Gamma}\int_{
{\cal F}_N}\chi(\gamma)k(u(X,\gamma X))dX
=\sum_{\{\gamma\}}\chi(\gamma)\int_{ {\cal F}_{\gamma}}k(u(X,\gamma
X)dX \:.\label{hrj} \end{eqnarray} Once more, each number $r_j$ is the
root of $r_{j}^{2}=\lambda_j -\rho_N^2$ in the upper half complex
plane and the sum runs over the eigenvalues of $A_{\Gamma,\chi}$. We
also denoted by  $dX=r^{-N}drdx_1...dx_{N-1}$ the invariant volume
element and the integral extends over a fundamental domain for
$\Gamma$. Moreover, ${ {\cal F}}_{\gamma}$ is a fundamental domain for
$C(\gamma)$, the centralizer of $\gamma$ in $\Gamma$. In addition, the
formula \begin{eqnarray} \int_{H^N}K(u(X,X))dX=\int_0^\infty
h(r)\Phi_N(r)dr \nonumber\end{eqnarray} holds for any trace-class
integral operator acting on square integrable functions on $H^N$.

The connection $k(u)\leftrightarrow h(r)$ is known as the Selberg
transform \cite{selb56-20-47}. In order to introduce such a transform,
we define the function $p(z)=\,\mbox{cosh$^{-1}$}\,(1+2z)$ and its
inverse $z(p)=(\cosh p-1)/2$. For $N=2n+1$, the Selberg transform can
be written as \begin{eqnarray} k(u)&=&\frac{(-1)^n}{(4\pi)^n}\hat
h^{(n)}(p(u))\:,\nonumber\\ \hat h(p)&=&\frac{(4\pi)^n}{\Gamma(n)}
\int_{z(p)}^{\infty}k(x)[x-z(p)]^{n-1}dx \:,\nonumber\end{eqnarray}
while if $N=2n$ \begin{eqnarray}
k(u)&=&\frac{2(-1)^n}{(4\pi)^n}\int_u^\infty \hat
h^{(n)}(p(z))(z-u)^{-1/2}dz\:,\nonumber\\ \hat
h(p)&=&\frac{(4\pi)^n}{2\pi^{1/2}\Gamma(n-1/2)}
\int_{z(p)}^{\infty}k(x)[x-z(p)]^{n-\frac{3}{2}}dx
\:,\nonumber\end{eqnarray} from which we can reconstruct $h(r)$ by the
inverse Fourier transform. For a simple derivation of both the theorem
and the formulae see Ref.~\cite{chav84b}. By $\hat h^{(n)}(p(u))$ we
indicate the $n^{th}$ derivative of $\hat h(p(u))$ with respect to the
variable $u$. It should be noted that these formulae are a particular
case of the theorem on the expansion of an arbitrary spherical
function in zonal functions (the Fourier-Harish-Chandra transform),
which is valid in more general cases than the ones considered here.

Evaluating the integrals in Eq.~(\ref{hrj}), using Selberg transform,
one finally gets the Selberg trace formula. For the proof, we refer
the reader to the literature (see for example
Refs.~\cite{hejh76b,chav84b}. The central point is that the domain ${
{\cal F}_{\gamma}}$ is very simple and this makes possible to perform
the integral. Here, we explicitly compute the contribution due to the
identity element of the isometry group, that is the term $\Omega(
{\cal F}_N)k(0)$ in Eq.~(\ref{hrj}). The integral operator determined
by $k(u(X,Y))$ is the operator function $\tilde h(\Delta)$ of the
Laplace operator in $H^N$, $k(u)$ being related to $h(r)$ by Selberg
transform. Thus we can expand $k(u(X,Y))$ over the eigenfunctions of
the continuous spectrum, which have been computed previously in the
hyperboloid model. We choose $Y=0$ to be the bottom of the
hyperboloid. In this way $u(X,0)=(\cosh\sigma-1)/2$ and $k(u(X,0))$ is
a radial function. Thus, the expansion contains only zero angular
momentum eigenfunctions and reads \begin{eqnarray}
k(u(X,0))=\int_{0}^{\infty}
\Phi_N(r)h(r)f_{\lambda(r)}(X)f_{\lambda(r)}(0)\,dr
\:,\nonumber\end{eqnarray} since the radial eigenfunctions have
density $\Phi_N(r)$. They were defined so that $f_{\lambda}(0)=1$.
Hence, setting $X=0$ we obtain the formula \begin{eqnarray}
k(0)=\int_{0}^{\infty}drh(r)\Phi_N(r) \nonumber\end{eqnarray} between
Selberg transform pairs determining the contribution of the identity.

\subsubsection{The Selberg $Z$-function} \label{S:SZF}

This section is devoted to a brief discussion of Selberg $Z$-function
and its associated $\Xi$-function, with a heuristic derivation of
their main properties. Let $C$ be a contour in the complex  plane from
$+i\infty$ to $+i\infty$ around the positive imaginary axis with
anti-clockwise orientation and including all the numbers $ir_j$. Then
for any $h(r)$ for which the Selberg formula holds, we may rewrite
\begin{eqnarray} \sum_j\int_C\frac{2z\,h(-iz)}{z^2+r_j^2}\,dz
&=&\Omega( {\cal F}_N) \int_{-\infty}^\infty dr\,\Phi_N(r)
\int_C\frac{z\,h(-iz)}{z^2+r^2}\,dz\nonumber\\
&&+\sum_{\{\gamma\}}\sum_{n=1}^{\infty} \frac{\chi^n(\gamma)l_{\gamma}}
{S_N(n;l_{\gamma})}\:\int_{-\infty}^\infty dr\,\frac{e^{i nl_\gamma
r}}{2\pi} \int_C\frac{2z\,h(-iz)}{z^2+r^2}\,dz \:,
\nonumber\end{eqnarray} The integration with respect to the variable
$r$, in the second term on the right hand side of the equation above,
can be done without any problem. Since the function $\Phi_N(r)$ goes
like $|r|^{N-1}$ for $|r|\to\infty$, in order to exchange and perform
the integration with respect to $r$ also in the first term, we
introduce a regularization function $f(r,\varepsilon)$, which is
analytic for $\,\mbox{Re}\, r\geq0$, goes to zero faster than
$|r|^{N-2}$ for $|r|\to\infty$ and goes to 1 for $\varepsilon\to0$ (an
example is $f(r,\varepsilon)=(r+i)^{-\varepsilon}$ with
$\varepsilon>N-2$). Then we can easily make the integration in the $r$
complex plane and, at the end, take the limit $\varepsilon\to0$. The
final result reads \begin{eqnarray}
\sum_j\int_C\frac{2z\,h(-iz)}{z^2+r_j^2}\,dz =\pi\,\Omega( {\cal
F}_N)\int_C\Phi_N(iz)h(-iz)\,dz +\int_C\Xi(z+\rho_N)\,h(-iz)dz
\:,\nonumber\end{eqnarray} where $\Xi(z)$ is the Selberg
$\Xi$-function, defined for $\,\mbox{Re}\, z>2\rho_N=N-1$ by
\begin{eqnarray} \Xi(z)=\sum_{\{\gamma\}}\sum_{n=1}^{\infty}
\frac{\chi^n(\gamma)l_{\gamma}}{S_N(n;l_{\gamma})}
e^{-(z-\rho_N)nl_{\gamma}} \:.\label{SXF}\end{eqnarray} This shows that
\begin{eqnarray} \sum_j\frac{2z}{z^2+r_j^2} =\pi\,\Omega( {\cal
F}_N)\Phi_N(iz)+\Xi(z+\rho_N) \label{traxi} \:,\end{eqnarray} where
$z\to\Phi_N(iz)$ has to be considered as a regular analytic
distribution and the series converge in the space of regular analytic
distributions \cite{guel72b}.

The asymptotic behaviour
$S_N(n;l_{\gamma})\simeq\exp(n\rho_Nl_{\gamma})$ for large $n$ and the
known fact that the number of conjugacy classes with a given length
$l_{\gamma}$ is asymptotically
$l_{\gamma}^{-1}\exp(2\rho_Nl_{\gamma})$, show that the $\Xi$-function
is analytic in the half-plane $\,\mbox{Re}\, z>N-1$. Thus, from
Eq.~(\ref{traxi}) we deduce that $\Xi(z)$ extends to a meromorphic
function with simple poles at $z_j=\rho_N\pm ir_j$ whose residues are
$d_j$ (the eigenvalues multiplicity), if the eigenvalue
$\lambda_j\neq\rho_N$ or $2d_j$, if $\lambda_j=\rho_N$. Moreover,
since $\Phi_N(z)$ is an even function, while the left hand side of
Eq.~(\ref{traxi}) is an odd function of $z$, we have the functional
equation \begin{eqnarray} \Xi(z+\rho_N)+\Xi(-z+\rho_N)=-2\pi \Omega(
{\cal F}_N)\Phi_N(iz) \:.\label{funxi}\end{eqnarray} It can be proved
that this holds true in any dimensions \cite{gang77-21-1}. In even
dimensions there are also the additional poles of the density
$\Phi_N(iz)$ at $z_k=\pm(\rho_N+k)$, $k\geq 0$ with certain real
residues, say $R_k$. Since $\Xi(z)$ is analytic for $\,\mbox{Re}\,
z>N-1$, it follows that $\Xi(z)$ in even dimensions has additional
simple poles at $z_k=-k$, $k\geq 0$.

As shown by Selberg, the analytic properties of $\Xi(z)$ permit the
existence of a function $Z(s)$ such that \begin{eqnarray} \ln
Z(s)=-\sum_{\{\gamma\}}\sum_{n=1}^{\infty}\frac{\chi^n(\gamma)}
{S_N(n;l_{\gamma})n}e^{-(s-\rho_N)nl_{\gamma}}
\:,\nonumber\end{eqnarray} then obviously $\Xi(s)=Z'(s)/Z(s)$ and
Eq.~(\ref{funxi}) becomes a functional equation for $Z(s)$ which reads
\begin{eqnarray} Z(-s+\rho_N)=Z(s+\rho_N)\exp\int_{0}^{s}2\pi \Omega(
{\cal F}_N)\Phi_N(iz)dz \:,\nonumber\end{eqnarray} where the complex
integral in the exponent is over any contour not crossing the poles of
$\Phi_N(iz)$. It can be shown that the residues of the function $2\pi
\Omega( {\cal F}_N)\Phi_N(iz)$ are multiples of the Euler number of
the manifold, so the integral is well defined. The poles of $\Xi(s)$
become poles or zeroes of $Z(s)$. More precisely, if $N$ is odd then
$Z(s)$ is an entire function of order $N$ and all the poles of
$\Xi(s)$ are zeroes of $Z(s)$. If $N$ is even there are the additional
poles of the density $\Phi_N(is)$. When the residues $R_k>0$, $Z(s)$
has poles at all negative integers. This occur precisely when $N=4n$,
as can be seen with some calculations. In all the other cases, the
negative integers are zeroes of $Z(s)$ \cite{gang77-21-1}. The point
$s=0$ is special. In odd dimensions it is always a zero, if
$\chi(\gamma)=1$. In even dimensions it can also be a non simple pole,
depending on the Euler characteristic of the manifold
\cite{gang77-21-1}.

\subsubsection{The Selberg trace formula for compact $H^3/\Gamma$}
\label{S:STF3}

{}From a physical point of view, this is certainly one of the most
important cases. One peculiarity is the explicit realization of the
isometry group as
$PSL(2,\mbox{$I\!\!\!\!C$})=SL(2,\mbox{$I\!\!\!\!C$})/\{-1,1\}$ and
the possibility to make use of complex numbers formalism. A second
fact is the possibility to include elliptic elements into the trace
formula. When $\Gamma$ is a discrete subgroup of
$PSL(2,\mbox{$I\!\!\!\!C$})$ containing elliptic elements, the space
$H^3/\Gamma$ is called the associated 3-orbifold. It is known that
$H^3/\Gamma$ is always a manifold, but the Riemannian metric is
singular along the axis of rotation of elliptic elements.

Using complex numbers, we can represent
$H^3\equiv\{X=(z,r)|z=x^1+ix^2\in
\mbox{$I\!\!\!\!C$},r=x^3\in(0,\infty)\}$, with the Riemannian metric
$dl^2=(d\bar zdz+dr^2)/r^2$. The group  $SL(2,\mbox{$I\!\!\!\!C$})$
acts on $H^3$ in the following way. Given a matrix $\sigma\in
SL(2,\mbox{$I\!\!\!\!C$})$ \begin{equation} \sigma=\left(
\begin{array}{cc} a&b\\c&d \end{array}\right) \nonumber\end{equation}
and $X\in H^3$, one defines the following orientation preserving
isometry \cite{elst85-17-83,elst87-277-655}: \begin{equation} \sigma
X=\left(\frac{(az+b)(\bar c\bar z+\bar d) +a\bar
cr^2}{|cz+d|^2+|c|^2r^2},\frac{r}{|cz+d|^2+|c|^2r^2}\right)
\:.\label{isom} \end{equation} It is evident that the isometry
determines $\sigma$ up to sign. Eq.~(\ref{isom}) is simply a special
way to rewrite the generators of $G_3$ (see Sec.~\ref{S:CHM}).

As in the general case, all elements of $PSL(2,\mbox{$I\!\!\!\!C$})$
belong to one of the following conjugacy classes: elliptic, loxodromic
and parabolic and, in this case,  we also discuss the case
corresponding to  $\Gamma$ containing elliptic elements. From the
normal form of the isometries we have discussed in Sec.(\ref{S:CHM}),
we see that every loxodromic element $\gamma\in\Gamma$ is conjugate in
$PSL(2,\mbox{$I\!\!\!\!C$})$ to a unique element $D(\gamma)$ of the
form \begin{equation} D(\gamma)=\left(\begin{array}{cc}
a_{\gamma}&0\\0&a_{\gamma}^{-1} \end{array}\right) \:,\qquad\qquad
|a_{\gamma}|>1 \:.\label{hele} \end{equation} From Eq.~(\ref{isom}) we
see that $D(\gamma)$ is a dilatation in the vertical line $H^1$ with
dilatation factor $|a_{\gamma}|^2$. The number
$N_{\gamma}=|a_{\gamma}|^2$ is called the norm of $\gamma$ and
$l_{\gamma}=\ln N_{\gamma}$. The phase of $a_{\gamma}$ is an $SO(2)$
rotation in $\mbox{$I\!\!\!\!C$}$ around $H^1$. Every elliptic element
$\alpha\in\Gamma$ is conjugate in $PSL(2,\mbox{$I\!\!\!\!C$})$ to a
unique element $D(\alpha)$ of the form \begin{equation}
D(\alpha)=\left(\begin{array}{cc} \xi_{\alpha}&0\\0&\xi_{\alpha}^{-1}
\end{array}\right) \:,\qquad\qquad |\xi_{\alpha}|=1.\label{eele}
\end{equation} From Eq.~(\ref{isom}), we see that $D(\alpha)$ is a
rotation in $\mbox{$I\!\!\!\!C$}$ around the origin with angle
$\theta=2\arg\xi_{\alpha}$. Let $C_{\gamma}$ be the centralizer of
$\gamma$ in $\Gamma$. Again $C_{\gamma}$ in infinite cyclic but this
time it may contain a discrete group of rotations around the axis of
$\gamma$ (since these commute with $\gamma$). We denote by
$R_{\gamma}$ the one with minimal rotation angle and by $m_{\gamma}$
its order, so that $(R_{\gamma})^{m_{\gamma}}=1$. Similarly, the
centralizer $C_{\alpha}$ of any elliptic element is infinite cyclic so
it must contain loxodromic elements \cite{elst85-17-83,gonc90-19-73}.
The reason is very simple. From the normal form of $\alpha$ and
$\gamma$ given by Eqs.~(\ref{hele}) and (\ref{eele}), it follows that
the dilatations in $H^1$ commute with $D(\alpha)$. Hence, $C_{\alpha}$
will contain a discrete group of dilatations on $H^1$ and it is a
known fact that such a group is discontinuous if and only if it is
infinitely cyclic. Let us choose a primitive element among the
loxodromic elements in $C_{\alpha}$ and call $N_{\alpha}$ its norm.
This is uniquely determined by $\alpha$, since the square modulus
eliminates the rotational part. Moreover, each elliptic $\alpha$ is
the power of a primitive elliptic element whose order is denoted by
$m_{\alpha}$ (simply take the rotation with minimal rotation angle
around the axis of $\alpha$). Finally, let we choose a character
$\chi$ for $\Gamma$.

{}From Eq.~(\ref{phi23}) we have $2\pi^2\Phi_3(r)=r^2$ and $\rho_3=1$.
Moreover Eq.~(\ref{factor}) gives \begin{eqnarray}
S_3(n;l_{\gamma})=|a_{\gamma}^{n}-a_{\gamma}^{-n}|^2
\:,\nonumber\end{eqnarray} where the determinant has been written
using the $SL(2,\mbox{$I\!\!\!\!C$})$ representation. Nevertheless,
the contribution of primitive loxodromic elements to the trace is not
yet determined if there are elliptic conjugacy classes. This is
because in the strictly loxodromic case all the conjugacy classes have
been parametrized by the powers $\gamma^n$, with $\gamma$ primitive.
In the present case, they are parametrized by the powers
$\gamma^n(R_{\gamma})^j$, where $0\leq j<m_{\gamma}$ and $R_{\gamma}$
is the rotation of minimal angle around the axis of $\gamma$ whose
order was denoted by $m_{\gamma}$. In fact, two different elements
among these powers are not conjugate in $\Gamma$ and it is not hard to
see that they form a complete representative system of all conjugacy
classes, if $\gamma$ runs through all primitive loxodromic elements.

The Selberg trace formula for scalar fields, in the presence of
elliptic elements is now \cite{elst85-17-83} \begin{eqnarray}
\sum_{j}h(r_j)&=& \frac{\Omega( {\cal F}_3)}{2\pi^2}
\int_{0}^{\infty}h(r)r^2dr \nonumber \\
&&\qquad\qquad+\sum_{\{\alpha\}}\sum_{n=0}^{m_{\alpha}-1}
\frac{\chi^n(\alpha)\ln N_{\alpha}}
{m_{\alpha}|(\xi_{\alpha})^n-(\xi_{\alpha})^{-n}|^2} \hat h(0)
\nonumber \\ &&\qquad\qquad+\sum_{\{\gamma\}}\sum_{n=1}^{\infty}
\sum_{j=0}^{m_{\gamma}-1}
\frac{\chi^n(\gamma)l_{\gamma}}{m_{\gamma}|(a_{\gamma})^n
\zeta^j-(a_{\gamma})^{-n} \zeta^{-j}|^2} \hat h(nl_{\gamma})
\:.\nonumber\end{eqnarray} The summations are extended over the
primitives elliptic $\{\alpha\}$ and hyperbolic $\{\gamma\}$ conjugacy
classes in $\Gamma$, as we have done in arbitrary dimensions. All
symbols have been defined above but the complex numbers $\zeta$,
$\zeta^{-1}$. They are the eigenvalues of $R_{\gamma}$ and have
modulus one. Because $R_{\gamma}$ has finite order $m_{\gamma}$,
$\zeta^{2m_{\gamma}}=1$ so $\zeta$ is a primitive $2m_{\gamma}$-th root
of unity. Once more, the integral is the contribution of the identity
class with the density $\Phi_3(r)$.

The general discussion we have presented in Sec.(\ref{S:STFN}) about
the trace formula remains valid in this case, except for the geometric
description. In particular, Eq.~(\ref{hrj}) is the starting point for
the trace formula also if elliptic elements are allowed.

For use later on, it is convenient to introduce the compact notation
\begin{eqnarray} E=\sum_{\{\alpha\}}\sum_{n=0}^{m_{\alpha}-1}
\frac{\chi^n(\alpha)\ln N_{\alpha}}
{m_{\alpha}|\xi_{\alpha}^{n}-\xi_{\alpha}^{-n}|^2}
\:,\nonumber\end{eqnarray} \begin{eqnarray}
H(\gamma;n)=\sum_{j=0}^{m_{\gamma}-1}
\frac{\chi^n(\gamma)l_{\gamma}}{m_{\gamma}|a(\gamma)^n\zeta^j-a^{-n}(\gamma)\zeta^{-j}|^2}
\:.\nonumber\end{eqnarray} The real number $E$ is called the elliptic
number of the manifold $H^3/\Gamma$. Its meaning in terms of the
geometric properties of $H^3/\Gamma$ is unclear. However, if any two
associated 3-orbifolds have the same eigenvalue spectrum, then the
volumes and the elliptic numbers are the same. This is known as (part)
of Huber theorem.

The Selberg $\Xi$-function is here defined as in $N$-dimensions by
\begin{eqnarray}
\Xi(z)=\sum_{\{\gamma\}}\sum_{n=1}^{\infty}\sum_{j=0}^{m_{\gamma}-1}
\frac{\chi^n(\gamma)l_{\gamma}}{m_{\gamma}|(a_{\gamma})^n\zeta^j-(a_{\gamma})^{-n}
\zeta^{-j}|^2} e^{-(z-1)nl_{\gamma}} \nonumber\end{eqnarray} and
again, with indefinite integration $Z(s)=\exp\int\Xi(z)dz$. This
function is an entire function of order three \cite{elst85-17-83} and
satisfies the functional equation \begin{eqnarray}
Z(-s+1)=Z(s+1)\exp\left(-\frac{\Omega( {\cal F}_3)s^3}{3\pi}+2Es\right)
\:.\nonumber\end{eqnarray} It may be of interest to mention that if
$\lambda=1$ is an eigenvalue of multiplicity $m$, then $s=1$ is a zero
of $Z(s)$ of multiplicity $2m$.

\subsubsection{The Selberg trace formula for compact $H^2/\Gamma$}
\label{S:STF2}

Here again the use of complex numbers is  useful. We take $H^2
=\{z\in\mbox{$I\!\!\!\!C$}| \,\mbox{Im}\, z>0\}$. We shall be
interested in the so called Fuchsian group of the first kind, namely a
discrete group of motion of $H^2$ for which there exists a fundamental
domain with a finite invariant measure. The group of all orientation
preserving isometries is
$PSL(2,\mbox{$I\!\!R$})=SL(2,\mbox{$I\!\!R$})/\{1,-1\})$, acting on
$H^2$ by fractional linear transformations $z\to(az+b)/(cz+d)$.
According to a classical result due to Fricke, any Fuchsian group of
the first kind is finitely generated and a certain standard system of
generators and relations may be given (see for example
Ref.~\cite{venk90b}). The measure of the fundamental domain can be
computed in terms of signature $(g,m_1,...,m_l,h)$ \begin{eqnarray}
\Omega({ {\cal F}_2})=2\pi\left[2g-2
+\sum_{j=1}^{l}\left(1-\frac{1}{m_j}\right)\right]
\:,\nonumber\end{eqnarray} where $g$ is the genus and the numbers
$m_j$ and $h$ are associated with elliptic and parabolic generators
respectively. If we allow $\Gamma$ to contain elliptic, but no
parabolic elements, the orbifold $H^2/\Gamma$ will be compact, but the
Riemannian metric will be singular at the fixed points (not lines as
in $H^3/\Gamma$) of the elliptic elements. The canonical form of a
loxodromic element is easily seen to be a dilatation \begin{eqnarray}
\gamma=\left(\begin{array}{cc} a&0\\0&a^{-1} \end{array}\right)
\:,\qquad\qquad a>1 \:.\nonumber\end{eqnarray} The number
$N_{\gamma}=a^2$ is the dilatation factor of $\gamma$ or the norm of
$\gamma$. Since the matrix trace is invariant under conjugation, we
see that $\gamma$ is loxodromic if and only if
$(\,\mbox{Tr}\,\gamma)^2>4$. Every hyperbolic element $\gamma$ is
always a power of a primitive one, which is uniquely determined by
$\gamma$ itself. This is the dilatation with the smallest possible
dilatation factor.

The canonical form of an elliptic element is \begin{eqnarray}
\alpha=\left(\begin{array}{cc} \cos\theta &-\sin\theta\\ \sin\theta
&\cos\theta \end{array}\right) \:.\nonumber\end{eqnarray} This is a
non linear rotation around the point $i=(0,1)$. We call $\theta\in
[0,\pi]$ the angle of rotation. Thus $\alpha$ is elliptic if and only
if $(\,\mbox{tr}\,\alpha)^2<4$. This non linear realization of
rotations is peculiar of $H^2$, since there are no true linear
rotations around the vertical line  $H^1$. Every elliptic element
$\alpha$ is also a power of a uniquely determined primitive one. This
is also the non linear rotation around the fixed point of $\alpha$
with the smallest possible angle. The angle is $\theta_0=\pi/m$ for
some integer $m$ (otherwise the group is not discrete). The order of
the corresponding rotation is thus $m$ (actually, the $m$-th power of
this primitive rotation is $-1$, but this is the identity in
$PSL(2,\mbox{$I\!\!R$})$). Finally, an element $\beta$ is parabolic if
and only if $(\,\mbox{tr}\,\beta)^2=4$.

Eq.~(\ref{phi23}) gives $2\pi\Phi_2(r)=r\tanh\pi r$ and $\rho_2=1/2$.
Furthermore, in $2$-dimensions the rotation $\Lambda_{\gamma}=1$ for
all $\gamma$. Then Eq.~(\ref{factor}) simply gives \begin{eqnarray}
S_2(n;l_{\gamma})=2\sinh\left(\frac{nl_{\gamma}}{2}\right)
\:.\nonumber\end{eqnarray} The contribution of hyperbolic elements is
thus determined as in $N$-dimensions. But now we have also elliptic
elements. The Selberg trace formula for scalar fields on compact
$H^2/\Gamma$ including elliptic conjugacy classes is
\cite{hejh76b,venk79-34-79} \begin{eqnarray}
\sum_jh(r_j)&=&\frac{\Omega({ {\cal F}_2})}{2\pi} \int_0^\infty dr
h(r)r\tanh\pi r \nonumber \\
&&\qquad\qquad+\sum_{\{\gamma\}}\sum_{n=1}^{\infty}
\frac{\chi^n(\gamma)l_{\gamma}}
{2\sinh\left(\frac{nl_{\gamma}}{2}\right)} \,\hat h(nl_{\gamma})
\nonumber \\ &&\qquad\qquad+\sum_{\{\alpha\}}\sum_{j=1}^{m_{\alpha}-1}
\frac{\chi^j(\alpha)}{2m_{\alpha}\sin\frac{j\pi}{m_{\alpha}}}
\int_{-\infty}^{\infty} \frac{\exp\left(-\frac{2\pi
rj}{m_{\alpha}}\right)} {1+\exp(-2\pi r)}h(r)dr
\nonumber\:.\end{eqnarray} The summations are extended over the
primitive elliptic $\{\alpha\}$ and hyperbolic $\{\gamma\}$ conjugacy
classes in $\Gamma$ and each number $m_{\alpha}$ is the order of the
primitive representative $\alpha$.

In 2-dimensions and without elliptic elements, the Selberg
$Z$-function admits a simple product representation \begin{eqnarray}
Z(s)=\prod_{\{\gamma\}}\prod_{k=0}^{\infty}
\left(1-\chi(\gamma)e^{-(s+k)l_{\gamma}}\right) \nonumber\end{eqnarray}
and $\Xi(s)=Z'(s)/Z(s)$. Such infinite product representations also
exist in any dimension but they are much more complicated. $Z(s)$ is
an entire function of order two, with trivial zeroes at $s_k=-k$,
$k=1,2,..,\infty$, of multiplicity $(2g-2)(2k+1)$, where $g$ is the
genus of the Riemann surface $H^2/\Gamma$. For trivial character,
$s=0$ is a zero of multiplicity $2g-1$ and $s=1$ is a simple zero.
Moreover, there are non trivial zeroes at $s_j=1/2\pm ir_j$, in
agreement with the general discussion. The functional equation is
\begin{eqnarray} Z(1/2-s)=Z(1/2+s)\exp\left[-\Omega( {\cal F}_2)
\int_{0}^{s}z\tan(\pi z)dz\right] \:.\nonumber\end{eqnarray} Since
$Z(s)$ has a simple zero at $s=1$ for trivial character, it follows
that $Z'(1)$ is non zero. This quantity is important because it
determines the functional determinant of the Laplace operator on a
Riemann surface.

\subsubsection{Recurrence relations for heat kernel and
$\zeta$-function} \label{S:HKZFHM}

We know that for physical applications heat kernel and
$\zeta$-function related to the Laplace-Beltrami operator $\Delta_N$
acting on fields in $H^N/\Gamma$ are relevant. Then we choose
$h(r)=\exp(-t(r^2+\varrho_N^2))$, compute the trace of the heat kernel
using Eq.~(\ref{STFN}) and then, using Eq.~(\ref{ZFdef}) we shall also
get the $\zeta$-function. Alternatively, we can obtain the
$\zeta$-function by choosing $h(r)=(r^2+\varrho_N^2)^{-s}$ and
directly apply the Selberg trace formula. It is convenient to
distinguish between the contribution to these quantities coming from
the identity and the hyperbolic elements of the isometry group. In 2
and 3-dimensions also the contribution of elliptic elements shall be
taken into account.

By a simple calculation we obtain \begin{eqnarray} \,\mbox{Tr}\,
e^{-t\L_N}&=&\Omega({ {\cal F}_N})e^{-t\alpha^2} \int_0^\infty
e^{-tr^2}\Phi_N(r)\;dr +\frac{e^{-t\alpha^2}}{(4\pi t)^{1/2}}
\sum_{\{\gamma\}}\sum_{n=1}^{\infty}
\frac{\chi^n(\gamma)l_\gamma}{S_N(n;l_{\gamma})}
e^{-(nl_\gamma)^2/4t}\:,\nonumber\\ &=&K_I(t|\L_N)+K_H(t|\L_N)
\:,\label{KI+KH}\end{eqnarray} \begin{eqnarray}
\zeta(s|\L_N)&=&\Omega({ {\cal F}_N}) \int_0^\infty
(r^2+\alpha^2)^{-s}\Phi_N(r)\;dr \nonumber\\
&&\qquad\qquad+\frac1{\sqrt\pi\Gamma(s)}
\sum_{\{\gamma\}}\sum_{n=1}^{\infty}
\frac{\chi^n(\gamma)l_\gamma}{S_N(n;l_{\gamma})}
\left(\frac{2\alpha}{nl_{\gamma}}\right)^{\frac12-s}
K_{\frac12-s}(nl_{\gamma}\alpha)\:, \nonumber\\
&=&\zeta_I(s|\L_N)+\zeta_H(s|\L_N) \:.\label{zeI+zeH} \end{eqnarray}
For the sake of generality and for future applications, we again
consider the operator $\L_N=-\Delta_N+\alpha^2+\kappa\varrho_N^2$.
Note that in contrast with the spherical case now $\alpha^2$ is always
positive, since $\kappa<0$. $\alpha^2=0$ and
$\alpha^2=m^2-\kappa\varrho_N^2$ give the conformal and minimal
couplings respectively, while $\alpha^2=-\kappa\varrho_N^2$
corresponds to the massless case (note that we think of $H^N$ as the
spatial section of an ultrastatic $N+1$ dimensional manifold. Only in
this case $\alpha^2=0$ corresponds to the conformal coupling).

Using recurrence relation (\ref{RRPhiN}) we easily get a recurrence
formula for $K_I(t|\L_N)$ which is very similar to the corresponding
equation on $S^N$, Eq.~(\ref{RRK-SN}). It reads \begin{eqnarray}
\frac{K_I(t|\L_{N+2})}{\Omega( {\cal F}_{N+2})} =-\frac1{2\pi N\Omega(
{\cal F}_N)}\left[\partial_t +\alpha^2+\kappa\varrho_N^2\right]
K_I(t|\L_N) \:.\nonumber\end{eqnarray} We also have recurrence
formulae for the spectral coefficients and also for $\zeta_I(s|\L_N)$,
that is \begin{eqnarray} \frac{K_n(\L_{N+2})}{\Omega( {\cal F}_{N+2})}
=-\frac1{2\pi N\Omega( {\cal F}_N)}\left[
\frac{n+N}2K_n(\L_N)+(\alpha^2+\kappa\varrho_N^2)K_{n-2}(\L_N)\right]
\:,\nonumber\end{eqnarray} \begin{eqnarray}
\frac{\zeta_I(s|\L_{N+2})}{\Omega( {\cal F}_{N+2})} =-\frac{1}{2\pi
N\Omega( {\cal F}_N)} \left[(\alpha^2+\kappa\varrho_N^2)
\tilde\zeta_I(s|\L_N)-\tilde\zeta_I(s-1|\L_N)\right]
\:.\label{RRze}\end{eqnarray} Knowing the partition function on $H^2$
and $H^3$ and using this latter equation we can obtain the partition
function for a scalar field on $H^N$. It is easy to see that
Eq.~(\ref{RRze}) is valid also for the corresponding regularized
functions according to Eq.~(\ref{ZF-r}). Equations above are formally
the same ones that we had in Sec.~\ref{S:ZFRR} for the spherical case.
The only difference is the sign of the curvature. The specific
computations for $H^2/\Gamma$ and $H^3/\Gamma$ are collected in
Appendix \ref{S:ZF-ExComp}.

The topological contribution to $\zeta$ due to hyperbolic elements
(the last term in Eq.~(\ref{zeI+zeH})) can be conveniently expressed
in terms of the logarithmic derivative of Selberg $Z$-function. In
fact, using a suitable integral representation for $K_\nu$ (see
Appendix \ref{S:UR}) and definition (\ref{SXF}) we have
\begin{eqnarray} \zeta_H(s|\L_N)= \frac{\alpha^{1-2s}\sin\pi s}{\pi}
\int_1^\infty(u^2-1)^{-s}\; \Xi(u\alpha+\varrho_N)\;du
\:,\label{ZF-H-Z} \end{eqnarray} which gives an integral
representation of $\zeta_H(s|\L_N)$ valid for $\,\mbox{Re}\, s<1$.
Deriving the latter expression at the point $s=0$, we obtain the
expression, which is relevant in the discussion of the factorization
formulae of the next subsection \begin{eqnarray} \zeta'_H(0|\L_N)=-\ln
Z(\alpha+\varrho_N) \:.\label{dzeHN} \end{eqnarray}

In the particular but important cases in which the hyperbolic section
has 2 or 3 dimensions, we are able to compute also the elliptic
contributions to the $\zeta$-function. The results are reported in
Appendix \ref{S:ZF-ExComp}, where for reader convenience we collect
some representations for $\zeta$-function on $H^3/\Gamma$ and
$H^2/\Gamma$.

\subsubsection{A factorization formula and zero modes} \label{S:HM-ZM}

We finish the section with some considerations on the possible
presence of zero modes. First of all we define the numbers $b_k^N$ by
means of \begin{eqnarray} \Phi_N(r)=\sum_{k=0}^{(N-3)/2}b_k^Nr^{2k}\:
\frac{r^2}{2\pi^2}\:, \qquad\qquad\mbox{for odd }N\geq 3\:,
\nonumber\end{eqnarray} \begin{eqnarray}
\Phi_N(r)=\sum_{k=0}^{(N-2)/2}b_k^Nr^{2k}\: \frac{r\tanh\pi r}{2\pi}\:,
\qquad\qquad\mbox{for even }N\geq 2\:. \nonumber\end{eqnarray} These
can be easily computed making use of Eq.~(\ref{RRPhiN}). For example
we have \begin{eqnarray} a_0^3=1\:,\qquad\qquad
a_0^5=a_1^5=\frac1{6\pi} \:,\nonumber\end{eqnarray} \begin{eqnarray}
a_0^2=1\:,\qquad\qquad a_0^4=\frac1{16\pi}\:,\qquad\qquad
a_1^4=\frac1{4\pi} \:.\nonumber\end{eqnarray} In this way we can write
\begin{eqnarray} \zeta_I(s|\L_N)&=&\Omega( {\cal
F}_N)\sum_{k=0}^{(N-3)/2}
b_k^N\frac{\Gamma(k+\frac32)\Gamma(s-k-\frac32)}
{4\pi^2\Gamma(s)}\,\alpha^{3+2k-2s} \:,\nonumber\end{eqnarray}
\begin{eqnarray} \zeta_I(s|\L_N)&=&\Omega( {\cal
F}_N)\sum_{k=0}^{(N-2)/2}b_k^N \left[\frac{\Gamma(k+1)\Gamma(s-k-1)}
{4\pi\Gamma(s)}\,\alpha^{2+2k-2s}\right.\nonumber\\
&&\left.\qquad\qquad\qquad\qquad\qquad\qquad-\frac1\pi\int_0^\infty
\frac{(r^2+\alpha^2)^{-s}\,r^{2k+1}}{1+e^{2\pi r}}\,dr\right]
\:,\nonumber\end{eqnarray} which are valid for odd $N\geq3$ and even
$N\geq2$ respectively. By setting $\zeta_I(s|\L_N)
=\frac{K_N(\L_N)}{\Gamma(s+1)}+\frac{I(s|\L_N)}{\Gamma(s)}$, for odd
$N\geq3$ we easily obtain \begin{eqnarray} K_N(\L_N)=0\:,\qquad\qquad
I(0|\L_N)=\Omega( {\cal F}_N)\sum_{k=0}^{(N-3)/2}
b_k^N\frac{\Gamma(k+\frac32)\Gamma(-k-\frac32)} {4\pi^2}\,\alpha^{3+2k}
\:,\label{IN-odd}\end{eqnarray} while for even $N\geq2$, after some
calculations we get \begin{eqnarray} K_N(\L_N)&=&\Omega( {\cal
F}_N)\sum_{k=0}^{(N-2)/2}b_k^N
\left[\frac{(-\alpha^2)^{k+1}}{4\pi(k+1)}
-\frac{\Gamma(2k+2)(1-2^{-2k-1})\zeta_R(2k+2)}
{\pi\,(2\pi)^{2k+2}}\right] \:,\nonumber\end{eqnarray} \begin{eqnarray}
I(0|\L_N)&=&-\gamma K_N(\L_N)+ \Omega( {\cal
F}_N)\sum_{k=0}^{(N-2)/2}b_k^N
\left[\frac{(-\alpha^2)^{k+1}}{4\pi(k+1)} \left(
C_{k+1}-\ln\alpha^2\right)\right.\nonumber\\
&&\left.\qquad\qquad\qquad\qquad\qquad\qquad\qquad\qquad+\frac1\pi\int_0^\infty
\frac{\ln(r^2+\alpha^2)\,r^{2k+1}}{1+e^{2\pi r}}\,dr\right]
\:,\label{IN-even}\end{eqnarray} where $\gamma$ is the
Euler-Mascheroni constant and $C_n=\sum_{k=1}^n1/k$.

Now we consider the pure Laplacian $L=-\Delta_N$ in $H^N/\Gamma$ and
take into consideration the possible presence of $ {\cal N}$
zero-modes. In order to deal with them, we have already pointed out in
Sec.~\ref{S:ZFR} that it is convenient to make use of the operator
$L_\lambda=L+\lambda$, with $\lambda$ a suitable non negative
parameter and define the determinant of $L$ by means of
Eq.~(\ref{logdetZM}), namely \begin{eqnarray} \det L=\frac{e^{-\gamma
K_N(L)}}{ {\cal N}!} \frac{d^{ {\cal N}}}{d\lambda^{ {\cal
N}}}\exp[-G(0|L_\lambda)] \label{gen}\end{eqnarray} Recalling the
definition of $G(s|L)=\Gamma(s)\zeta(s|L)-K_N(L)/s$
(Eq.~(\ref{ZFpoles})), Eqs.~ (\ref{zeI+zeH}) and (\ref{dzeHN}) we can
write \begin{equation} G(0|L_\lambda)=I(0|L_\lambda) -\ln
Z(\rho_N+\sqrt{\rho^2_N+\lambda})\,. \nonumber\end{equation}

Looking at Eqs.~(\ref{IN-odd}) and (\ref{IN-even}), we see that
$I(0|\L_N)$ is well defined also in the limit
$\alpha^2\to\varrho_N^2$, which corresponds to the pure Laplacian.
This means that possible singularities for $\lambda\to0$ in the
determinant of the Laplacian come from Selberg $Z$-function and this
is in agreement with the fact that zero-modes come from non trivial
topology. In fact we have the factorization formula \begin{equation}
Z(\rho_N+\sqrt{\rho_N^2+\lambda})=\lambda^{ {\cal N}} \exp[\gamma
K_N(L_\lambda)+I(0|L_\lambda)-f'(0|L_\lambda)] \,,\label{3.133}
\end{equation} where $f(s|L_\lambda)=\zeta(s|L_\lambda)- {\cal
N}\lambda^{-s}$ is well defined in the limit $\lambda\to0$.
Eq.~(\ref{3.133}) tells us that if $ {\cal N}$ zero-modes are present,
the Selberg $Z(s)$-function has a zero of multiplicity $ {\cal N}$ at
$s=2\rho_N=N-1$, in agreement with the general results reported in
Ref.~\cite{gang77-21-1} and discussed in Sec.~\ref{S:SZF}. However,
the presence of zero modes depends on the characters $\chi$.

Finally, by Eq.~(\ref{gen}) we get the regularized determinant for the
Laplace operator in the form \begin{equation} \det L=\frac1{ {\cal
N}!}\exp[-\gamma K_N(L)-I(0|L)] \,Z^{( {\cal N})}(2\rho_N)\,.
\label{detL-ZM} \end{equation}

\newpage
\setcounter{equation}{0}
\section{Zero temperature quantum properties on ultrastatic manifolds
with constant curvature section}

Here we recall some basic properties concerning quantum field theory in
ultrastatic space-times. It has been proved in
Ref.~\cite{brow82-26-1881} that any conformally flat, ultrastatic
space-time is locally the Minkowski space, the Einstein static
universe or the open Einstein universe, namely $\mbox{$I\!\!R$}\times
H^3$ with the product metric. Actually this holds true in arbitrary
dimensions, in which case we denote by $\cal M^N$ the corresponding
spatial section. The global topology of such spaces however can be
very involved and firstly only simply connected space-times have been
considered. Later, in
Refs.~\cite{isha78-362-383,dowk78-11-2255,bana79-12-2527} field
theories on space-times carrying non-trivial topology have been
considered. When $\cal M^N$ has constant curvature, the spaces
$\mbox{$I\!\!R$}\times\cal M^N$ are all conformally flat. The question
arises whether they are globally hyperbolic, due to the importance
that this property has for the quantum theory. One can easily see that
this cannot be the case if the spatial section is incomplete as a
Riemannian manifold. It is a known fact that any ultrastatic
space-time is globally hyperbolic provided the spatial section is a
complete Riemannian manifold \cite{kay78-62-55}. Furthermore, in the
physical $4$-dimensional case, we have the Fulling-Narcowich-Wald
theorem \cite{full81-136-243}, stating that the Hadamard elementary
solution of the wave equation has the singularity structure of the
Hadamard form. The consequences of this for the renormalization
program are well known. When the spatial section of space-time is
simply connected and maximally symmetric, it can be shown that the
renormalized vacuum stress tensor takes the general form
\begin{eqnarray} T_{ij}=\frac{1}{3}\rho_v[g_{ij}+4\xi_i\xi_j]
\label{genstr} \end{eqnarray} and thus it is  completely determined by
the vacuum energy density. Here $\xi_i$ is the global time-like
killing field which defines the given ultrastatic space-time. Now we
discuss these issues for the three possible ultrastatic geometries but
admitting non trivial global topology.

\subsection{The vacuum energy for massive scalar fields on
$\mbox{$I\!\!R$}\times T^N$}

This is a well studied case, see for example
Ref.\cite{bana79-12-2545}. The regularized vacuum energy for a scalar
field on an ultrastatic manifold can be obtained using
Eq.~(\ref{VEZreg}). Here we have to deal with an operator of the kind
$L_D=-\partial_\tau^2+\L_N$ with $D=N+1$ and $\L_N=-\Delta_N+m^2$
represents an operator acting on fields in $T^N$. As in
Sec.~\ref{S:ZFRR}, we call $r_i$ ($i=1,\dots,N$) the radii of the
circles. Then, using Eqs.~(\ref{ZF-r}), (\ref{VEZreg}) and
(\ref{ZFTN}) for the vacuum energy $E_v$ we obtain, for even $D$
\begin{eqnarray} E_v&=&\nu\left(\frac{-1}{4\pi}\right)^{\frac D2}
\frac{\Omega_Nm^D}{(D/2)!} \left[\ln(m^2\ell^2)-C_{D/2}\right]
\nonumber\\&&\qquad\qquad\qquad\qquad -\frac{\nu\Omega_Nm^D}
{(4\pi)^{\frac N2}\Gamma(\frac{D+1}2)} \sum_{\vec
n\neq0}\int_{1}^{\infty}(u^2-1)^{\frac N2} e^{-2\pi m|\vec n\cdot\vec
r|u}\,du \:,\label{vaene} \end{eqnarray} while for odd $D$
\begin{eqnarray} E_v&=&-\frac{\nu\Omega_N\Gamma(-D/2)m^D}
{(4\pi)^{\frac D2}} -\frac{\nu\Omega_Nm^D} {(4\pi)^{\frac
N2}\Gamma(\frac{D+1}2)} \sum_{\vec
n\neq0}\int_{1}^{\infty}(u^2-1)^{\frac N2} e^{-2\pi m|\vec n\cdot\vec
r|u} \,du \:,\label{enetor}\end{eqnarray} with $C_D=\sum_{k=1}^D1/k$
and $\vec n\in\mbox{$Z\!\!\!Z$}^N$.

Now we specialize this result to a $3$-torus universe and describe the
stress tensor for the untwisted, massive scalar field. With the three
radii of the torus let us form the four vector
$N^i=2\pi(0,r_1,r_2,r_3)$, the integer vector $\vec{n}=(n_1,n_2,n_3)$
and let $\xi^i$ be the globally time-like Killing field of
$\mbox{$I\!\!R$}\times T^3$. In the natural coordinate system
$(t,\vec{x})$, $\vec{x}\in T^3$ it takes the form $\xi^i=(1,0,0,0)$.
Moreover we denote $\varrho^i=(x-y)^i$, $x,y\in\mbox{$I\!\!R$}\times
T^3$. The Hadamard elementary function on the torus is then given as
the images sum \begin{eqnarray}
G_T^{(1)}(x,y)=:<0|\{\phi(x),\phi(y)\}|0>
=\sum_{\vec{n}}G^{(1)}(g_{ij}(\varrho^i+N^i)(\varrho^j+N^j))\,.
\nonumber\end{eqnarray} Here, $\{\phi(x),\phi(y)\}$ is the
anticommutator and $G^{(1)}(x,y)$ is the Hadamard function for flat
Minkowski space (see for example Ref.~\cite{full89b} and references
therein). This function then defines a global vacuum state for the
torus universe. For a minimally coupled scalar field, the quantum
expectation value of the stress tensor in this vacuum is defined by
\begin{eqnarray} T_{ij}=\frac{1}{2}\lim_{y\rightarrow x}
\left[\nabla_{x^i}\nabla_{y^j}-\frac{1}{2}g_{ij}
\left(\nabla_{x^k}\nabla^{y^k}+m^2\right)
\right]G_s^{(1)}(x,y)\nonumber\,, \nonumber\end{eqnarray} where the
Minkowski $(\vec{n}=0)$ term has been subtracted from $G^{(1)}_T(x,y)$
in order to define the singularity free function $G_s^{(1)}(x,y)$. As
a result, the stress tensor takes the form \begin{eqnarray}
T_{ij}=Ag_{ij}+Bg_{i\mu}g_{j\nu}N^{\mu}N^{\nu}
\:,\nonumber\end{eqnarray} in which \begin{eqnarray}
A=-\frac{m^4}{4\pi^2}\sum_{\vec{n\neq0}} \frac{1}{U_n^2}K_2(U_n)
\:,\label{enea} \end{eqnarray} \begin{eqnarray}
B=-\frac{m^6}{4\pi^2}\sum_{\vec{n\neq0}} \left[\frac{3}{U_n^4}K_2(U_n)+
\frac{1}{U_n^3}K^{''}_1(U_n)\right] \:,\nonumber\end{eqnarray} where
$U_n=m\sqrt{g_{ij}N^iN^j}$. For $m=0$, the stress tensor becomes
\begin{eqnarray} T_{ij}=\frac{1}{2\pi^2}\sum_{\vec{n}}
\frac{g_{ij}N^2-4g_{ik}g_{jl}N^{k}N^{l}}
{(g_{\mu\nu}N^{\mu}N^{\nu})^3}\:, \qquad\qquad N^2=:g_{ij}N^iN^j\,.
\label{strtor} \end{eqnarray}

Another important case is a massless field on the equilateral torus
for which $r_1=r_2=r_3$. The stress tensor takes the form announced
above, i.e. \begin{eqnarray}
T_{ij}=\frac{1}{3}\rho_v[g_{ij}+4\xi_i\xi_j]\,, \nonumber\end{eqnarray}
where $\rho_v=\Omega_3^{-1}E_v$ is actually the limit $m\to0$ in
Eq.~(\ref{enetor}) and is also given by Eq.~(\ref{strtor}) by means of
Epstein function (see Appendix~\ref{S:UR}) \begin{eqnarray}
\rho_v=-\frac{1}{2\pi^2}\sum_{\vec{n}} \frac{1}{(g_{ij}N^iN^j)^2}\,.
\nonumber\end{eqnarray} Hence, the (negative) vacuum energy density
determines the stress tensor.

For the non-equilateral torus, one can compute the three principal
pressures $P_{(i)}$, $i=1,2,3$. With a simple calculation we get (here
the sum is excluded) \begin{eqnarray}
P_{(i)}=-\rho_v-r_i\frac{\partial\rho_v}{\partial r_i}\,,
\nonumber\end{eqnarray} where $\rho_v=\Omega_3^{-1}A$, $A$ given by
Eq.~(\ref{enea}). Thus \begin{eqnarray}
\rho_v-\sum_{i=1}^{3}P_{(i)}\geq 0 \nonumber\end{eqnarray} and also in
this case the energy density determines the stress tensor. We see that
the above renormalization ansatz is equivalent to the $\zeta$-function
prescription of removing the first term in Eq.~(\ref{vaene}).

\subsection{The vacuum energy for massive scalar fields on
$\mbox{$I\!\!R$}\times S^N$}

Now we review Einstein-like static universes, which have been
extensively studied, for example, in
Refs.~\cite{ford76-14-3304,dowk84-1-359,camp90-196-1}. As in the
previous section, we consider an operator of the kind
$L_D=-\partial_\tau^2+\L_N$, but now
$\L_N=-\Delta_N+\alpha^2+\kappa\varrho^2_N$ acts on fields in $S^N$.
We have $\kappa\varrho^2_N=\kappa(N-1)^2/4=R(N-1)/2N$. We would like
to compute vacuum energy $E_v^S(\L_N)$ for such systems. As a first
step, we carry on the computation for the two cases in which the
manifold is $\mbox{$I\!\!R$}\times S^1$ and $\mbox{$I\!\!R$}\times
S^2$ with the assumption that $\alpha^2+\kappa\varrho_N^2$ is not
vanishing (no zero-modes). Using Eqs.~(\ref{ZF-r}), (\ref{VEZreg}),
(\ref{ZFS1-I}) and (\ref{ZFs2}) we obtain the results \begin{eqnarray}
E_v^S(\L_1)&=&-\frac{\nu\Omega_1\alpha^2}{4\pi}
\left(\ln(\ell^2\alpha^2)-1\right)
-\frac{2\nu\Omega_1\alpha^2}{\pi}\int_1^\infty
\frac{\sqrt{u^2-1}}{e^{2\pi\alpha ru}-1}\,du
\:,\label{EvS1}\end{eqnarray} \begin{eqnarray}
E_v^S(\L_2)&=&-\frac{\nu\Omega_2\alpha^3}{12}\int_{\Gamma}
\left(1+\frac{|\kappa|z^2}{\alpha^2}\right)^{3/2} \frac{dz}{\cos^2\pi
z} \label{EvS2} \:.\end{eqnarray} where in Eq.~(\ref{EvS1}) $r$ is the
radius of $S^1$.

When $\alpha^2+\kappa\varrho_N^2=0$, zero-modes are present. In this
case, for $S^1$ we have to consider only the right hand side of
Eq.~(\ref{ZFS1-I}), so that the $\zeta$-function becomes
$2r^{2s}\zeta_R(2s)$-function and the vacuum energy
$E_v(-\Delta_{S^1})=2\nu\zeta_R(-1)/r$, while for $S^2$ we have to
exclude the branch point $z=|\kappa|^{1/2}/2$ from the path of
integration in Eq.~(\ref{EvS2}).

Using Eq.~(\ref{ZFRR}) in any $D+2$ dimensional manifold of the kind
$\mbox{$I\!\!R$}\times S^{N+2}$ we get \begin{eqnarray}
\frac{E_v^S(\L_{N+2})}{\Omega_{N+2}} &=&-\frac{1}{2\pi N\Omega_N}
\left[(\alpha^2+\kappa\varrho_N^2)E_v^S(\L_N)
-\nu\zeta^{(r)}(-3/2|\L_N)\right] \:.\nonumber\end{eqnarray} Using
Eq.~(\ref{ZFRR}) we also see that vacuum energies $E_v^S(\L_N)$ are
determined by the values of $\zeta^{(r)}(s|\L_1)$ or
$\zeta^{(r)}(s|\L_2)$ at the points $s=-n/2$, with $n$ any odd integer
smaller than or equal to $N$. Note that $\zeta(s|\L_2)$ has only a
pole at $s=1$ and therefore (as we know) the vacuum energy does not
need renormalization.

For the sake of completeness, we report also the expression of vacuum
energy in $3+1$ dimensions. This is the Einstein static universe first
treated in Ref.~\cite{ford76-14-3304}. It reads \begin{eqnarray}
E_v^S(\L_3)&=&-\frac{\nu\Omega_3\alpha^4}{32\pi^2}
\left(\ln(\ell^2\alpha^2)-\frac32\right)
+\frac{\nu\Omega_3\alpha^{4}}{\pi^2}
\int_1^\infty\frac{u^2\sqrt{u^2-1}} {e^{2\pi\alpha|\kappa|^{-1/2}
u}-1}\,du \:.\label{enesph}\end{eqnarray} The renormalization
procedure employed in Ref.~\cite{ford76-14-3304} amounts, in the
present treatment, to omit the first term in the formula. The
renormalized stress tensor is again given by Eq.~(\ref{genstr}) with
$\rho_v=\Omega_3^{-1}E_R$, where $E_R$ is the last term in
Eq.~(\ref{enesph}).

\subsection{The vacuum energy for massive scalar fields on
$\mbox{$I\!\!R$}\times H^N/\Gamma$}

This case has recently been studied in
Refs.~\cite{byts91-8-2269,cogn92-33-222,byts92-9-1365}. Here
$\L_N=-\Delta_N+\alpha^2+\kappa\varrho_N^2$ acts on fields in
$H^N/\Gamma$. Note that in contrast with the case treated above, now
$R=N(N-1)\kappa$ is negative. First of all we concentrate on the
contribution to vacuum energy due to the identity element of the group
$\Gamma$. We shall indicate it by $E_v^I(\L_N)$. As a first step we
carry on the computation for the two cases in which the manifold is
$\mbox{$I\!\!R$}\times H^3/\Gamma$ and $\mbox{$I\!\!R$}\times
H^2/\Gamma$ and then we extend the results to any dimension by using
the recurrence relations (\ref{RRze}).

In the first case we have the heat coefficients in a closed form and
so, using Eqs.~(\ref{ZF-r}), (\ref{VEZreg}) and (\ref{ze3}) we obtain
\begin{eqnarray} E_v^I(\L_3)&=&-\frac{\nu \Omega( {\cal
F}_3)\alpha^4}{32\pi^2} \left(\ln(\ell^2\alpha^2)-\frac32\right)
\:,\nonumber\end{eqnarray} which differs from the analog expression in
flat space-time for a modification of the mass term by the constant
curvature. In the zeta-function treatment, the usual renormalization
ansatz (see Ref.~\cite{bunc78-18-1844} and references therein) for the
open Einstein universe amounts to remove the entire expression, thus
getting a vanishing vacuum energy. The renormalized stress tensor is
given by Eq.~(\ref{genstr}) with $\rho_v=0$, hence it is zero. The
same conclusion is reached in Ref.~\cite{cand79-19-2902}.

In the second case, since the manifold without boundary has odd
dimension, all odd coefficients of the heat kernel are vanishing. As a
result, there is no $\ell$ ambiguity. Using Eq.~(\ref{ze2}), we get
\begin{eqnarray} E_v^I(\L_2)&=-&\frac{\nu \Omega( {\cal
F}_2)\alpha^3}{6}
\int_{0}^{\infty}\left(1+\frac{|\kappa|r^2}{\alpha^2}\right)^{3/2}
\frac{dr}{\cosh^2\pi r} \:.\nonumber\end{eqnarray}

Now, using Eq.~(\ref{RRze}) in any $D+2$ dimensional manifold of the
kind $\mbox{$I\!\!R$}\times H^{N+2}/\Gamma$ we have \begin{eqnarray}
\frac{E_v^I(\L_{N+2})}{\Omega( {\cal F}_{N+2})} &=&-\frac{1}{2\pi N
\Omega( {\cal F}_N)} \left[(\alpha^2+\kappa\varrho_N^2)E_v^I(\L_N)
-\nu\zeta_I^{(r)}(-3/2|\L_N)\right] \:,\nonumber\end{eqnarray} from
which we see that in order to compute the contribution to vacuum
energy due to the identity of the group $\Gamma$, it is sufficient to
know $\zeta(s|\L_2)$ and $\zeta(s|\L_3)$ at the negative half-integers.

The topological contribution due to hyperbolic elements only, can be
easily expressed in terms of Selberg $\Xi$-function making use of
Eq.~(\ref{ZF-H-Z}). One has \begin{eqnarray}
E_v^H(\L_N)=-\frac{\nu\alpha^2}{\pi|\kappa|^{1/2}} \int_1^\infty
\sqrt{u^2-1}\; \Xi_N(u\alpha|\kappa|^{-1/2}+\varrho_N)\;du
\:.\label{EvN-H}\end{eqnarray} Due to the properties of Selberg
$\Xi$-function, this contribution to vacuum energy is always negative
if one is dealing with untwisted fields, i.e. $\chi=1$
\cite{cogn92-33-222,byts92-9-1365}.

Finally, in 4 ($N=3$) and 3 ($N=2$) dimensions we are able to evaluate
also the contribution to vacuum energy due to elliptic elements of the
$\Gamma$ isometry group. In fact, using Eqs.~(\ref{ZF3-E}) and
(\ref{ZF2-E}) we obtain \cite{byts92-33-3108} \begin{eqnarray}
E_v^E(\L_3)=-\frac{\nu E\alpha^2}{4\pi}
\left(\ln(\ell^2\alpha^2)-\frac{1}{2}\right)\,, \nonumber\end{eqnarray}
\begin{eqnarray} E_v^E(\L_2)=4\nu\alpha\int_0^\infty
\sqrt{1+\frac{|\kappa|r^2}{\alpha^2}}\; E_2(r)\;dr
\nonumber\end{eqnarray} and we see that the above renormalization
ansatz is not sufficient to get rid of all the arbitrary scale
dependences of the vacuum energy.

In the conformal case ($\alpha=0$) the vacuum energy is trivially
vanishing (in 3+1-dimensions). Note however that if $\lambda=1$ is an
eigenvalue of $\L_3$, the limit is trickier since the integrand in
Eq.~(\ref{EvN-H}) has a simple pole just at $\alpha=0$.

\subsection{The self-interacting scalar field} \label{S:SISF-R}

Here we concentrate our attention on a self-interacting scalar field
defined on an ultrastatic space-time $\cal M^4$ in which the spatial
section is a manifold with constant curvature. For the sake of
generality, we shall derive a general expression for the one-loop
effective potential on such kind of space-times and then we shall
discuss in detail the case in which the spatial section is
$H^3/\Gamma$, the discrete group of isometries containing hyperbolic
elements only. When $\Gamma$ contains elliptic elements, conical-like
singularities appear in the manifold, and in this case the
renormalization becomes quite delicate. For a discussion of similar
situations see Ref.~\cite{cogn94-49-1029}. We analyze the possibility
of symmetry breaking (or symmetry restoration) due to topology. We
shall see that the sign of the square of the topological mass will
depend on the character of the field, being positive for trivial
character (untwisted fields-possible symmetry restoration). For
twisted fields there could exist a mechanism for symmetry breaking.
Here we shall use $\zeta$-function regularization but, as we already
have pointed out, other regularizations are equally good.

We start with the classical Euclidean action for the scalar field
\begin{eqnarray} S[\phi,g]=\int_{{\cal
M}^4}\left[-\frac12\phi\Delta\phi +V_c(\phi,R)\right]\sqrt{g}d^4x
\:,\label{Sc}\end{eqnarray} where the classical potential reads
\begin{eqnarray} V_c(\phi,R)=\frac{\lambda\phi^4}{24}
+\frac{m^2\phi^2}2+\frac{\xi R\phi^2}2 \:.\nonumber\end{eqnarray} From
Eq.~(\ref{Sc}) we obtain the small disturbance operator
\begin{eqnarray} L=-\Delta+V_c''(\phi_c,R)= -\Delta+m^2+\xi
R+\frac{\lambda\phi_c^2}{2} \:,\nonumber\end{eqnarray} the prime
representing the derivative with respect to $\phi$.

Using $\zeta$-function regularization and Eq.~(\ref{OLEP}), we write
the one-loop quantum corrections to the effective potential in the
convenient form \begin{eqnarray} V^{(1)}(\phi_c,R)&=&-\frac1{2V(\cal
M)}\zeta'(0|L\ell^2)\nonumber\\ &=&\frac1{64\pi^2}\left[
2a_2\ln(M^2\ell^2)-\frac32M^4\right] +f(\phi_c,R)
\:,\nonumber\end{eqnarray} where $a_2$ is the spectral coefficient
defined in Appendix~\ref{S:SC} and $M^2=-a_1$ is a function of
(constant) $\phi_c$ and $R$. We have introduced the quantity
\begin{eqnarray}
f(\phi_c,R)=\frac{3M^4}{128\pi^2}-\frac{\zeta'(0|L/M^2)}{2V(\cal M)}
\:,\nonumber\end{eqnarray} which does not depend on the
renormalization parameter $\ell$. Its utility shall become clear in
the following.

Specializing Eqs.~(\ref{a1}) and (\ref{a2}) to ultrastatic manifolds
with constant curvature spatial section and recalling that, since we
are computing the effective potential, $V(x)=V''_c(\phi_c,R)$  is a
constant, we obtain \begin{eqnarray} V^{(1)}(\phi_c,R)=\frac1{64\pi^2}
\left[\left( M^4-\frac{c_0R^2}2\right) \ln(M^2\ell^2)-\frac32
M^4\right]+f(\phi_c,R) \:,\nonumber\end{eqnarray} with \begin{eqnarray}
M^2=V_c''(\phi_c,R)-\frac R6= m^2+\left(\xi-\frac16\right)
R+\frac{\lambda\phi_c^2}{2} \:.\nonumber\end{eqnarray} In this way,
only $c_0$ and $f(\phi_c,R)$ depend on the chosen manifold. Note that
only $R^2$ appears in the expression for $a_2$, since for the
manifolds we are considering, all curvature invariants are
proportional to $R$. For a more general treatment see
Ref.~\cite{cogn94-50-909}.

It follows that the one-loop quantum corrections to the classical
action generate quadratic terms in the curvature  \cite{utiy62-3-608},
so for the gravitational contribution we have to take the expression
\begin{eqnarray} V_g=\Lambda+\alpha_1R+\frac{\alpha_2 R^2}2
\nonumber\end{eqnarray} and we must add to the classical potential a
counterterm contribution, which reflects the structure of the one-loop
divergences. It has the form \begin{equation} \delta
V(\phi_c,R)=\delta\Lambda+\frac{\delta\lambda\phi_c^4}{24}
+\frac{\delta m^2\phi_c^2}{2}+\frac{\delta\xi R\phi_c^2}{2}
+\delta\alpha_1R+\frac{\delta\alpha_2R^2}{2} \:,\nonumber\end{equation}
$\Lambda\sim0$, $\alpha_1$ and $\alpha_2$ being the cosmological and
coupling constants respectively. As a result, the renormalized
one-loop effective potential reads \begin{eqnarray}
V_{eff}=V_c+V_g+V^{(1)}+\delta V=V_c+V_g+V^{(1)}_{eff}
\:.\nonumber\end{eqnarray}

When it will be convenient, we shall denote by
$\delta\eta_q\equiv(\delta\lambda,\delta
m^2,\delta\xi,\delta\alpha_1,\delta\alpha_2)$ the set of whole
counterterm coupling constants.

\subsubsection{Renormalization conditions.} The quantities
$\delta\eta_q$, which renormalize the coupling constants, are
determined by the following renormalization conditions (see for
example Refs.~\cite{hu84-30-743,berk92-46-1551}) \begin{eqnarray}
\frac{\partial^4V^{(1)}_{eff}(\phi_1,R_1)}{\partial\phi_c^4}
=0\:,\qquad\qquad \frac{\partial V^{(1)}_{eff}(0,0)}{\partial R}
=0\:,\nonumber\end{eqnarray} \begin{eqnarray}
\frac{\partial^2V^{(1)}_{eff}(0,0)}{\partial\phi_c^2} =0\:,\qquad\qquad
\frac{\partial^2V^{(1)}_{eff}(\phi_5,R_5)}{\partial R^2}
=0\:,\label{RC}\end{eqnarray} \begin{eqnarray}
\frac{\partial^3V^{(1)}_{eff}(\phi_3,R_3)}{\partial R\partial\phi_c^2}
=0\:,\qquad\qquad V^{(1)}_{eff}(\phi_0,0) =0\nonumber\:.\end{eqnarray}
The latter equation is equivalent to say that the the cosmological
constant is equal to $V_c(\phi_0,0)$, $\phi_0=<\phi_c>$ being the mean
value of the field. The choice of different points
$(\phi_c,R)\equiv(\phi_q,R_q)$ in the renormalization conditions which
define the physical coupling constants, is due to the fact that in
general they are measured at different scales, the behaviour with
respect to a change of scale being determined by the renormalization
group equations (\ref{IRGE}), discussed in Sec.~\ref{S:OLEA}. By a
suitable choice of $(\phi_q,R_q)$, Eqs.~(\ref{RC}) reduce to the
conditions chosen by other authors (see for example
Refs.~\cite{hu84-30-743,berk92-46-1551}).

{}From such conditions, after straightforward calculations we get the
counterterms \begin{eqnarray}
64\pi^2\left(\delta\Lambda+f(\phi_0,0)\right) &=& 64\pi^2\left[
\left(\frac{\partial^2f(0,0)}{\partial\phi_c^2}
-m^2\right)\frac{\phi_0^2}{2}+\left(
\frac{\partial^4f(\Phi_1)}{\partial\phi_c^4}
-\lambda\right)\frac{\phi_0^4}{24}\right]\nonumber\\
&&\qquad\qquad-\lambda m^2\phi_0^2\left[1-   \ln(m^2\ell^2)\right]
+M_0^4\left[\frac32-\ln(M_0^2\ell^2)   \right]\nonumber\\
&&\qquad\qquad\qquad\qquad
+\frac{\lambda^2\phi_0^4}{4}\ln(M_1^2\ell^2)
-\frac{\lambda^4\phi_0^4\phi_1^4}{12M_1^4}
+\frac{\lambda^3\phi_0^4\phi_1^2}{2M_1^2} \:,\nonumber\end{eqnarray}
\begin{eqnarray} 64\pi^2\left[\delta\lambda
+\frac{\partial^4f(\phi_1,R_1)}{\partial\phi_c^4}\right]
&=&-6\lambda^2\ln(M_1^2\ell^2)   +\frac{2\lambda^4\phi_1^4}{M_1^4}
-\frac{12\phi_1^2\lambda^3}{M_1^2} \nonumber\:,\end{eqnarray}
\begin{eqnarray} 64\pi^2\left[\delta m^2
+\frac{\partial^2f(0,0)}{\partial\phi_c^2}\right] &=&2\lambda
m^2\left[1-\ln(m^2\ell^2)\right] \nonumber\:,\end{eqnarray}
\begin{eqnarray} 64\pi^2\left[\delta\xi
+\frac{\partial^3f(\phi_3,R_3)}{\partial R\partial\phi_c^2}\right]
&=&-2\lambda\left(\xi-\frac16\right)    \ln(M_3^2\ell^2)
-\frac{2\lambda^2\left(\xi-\frac16\right)
\phi_3^2}{M_3^2}\nonumber\\
&&\qquad\qquad\qquad\qquad+\frac{c_0\lambda R_3^2}{M_3^2}
\left(1-\frac{\lambda\phi_3^2}{M_3^2}\right) \:,\nonumber\end{eqnarray}
\begin{eqnarray} 64\pi^2\left[\delta\eta_4   +\frac{\partial
f(0,0)}{\partial R}\right] &=& 2m^2\left(\xi-\frac16\right)
\left[1-\ln(m^2\ell^2)\right] \nonumber\:,\end{eqnarray}
\begin{eqnarray} 64\pi^2\left[\delta\eta_5
+\frac{\partial^2f(\phi_5,R_5)}{\partial R^2}\right]
&=&-\left[2\left(\xi-\frac16\right)^2-c_0\right]    \ln(M_5^2\ell^2)+
  \frac{2\left(\xi-\frac16\right)c_0R_5^2}{M_5^2}
\nonumber\:,\end{eqnarray} where we introduced the constants
$M_n^2=m^2+\left(\xi-\frac16\right)R_n+\frac{\lambda}2 \phi_n^2$.

The renormalized one-loop contribution to the effective potential
looks very complicated. We write it in the form \begin{eqnarray}
64\pi^2V_{eff}^{(1)}&=&
-32\pi^2m^2\phi_0^2-\frac{8\pi^2\lambda\phi_0^4}{3} +\lambda
m^2\phi_0^2\left(\ln\frac{m^2}{M_0^2} +\frac12\right) \nonumber\\
&&\qquad\qquad-\frac{\lambda^2\phi_0^4}4
\left[\ln\frac{M_0^2}{M_1^2}-\frac32
-\frac{4(M_1^2-m^2)(2M_1^2+m^2)}{3M_1^4}\right]\nonumber\\
&&\hspace{-1.5cm}+m^4\ln\frac{M^2}{M_0^2} +2m^2\left(\xi-\frac16\right)
R\left(\ln\frac{M^2}{m^2}-\frac12\right)
+\left(\xi-\frac16\right)^2R^2\left(\ln\frac{M^2}{M_5^2}
-\frac32\right)\nonumber\\ &&+\left\{\left(\xi-\frac16\right)
R\left[\ln\frac{M^2}{M_3^2}-\frac32
-\frac{\lambda\phi_3^2}{M_3^2}\right]
+m^2\left[\ln\frac{M^2}{m^2}-\frac12\right]\right\}
\lambda\phi_c^2\nonumber\\
&&+\left\{\left(\ln\frac{M^2}{M_1^2}-\frac{25}6\right)
+\frac{4m^2(m^2+M_1^2)}{3M_1^4}\right\}
\frac{\lambda^2\phi_c^4}4\nonumber\\ &&-c_0\left\{\frac{\lambda
R_3^2}{2M_3^2}     \left(\frac{\lambda\phi_3^2}{M_3^2}-1\right) R
+\left[\ln\frac{M^2}{M_5^2}-
\frac{2\left(\xi-\frac16\right)R_5^2}{M_5^2}
\right]\frac{R^2}2\right\}\nonumber\\ &&+64\pi^2F(f)
\:,\nonumber\end{eqnarray} where $F(f)$ contains all terms depending
on $f(\phi_c,R)$. It reads \begin{eqnarray}
F(f)&=&f(\phi_c,R)-f(\phi_0,0)
+\frac{\partial^4f(\phi_1,R_1)}{\partial\phi_c^4}
\frac{\phi_c^4-\varphi^4_0}{24}
+\frac{\partial^2f(0,0)}{\partial\phi_c^2} \frac{\phi_c^2-\varphi^2_0}2
\nonumber\\&&\qquad\qquad
-\frac{\partial^3f(\phi_3,R_3)}{\partial\phi_c^2\partial R}
\frac{R\phi_c^2}2-\frac{\partial f(0,0)}{\partial R}R
-\frac{\partial^2f(\phi_5,R_5)}{\partial R^2}\frac{R^2}2
\:.\label{Ff}\end{eqnarray} We see that the computation of the
one-loop effective Lagrangian reduces to the determination of the
function $f(\phi_c,R)$.

If we consider a flat space-time and take the limit $m\to0$, of course
we get the Coleman-Weinberg result \cite{cole73-7-1888}
\begin{eqnarray} V_{eff}&=&\frac{\lambda\phi_c^4}{24}
+\frac{\lambda^2\phi_c^4}{256\pi^2}
\left(\ln\frac{\lambda\phi_c^2}{2M_1^2}-\frac{25}6\right)
\:.\nonumber\end{eqnarray}

What is relevant for the discussion of the phase transition of the
system, is the second derivative of $V_{eff}$ with respect to the
background field $\phi_c$. Then we define \begin{eqnarray}
V_{eff}=\Lambda_{eff}(R)+\frac{m^2_{eff}(R)\phi_c^2}2+O(\phi_c^4) \:,
\label{veff} \end{eqnarray} where $\Lambda_{eff}(R)$ and
$m^2_{eff}(R)$ are complicated expressions not depending on $\phi_c$.
They contain curvature and topological contributions to $\Lambda$ and
$m^2$. By a straightforward computation we get \begin{eqnarray}
m^2_{eff}&=&m^2+\xi R\nonumber\\&& +\frac{\lambda}{32\pi^2}
\left\{m^2\ln \frac{m^2+\left(\xi-\frac16\right)R}{m^2}
+\left(\xi-\frac16\right)R
\left[\ln\frac{m^2+\left(\xi-\frac16\right)R}{M_3^2}
-\frac{\lambda\phi_3^2}{M_3^2}-1\right] \right\}\nonumber\\&&
-\frac{\lambda c_0R^2} {128\pi^2\left[
m^2+\left(\xi-\frac16\right)R\right]}
+\frac{\partial^2f(0,R)}{\partial\phi_c^2}
-\frac{\partial^2f(0,0)}{\partial\phi_c^2}
-\frac{\partial^3f(0,R_3)}{\partial\phi_c^2\partial R}R
\:.\label{meff}\end{eqnarray} When $\phi_c=0$ is a minimum for the
classical potential, Eqs.~(\ref{veff}) and (\ref{meff}) are useful in
the discussion of phase transition.

\subsubsection{The ultrastatic space-time $\mbox{$I\!\!R$}\times
H^3/\Gamma$.} \label{S:OLEP}

For this case we shall give a detailed treatment and in particular, we
shall analyze the possibility of symmetry breaking due to topology. At
the end of the section, other examples shall be briefly discussed.

For these kind of manifolds we have $c_0=0$ ($a_2=M^4/2$) and
moreover, as one can easily check by using Eqs.~(\ref{ZFfact}) and
(\ref{ze2}), the identity of the isometry group $\Gamma$ does not give
contributions to the function $f(\phi_c,R)$. This means that it has
only a topological contribution given by \begin{eqnarray}
V_{top}(\phi_c,R)=f(\phi_c,R)= -\frac{M^2|\kappa|^{-1/2}}{2\pi \Omega(
{\cal F}_3)} \int_{1}^{\infty}\sqrt{u^2-1}\,\Xi_3
\left(1+uM|\kappa|^{-1/2}\right)\,du \:,\label{Vtop}\end{eqnarray}
where we have used Eqs.~(\ref{ZFfact}) and Eq.~(\ref{ZF-H-Z}) with
$\alpha=M$. The latter equation represents the unrenormalized
topological contribution to the one-loop effective potential. Using
the properties of the $\Xi$-function one can see that for $M>0$,
$V_{top}$ is exponentially vanishing when $\kappa=R/6$ goes to zero.
Then, from Eq.~(\ref{Ff}) we obtain \begin{eqnarray}
F(f)&=&V_{top}(\phi_c,R)
+\frac{\partial^4V_{top}(\phi_1,R_1)}{\partial\phi_c^4}
\left(\frac{\phi_c^4-\phi_0^4}{24}\right) \nonumber\\&&\qquad\qquad
-\frac{\partial^3V_{top}(\phi_3,R_3)} {\partial\phi_c^2\partial R}
\frac{R\phi_c^2}2 -\frac{\partial^2V_{top}(\phi_5,R_5)} {\partial
R^2}\frac{R^2}2 \label{OLEPtop}\end{eqnarray} and this is the
renormalized topological contribution to the one-loop effective
potential. We also have from Eq.~(\ref{meff}) \begin{eqnarray}
m^2_{eff}&=&m^2+\xi R+m^2_{top} +\frac{\lambda}{32\pi^2} \left\{m^2\ln
\frac{m^2+\left(\xi-\frac16\right)R}{m^2}\right.\nonumber\\
&&\left.\qquad\qquad\qquad\qquad+\left(\xi-\frac16\right)R
\left[\ln\frac{m^2+\left(\xi-\frac16\right)R}{M_3^2}
-\frac{\lambda\phi_3^2}{M_3^2}-1\right]\right\}
\:,\nonumber\end{eqnarray} where the topological contribution
$m^2_{top}$ has been introduced. To simplify the discussion, from now
on we choose $R_1=R_3=R_5=0$ like in Ref.~\cite{berk92-46-1551}. In
this way $m^2_{top}$ assumes the form \begin{eqnarray}
m^2_{top}=\frac{\sqrt{3}\lambda|R|^{-1/2}} {4\pi \Omega({\cal F}_3)}
\int_1^{\infty} \Xi_3(1+u\sqrt{6m^2|R|^{-1}+1-6\xi})\:
\frac{du}{\sqrt{u^2-1}} \:.\nonumber\end{eqnarray}

\paragraph{The symmetry breaking.} For the discussion on the physical
implications of the one-loop effective potential, let us specialize to
different cases and to several limits.

Let us first concentrate on the regime $m^2+\xi R>0$. Then $\phi_c=0$
is a minimum of the classical potential and we can use expansion
(\ref{veff}) to carry on the analysis. We consider the small and large
curvature limits separately. As mentioned before, the topological
contribution is negligible when $R\to0$ and so the leading orders to
$m^2_{eff}$ only including linear curvature terms are easily obtained
from Eq.~(\ref{meff}). We have \begin{equation} m^2_{eff}(R)=m^2+\xi R
-\frac{\lambda\left(\xi-\frac16\right) R}{32\pi^2}
\left(\frac{\lambda\phi_3^2}{M_3^2} -\ln\frac{m^2}{M_3^2}\right)
\:.\nonumber\end{equation} This result is effectively true for any
smooth Riemannian manifold. Due to $R<0$ in our example, for
$\xi<\frac16$ the one loop term will help to break symmetry, whereas
for $\xi>\frac16$ the quantum contribution acts as a positive mass and
helps to restore symmetry. The quantum corrections should be compared
to the classical terms which in general dominate, because the one loop
terms are suppressed by the square of the Planck mass. But if $\xi$ is
small enough the one loop term will be the most important and will
then help to break symmetry.

Let us now concentrate on the opposite limit, that is $|R|\to\infty$
with the requirement $\xi<0$. In that range the leading order of the
topological part reads \begin{eqnarray}
m^2_{top}=\frac{\sqrt{3}\lambda|R|^{-1/2}}{4\pi \Omega({\cal F}_3)}
\int_1^{\infty}\Xi_3(1+u\sqrt{1-6\xi})\, \frac{du}{\sqrt{u^2-1}}
+O(R^0)\,. \nonumber\end{eqnarray} It is linear in the scalar
curvature $R$ because $\Omega({\cal F}_3)\sim |R|^{-3/2}$ . The sign
of the contribution depends on the choice of the characters
$\chi(\gamma))$. For trivial character $\chi=1$ we can say that the
contribution helps to restore symmetry, whereas for nontrivial
character it may also serve as a symmetry breaking mechanism. So, if
the symmetry is broken at small curvature, for $\chi=1$ a symmetry
restoration at some critical curvature $R_c$, strongly depending on
the non trivial topology, will take place.

Finally, let us say some words about the regime $m^2+\xi R<0$, which
includes for example the conformal invariant case. The classical
potential has two minima at $\pm\sqrt{6|m^2+\xi R|/\lambda}$. So even
for $\xi>0$, due to the negative curvature in the given space-time,
the classical starting point is a theory with a broken symmetry. As is
seen in the previous discussion, in order to analyze the influence of
the quantum corrections on the symmetry of the classical potential, a
knowledge of the function $\Xi_3(s)$ for values $\,\mbox{Re}\, s<2$ is
required. But $\Xi_3(s)$ has a simple pole at $s=2$
\cite{venk79-34-79}. As may then be seen in equation (\ref{Vtop}), the
topological contribution contains a part which behaves like
$\sqrt{m^2+\xi R}$ near $m^2+\xi R=0$ (resulting from the range of
integration near $u=1$). So for $m^2+\xi R<0$ the effective potential
becomes complex, which reflects the well known failure of the loop
expansion in this range of parameters \cite{dola74-9-3320,rive87b}.

\paragraph{Some remarks.} In new inflationary models, the effective
cosmological constant is obtained from an effective potential, which
includes quantum corrections to the classical potential of a scalar
field \cite{cole73-7-1888}. This potential is usually calculated in
Minkowski space-time, whereas to be fully consistent, the effective
potential must be calculated for more general space-times. For that
reason an intensive research has been dedicated to the analysis of the
one-loop effective potential of a self-interacting scalar field in
curved space-time. Especially to be mentioned are the considerations
on the torus
\cite{ford80-21-933,toms80-21-2805,dena80-58-243,acto90-7-1463,eliz90-244-387},
which however do not include nonvanishing curvature, and the
quasi-local approximation scheme developed in Ref.~\cite{hu84-30-743}
(see also Ref.~\cite{kirs93-48-2813}), which however fails to
incorporate global properties of space-time. To overcome this
deficiency, we were naturally led to the given considerations on the
space-time manifold $\mbox{$I\!\!R$}\times H^3/\Gamma$. Apart from its
physical relevance \cite{elli71-2-7}, this manifold combines
nonvanishing curvature with highly nontrivial topology, still
permitting the exact calculation of the one-loop effective potential
by the use of the Selberg trace formula. So, at least for $m^2+\xi
R>0$ and trivial line bundles ($\chi=1$), we were able to determine
explicitly the influence of the topology, namely the tendency to
restore the symmetry. Furthermore, this contribution being
exponentially damped for small curvature, we saw, in that regime, that
for $\xi<\frac16$ the quantum corrections to the classical potential
can help to break symmetry.

\subsubsection{The ultrastatic space-times $\mbox{$I\!\!R$}\times
T^3$, $\mbox{$I\!\!R$}\times S^3$, $\mbox{$I\!\!R$}^2\times S^2$,
$\mbox{$I\!\!R$}^2\times H^2/\Gamma$}

For the sake of completeness we shall consider also these examples,
but for all of them we shall simply give the value of $c_0$ and the
explicit expression of the function $f(\phi_c,R)$, which is what one
needs for the evaluation of $V_{eff}$.

\paragraph{Example: $\mbox{$I\!\!R$}\times T^3$.} For simplicity we
consider an equilateral torus $T^3$ with radius $r$. $\cal M^4$ is a
flat manifold and so we have $R=0$ in all formulae. Using
Eqs.~(\ref{ZFfact}) and (\ref{ZFTN}) for $f(\phi_c)$ we get
\begin{eqnarray} f(\phi_c)=\frac{M^4}{3\pi^2}
\int_1^{\infty}\sqrt{u^2-1}\, \sum_{\vec n\in\mbox{$Z\!\!\!Z$}^3,\vec
n\not=0} e^{-2\pi Mr|\vec n|u}\,du \:.\nonumber\end{eqnarray}

\paragraph{Example: $\mbox{$I\!\!R$}\times S^3$.} It is easy to see
that if the spatial section is a maximally symmetric space, then
$a_2=M^4/2$, that is $c_0=0$. Again, using Eqs.~(\ref{ZFfact}) and
(\ref{ZFS3}) with $\alpha=M>0$ we obtain \begin{eqnarray}
f(\phi_c,R)=\frac{M^4}{2\pi^2} \int_{1}^{\infty}\frac{u^2\sqrt{u^2-1}}
{e^{2\pi M|\kappa|^{-1/2}u}-1}\,du \:,\nonumber\end{eqnarray} where
$\kappa=R/6$.

\paragraph{Example: $\mbox{$I\!\!R$}^2\times S^2$.} For this case we
have $c_0=1/180$. Using Eqs.~(\ref{ZFfact}) and (\ref{ZFS2}) with
$\alpha^2=M^2+\kappa/12=M^2+R/24$, we get \begin{eqnarray}
f(\phi_c,R)&=&\frac1{64\pi^2}\left\{ -\frac{\kappa^2}{30}+\pi M^4
\int_{0}^{\infty}\left[1-\frac{\kappa(u^2-1/12)}{M^2}\right]^2
\right.\nonumber\\&&\qquad\qquad\qquad\qquad\left.\times\,
\ln\left|1-\frac{\kappa(u^2-1/12)}{M^2}\right| \,\frac{du}{\cosh^2\pi
u} \right\}\:.\label{EPS2}\end{eqnarray}

\paragraph{Example: $\mbox{$I\!\!R$}^2\times H^2/\Gamma$.} We again
have $c_0=1/180$. Using Eqs.~(\ref{ZFfact}) and (\ref{ze2}) with
$\alpha^2=M^2+\kappa/12=M^2+R/24$, the identity contribution is given
by Eq.~(\ref{EPS2}), however  now $\kappa<0$. If the isometry group
$\Gamma$ contains only hyperbolic elements, then topological
contributions  can be easily computed using Eq.~(\ref{ZF-H-Z}). As a
result \begin{eqnarray} f(\phi_c,R)&=&\frac1{64\pi^2}\left\{
-\frac{\kappa^2}{30}+\pi M^4
\int_{0}^{\infty}\left[1-\frac{\kappa(u^2-1/12)}{M^2}\right]^2
\right.\nonumber\\&&\qquad\qquad\qquad\qquad\left.\times\,
\ln\left|1-\frac{\kappa(u^2-1/12)}{M^2}\right| \,\frac{du}{\cosh^2\pi
u} \right\}\nonumber\\&& -\frac{\left(
M^2+k/12\right)^{3/2}|\kappa|^{-1/2}}{8\pi \Omega( {\cal F}_2)}
\int_{1}^{\infty}(u^2-1)\,\Xi_2
\left(\frac12+u\sqrt{\frac{M^2}{\kappa}+\frac1{12}}\right)\,du
\:.\nonumber\end{eqnarray}

\newpage
\setcounter{equation}{0}
\section{Quantum $p$-branes in curved space-times}

It is well known the great interest that has been recently arisen in
regarding string theory as the Theory of Everything, including
possibly also the gravitational interaction \cite{gree87b}. In spite
of his successes, soon after it has been proposed a generalization to
higher dimensional extended objects, the $p$-branes  (strings for
$p=1$, membranes for $p=2$, etc.).

There are several reasons for considering relativistic extended
objects or $p$-branes ($p>2$). To begin with, they are the natural
generalizations of strings. Furthermore the  study of
(super)$p$-branes is a way to better understand, in particular,
(super)strings. By the simultaneous reduction of the world-volume and
space-time of a supermembrane model (modified in an appropriate way in
order to account for the 11-dimensional supergravity background) it is
possible to derive the type II A superstring in 10-dimensions
\cite{duff88-5-189}. Furthermore, the field theory limit of the
underlying string theory is just the supergravity theory which links
string and particle physics at low energy. In fact
(D=11,N=1)-supergravity cannot be obtained from string theories, but
it can be obtained from suitable $p$-brane models. These are some of
the reasons advocated in starting the study of these higher
dimensional extended objects.

The interest in the quantum theory of (super)$p$-branes, has also been
motivated by their mathematical structure as well as  by their
possible significance for the unification of fundamental interactions
\nocite{kikk86-76-1379,fuji87-199-75,mezi88-309-317,lind88-3-2401,duff89-6-1577,barc89-228-193,amor90-5-2667,barc90-245-26,odin90-7-1499}.
One of the central issues is the nature of the mass spectrum in the
effective low energy limit. In order to investigate the massive state
one needs the knowledge of the quantum properties of the theory.
Unfortunately, within the quantum theory of $p$-branes, there exist
many unsolved questions up to now. The main difficulty is a
consequence of the inherent non-linearity of the field equations as
well as the related quantization of the theory.

In the following, we shall discuss some aspects of quantization of
extended objects. These brief considerations do not claim to be a
complete description and the aspects of the quantization which will be
considered, reside mainly in the interests of the authors. As far as
the issue of quantization is concerned, we say again that the main
problems come from the non-linearity of $p$-branes models as well as
the non-renormalizability in $(p+1)$-dimensions. Also the connection
between the dimension of the embedding space ($D$) and the quantum
consistency has to be stressed. We shall aside the problem of the
renormalizability and we shall concentrate only on some aspects of the
quantizations of these systems, which have some similarity with string
theory.

\subsection{Classical properties of (super)$p$-branes}

To begin with we recall some properties of $p$-branes. There are
several proposals for the $p$-brane action, which are equivalent at
the classical level. The most popular among them are the Dirac
\cite{dira62-268-7}, Howe-Tucker \cite{howe77-10-155} and conformally
invariant actions \cite{duff89-6-1577,alve88-7-395} which look as
follows \begin{equation} S_D=k\int\left[ -\det
(\partial_iX^{\mu}\partial_jX^{\nu} g_{\mu\nu})\right]^{\frac{1}{2}}
d^{p+1}{\xi}\,, \label{1a} \end{equation} \begin{equation}
S_{HT}=\frac{k}{2}\int \sqrt{|\gamma|}\left[
\gamma^{ij}\partial_iX^{\mu} \partial_jX^{\nu}g_{\mu\nu}-(p-1)\right]
d^{p+1}{\xi}\,, \label{2a} \end{equation} \begin{equation} S_C=k\int
\sqrt{|\gamma|}\left( \frac{1}{p+1}{\gamma}^{ij}\partial_i
X^{\mu}\partial_jX^{\nu}g_{\mu\nu} \right)^{\frac{p+1}{2}}
d^{p+1}{\xi} \label{553} \end{equation} respectively. Here $k$ is the
$p$-brane tension parameter with dimensions of $(mass)^{p+1}$ and
$\gamma$ is the determinant of the world metric
$\gamma_{ij}=\partial_iX^{\mu}\partial_jX^{\nu}g_{\mu\nu} $. The
indices $i,j,...$ label the coordinates $ \xi^i$  on the world volume
(one time-like, $p$ space-like) and the fields $X^{\mu}
(\mu=0,...,D-1)$ are the coordinates of space-time. The equations  of
motion derived from action (\ref{1a}) are \begin{eqnarray}
\partial_i(\sqrt{\gamma}\gamma^{ij}\partial_jX^{\nu}g_{\mu\nu})=
\frac{1}{2}\sqrt{\gamma}\gamma^{ij}\partial_iX^{\nu}\partial_jX^{\lambda}
\frac{\partial g_{\nu\lambda}}{\partial X^{\mu}} \:,\qquad\qquad
\gamma^{im}\gamma_{jm}=\delta^i_j\,. \nonumber\end{eqnarray} The Dirac
action enjoys general coordinate (reparametrization) invariance.
Therefore the Hamiltonian of the model is a linear combination of the
$(p+1)$ first class constraints \cite{henn83-120-179} \begin{eqnarray}
H_T= P^{\mu}P_{\mu}+k^2 G \:,\qquad\qquad H_i= P_{\mu} \partial_i
X^{\mu}\,, \nonumber\end{eqnarray} where $G$ is the determinant of the
space-like part of the metric
$\gamma_{ij}$.

For the Howe-Tucker action (\ref{2a}) we have equations equivalent to
the above ones. It is clear that the world volume  cosmological
constant $ \Lambda=p-1$ vanishes only for the special  case $p=1$.
This means  that only the string action is invariant under the
conformal symmetry. Finally, the equations of motion generated by the
conformally invariant action (\ref{553}) are \begin{eqnarray}
(p+1)\sqrt{\gamma}\gamma^{ij}
(\gamma^{kl}\partial_kX^{\mu}\partial_lX^{\nu}
g_{\mu\nu})^{\frac{p+1}{2}}&=&\nonumber\\&&\hspace{-2cm}
(\gamma^{kl}\partial_kX^{\mu}
\partial_lX^{\nu}g_{\mu\nu})^{\frac{p-1}{2}}
\sqrt{\gamma}\gamma^{im}\gamma^{jn}
\partial_mX^{\rho}\partial_nX^{\sigma}g_{\rho\sigma}
\,,\nonumber\end{eqnarray} \begin{eqnarray} \partial_i\left(
\sqrt{\gamma} (\gamma^{kl}\partial_kX^{\mu}\partial_lX^{\sigma}
g_{\mu\sigma})^{\frac{p-1}{2}}
\gamma^{ij}\partial_jX^{\nu}g_{\mu\nu}\right)&=&
\nonumber\\&&\hspace{-3cm} \frac{\sqrt\gamma}{2}\left(
\frac{\gamma^{kl}}{p+1}\partial_kX^{\rho}
\partial_lX^{\sigma}g_{\rho\sigma}
\right)^{\frac{p-1}{2}}\gamma^{ij}\partial_i
X^{\nu}\partial_jX^{\lambda} \frac{\partial g_{\nu\lambda}}{\partial
X^{\mu}} \,.\nonumber\end{eqnarray} In fact, the latter equations are
equivalent to the other two ones provided that $ \gamma_{ij}$ is a
function of the space-time parameter $\xi^i$, namely
$\gamma_{ij}=F(\xi^i)$ \cite{alve88-7-395}.

\paragraph{A novel $p$-brane action} From the above discussion, one
can see that there are several reasonable generalizations of the
string action. We only mention the Lagrangian  for an extended
$p$-dimensional object which has been proposed in
Ref.~\cite{stru75-13-337}. Another example will be discussed here and
we propose it only for the sake of completeness. The idea is to
generalize to extended objects  the following reparametrization
invariant action for a relativistic point particle (Euclidean
signature): \begin{equation} S=\int \left(
\frac{g_{\mu\nu}\dot{X}^{\mu}\dot{X}^{\nu}}{2\dot{f}}+
\frac{m^2\dot{f}}{2}\right) d\xi\, . \label{22.7} \end{equation} Note
that in contrast with  the well known einbein action
\cite{brin76-65-471}, where the einbein plays the role of a Lagrangian
multiplier, here the additional degree of freedom is represented by
means of an auxiliary field  $f$, appearing at the same level of the
coordinates $X^{\mu} $. Note that  the the extended object version of
the einbein action is given by the Polyakov-Howe-Tucker action. Coming
back to action (\ref{22.7}), the conjugate momentum of the variable
$f$ must vanish. As a result, making use of the equation of motion,
the above action reduces to the familiar square root action. Note also
that the reparametrization invariance of the action is a direct
consequence of the structure of the Lagrangian, since it is a
homogeneous function of degree one in the velocities.

For an extended object the natural generalization of the above action
requires an enlargement of the configuration space, namely
$(X^\mu,f^a)$, with $a=0,1,2,\ldots p$. The $p+1$ additional
"coordinates " $f^a$, are scalar functions. Therefore one may write
down an action which is linear in $\gamma$ and which is a
generalization of Schild string action \cite{schi77-16-1722}
\begin{eqnarray} S=\int  \left[ \frac{\gamma }{2F}+\frac{k^2F}{2}
d^{p+1}\xi \right]\, , \label{2.8} \end{eqnarray} with $\gamma$ as
above $f\equiv(f^0,f^1,...f^p)$ and $F=\left|\frac{\partial
f}{\partial\xi}\right|$. The presence of the Jacobian $F$ makes the
Lagrangian a scalar density of weight $-1/2$, rendering the action
reparametrization invariant.

A classical canonical analysis can be done, but we stop here because
we shall be interested in path integral quantization. Here, we shall
limit ourselves only to few remarks. The conjugate momenta associated
with $X_\mu$ and $f^a$ are respectively \begin{equation}
P_\mu=\frac{\partial L}{\partial \dot{X_\mu}}=\frac{\gamma}{F}
\gamma^{m0}\partial_m X_\mu\,, \label{2.12} \end{equation}
\begin{equation} \pi_a=\frac{\partial L}{\partial {\dot f^a}}=
\left(-\frac{ \gamma}{2F^2}+\frac{k^2}{2}\right) \left|\frac{\partial
f}{\partial\xi} \right|_a\,, \nonumber\end{equation} where the
time-like variable $f^0$ and the generic coordinates $\xi^a$ are
missing in the Jacobian. At  $F^2=\gamma k^{-2}$, we obtain  $p+1$
primary constraints. Further, for $j\neq 0$ and taking
Eq.~(\ref{2.12}) into account, we obtain further $p$ constraints
$\partial_jX^{\mu}P_\mu=0$. As a result, we arrive at the secondary
Hamiltonian constraint $P^2+k^2G=0$. Since the Lagrangian density is a
homogeneous function of degree one in the velocities, the canonical
Hamiltonian is identically vanishing. This is a general feature of
theories which are reparametrization invariant. A direct calculation
shows that all the constraints we have obtained are identically
preserved in time. As a consequence there are no other constraints and
the number of independent degrees of freedom is given by the number of
initial Lagrangian coordinates $ D+p+1$ minus the number of
constraints which are $2p+2$. As a result we have $ D-p-1$ independent
degrees of freedom. Furthermore, the constraint algebra of the action
(\ref{2.8}) and the Dirac action are the same and therefore describe
the same physics. It should be noted that the action (\ref{2.8})
allows one to take the limit $k$ goes to zero and permits the study of
null $p$-branes which are the higher dimensional counterpart of
massless particles.

\subsection{Some remarks on $p$-brane quantization}

We have shown that at a classical level the $p$-brane theory has some
problems. In fact, the action of the theory contains a world-volume
cosmological constant (the cosmological term) and this constant is
vanishing only for $p=1$. As a consequence only the string action is
invariant under conformal symmetry. In Ref. \cite{dola87-198-447} an
alternative bosonic $p$-brane action with an independent world volume
metric and without the cosmological term has been proposed, but
difficulties associated with the corresponding tensor calculus are
present in this model, even for $p=2$.

We shall briefly consider later the path integral quantization (or
Polyakov approach \cite{poly81-103-207}). Here we only limit ourselves
to some remarks.  One issue is the difficulty related to the
connection between any topological characterization of
$(p+1)$-manifolds (largely incomplete) and loop diagrams. Another one
is the generalization of (super)strings  geometrical quantization
\cite{bowi87-58-535,bowi87-293-348} based on the reparametrization
invariance (super)diff$\:S^1/S^1$. In a pionering work \cite{hopp87b},
Hoppe has showed that the algebra diff$:S^2/S^2$, in the limit
$N\rightarrow\infty$ is isomorphic to $SU(N)$. In the linear
approximation the simplest possibility is to consider $p$-branes with
the topology of the $p$-torus.

The supersymmetric version of Hoppe construction is contained in
\cite{dewi88-305-545}. The gauge theory of the area-preserving
diffeomorphisms of the supermembrane can be obtained as a limit of
supersymmetric quantum mechanics. In contrast with the perturbative
approach, massless modes are absent in the spectrum of the
11-dimensional supermembrane.

The BRST formalism for bosonic $p$-branes has been developed in
\cite{evan88-298-92}. The covariant BRST quantization of the
Green-Schwartz superstring has been carried out in
\cite{kall87-195-369,carl87-284-365} and investigations along these
lines for $p>1$ are in progress. With regard to the connection between
$D$ and the absence of anomalies (quantum consistency), it is well
known that for string $D=26$ and for super-string $D=10$. In Ref.
\cite{marq89-227-227}, it is claimed that for the bosonic membrane,
$D=27$ is necessary for the quantum consistency (see however Ref.
\cite{bars90-343-398}). We also would like to mention the quantization
scheme based on $p$-volume functionals \cite{mars75-85-375}. Within
this approach, there exists the possibility to find a functional
diffusion equation in which the measure of the $p$-volume plays the
role analogue of the proper time in the point particle dynamics
\cite{namb79-80-372,poly79-82-247,poly80-164-171}. This formalism has
been introduced for the bosonic string in
Refs.~\cite{eguc80-44-126,Ogie80-22-2407} and for any $p$ in
Ref.~\cite{gamb88-205-245}. This issue is related to the path integral
quantization of Sec.~\ref{S:PI-EO}.

Now we shall consider the supersymmetric $p$-brane. The non-linear
action for the supermembrane in 11-dimensional flat space-time may be
written as \cite{duff88-5-189} \begin{eqnarray} S=-\int d^3\xi [\sqrt
{-g}(g^{ij}\Pi_i^\mu \Pi_{j\mu}-1)+
2\varepsilon^{ijk}\Pi_i^A\Pi_j^B\Pi_k^C B_{CBA}]\,.
\nonumber\end{eqnarray} Here the first  term is the straightforward
supersymmetrization of the action for a bosonic membrane, while the
second term may be conveniently understood as a Wess-Zumino-Witten term
\begin{eqnarray} \Pi_i^A=(\Pi_i^\mu,\Pi_i^\alpha)\:,\qquad\qquad
\Pi_i^\mu= \partial_i X^\mu-i \bar{\psi}\Gamma^\mu \partial_i
\psi\:,\qquad\qquad \Pi_i^\alpha= \partial_i \psi^\alpha\:.
\nonumber\end{eqnarray} Our conventions are: $X^\mu$ is the bosonic
variable ($\mu=0,1,...,10$), $\psi^\alpha$ is a 32-component Majorana
spinor, the tension membrane is equal to unity,
$\{\Gamma^\mu,\Gamma^\nu\}=-2\eta^{\mu\nu}$ with
$\eta^{\mu\nu}=diag(-1,1,...,1)$, $\varepsilon^{012}=-1$. The
11-dimensional charge conjugation matrix $C$ is given by $C=1\otimes
\sigma_2\otimes \sigma_1$,  $\bar{\psi_\alpha}=-\psi^{\beta}
C_{\beta\alpha}$ ($C_{\alpha \beta}=-C_{\beta \alpha}$) and the super
3-form $B_{CBA}$ satisfies $dB=H$, with all  components of $H$
vanishing except $H_{\mu\nu\alpha\beta}=-\frac{1}{3}
(C\Gamma_{\mu\nu})_{\alpha\beta}$. In addition to Poincar\'e
invariance and world-sheet diffeomorphism invariance, the above action
is also invariant under rigid space-time supersymmetry transformations
and local Siegel transformations \cite{sieg83-128-397,hugh86-180-370}.

Solving for $B_{CBA}$, the action may be rewritten in terms of
$g^{ij},\Pi_i^\mu$ and $\psi$, i.e. \begin{eqnarray} I&=&-\int d^3\xi
[\sqrt {-g}(g^{ij}\Pi_i^\mu \Pi_{j\mu}-1)+
\varepsilon^{ijk}\Pi_i^\mu\Pi_j^\nu\bar{\psi}
\Gamma_{\mu\nu}\partial_k \psi \nonumber\\
&&-i\varepsilon^{ijk}\Pi_i^\mu\bar{\psi} \Gamma_{\mu\nu}\partial_j
\psi \bar{\psi} \Gamma^{\nu}\partial_k \psi
-\frac{1}{3}\varepsilon^{ijk}\bar{\psi} \Gamma_{\mu\nu}\partial_i \psi
\bar{\psi} \Gamma^{\mu}\partial_j\psi
\bar{\psi}\Gamma^{\nu}\partial_k\psi]\,. \nonumber\end{eqnarray} The
corresponding equations of motion read \begin{eqnarray}
\partial_i(\sqrt {-g}g^{ij}\Pi_j^\mu)+ \varepsilon^{ijk}\Pi_i^\nu
\partial_j \bar{\psi} \Gamma_{\nu}^{\mu}\partial_k\psi=0\,,
\qquad\qquad(1-\Gamma)g^{ij}\Pi_i^\mu\Gamma_{\mu}\partial_j\psi=0
\,,\label{6.4} \end{eqnarray} in which
$\Gamma=i\varepsilon^{ijk}\Pi^\mu_i\Pi^\nu_j\Pi^\rho_k
\Gamma_{\mu\nu\rho}/6\sqrt{-g}$.

Now let us make some considerations for generic $p$. The variation of
the Wess-Zumino-Witten term \begin{eqnarray}
\varepsilon^{i_1...i_{p+1}}
\Pi_{i_1}^{A_1}...\Pi_{i_{p+1}}^{A_{p+1}}B_{A_{p+1}...A_1}
\nonumber\end{eqnarray} of the global supersymmetric $p$-brane action
is a total derivative if and only if the $\Gamma$-matrix identity
holds for arbitrary spinors, namely \begin{eqnarray}
\varepsilon^{ijk}[\Gamma^{\mu_1}\psi_i(\bar{\psi}_j\Gamma_{\mu_1...\mu_p}
\psi_k)+ \Gamma_{\mu_1...\mu_p}\psi_i(\bar{\psi}_j\Gamma^{\mu_1}
\psi_k)]=0\,. \nonumber\end{eqnarray} There exists only a finite
number of admissible pairs $(p,D)$ for each spinor type. The
$\Gamma$-matrix identity holds for arbitrary Majorana spinors only if
$(p,D)$ is equal to one of the following pairs: $(1,3)$, $(1,4)$,
$(2,4)$, $(2,5)$, $(2,7)$, $(2,11)$, $(3,8)$ or $(4,9)$. If the
spinors are both Majorana and Weyl we also have $(p,D)=(1,10)$ and
$(p,D)=(5,10)$. Finally for Weyl spinors we have $(p,D)=(1,4)$,
$(1,6)$, $(2,4)$ and $(3,6)$. Note that for Dirac spinors the
$\Gamma$-matrix identity is never satisfied. The Fermionic degrees of
freedom of a $p$-brane are described by spinor of minimal size at each
allowed $D$. When $p=1$, it follows the well known result that the
classical Green-Schwarz superstring may be formulated only in
$D=3,4,6,10$. Each of these string cases extends to other allowed
$(p,D)$ pairs by simultaneously increasing  $p$ and $D$ in the $(p,D)$
plane up to $D=11$. The four discrete series are related to the four
composition-division algebras $R$, $C$, $H$ and $O$
\cite{achu87-198-441,sier87-4-227,evan88-298-92} as shown in Table
\ref{tab1}. \begin{table} \begin{center}
\begin{tabular}{||c|lllll|c||}\hline Algebra
&\multicolumn{5}{c|}{$(p,D)$}&Codimension $D-p-1$\\  & & & & & &\\
\hline $R$ & (1,3) & (2,4) & & & &  1\\ $C$ & (1,4)& (2,5)& (3,6)  & &
&  2\\ $H$ & (1,6)&(2,7)&(3,8)&(4,9)&(5,10) &   4\\ $O$ &
(1,10)&(2,11) && &&  8\\ \hline \end{tabular} \end{center} \caption{}
\label{tab1} \end{table} The case $p=0$  (superparticle) has been
excluded. In this case, the Wess-Zumino-Witten term can be interpreted
as a mass term \cite{deaz88-38-509}. Note that the codimension
$(D-p-1)$ of the allowed $p$-branes equals the dimension of the
related composition-division algebra. The $p$-branes in each series
can be obtained from the highest-$p$ one  (maximal superimmersion) by
simultaneous dimensional reduction of the space-time and world-volume
\cite{duff88-5-189}.

In order to obtain the simplest stable classical solution of the
non-linear equations of motion (\ref{6.4}), one can consider the
supermembrane in a space-time with topology $S^1\times
S^1\times\mbox{$I\!\!R$}^9$ \cite{duff88-297-515}. As a result the
stabilization is carried out by the supermembrane stretching over the
2-torus. The classical solution takes the form of a purely bosonic
background  with $\psi=0 $ (for more details see
Ref.~\cite{duff88-297-515}). Then one can quantize the linearized
fluctuations around this background.

This can be generalized to a compactified (super)$p$-brane. Such a
semiclassical quantization leads to the algebra of number operators
$N_{\vec n} $ and (anti)commutation relations
\cite{duff88-297-515,byts92-9-391,byts93-41-1}. Finally, in
Ref.~\cite{marq89-227-234} the critical dimension for the
supermembrane is reported to be $D=11$, namely one of the classically
admitted dimensions, which has been described above.

\subsection{Classification of 3-geometries}

The partition function as the fundamental object in the $p$-brane
quantization may be expressed by means of a path integral evaluated
over all the $p$-dimensional manifolds and the metrics on them. In
particular, the  functional integration over the 3-dimensional metric
can be separated into an integration over all metrics for a 3-volume
of definite topology, followed by a sum over all topologies
\cite{berg88-185-330}. But even for a 3-dimensional manifold of fixed
topology, the moduli space of all metrics, modulo 3-dimensional
diffeomorphisms, is infinite dimensional. Here we shall present a
necessarily brief description of the classification (uniformization)
and sum over the topology for 3 and 4-dimensional manifolds.

The uniformization concept is one of the main concepts in complex
analysis and other areas of mathematics. Here we shall discuss mainly
uniformization of complex algebraic or more general analytic curves,
i.e. Riemann surfaces and also multi-dimensional real manifolds
admitting a conformal structure.

It should be recall that all curves of genus zero can be uniformized
by rational functions, all those of genus one can be uniformized by
elliptic functions, and all those of genus $g>1$, can be uniformized
by meromorphic functions, defined on proper open subsets of
$\mbox{$I\!\!\!\!C$}$, for example in the disk. This result, due to
Klein, Poincar\'e and Koebe, is one of the deepest achievements in
mathematics as a whole. A complete solution of the uniformization
problem has not yet been obtained (with the exception of the
1-dimensional complex case). However, there have been essential
advances in this problem, which have brought to foundations for
topological methods, covering spaces, existence theorems for partial
differential equations, existence and distorsion theorems for
conformal mappings and so on.

With regard to one-dimensional complex manifolds, in accordance with
Klein-Poincar\'e uniformization theorem, each Riemann surface can be
represented (within a conformal equivalence) in the form
$\Sigma/\Gamma$, where $\Sigma$ is one of the three canonical regions,
namely the extended plane $\bar{\mbox{$I\!\!\!\!C$}}$ (the sphere
$S^2$), the plane $\mbox{$I\!\!\!\!C$}$ ($\mbox{$I\!\!R$}^2$), or the
disk, and $\Gamma$ is a discrete group of M\"{o}bius automorphisms of
$\Sigma$ acting freely there (without fixed points). Riemann surfaces
with such coverings are elliptic, parabolic and hyperbolic type
respectively. The theorem given above admits generalization also to
surfaces with branching.

A different approach to the solution of the uniformization problem was
proposed by Koebe. The general uniformization principle of Koebe
asserts that if a Riemann surface $\tilde{\Sigma}$ is topologically
equivalent to a planar region $P$, then there also exists a conformal
homeomorphism of $\tilde{\Sigma}$ onto $P$. The same problem of
analytic uniformization reduces to the topological problem of finding
all the (branched, in general) planar coverings  $\tilde{\Sigma}
\mapsto \Sigma$ of a given Riemann surface $\Sigma$. The solution of
this topological problem is given by the theorem of Maskit.

It should be noted that, with the help of standard uniformization
theorems and decomposition theorems \cite{mask73-130-243}, one can
construct and describe all the uniformizations of Riemann surfaces by
Kleinian groups. Furthermore, by using the quasiconformal mappings, it
is possible to obtain an uniformization theorem of more general
character (this fact is related to Techm\"{u}ller spaces), namely it
is possible to prove that several surfaces can be uniformized
simultaneously (see for example Ref.~\cite{apan76-17-1670}).

\paragraph{The Thurston classification} In the path integral approach
to membranes the two following problems arise. Should one include all
3-dimensional manifolds or only orientable ones? What is known about
the classification of manifolds?

{}From the physical point of view, we shall restrict ourselves only to
orientable 3-dimensional manifolds  (see for example
\cite{berg88-185-330}). Furthermore, in dealing  with the evaluation
of the vacuum persistence amplitude, we shall consider the sum over
all compact orientable manifolds without boundaries. With regard to
the classification of the 3-dimensional manifolds, this is a difficult
problem but important progress has been made by Thurston
\cite{thur82-6-357}.

It is well known that for any closed orientable 2-dimensional manifold
$\cal M$ the following result holds: every conformal structure on
$\cal M$ is represented by a constant curvature geometry. The only
simply connected manifolds with constant curvature are $\Sigma=S^2$ or
$\mbox{$I\!\!R$}^2$ or $H^2$ and $\cal M$ can be represented as
$\Sigma/{\Gamma}$, where $\Gamma $ is a group of isometries.

Let us now turn to the classification of the 3-geometries following
the presentation of Ref.~\cite{scot83-15-401}. By a geometry  or a
geometric structure we mean a pair $(\Sigma,\Gamma)$, that is a
manifold $\Sigma$ and a group  $\Gamma$ acting transitively on
$\Sigma$ with compact point stabilizers. Two geometries
$(\Sigma,\Gamma)$  and  $(\Sigma',\Gamma')$ are equivalent if there is
a diffeomorphism of $\Sigma$ with $\Sigma '$ which throws the action
of $\Gamma$ onto the action of $\Gamma'$. In particular, $\Gamma$ and
$\Gamma'$ must be isomorphic. We shall assume: \begin{description}
\item{i)} The manifold $\Sigma$ is  simply connected. Otherwise it
will be sufficient to consider a natural geometry
$(\tilde{\Sigma},\tilde{\Gamma})$, $\tilde{\Sigma}$ being the
universal covering of $\Sigma$ and $\tilde{\Gamma}$ denoting the group
of all diffeomorphisms of $\tilde{\Sigma}$  which are lifts of
elements of $\Gamma$. \item{ii)} The geometry admits a compact
quotient. In another words, there exists a subgroup $G$ of $\Gamma$
which acts on $\Sigma$ as covering group and has compact quotient.
\item{iii)} The group  $\Gamma$ is maximal. Otherwise, if
$\Gamma\subset\Gamma'$ then any geometry $(\Sigma,\Gamma)$ would be
the geometry $(\Sigma,\Gamma')$ at the same time. \end{description}
After these preliminaries we can state the classification theorem.
\begin{Theorem}[Thurston] Any maximal, simply connected, 3-dimensional
geometry admitting a compact quotient is equivalent to one of the
geometries $(\Sigma,\Gamma)$, where $\Sigma$ is one of the eight
manifolds
$\mbox{$I\!\!R$}^3\,,S^3\,,H^3\,,S^2\times\mbox{$I\!\!R$}\,,H^2\times\mbox{$I\!\!R$}\,,
\widetilde{SL(2,\mbox{$I\!\!R$})}\,,Nil\,,Sol$. \end{Theorem} The
group properties and more details of these manifolds may be found in
Ref.~\cite{scot83-15-401}. The first five geometries are  familiar
objects, so we explain metric and isometry group of the last three
ones.

\paragraph{The geometry of $\widetilde{SL(2,\mbox{$I\!\!R$})}$.} The
group $\widetilde{SL(2,\mbox{$I\!\!R$})}$ is the universal covering of
$SL(2,\mbox{$I\!\!R$})$, the 3-dimensional Lie group of all $2\times2$
real matrices with determinant $1$. The standard metric on
$\widetilde{SL(2,\mbox{$I\!\!R$})}$ is one of the left
(right)-invariant metrics. It is well known that for a Riemannian
N-dimensional manifold $\cal M$ there is a natural 2N-dimensional
metric on the tangent bundle $T\cal M$ of $\cal M$. If $f:\cal
M\to\cal M$ is an isometry, then $df: T\cal M\to T\cal M$ is also an
isometry. We shall use this argument for the hyperbolic plane $\cal
M=H^2$. The unit tangent bundle $UH^2$ of $H^2$ has a metric induced
from the base manifold $TH^2$. Since there is a natural identification
of $UH^2$ with $PSL (2,\mbox{$I\!\!R$})$, the orientation preserving
isometry group of $H^2$, then we have a metric on $PSL
(2,\mbox{$I\!\!R$})$. Note that $PSL (2,\mbox{$I\!\!R$})$ is doubly
covered by $SL(2,\mbox{$I\!\!R$})$, therefore its universal covering
is $\widetilde {SL(2,\mbox{$I\!\!R$})}$ and the induced metric on
$\widetilde {SL(2,\mbox{$I\!\!R$})}$ is the one in which we are
interested. Finally $\widetilde {SL(2,\mbox{$I\!\!R$})}$ is naturally
a line bundle over $H^2$ since the bundle $UH^2$ is a circle bundle
over $H^2$. The 4-dimensional isometry group of $\widetilde
{SL(2,\mbox{$I\!\!R$})}$ preserves this bundle structure and has two
components both orientation preserving \cite{scot83-15-401}.

\paragraph{The geometry of $Nil$.} $Nil$ is the 3-dimensional Lie
group of all $3 \times 3$ real upper triangular matrices of the form
\begin{eqnarray}\left( \begin{array}{ccc} 1 & x & z \\ 0 & 1 & y \\ 0
& 0 & 1 \end{array} \right)\,,\nonumber\end{eqnarray} with ordinary
matrix multiplication, $x,y$ and $z$ being real numbers. It is also
known as the nilpotent Heisenberg group. It is easy to write down a
metric which is invariant under left multiplication for $Nil$. A basis
of left-invariant 1-forms is \begin{eqnarray} \sigma^1=
dz-xdy\,,\qquad\qquad \sigma^2=dx\,, \qquad\qquad\sigma^3=dy\,.
\nonumber\end{eqnarray} Therefore, the standard metrics reads
\begin{eqnarray} ds^2=(\sigma^1)^2 + (\sigma^2)^2 + (\sigma^3)^2 =dx^2
+ dy^2 + (dz-xdy)^2 \,. \nonumber\end{eqnarray} The isometry group has
$Nil$ as its subgroup. There is an additional one-parameter family of
isometries isomorphic to $U(1)$ which can be written as
($0\leq\theta<2\pi$) \begin{eqnarray} S_{\theta}: \left(
\begin{array}{c} x\\ y \\ z \end{array} \right) \to \left(
\begin{array}{c} x\cos\theta+y\sin\theta \\ -x\sin \theta+y\cos\theta
\\ z + \frac{1}{2} \left[(x^2 - y^2)\cos\theta-2xy\sin\theta\right]
\sin \theta \end{array} \right) \,.\nonumber\end{eqnarray} Here the
2-dimensional rotation matrix of angle $\theta$ appears. The
4-dimensional isometry group has two components. A discrete isometry
is given by $(x, y, z) \to (x, - y, - z)$ and besides, all isometries
preserve the orientation of $Nil$.

\paragraph{The geometry of $Sol$.} $Sol$ is the 3-dimensional
(solvable) group with the following multiplication rule
\begin{eqnarray} \left(\begin{array}{c} x\\y\\z \end{array} \right)
\left( \begin{array}{c} x'\\y'\\z ' \end{array} \right)= \left(
\begin{array}{c} x+e^{- z}  x' \\ y+e^z  y' \\ z+z' \end{array}
\right) \,.\nonumber\end{eqnarray} A basis of left-invariant 1 - forms
is \begin{eqnarray}  \sigma^1  =  e^z dx\,, \qquad\qquad  \sigma^2  =
e^{- z} dy\,, \qquad\qquad  \sigma^3  =  dz \,, \nonumber\end{eqnarray}
while the standard left-invariant metric reads \begin{eqnarray}
ds^2=(\sigma^1)^2 + (\sigma^2)^2 + (\sigma^3)^2 =e^{2 z} dx^2 + e^{- 2
z} dy^2 + dz^2 \,. \nonumber\end{eqnarray} The discrete isometries are
\begin{eqnarray} (x,y,z)\to \left\{ \begin{array}{c} (\pm x, \pm y,
z)\\ (\pm y, \pm x, - z)\\ \end{array} \right.\, ,
\nonumber\end{eqnarray} so the group $\Gamma$ of $Sol$ has eight
components. Moreover, four of them, connected to the following
elements, are orientation preserving: \begin{eqnarray} (x,y,z)\to
\left\{ \begin{array}{c} (x, y, z) \\ (- x, - y, z) \\ (y, x, - z) \\
(-y, -x, - z) \end{array} \right.\,. \nonumber\end{eqnarray}

As for the manifolds modelled on  $H^2\times\mbox{$I\!\!R$}$,
$S^1\times H^2/\Gamma$ contains a compact Riemann surface and these
are relevant for string theory. The manifold modelled on $
\widetilde{SL(2,\mbox{$I\!\!R$})}$ or $Nil$ are Seifert fibre spaces
and those modelled on $Sol$ are bundles over $S^1$ with fibers the
torus or the Klein bottle. A compact 3-manifold without boundary
modelled on
$\mbox{$I\!\!R$}^3,S^3,S^2\times\mbox{$I\!\!R$},H^2\times\mbox{$I\!\!R$},
\widetilde{SL(2,\mbox{$I\!\!R$})},Nil$ is a Seifert fibre space and
vice-versa \cite{scot83-15-401}.

\subsection{Classification of 4-geometries}

Unlike the case of compact Riemann surfaces or 3-dimensional
manifolds, very little is known about the uniformization of
$N$-dimensional manifolds $(N>3)$ by Kleinian groups. The reader can
find some results along these lines for  conformal manifolds in
Ref.~\cite{krus81b}.

The Donaldson theorem \cite{dona83-18-316} for smooth structure on
$\mbox{$I\!\!R$}^4$ shows that in the theory of differentiable
4-dimensional manifolds is necessary to use low dimensional methods,
in particular geometrical methods. Although in this case there are no
decomposition theorems which permit the use of Thurston methods,
nevertheless there is a classification of the 4-dimensional geometries
$(\Sigma,\Gamma)$ \cite{wall86b}. The reader can find some necessary
informations about 4-geometries from the point of view of homogeneous
Riemannian manifolds and Lie groups in
Refs.~\cite{ishi55-7-345,bess87b}.

The list of Thurston 3-geometries can be organized in terms of the
compact stabilizers $\Gamma_\sigma$ of $\sigma\in\Sigma$ isomorphic to
$SO(3)$, $SO(2)$ or trivial group $SO(1)$. The analogue list of
4-geometries can be organized (using only connected groups of
isometries) as in in Table \ref{tab2}. \begin{table} \begin{center}
\begin{tabular}{||l|l||}\hline stabilizer $\Gamma_\sigma$ & manifold
$\Sigma$ \\  & \\ \hline $SO(4) $&$ S^4,\mbox{$I\!\!R$}^4, H^4$\\
$U(2) $&$ \mbox{$I\!\!\!\!C$} P^2,\mbox{$I\!\!\!\!C$} H^2   $\\
$SO(2)\times SO(2) $&$ S^2\times \mbox{$I\!\!R$}^2,S^2\times
S^2,S^2\times H^2, H^2\times \mbox{$I\!\!R$}^2,H^2\times H^2$\\ $SO(3)
$&$ S^3\times \mbox{$I\!\!R$}, H^3\times \mbox{$I\!\!R$}$\\ $SO(2) $&$
Nil^3\times
R,\widetilde{PSL(2,\mbox{$I\!\!R$})}\times\mbox{$I\!\!R$},Sol^4 $\\
$S^1 $&$ F^4 $\\ trivial &$ Nil^4,Sol^4_{m,n}$ (including
$Sol^3\times\mbox{$I\!\!R$}),Sol^4_1$ \\ \hline \end{tabular}
\end{center} \caption{} \label{tab2} \end{table} Here we have the four
irreducible 4-dimensional Riemannian symmetric spaces: sphere $S^4$,
hyperbolic space $H^4$, complex projective space $\mbox{$I\!\!\!\!C$}
P^2$ and complex hyperbolic space $\mbox{$I\!\!\!\!C$} H^2$ (which we
may identify with the open unit ball in $\mbox{$I\!\!\!\!C$}^2$ with
an appropriate metric). The other cases are more specific and for the
sake of completeness we shall illustrate them.

\paragraph{The geometry of $Nil^4$, $Sol^4_{m,n}$, $Sol^4_1$ and
$F^4$.} The nilpotent Lie group $Nil^4$ can be presented as the split
extension $\mbox{$I\!\!R$}^3 \oslash _U\mbox{$I\!\!R$}$ of
$\mbox{$I\!\!R$}^3$ by $\mbox{$I\!\!R$}$ (the symbol $\oslash$ denotes
semidirect product). The quotient $\mbox{$I\!\!R$}$ acts on the
subgroup $\mbox{$I\!\!R$}^3$ by means of $U(t)=\exp (tB)$, where
\begin{eqnarray} B=\left( \begin{array}{ccc} 0 & 1 & 0\\ 0 & 0 & 1\\ 0
& 0 & 0 \end{array} \right)\,.\nonumber\end{eqnarray}

In the same way, for the soluble Lie groups one has
$Sol_{m,n}^4=\mbox{$I\!\!R$}^3 \oslash _{T_{m,n}}\mbox{$I\!\!R$}$,
where $T_{m,n}(t)=\exp (tC_{m,n})$ and \begin{eqnarray} C_{m,n}=\left(
\begin{array}{ccc} \alpha & 0 & 0\\ 0 & \beta & 0\\ 0 & 0 & \gamma
\end{array} \right)\,,\nonumber\end{eqnarray} with the real numbers
$\alpha>\beta>\gamma$ and $ \alpha+\beta+\gamma=0 $. Furthermore
$e^{\alpha}$, $e^{\beta}$ and $e^{\gamma}$ are the roots of
$\lambda^3-m\lambda^2+n\lambda-1=0$, with $m,\;n$ positive integers.
If $m=n$, then $\beta=0$ and  $Sol_{m,n}^4=Sol^3\times \mbox{$I\!\!R$}
$. In general, if $C_{m,n}\propto C_{m',n'}$, then $Sol_{m,n}^4\cong
Sol_{m',n'}^4$. When $m^2n^2+18=4(m^3+n^3)+27$, one has a new
geometry, $Sol^4_0$, associated with group $SO(2) $ of isometries
rotating the first two coordinates.

The soluble group $Sol^4_1$, is most conveniently represented as the
matrix group \begin{eqnarray} \left(\begin{array}{ccc} 1 & b & c\\ 0 &
\alpha & a\\ 0 & 0 & 1 \end{array} \right)\,,\nonumber\end{eqnarray}
with $\alpha,a,b,c\in\mbox{$I\!\!R$}$, $\alpha>0$.

Finally the geometry $F^4$, related to the isometry group $
\mbox{$I\!\!R$}^2 \oslash PSL(2,\mbox{$I\!\!R$})$ and stabilizer
$SO(2)$, is the only geometry which admits no compact model. A
connection of these geometries with complex and K\"ahlerian structures
(preserved by the stabilizer $\Gamma_{\sigma}$) can be found in
Ref.~\cite{wall86b}

We conclude this subsection with some remarks. It is well known that
there are only a finite number of manifolds of the form
$\mbox{$I\!\!R$}^N/\Gamma$, $S^N/\Gamma$ for any $N$ \cite{wolf77b}. A
fortiori this holds also for $S^2\times \mbox{$I\!\!R$}^2,S^2\times
S^2, S^3\times \mbox{$I\!\!R$} $ manifolds. Besides, if we make the
intuitive requirement that only irreducible manifolds have to be taken
into account, then the manifolds modelled on
$S^2\times\mbox{$I\!\!R$}$, $H^2\times\mbox{$I\!\!R$}$ have to be
excluded in 3-dimensions, while the ones modelled on \begin{eqnarray}
S^2\times\mbox{$I\!\!R$}^2,\:S^2\times S^2,\:S^2\times H^2,\:
H^2\times\mbox{$I\!\!R$}^2,\:H^2\times H^2, \nonumber\end{eqnarray}
\begin{eqnarray} S^3\times \mbox{$I\!\!R$},\:H^3\times
\mbox{$I\!\!R$},\: Nil^3 \times \mbox{$I\!\!R$},\:
\widetilde{PSL(2,\mbox{$I\!\!R$})}\times
\mbox{$I\!\!R$},\:Sol^3\times\mbox{$I\!\!R$} \nonumber\end{eqnarray}
have to be neglected in 4-dimensions. As a consequence it seems that
the more important contribution to the vacuum persistence amplitude
should be given  by the compact hyperbolic geometry, the other
geometries appearing only for a small number of exceptions
\cite{bess87b}. It has to be noted that gluing of the above
geometries, characterizing different coupling  constants, by a
complicated set of moduli, is a very difficult task (for more details
see Refs.~\cite{alva92b,alva93-2-1}). Therefore, in the following
physical applications, we shall consider the compact hyperbolic
manifolds $H^N/\Gamma$.

\subsection{The path integral associated with loop expansion}
\label{S:PI-EO}

In this subsection a path integral technique for the closed quantum
$p$-brane will be considered. Such an approach has been pioneered for
the string case ($p=1)$ in Ref.~\cite{eguc80-44-126}. More recently
(see Refs.~\cite{gonc90-19-73,byts91-6-669,byts93-8-1573}) it has been
used to discuss the quantization of a closed $p$-brane when the
extended object sweeps out a compact $(p+1)$-dimensional manifold
without boundary.

In the framework of the path integral approach, the idea is to find a
classical solution of the equation of motion, expand the action up to
quadratic terms in fluctuations around the classical solution and
compute the determinants of second order elliptic operators which
arise in the Gaussian functional integration. Such operators are
always Laplace-Beltrami type operators, acting in different bundles
over the above compact manifold. They can be considered as the main
building blocks of the $p$-brane path integral.

Note that the usual perturbative expansion methods cannot be applied.
It would be nice to quantize covariantly the model, but from the
experience with the string case, it is known that this may be
extremely difficult. It is also less satisfactory to fix the
reparametrization invariance in a particular gauge, checking the
covariance afterwards. As far as this issue is concerned, it is known
\cite{coll76-112-150} that for $p>1$ there exists no gauge in which
the model can be cast in a linear form. As a consequence, as a first
step to quantization of a non-linear theory, one can attempt a
semiclassical (one-loop) approximation \cite{frad82-143-413}.

\subsubsection{The free relativistic point particle}

With regard to the path integral quantization, let us show how action
defined by Eq.~(\ref{22.7}) leads quite naturally to the point
particle Euclidean relativistic propagator. Indeed it is well known
that the propagator related to the relativistic point particle can be
written as \begin{eqnarray} A(X,f;X',f')= N\int d[X]d[f]\,\exp\left[-
\int_0^{1}\left(\frac{g_{\mu\nu} \dot{X}^{\mu}\dot{X}^{\nu}}{2\dot{f}}+
\frac{m^2\dot{f}}{2}\right)\,d\xi\right] \,.\nonumber\end{eqnarray} We
may choose $f(0)=0$ and $f(1)=c >0$, the meaning of $c$ being  the
length of the trajectory. Note that \begin{equation} c = \int_0^1
\frac{df}{d\xi}d\xi\:,\qquad\qquad \int_0^{1}
\frac{g_{\mu\nu}\dot{X}^{\mu}\dot{X}^{\nu}}{\dot{f}} d\xi= \int_0^c
g_{\mu\nu}\frac{dX^\mu}{df}\frac{dX^\nu}{df} df\, .
\nonumber\end{equation} Let us introduce $f=c\rho$ so that
$d[f]=dc\,d[\rho]$. Thus, we arrive at \begin{equation} A(X,f;X',f')=N
\int d[\rho]\int_0^{\infty} \left[
e^{-\frac{m^2c}{2}}\int\,d[X]\exp\left(-\int_0^c
g_{\mu\nu}\frac{dX^\mu}{df}\frac{dX^\nu}{df}\,df\right)\right]\,dc
\:.\nonumber\end{equation} The true propagator can be obtained
factorizing out the infinite measure due to reparametrization
invariance of the action. Thus, we get the well known result
\begin{equation} A(X;X')=\frac1{(4\pi)^{D/2}}
\int_0^{\infty}c^{-\frac{D}{2}}\,
e^{-\frac{m^2c}{2}-\frac{(X-X')^2}{4c}}\, dc \, .
\nonumber\end{equation}

\subsubsection{The $p$-brane model}

Here we shall mimic the approach used in the point-like case. To start
with, the partition function which describes the quantized extended
object may be written as \begin{equation} Z=N \int d[f]\,d[X]
\exp\left[-\int\left(\frac{\gamma }{2F}+\frac{k^2F}{2}\right)
d^{p+1}{\xi}\right]\,. \nonumber\end{equation} An argument similar to
the one given above for the point particle leads to \begin{equation}
Z=N \int d[\rho]\int_0^{\infty}\left[
\exp\left(-\frac{k^2\Omega}2\right)\int d[X]
\exp\left(-\int\gamma\,d^{p+1}f\right)\right]\,d\Omega
\,,\nonumber\end{equation} where $\Omega$ is the volume of the
$(p+1)$-dimensional closed  manifold and the infinite integration
associated with reparametrization invariance of the model has been
factorized out. This approach to quantization of extended objects has
been proposed for strings in Ref.~\cite{eguc80-44-126} (see also
Ref.~\cite{gamb88-205-145}). The evaluation of the above functional
integral is a formidable problem. For $p>2$, one is forced to make use
of the Gaussian approximation. With regard to this, we would like to
recall that the main issue one has to deal with, is the classification
of $(p+1)$-dimensional closed manifolds  and the related  evaluation
of determinant of Laplacian operators. To our knowledge, such a task
is far from being solved. We have argued previously that the compact
hyperbolic geometries seem to play a significant role among all the
possible ones. As a consequence, it seems reasonable to consider
within the semiclassical approximation, that the bosonic contribution
of the $p$-brane is represented by factors like $(\det
L)^{-(D-p-1)/2}$, $L$ being a suitable Laplace-Beltrami operator
acting over $\cal M^N=H^N/\Gamma$ Furthermore one should observe that
in general, there exist a number of different topologically
inequivalent real bundles over $\cal M^N$, this number being given by
the number of elements of $H^1(\cal M^N;\mbox{$Z\!\!\!Z$}_2)$, the
first cohomology group of $\cal M^N$ with  coefficients in
$\mbox{$Z\!\!\!Z$}_2$. Thus, one has to try to evaluate such
determinants on compact manifolds, the hyperbolic ones, i.e. $\cal
M^N=H^N/\Gamma$ giving the most important contributions.

At the end of Sec.~\ref{S:HM-ZM} we have evaluated the regularized
determinant on hyperbolic manifolds by taking into account also the
possible presence of zero modes, which strictly depends on the
characters $\chi$. We have derived the  equation (see
Eq.~(\ref{detL-ZM})) \begin{eqnarray} \det L_N=\frac1{ {\cal N}!}
\exp[-\gamma K_N(L_N)-I(0|L_N)]\:Z^{( {\cal N})}_N(2\varrho_N)
\:,\label{detLN5}\end{eqnarray} where $L_N=-\Delta_N$ and $ {\cal N}$
is the number of zero-modes. For example, for $N=2$ and trivial real
line bundle $(\chi=1)$, $s=1$ is a zero with multiplicity 1 of
$Z_2(s)$. In this case (string model), a formula similar to
Eq.~(\ref{detLN5}) has been discussed in
Refs.~\cite{dhok86-104-537,sarn87-110-113,cogn92-33-222}.

For $N=3$ (membrane model), the situation is quite similar and the
evaluation of the Laplace determinant has been done in
Ref.~\cite{gonc90-19-73}, but with a different technique. Also in this
case, for untwisted fields ($\chi=1$) there exists a zero mode.
Moreover, $K_3(L_3)=0$ (odd dimension) and from Eq.~(\ref{IN-odd}),
$I(0|L_3)=\Omega( {\cal F}_3)/6\pi$ easily follows. As a consequence
we obtain the simple result for the determinant of the Laplacian
\begin{equation} \det L_N=Z_3'(2)\,\exp\left[-\frac{\Omega_( {\cal
F}_3)}{6\pi}\right] \,.\nonumber\end{equation} As a result, it follows
that the leading contribution seems to come from 3-dimensional compact
hyperbolic manifolds having the smallest volume.

In this Section, we have discussed in some detail hyperbolic
contribution to the one-loop approximation of closed $p$-branes. It
has also been proposed a slight variation of Dirac $p$-brane action
which however involve a set of new scalar fields necessary for the
reparametrization invariance of the action. Even though at the
classical level this  action leads to some simplification, it is not
clear if, at the quantum level, such simplifications still remain.
However, within the one-loop approximation the bosonic sectors are all
equivalent. The main issue to be solved is the evaluation of a Laplace
determinant for scalar fields on a compact $(p+1)$-dimensional
manifold swepts out by the $p$-brane.

As far as the extension of these results to super $p$-branes is
concerned, we note that at the classical level the fermionic sector
may present some difficulties which can be overcome (see
Ref.~\cite{duff89-6-1577}). At the quantum level, one should deal with
determinants of the square of the Dirac operator on
$(p+1)$-dimensional compact hyperbolic manifolds. To our knowledge
only the string case has been successfully considered (see
Ref.~\cite{dhok86-104-537,sarn87-110-113}).

\subsection{The Casimir energy for $p$-branes in space-times with
constant curvature}

The physical properties of a $p$-brane in the quantum regime may be
obtained from a study of the effective action for various $p$-brane
configurations. The first attempts along these lines have been
performed for bosonic and supersymmetric membranes in
Ref.~\cite{kikk86-76-1379,fuji87-199-75,mezi88-309-317}, for bosonic
membrane in a 1/D approximation in
Refs.~\cite{flor89-220-61,flor89-223-37,odin89-10-439} and for bosonic
$p$-branes in Ref.~\cite{odin90-7-1499}. Open, toroidal and spherical
$p$-branes have been considered \cite{sawh88-202-505} with the
interesting result that the Casimir energy provides a repulsive force
which stabilizes the membrane at non zero radius, but the net energy
of this stabilized membrane is negative, suggesting that the membrane
ground state is tachyonic. Since the Casimir energy is likely to
vanish for the supermembrane, it seems unlikely that these results
will carry over to that case.

Here we will present a general expression for the static potential
(Casimir energy) of $p$-branes compactified on constant curvature
Kaluza-Klein space-times \cite{byts92-9-391,byts92-9-1275}. Thus we
shall consider $p$-branes which evolve in space-times of the kind
$\cal M=\cal M^D=\cal M^p\times\mbox{$I\!\!R$}^{D-p}$ ($D>p$) with
$\cal M^p=T^p$,  $\cal M^p=T^K\times S^Q$ ($p=K+Q$), $\cal
M^p=T^K\times S^Q\times H^N/\Gamma$ ($p=K+Q+N$), $\Gamma$ being a
discrete group of isometries of $H^N$.

\paragraph{Classical solutions and gauge conditions.}

In the case of toroidal $p$-brane configurations, i.e. $\cal M^p=T^p$,
we will consider the following classical solutions of the equation of
motion \begin{eqnarray} \vec X^0_{cl} \equiv\tau&=&\xi_0,\,
\qquad\qquad\vec X^\bot_{cl}= 0\,, \qquad\qquad\vec X^ {D - 1} _{cl}
\equiv \theta_1 = \xi_1, \ldots, X^{D-p}_{cl} = \xi_p \,,\nonumber\\
\left( \gamma_{cl} \right)_{ij}& =& \eta_{ij}\,,
\nonumber\end{eqnarray} where $\vec
X^\bot_{cl}=(X^1,\ldots,X^{D-p-1})$ and $(\xi_1,\ldots,\xi_p)\in U
=[O,r_1]\times\ldots\times[O,r_p]$, $r_i$ being the circle radii of
the space $T^p$. In all cases examined below, the fields are taken to
be periodic in the imaginary time with period $T$, that is
\begin{eqnarray} \vec X^\bot(0,\xi_1,...,\xi_p) =\vec X^\bot(T,
\xi_1,...,\xi_p)\,. \nonumber\end{eqnarray}

The nontrivial topology of a space-time leads to the existence of the
topologically inequivalent field configurations of fields
\cite{isha78-362-383,isha78-364-591}. The number of such
configurations is equal to the number of non-isomorphic linear real
vector bundles over $\cal M$, i.e. the number of elements in $H^1(\cal
M;\mbox{$Z\!\!\!Z$}_2)$, the first cohomology group of $\cal M$ with
coefficients in $\mbox{$Z\!\!\!Z$}_2$
\cite{isha78-362-383,isha78-364-591,avis79-156-441,choc80-13-2723}.
Each field sector is characterized by some quantum number, the
M\"obius character or twist $h\epsilon H^1(\cal
M;\mbox{$Z\!\!\!Z$}_2)$. Since $H^1(\cal M;\mbox{$Z\!\!\!Z$}_2)$ is
always an abelian group, there exists the obvious condition
$h^2=h+h=0\in H^1(\cal M;\mbox{$Z\!\!\!Z$}_2)$. We introduce the
vector $\vec g=(g_1, \ldots, g_p)$ which defines the type of field
(i.e. the corresponding twist h). Depending on the field type chosen
in $\cal M$, we have $g_i=0$ (untwisted field) or 1/2 (twisted field).
In our case $H^1(T^p;\mbox{$Z\!\!\!Z$}_2)= \mbox{$Z\!\!\!Z$}_2^p$ and
the number of configurations of real scalar field is $2^p$. Therefore
we take as remaining boundary conditions, the $p$ equations
\begin{eqnarray} \vec X^\bot(\xi_0,\xi_1,\ldots,\xi_i=0,\ldots,\xi_p)
&=&(1-4g_i) \vec X^\bot(\xi_0,\xi_1,\ldots,R_{Ti},\ldots,\xi_p)
\label{(103)} \:,\end{eqnarray} where the index $i$ runs from 1 to $p$.
Eqs. (\ref{(103)}) generalize the corresponding boundary conditions
for the toroidal $p$-brane \cite{odin90-7-1499} (see also
Refs.~\cite{flor89-223-37,odin89-10-439}).

In the case of space-times with topology $T^K \times S^Q$ the only
non-zero elements of the metric are given by \begin{eqnarray}
g_{D-1,D-1}& =& R^2_S, \qquad\qquad g_ {D-j, D-j} = R^2_S \prod^{j-1}_
{l = 1} \sin^2 (\theta_l), \qquad\qquad j = 2, \ldots, Q\,,\nonumber
\\ g_{ii}& =& 1, \qquad\qquad i = 0,1,\ldots,D-(Q+1)\,,
\nonumber\end{eqnarray} where $R_S$ is the fixed radius of the
hypersphere. We shall generalize the spherical $p$-brane results of
Refs.~\cite{flor89-223-37,odin89-10-439,odin90-7-1499} and we shall
find the classical solution in the form \begin{eqnarray} \vec X^0_{cl}
= \xi_0\:, \qquad\qquad\vec X^ \bot_ {cl} = 0\:, \qquad\qquad
X^{D-1}_{cl} = \xi_1, \dots, X^{D-p}_ {cl} = \xi_p
\:,\nonumber\end{eqnarray} \begin{eqnarray} (\gamma _{cl})_{ij} d
\xi^i d \xi^j = d \tau^2 + R_S^2 d \Omega^2_Q + \sum^K_{l= Q + 1} d
\xi^l d \xi^l. \nonumber\end{eqnarray} There are two types of boundary
conditions for the function $\vec
X^\bot(\xi_0,\xi_1,\ldots,\xi_K,\ldots,\xi_p)$. The first of them
(with respect to space-time parameters $\xi_i, i = Q + 1,..., K)$
looks like Eqs. (\ref{(103)}), while the second one (related to
$\xi_i, i=1,..., Q$) is the appropriate boundary conditions for a
hypersphere.

Finally in the case $T^K \times S^Q \times (H^N / \Gamma)$ the nonzero
elements of the metric are given by \begin{eqnarray} g_{D-1, D-1}& =&
R^2_S, \qquad\qquad g_ {D-j, D-j} =
R^2_S\prod^j_{l=1}\sin^2(\theta_l), \qquad\qquad j = 2,\dots, Q
\:,\nonumber\\ g_{D-i, D-i}&=& R^{-2}_H\xi^{-2}_2\,, \qquad\qquad
i=Q+1,\dots, Q+1+N\:,\label{6.2.106}\\ g_{l,l}&=&1, \qquad\qquad l =
0,1,..., D - (Q + 1 + N) \,,\nonumber\end{eqnarray} where $-R^{-2}_H$
is the curvature of hyperbolic metric. The classical solution of Eqs.
(\ref{6.2.106}) has the form \begin{eqnarray} \vec X^0_{cl} \equiv
\tau=\xi_0\:, \qquad\qquad\vec X^ \bot_{cl}=0\:, \qquad\qquad
X^{D-1}_{cl}=\xi_1,\dots, X^{D-p}_{cl}=\xi_p \:,\nonumber\end{eqnarray}
\begin{eqnarray} (\gamma_{cl})_{ij} d \xi^i d \xi^j=d\tau^2+ R^2_S d
\omega^2_Q+\xi_2^{- 2} R^{- 2}_H d\Omega^2_N +\sum^{Q + N}_{l=K+1}
d\xi^l d\xi^l\,. \nonumber\end{eqnarray}

Let us make some considerations necessary for further calculations.
Since the fundamental group of the manifold $\cal M$ now is
$\mbox{$Z\!\!\!Z$}^K\times\Gamma$, it follows that the real bundles
over $\cal M$ correspond to  multiplets $(\vec g,\vec\chi)$. Here
$\vec\chi$ is a character of the group $\Gamma$. For example a scalar
Laplacian $L_D$ in such a bundle is the Kronecker sum of the following
Laplacians: a Laplacian $L_K$ in the real line bundle labelled by
$\vec g$ over torus $T^K$, the standard Laplacian $L_Q$ on the
$Q$-dimensional sphere and a Laplacian $L_N$ in the real line bundles
$\vec\chi$ over $H^N/\Gamma$. Therefore we have three types of
boundary conditions for the functions $\vec X^\bot (\xi_0, \ldots,
\xi_p)$. The first of them is the boundary conditions for torus (it
looks like Eq.~(\ref{(103)})). The second and the third types,
concerning the parameters $\xi_i$ for $i=1,\ldots,Q$ and
$i=Q+1,\ldots,Q+N$, are the appropriate boundary conditions for
hyperspheres and compact hyperbolic spaces correspondingly. Note that
in all cases examined below we shall use the background gauge as in
Refs.~\cite{flor89-220-61,flor89-223-37,odin89-10-439,odin90-7-1499,byts92-9-1275}
\begin{eqnarray} X^0 = X^0_{cl}\:,\: X^{D-1} =
X^{D-1}_{cl}\:,\dots\:,X^{D-p}= X^{D-p}_{cl} \:,\nonumber\end{eqnarray}
in which there are no Faddeev-Popov ghosts.

\subsubsection{The semiclassical approximation}

Here we shall derive the expression of the static potential keeping
only the quadratic quantum fluctuations around a static classical
solution and making use of the background field gauge. It is easy to
show that in this approximation all the actions we have considered
reduce to \begin{eqnarray} S=kT\Omega_p+\frac{1}{2}\int_0^T d\xi_0
\int_{{\cal M}^p} \vec {\eta}^{\bot}\cdot L \vec {\eta}^{\bot}d\xi\,,
\nonumber\end{eqnarray} where $L$ is the Laplace operator acting on
the transverse fluctuation fields $\vec{\eta}^{\bot}$. The Euclidean
vacuum-vacuum amplitude reads \begin{eqnarray} Z&=&\int d[\vec
{\eta}^{\bot}]e^{-S} =e^{-kT\Omega_p} \int d[\vec {\eta}^{\bot}]\exp
\left({-\frac{1}{2}\int_0^T d\xi_0\int_{{\cal M}^p}
\vec{\eta}^{\bot}\cdot kL\vec{\eta}^{\bot}}\,d\xi\right) \nonumber\\&=&
e^{-kT\Omega_p}\left(\det kL\ell^2\right)^{-\frac{D-p-1}2}
\,.\nonumber\end{eqnarray} So one has \begin{eqnarray} \ln
Z=-kT\Omega_p+\frac{D-p-1}{2}\zeta'(0|k L\ell^2) \,.
\nonumber\end{eqnarray} The static potential is defined by
\begin{eqnarray} V=-\lim _{T\to\infty}\frac{\ln Z}{T}
=k\Omega_p-\frac{D-p-1}{2}\lim _{T \to \infty}
\frac{\zeta'(0|kL\ell^2)}{T} \:.\nonumber\end{eqnarray} A direct
computation gives (see Eq.~(\ref{ZFfact}) with $p=1$ and $N=p$)
\begin{eqnarray} \zeta(s|L)=\frac{T\,\Gamma(s-\frac12)}
{\sqrt{4\pi}\Gamma(s)}\,\zeta(s-\frac12|L_p) +O(e^{-T^2})
\:,\nonumber\end{eqnarray} in which $L_p$ is the Laplace operator on
$\cal M^p$. Using Eq.~(\ref{ZF1-N+1}) we can immediately write down
\begin{eqnarray}
V&=&k\Omega_p+\frac{D-p-1}2\zeta^{(r)}(-\frac12|kL_p\ell^2)\nonumber\\
&=&k\Omega_p+\frac{D-p-1}2
\left[\frac{\ln(4k\ell^2)-2}{\sqrt{4\pi}}K_{p+1}(L_p)
+\,\mbox{PP}\,\zeta(-\frac12|L_p)\right] \label{pbcpot}
\:,\end{eqnarray} which formally reduces to
\cite{flor89-220-61,flor89-223-37,odin89-10-439,odin90-7-1499,byts92-9-1275}
\begin{eqnarray} V=k\Omega_p+\frac{D-p-1}2\sum_i\lambda_i^{\frac12}\,,
\nonumber\end{eqnarray} where $\lambda_i$ run through the spectrum of
the operator $L_p$.

The above general formula for static potential, Eq. (\ref{pbcpot}),
has been obtained making use of zeta-function regularization. We only
remark that if $K_{p+1}(L_p)$ is not vanishing, the analytical
continuation of $\zeta(s|L_p)$ contains a simple poles at $s=-1/2$. As
a consequence, a contribution depending from the arbitrary scale
parameter $\ell$ appears. We know that such a term is always absent
for even $p$.

\subsubsection{The static potential on toroidal spaces}

As an application, we shall consider the Casimir energy of a toroidal
$p$-brane evaluated in a space-time with topology
$\mbox{$I\!\!R$}^{D-p}\times T^p$. In this case, it is well known that
the heat kernel expansion terminates to the Seeley-DeWitt coefficient
$K_p$ and so there is no $\ell$ ambiguity in the static potential.
Moreover, the $\zeta(s|L_p)$ reduces to the Epstein $Z$-function (see
Eq.~(\ref{zeTN-Epstein})). Making use of Eq. (\ref{pbcpot}) and the
functional relation for the Epstein $Z$-function (see
Eq.~(\ref{FR-Epstein}) in Appendix \ref{S:UR}) we may rewrite the
regularized potential as \begin{eqnarray} V_{reg}=\Omega_p\left[
k-\frac{D-p-1}2 \frac{\Gamma(\frac{p+1}2)}{(4\pi^3)^{\frac{p+1}2}} Z_{
{\cal R}_p}(\frac{p+1}p;0,-\vec g)\right] \:,\nonumber\end{eqnarray}
where $\Omega_p=(2\pi)^p\det {\cal R}_p^{1/2}=\prod_{i=1}^p\,l_i$ is
the volume of $T^p$.

For an equilateral torus $r_i=r$ ($i=1,\dots,p$), we find
\begin{eqnarray} V_{reg}&=&k\left[(2\pi r)^p-\frac{\alpha(\vec
g)}{r}\right]\nonumber\\ \alpha(\vec g)&=&\frac{D-p-1}{4\pi k}
\frac{\Gamma(\frac{p+1}2)}{\pi^{\frac{p+1}2}}
Z_{I_p}(\frac{p+1}p;0,-\vec g) \:,\nonumber\end{eqnarray} where $I_p$
is the identity matrix. For untwisted fields $\vec X^{\bot}$ ($\vec
g=0$) and for any $p\in\mbox{$I\!\!N$}$ one has
$Z_{I_p}(\frac{p+1}p;0,0)>0$ (see for example
Refs.~\cite{zuck74-7-1568,zuck75-8-1734,dowk78-11-2255}). This means
that for the untwisted toroidal $p$-brane $\alpha(0)>0$ (the Casimir
forces are attractive) and the $p$-brane tends to collapse
\cite{odin90-7-1499,byts92-9-391,byts92-9-1275}. The behavior of an
untwisted toroidal $p$-brane potential is similar to that of spherical
$p$-branes \cite{flor89-223-37,odin89-10-439,odin90-7-1499}. On the
other hand, for a twisted toroidal $p$-brane there are field sectors
for which $\alpha(\vec g)<0$. For example, if we choose $\vec
g\equiv(\frac12,\ldots,\frac12)$, then for any $p\in\mbox{$I\!\!N$}$
$Z_{I_p}(\frac{p+1}p;0,-\frac12)<0$
\cite{zuck74-7-1568,zuck75-8-1734,dowk78-11-2255}. In this case the
potential has a minimum at finite distance \begin{eqnarray}
r_0=\left[\alpha\left(\frac12,\ldots,
\frac12\right)\right]^{\frac1{p+1}} \nonumber\end{eqnarray} and its
behaviour is similar to the potential of the open $p$-brane
\cite{flor89-220-61,flor89-223-37,odin89-10-439,odin90-7-1499}.

Similar considerations hold for manifolds of the form  $\cal M^p=T^K
\times \cal M^Q$, with $p=K+Q$. It should be noted however that in
this case, the $\ell$ ambiguity term may be present. With regard to
this fact we  say only few things. Since the underlying theory is not
renormalizable, it is not possible to determine such parameter in the
usual way. Furthermore, we are only considering the semiclassical
(one-loop) approximation. Following Ref.~\cite{blau88-310-163}, the
dimensional parameter $\ell$ may phenomenologically summarize the
unknown physics related to the approximations made. In this context,
it should be determined experimentally.

\newpage
\setcounter{equation}{0}
\section{Asymptotic properties of $p$-brane quantum state density}

In order to study the statistical properties of extended objects (we
shall be mainly interested in (super)strings), it is necessary to have
informations on the asymptotics of the density of states. In the case
of field theory (point-like objects), these informations can be
obtained from the heat-kernel expansion, which we have illustrated in
some detail in Sec.~\ref{S:CPEO}, by making use of the Karamata
tauberian theorem. One has to consider the leading term (Weyl term)
and the result is the well known polynomial growing of the state
density when the energy is going to infinity. For an extended object,
the situation is more complicated and we need some preliminary
mathematical results, which, in turn, are relevant in number theory.
First, we shall discuss the analogue of heat-kernel expansion. Then,
the Meinardus theorem will give us the asymptotics of the level
degeneracy, which directly leads to the asymptotics of level state
density for a generic extended object.

\subsection{Asymptotic properties of generating functions}
\label{S:Meinardus}

We have seen  that the semiclassical quantization of a $p$-brane in
$(S^1)^p\times\mbox{$I\!\!R$}^{D-p}$, is equivalent to deal with to the
following "proper time Hamiltonian": \begin{eqnarray} L=\vec
{p}^2+M^2\, , \nonumber\end{eqnarray} where the mass operator $M^2$ is
linearly related to the total number operator \begin{eqnarray}
N=\sum_{i=1}^{d}\sum_{\vec n \in \mbox{$Z\!\!\!Z$}^p/\{0\}}
\omega_{\vec n}N_{\vec n}^i\,. \nonumber\end{eqnarray} Here $d=D-p-1$
and the frequencies are given by \begin{eqnarray} \omega_{\vec n}^2
=\sum _{i-1}^p \left(\frac{2\pi n_i}{l_i}\right)^2\,, \label{om-vec-n}
\end{eqnarray} with the compactification lenghts $l_i=2\pi r_i$,
$i=1,...,p$. The number operators $N_{\vec n}$ with $\vec n
=(n_1,...,n_p)\in \mbox{$Z\!\!\!Z$} ^p$ and the commutation relations
for the oscillators can be found for example in
Refs.~\cite{berg87-185-330,berg87-189-75,duff88-297-515,gand88-5-127,berg88-205-237,byts92-9-391}.
Due to the linear Regge trajectory relation, one may deal with the
number operator $N$. Furthermore, the flat particle Laplace operator
$\vec p^2$ commutes with the mass operator. As a consequence, one may
consider the trace of the heat number operator $\exp(-tN)$, the trace
being computed over the entire Fock space, namely \begin{eqnarray}
{\cal{Z}}(t)=\,\mbox{Tr}\, e^{-tN}=\prod_{j} \left[1-e^{-t\omega_{j}
}\right]^{-d} \:,\nonumber\end{eqnarray} where  $t>0$. For $p=1$
(string case) the function $ {\cal{Z}}(z)$ of the complex variable
$z=t+ix $ is known as the generating function of the partition
function, which is well studied in the mathematical literature
\cite{hard18-17-75}. The properties of this generating function have
been used to evaluate the asymptotic state density behaviour for $p=1$
\cite{huan70-25-895,gree87b,mitc87-294-1138,mats87-36-289}.

In the following, for the sake of completeness, we shall present some
mathematical results we shall use in investigating the heat-kernel
expansion of the operator $N$ \cite{byts93-304-235}. We shall be
interested in the asymptotics of the partition functions which admit
an infinite product as associated generating function. We shall
present a general theorem due to Meinardus
\cite{mein54-59-338,mein54-61-289} following Ref.~\cite{andr76b} (see
also Ref.~\cite{leve74b}) and in particular we shall discuss the so
called vector-like partition functions, which are relevant in the
determination of the asymptotic state density of quantum $p$-branes.

Let us introduce the generating function \begin{eqnarray} f(z)=\prod_{
n=1}^\infty[1-e^{-zn}]^{-a_n}=1+\sum_{n=1}^\infty v(n)e^{-zn},
\nonumber\end{eqnarray} where $\,\mbox{Re}\, z>0$ and $a_n$ are
non-negative real numbers. Let us consider the associated Dirichlet
series \begin{eqnarray} D(s)=\sum_{n=1}^\infty a_n n^{-s} \,
,\hspace{1cm} \hspace{1cm} s=\sigma+it, \nonumber\end{eqnarray} which
converges for $0<\sigma<p$. We assume that $D(s)$ can be analytically
continued in the region $\sigma\geq-C_0 $  ($0<C_0<1$) and here $D(s)$
is analytic except for a pole of order one at $s=p$ with residue $A$.
Besides we assume that $D(s)=O(|t|^{C_1})$ uniformly in
$\sigma\geq-C_0$ as $|t|\rightarrow\infty$, where $C_1$ is a fixed
positive real number. The following lemma
\cite{mein54-59-338,mein54-61-289} is useful with regard to the
asymptotic properties of $f(z)$ at $z=0$: \begin{Lemma} If $f(z)$ and
$D(s)$ satisfy the above assumptions and $z=y+2\pi ix$ then
\begin{equation} f(z)=\exp{\{A\Gamma(p)\zeta_R(1+p) z^{-p}-D(0)\ln
z+D'(0)+O(y^{C_0})\}} \label{Lemma}\end{equation} uniformly in $x$ as
$y\rightarrow0$, provided $|\arg z|\leq\pi/4$ and $|x|\leq1/2$.
Moreover there exists a positive number $\varepsilon$ such that
\begin{equation} f(z)=O(\exp{\{A\Gamma(p)\zeta_R(1+p)
y^{-p}-Cy^{-\varepsilon}\}}), \nonumber\end{equation} uniformly in $x$
with $y^{\alpha}\leq|x|\leq1/2$ as $y\rightarrow0$, $C$ being a fixed
real number and $\alpha=1+p/2-p\nu/4$, $0<\nu<2/3$. \end{Lemma} Here
is a sketch of the proof.  The Mellin-Barnes representation of the
function $\ln f(z)$ gives \begin{eqnarray} \ln f(z)=\frac{1}{2\pi i}
\int_{1+p-i\infty}^{1+p+i\infty}z^{-s} \zeta_R(s+1)\Gamma(s)D(s)ds\,.
\nonumber\end{eqnarray} The integrand in the above equation has a
first order pole at $s=p$ and a second order pole at $s=0$.  Therefore
shifting the vertical contour from $\,\mbox{Re}\, z=1+p$ to
$\,\mbox{Re}\, z=-C_0$ (due to the conditions of the Lemma the shift
of the line of integration  is permissible) and making use of the
theorem of residues one obtains \begin{eqnarray} \ln
f(z)&=&A\Gamma(p)\zeta_R(1+p)z^{-p}-D(0)\ln z+D'(0)\nonumber\\
&&\qquad\qquad+\frac{1}{2\pi i}\int_{-C_0-i \infty}^{-C_0+i \infty}
z^{-s} \zeta_R(s+1)\Gamma(s)D(s)ds \,. \label{6.16} \end{eqnarray} The
first part of the Lemma follows from  Eq.~(\ref{6.16}), since the
absolute value of the integral in the above equation can be estimated
to behave as $O(y^{C_0})$. In a similar way one can prove the second
part of the Lemma but we do not dwell on this derivation.

Now we are ready to state the main result, which  permits to know the
complete asymptotics of $v(n)$. \begin{Theorem}[Meinardus] For
$n\rightarrow\infty$ one has \begin{equation}
v(n)=C_pn^{\frac{2D(0)-p-2}{2(1+p)}} \exp\left\{\frac{1+p}{p}
[A\Gamma(1+p)\zeta_R(1+p)]^{\frac1{1+p}}
n^{\frac{p}{1+p}}\right\}(1+O(n^{-k_1}))\,, \nonumber\end{equation}

\begin{equation}
C_p=[A\Gamma(1+p)\zeta_R(1+p)]^{\frac{1-2D(0)}{2(1+p)}}
\frac{e^{D'(0)}}{[2\pi(1+p)]^{\frac12}} \,,\nonumber\end{equation}

\begin{equation} k_1=\frac{p}{1+p}\min
(\frac{C_0}{p}-\frac{\nu}{4},\frac{1}{2}-\nu)\,.
\nonumber\end{equation} \end{Theorem} The proof of this theorem relies
on the application of the saddle point method. Cauchy integral theorem
gives \begin{eqnarray} v(n)=\frac{1}{2\pi i}\int_{z_0}^{z_0+2\pi
i}f(z)e^{nz}dz =\int_{-1/2}^{1/2}f(y+2\pi ix)e^{n(y+2\pi ix)}dx\,.
\label{v}\end{eqnarray} Since the maximum absolute value of the
integral occurs for $x=0$, the Lemma implies that the integrand is
well approximated by \begin{eqnarray}
U=\exp[A\Gamma(p)\zeta_R(1+p)y^{-p}+ny] \:.\nonumber\end{eqnarray}
Within the saddle point method one has to minimize this expression,
i.e. $dU/dy=0$ and therefore \begin{eqnarray}
y=n^{-\frac{1}{1+p}}[pA\Gamma(1+p)\zeta_R(1+p)]^{\frac{1}{1+p}}\,.
\label{6.1.20} \end{eqnarray} The result of Meinardus theorem follows
by carefully making the estimation of the integral (\ref{v}) and
making use of Eq.~(\ref{6.1.20}) (for an extensive account of the
proof we refer the reader to Ref.~\cite{andr76b}).

Coming back to our problem, we note that the  quantity $\,\mbox{Tr}\,
e^{-tN}$ is a special kind of vector-like generating  function, we are
going to introduce. Let \begin{equation} F(z)=\prod_{\vec n \in
Z^{p}/\{0\}} [1-e^{-z\omega_{\vec n}}]^{-d}\,, \label{3p}
\end{equation} be a generating function, $\,\mbox{Re}\, z>0$,  $d>0$
and $\omega_{\vec n}$ given by Eq.~(\ref{om-vec-n}). The theorem of
Meinardus can be generalized to deal with a such vector valued
function. In the formulation of this theorem the Dirichlet series
$D(s)$ has been used (see the above discussion). In the case of the
generating function (\ref{3p}), the role of $D(s)$ is played by the
$p$-dimensional Epstein $Z$-function. More precisely, for $F(z)$ one
obtains Eq.~(\ref{Lemma}), but with $D(s)$ replaced by $d\,Z_{ {\cal
R}_p^{-1}}(s/p;0,0)$  and the residue $A$ replaced by $\pi^{p/2}\det
{\cal R}_p^{1/2}/\Gamma(1+p/2)$, where $ {\cal R}_p$ is the $p\times
p$ diagonal matrix $diag {\cal R}=(r_1^2,\ldots,r_p^2)$ (see Appendix
\ref{S:UR} for definitions and properties).

Recalling that $Z_{ {\cal R}}(0;0,0)=-1$, we get the following
asymptotic expansion for the function $ {\cal{Z}}(t)$ (for small $t$)
\begin{eqnarray}  {\cal{Z}}(t)\sim t^d \exp\left(\frac dpZ'_{ {\cal
R}_p^{-1}}(0;0,0)\right) \exp\left( B_pt^{-p}\right)\:,
\label{12zZ}\end{eqnarray} \begin{eqnarray} B_p=\frac{2^\frac
p2\pi^p\:d}p \Gamma(\frac p2)\zeta_R(p)\prod_{i=1}^p l_i
\,.\nonumber\end{eqnarray}

\subsection{Asymptotic density of $p$-brane quantum states}

In the following, the asymptotic behaviour of the degeneracy of the
state density level related to a generic $p$-brane will be discussed.
We have already mentioned that the knowledge of such asymptotic
behaviour is important in the investigation of thermodynamical
properties of extended objects. For ordinary matter fields, the
leading term in the heat-kernel expansion, the Weyl term, determines
the leading term of the density of states for large values of the
energy. Here we would like to present the analogue of it for extended
objects. For the sake of completeness we start with an elementary
derivation of such asymptotic behaviour for small $t$.

In terms of $ {\cal{Z}}$, the total number $q(n)$ of $p$-brane states
may be described by \begin{eqnarray}
{\cal{Z}}(t)=\sum_{n=0}^{\infty}q(n)e^{-zt}\,. \nonumber\end{eqnarray}
Making use of the "thermodynamical methods" of Ref.
\cite{fubi73-7-1732},  $ {\cal{Z}}(t)$ may be regarded as a "partition
function" and $t$ as the inverse "temperature". Thus, the related
"free energy" $F_t$, "entropy" $S_t$ and "internal energy" $N_t$ may
be written respectively as \begin{eqnarray}  F_t=-\frac{1}{t}\ln
{\cal{Z}}(t)\,, \qquad\qquad S_t=t^2\frac{\partial}{\partial t}F_t\,,
\qquad\qquad N_t=-\frac{\partial}{\partial t}\ln  {\cal{Z}}(t)\,.
\nonumber\end{eqnarray} The limit $n \to \infty $  corresponds to $t
\rightarrow 0$. Furthermore, in this limit the entropy may be
identified with $\ln q(n)$, while the internal energy is related to
$n$. Hence, one has from Eq.~(\ref{12zZ}) \begin{eqnarray} F_t \simeq
-B_p t^{-p-1}\,, \qquad\qquad S_t \simeq (p+1)B_p t^{-p}\,,
\qquad\qquad N_t=pB_p t^{-p-1}\,. \nonumber\end{eqnarray} Eliminating
the quantity $t$ between the two latter equations one gets
\begin{eqnarray} S_t \simeq \frac{p+1}{p}(pB_p)^{\frac{1}{p+1}}
N_t^{\frac{p}{p+1}}\,. \nonumber\end{eqnarray} As a result
\begin{eqnarray} \ln q(n) \simeq \frac{p+1}{p}
(pB_p)^{\frac{1}{p+1}}n^{\frac{p}{p+1}} \:.\nonumber\end{eqnarray} A
more complete evaluation based on the result of Meinardus gives
\begin{eqnarray} q(n) \sim C_pn^X\exp\left\{n^{\frac p{p+1}}
\left(1+\frac1p\right)(pB_p)^{\frac{1}{p+1}}\right\}\,, \label{133}
\end{eqnarray} with definitions \begin{eqnarray} C_p=
e^{\frac{d}{2}\zeta'(0|L)} \left(2\pi(1+p)\right)^{-\frac12}
(pB_p)^{\frac{1-2dK_0}{2(1+p)}}\,, \qquad\qquad
X=\frac{2dK_0-p-2}{2(p+1)}\,. \label{134} \end{eqnarray} In
Eq.~(\ref{133}) the complete form of the prefactor appears.

Some remarks are in order. First of all let us consider the
(super)string case. Then $p=1$ and for the open bosonic string
Eqs.~(\ref{133}) and (\ref{134}) give \begin{equation}
q(n)=C_1n^{-\frac{D+1}{4}} \exp{\{\pi \sqrt{\frac{2n(D-2)}{3}}\}}
(1+O(n^{-k_1}))\,, \label{16111} \end{equation} where the constant
$C_1$ is given by \begin{equation}
C_1=2^{-\frac12}\left(\frac{D-2}{24}\right)^{\frac{D-1}4} \label{17111}
\:.\end{equation} The formulae (\ref{16111}) and (\ref{17111})
coincide with previous results for strings (see for example
\cite{huan70-25-895,gree87b,mitc87-58-1577,mitc87-294-1138,mats87-36-289}).
The new feature of this considerations is that the constant $C_1$ has
been calculated now making use of Meinardus results.

The closed bosonic string can be dealt with by taking the constraint
$N=\tilde{N}$ into account. As a result the total degeneracy of the
level $n$ is simply the square of $q(n)$. In a similar way one can
treat the open superstring. Furthermore using the mass formula $M^2=n$
(for the sake of simplicity here and in the following we assume a
tension parameter, with dimensions of $(mass)^{p+1}$, equal to $1$) we
find for the number of string states of mass $M$ to $M+dM$
\begin{eqnarray} \nu_1 (M)dM\simeq2C_1M^{\frac{1-D}{2}}\exp(b_1M)dM
\,,\qquad\qquad b_1=\pi\sqrt{\frac{2(D-2)}3}\,. \nonumber\end{eqnarray}
One can show that the constant $b_1$ is the inverse of the Hagedorn
temperature. It is also clear that the Hagedorn temperature can be
obtained in a similar way for the other (super)string cases.

Now let us consider the (super)$p$-brane case, namely $p>1$. Using
again a linear mass formula $M^2=n$, Eqs.~(\ref{133}) and (\ref{134})
lead to an asymptotic density of states of the form \begin{eqnarray}
\nu_p (M) dM\simeq 2C_p M^{\frac{2p+1-2D}{1+p}}
\exp\left(b_pM^{\frac{2p}{1+p}}\right)\,, \label{20111}\end{eqnarray}
\begin{eqnarray} b_p=\frac{1+p}p\left[dA\Gamma(1+p)
\zeta_R(1+p)\right]^{\frac1{1+p}} \:.\nonumber\end{eqnarray} This
result has a universal character for all (super)$p$-branes. It was
presented  two decades ago in Refs.~\cite{fubi73-7-1732,stru75-13-337}
within the extended models for hadronic matter and more recently in
Ref.~\cite{alva92-7-2889}), but without the complete knowledge of the
prefactor. The complete derivation presented here is contained in
Ref.~\cite{byts93-304-235}.

\subsection{Asymptotic density of parabosonic string quantum states}

Here we shall use the Meinardus theorem to evaluate the asymptotics
for the level state density of parabosonic string. First, let us
briefly recall the parastatistic idea.

It is quite standard nowadays to describe the quantum field theory in
terms of operators obeying canonical commutation relations. However,
there exists the alternative logical possibility of para-quantum field
theory \cite{gree53-90-270,ohnu82b}, where parafields satisfy
tri-linear commutation relations. Later, the paraquantization proposal
was  investigated in Ref.~\cite{gree65-138-1155}. We also would like
to remind that, in a general study of particle statistics within the
algebraic approach to quantum field theory, parastatistics is one of
the possibilities found in Refs.~\cite{dopl71-23-199,dopl74-35-49}.
Despite the efforts to apply parastatistics for the description of
internal symmetries (for example in paraquark models \cite{ohnu82b})
or even in solid state physics for the description of quasiparticles,
no experimental evidence in favour of the existence of parafields has
been found so far. Nevertheless, parasymmetry can be of some interest
from the mathematical point of view. For example it can be considered
as formal extension of the supersymmetry algebra. Furthermore,
parasymmetry  may find some physical application in string theory,
where parastrings \cite{arda74-9-3341} have been constructed. It has
been shown there, that these parastrings possess some interesting
properties, like the existence of critical dimensions different from
the standard ones, i.e. $D=10$ and $26$.

In the following, we shall briefly review the paraquantization for
parabose harmonic oscillators, which are relevant to the parabosonic
string in the limit $\wp\to\infty$, where $\wp$ is the order of the
paraquantization. The Hamiltonian and the zero point energy for the
free parabose system has the form \begin{eqnarray}
\hat{H}=\sum_n\frac{\omega_n}{2} (a_n^{\dag}a_n +a_na_n^{\dag}
)-E_0\,, \qquad\qquad E_0 = \frac{\wp}{2}\sum_n\omega_n\,.
\nonumber\end{eqnarray} The operators $a_n$ and $a^{\dag} _n$ obey the
following tri-commutation relations \cite{gree53-90-270,ohnu82b}
\begin{eqnarray} \left[ a_n,\left\{ a^{\dag} _m,a_l\right\}\right]
=2\delta_{nm} a_l\,,\qquad\qquad \left[ a_n \left\{ a_m,a_l
\right\}\right]= 0\,. \nonumber\end{eqnarray} The vacuum will be
chosen to satisfy the relations \begin{eqnarray}
a_n|0>=0,\qquad\qquad\left\{ a^{\dag} _n,a_m\right\}\vert0>
=\wp\delta_{nm}|0>\,, \nonumber\end{eqnarray} so that
$\hat{H}|0>\,=0$. The paracreation operators $a_n^{\dag} $ do not
commute and therefore the Fock space is quite complicated
\cite{ohnu82b}. For $D$-dimensional harmonic oscillators $a_n^i$ of
parabosonic string with frequencies $\omega_n^i=n$ we have the
Hamiltonian \begin{eqnarray}
\hat{H}=\sum^D_{i=1}\sum^\infty_{n=1}\frac{n}{2} \left\{ a^{\dag i}_n,
a^i_n \right\} - E_0\,. \nonumber\end{eqnarray} A closed form for the
partition function $\hat{Z}(t)=\,\mbox{Tr}\, e^{-t\hat{H}}$, the trace
being computed over the entire Fock space, in the limit $\wp\to\infty$
reads \cite{hama92-88-149} \begin{equation}
\hat{Z}(t)=\left\{\prod_{n=1}^\infty\frac{1}{(1-e^{-tn})}\right\}^{D}
\left\{\prod_{n,m=1}^\infty\frac{1}{1-e^{-t(n+m)}}
\right\}^{\frac{D^2}{2}}
\left\{\prod_{n=1}^\infty(1-e^{-2nt})\right\}^{\frac{D}{2}}\,.
\nonumber\end{equation}

Our aim is to evaluate, asymptotically, the degeneracy or state level
density corresponding to a parabosonic string, in the limit of
infinite paraquantization order parameter. As a preliminary result, we
need the asymptotic expansion of the partition function for $t\to0$.
To this aim, it may be convenient to work with the quantity
\begin{eqnarray} F(t)=\ln \hat{Z}(t)=-DF_1(t)+\frac{D}{2}F_1(2t)
-\frac{D^2}{2}F_2(t)\,, \nonumber\end{eqnarray} where we have
introduced definitions \begin{eqnarray}
F_1(t)=\sum_{n=1}^\infty\ln(1-e^{-t n})\:,\qquad\qquad
F_2(t)=\sum_{n,m=1}^\infty\ln(1-e^{-t( n+m)})\,.
\nonumber\end{eqnarray} With regards to the contributions $F_1(t)$,
one may use a result, known in the theory of elliptic modular function
as Hardy-Ramanujan formula \cite{gree87b}, that is \begin{eqnarray}
F_1(t)=-\frac{\pi^2}{6t} -\frac{1}{2}\ln\frac{t}{2\pi}+\frac{t}{24}
+F_1\frac{4\pi^2}{t}\,. \nonumber\end{eqnarray}

Let us now consider the quantity $F_2(t)$. A  Mellin representation
gives \begin{eqnarray} \ln(1-e^{-ta})=-\frac{1}{2\pi i}
\int_{\,\mbox{Re}\, z=c>2} \Gamma(z)\zeta_R(1+z)(at)^{-z}\,dz\:.
\nonumber\end{eqnarray} As a result, \begin{eqnarray}
F_2(t)=-\frac{1}{2\pi i} \int_{\,\mbox{Re}\, z=c>2}
\Gamma(z)\zeta_R(1+z)\zeta_2(z)t^{-z}\,dz\, , \nonumber\end{eqnarray}
where $\zeta_2(z)\equiv \sum_{n,m=1}^\infty (n+m)^{-z}$. Now it is
easy to show that (see for example Ref.~\cite{acto87-20-927})
$\zeta_2(z)=\zeta_R(z-1)-\zeta_R(z)$ and so we have
$F_2(t)=G_2(t)-F_1(t)$, where we have set \begin{eqnarray}
G_2(t)=-\frac{1}{2\pi i} \int_{\,\mbox{Re}\, z=c>2}
\Gamma(z)\zeta(z+1)\zeta(z-1)t^{-z}\,dz=
\sum_{n=1}^\infty\ln\left(1-e^{-tn}\right)^n\,, \nonumber\end{eqnarray}
and the related generating function reads \begin{eqnarray}
g_2(t)=\prod_{n=1}^\infty (1-e^{-tn})^{n}\,. \nonumber\end{eqnarray}

For the estimation of the small $t$ behaviour, let us apply the
results of Sec.~\ref{S:Meinardus} to the generating function $g_2(t)$.
We have $D(s)=-\zeta_R(s-1)$, $\alpha=2$, and $A=-1$. According to
Meinardus lemma we arrive at the asymptotic expansions \begin{eqnarray}
G_2(t)\simeq-\zeta_R(3)t^{-2}-\frac{1}{12}\ln t -\zeta'_R(-1)+ O(t)\,,
\nonumber\end{eqnarray} \begin{eqnarray}
F(t)\simeq\frac{D^2}{2}\zeta_R(3)t^{-2}+
\ln\left(\pi^{\frac{D}{4}}(2\pi)^{\frac{D(D-2)}{4}}
t^{\frac{6D-5D^2}{24}}\right)+ \frac{D^2}{2}\zeta'_R(-1)+O(t^{-1})\,,
\nonumber\end{eqnarray} \begin{eqnarray} \hat{Z}(t)\simeq At^B \exp
{(Ct^{-2})}\,, \label{14.14} \end{eqnarray} where in the latter
equation \begin{eqnarray} A=\pi^{\frac{D}{4}}(2\pi)^{\frac{D(D-2)}{4}}
e^{\frac{D^2}{2}\zeta'_R(-1)}\:, \qquad\qquad B=\frac{6D-5D^2}{24}\:,
\qquad\qquad C=\frac{D^2}{2}\zeta_R(3) \,.\label{ABC}\end{eqnarray}
Note that in ordinary string theory, the asymptotic behaviour of
$Z_1(t)$ is of the kind $ \exp(ct^{-1})$.

Now, the degeneracy or density of levels can be easily calculated
starting from the above asymptotic behaviour. In fact the density
$\hat v(n)$ of levels for parabosonic strings (for a general
discussion on parastrings see Ref.~\cite{arda74-9-3341}) may be
defined by \begin{eqnarray} \hat{Z}(z)=\,\mbox{Tr}\, e^{-z\hat{H}}
=1+\sum_{n=1}^\infty \hat{v}(n)e^{-zn}. \nonumber\end{eqnarray} The
Cauchy theorem gives \begin{eqnarray} \hat{v}_n=\frac{1}{2\pi i}\oint
e^{zn}\hat Z(z)\,dz \, , \nonumber\end{eqnarray} where the contour
integral is a small circle about the origin. For $n$ very large, the
leading contribution comes from the asymptotic behaviour for $z\to0$
of $\hat Z(z)$. Thus, making use of the Eq. (\ref{14.14}) we may write
\begin{eqnarray} \hat{v}_n\simeq\frac{A}{2\pi i}\oint z^B
e^{zn+Cz^{-2}}\,dz\,. \nonumber\end{eqnarray} A standard saddle point
evaluation (or Meinardus main theorem) gives as $n\to\infty$
\begin{eqnarray} \hat{v}_n\simeq\hat{C}_1 n^{-\frac{B+2}{3}}
\exp{(\hat{b}_1 n^{\frac{2}{3}})}\:, \label{17.17} \end{eqnarray} with
$B$ as in Eq.~(\ref{ABC}),
$\hat{b}_1=\frac{3}{2}(D^2\zeta(3))^{\frac{1}{3}}$ and \begin{eqnarray}
\hat{C}_1=\frac1{\sqrt{6}}\, 2^{\frac{D(D-2)}4}\pi^{\frac{(D+1)(D-2)}4}
e^{\frac{D^2}{2}\zeta'_R(-1)} \left[
D^2\zeta_R(3)\right]^{\frac{6(D-2)-5D^2}{72}} \:.\label{19.19}
\end{eqnarray} Eqs.~(\ref{14.14}), (\ref{ABC}), (\ref{17.17}) and
(\ref{19.19}) have been obtained in \cite{byts94-35-2057}. The factor
$\hat{b}_1$ is in agreement with result in Ref.~\cite{hama92-88-149},
where however the prefactor $\hat{C}_1$ was missing. Here, with the
help of Meinardus techniques, we have been able to compute it.

The asymptotic behaviour given by Eq.~(\ref{17.17}) should be compared
with the one of the ordinary bosonic string and $p$-brane which we
have discussed in previous subsections. As a consequence, one may
conclude that the parabosonic string, in the limit of infinite
paraquantization parameter, behaves as an ordinary membrane ($p=2$).
We will see that there is some indication that canonical partition
function for $p$-branes does not exist. Thus, with regard to Hagedorn
temperature, the situation for parabosonic strings may be similar to
membranes. Hence, the concept of Hagedorn temperature is likely to be
meaningless for parastrings.

\subsection{Extented objects and black holes}

Recently, spacetimes with black $q$-brane solutions, namely singular
spacetimes for which the region of singularity assumes the shape of a
$q$-brane, have been constructed \cite{horo91-360-197}. Such solutions
have attracted much attention in view of the fact that they can
represent vacuum solutions of the 10-dimensional superstring for which
$q=10-D$ is the dimensionality of an embedded flat space. The problem
of finding black $q$-brane solutions of the 10-dimensional superstring
theory can be reduced to the problem of finding black hole solutions
to the Einstein equations in $D$-dimensions
\cite{horo91-360-197,gibb88-298-741}. These solutions have been used
to study the statistical mechanics of black holes using the
microcanonical ensemble prescription, this prescription being the
unique reasonable framework for analyzing the problem
\cite{hawk76-13-191,harm92-46-2334,harm93-47-2438}. In
Refs.~\cite{harm92-46-2334,harm93-47-2438} the approximate
semiclassical formula for the neutral black hole degeneracy $\rho(M)$
of states at mass level $M$ has been obtained, which reads
\begin{eqnarray} \rho(M)\sim B(M)\exp\left(S_E(M)\right)\,.
\nonumber\end{eqnarray} Here $S_E$ is the Euclidean action (the so
called Bekenstein-Hawking entropy) \begin{eqnarray} S_E(M)
=\sqrt{\pi}\left[\frac{2^{2D-3}
\Gamma(\frac{D-1}{2})G}{(D-2)^{D-2}}\right]
^{\frac{1}{D-3}}M^{\frac{D-2}{D-3}} \:,\label{02} \end{eqnarray} $G$
the generalized Newton constant and the prefactor $B(M)$ represents
general quantum field theoretical corrections to the state density. To
begin with, firstly, let us consider the quantity $S_E(M)$. For $D=4$,
it reduces to the well known result \cite{beke73-7-2333,hawk75-43-199}
\begin{eqnarray} S_E(M) =4\pi G M^2\, . \label{03} \end{eqnarray} The
exponential factor of the statistical mechanical density of states
(degeneracies) for black holes given in Eqs.~(\ref{03}) reveals great
similarities with the corresponding exponential factor of the density
$\nu_p(M)$ of states of quantum $p$-branes derived above. For this
reason it has been proposed that black holes might be considered as
quantum extended objects like $p$-branes
\cite{harm92-46-2334,harm93-47-2438}. In fact, the comparison of
Eqs.~(\ref{20111}) and (\ref{02}) yields \begin{eqnarray}
p=\frac{D-2}{D-4}\,. \label{05} \end{eqnarray} The only integer
solutions of Eq.~(\ref{05}) are $p=1$ ($D=\infty$), the string case
corresponds to an infinite dimensional black hole, $p=2$ ($D=6$),
$p=3$ ($D=5$) and the limit $p\to\infty$ corresponding to the
4-dimensional black hole. This last result has been  pointed out in
Ref.~\cite{alva92-7-2889}.

It should be noted that for any fixed $p$, the density of states of a
$p$-brane grows slower than the one for 4-dimensional black holes, but
faster than the one for strings. So the probability of a $p$-brane
being in an interval of mass ($M,M+dM$) increases with the mass and
the total probability diverges. In accordance with the argument of
Ref.~\cite{hawk76-13-191} this divergence indicates a breakdown of the
canonical ensemble.

For super $p$-branes the asymptotic behaviour of the state density
looks just in the same way. We have seen in the previous subsection
that the variation of the Wess-Zumino-Witten term of the globally
supersymmetric $p$-brane action is a total derivative if and only if
the $\Gamma$-matrix identity holds for arbitrary spinors. Then there
exist only several admissible pairs ($p,D$), $1\leq p\leq5$,
associated with the four composition-division algebras for each spinor
type \cite{achu87-198-441,sier87-4-227,evan88-298-92}. Furthermore,
the inclusion of winding modes can be done. The value of this function
at $s=0$ does not depend on winding numbers $g_i$, but its derivative
in general depends on them. Therefore the total degeneracy of the
level $n$ and the density of states change in the presence of winding
modes.

Finally a warning about the possible identification between $p$-branes
and black holes. The first trivial observation stems from the fact
that the asymptotic behaviour of two functions does not lead to the
conclusion that they are similar. The second one is more subtle. If
one naively assume that black holes share, asymptotically, some
properties related to quantum extended objects, it is a fact that the
nature of these extended object may not be  determined by their
asymptotic behaviour. In fact, in the previous subsection, we have
seen that a  parabosonic string, asymptotically, behaves as a
membrane. Thus the two original physical systems are quite different
from the physical view point, even though they asymptotically have the
same behaviour. Finally, there is a third serious fact against the
identification of black holes and $p$-branes. As we have shown, for
the $p$-branes, the prefactor $B_p(M)$ is computable and is finite.
For the black holes, however, as first pointed out in Ref.
\cite{thoo85-256-727}, the prefactor $B(M)$ seems untractable and
turns out to be divergent and this divergence may be regarded as the
first quantum correction to the Bekenstein-Hawking entropy (see, for
example,
\cite{suss94-50-2700,dowk94-11-55,furs94u-32,solo95-51-609,cogn95u-342}.)

These divergences appear also in the so called "entanglement or
geometric entropy"
\cite{bomb86-34-373,sred93-71-666,call94-333-55,kaba94-329-46} and are
peculiar of space-times with horizons and, in these cases,  a possible
physical origin of them can be traced back to the equivalence
principle \cite{isra76-57-107,scia81-30-327,barb94-50-2712}.

\newpage
\setcounter{equation}{0}
\section{Finite temperature quantum properties on
ultrastatic space-time with hyperbolic spatial section}
\label{S:FTQP}

In this section we are going to study some properties of free and self
interacting scalar fields on ultrastatic manifolds $\cal M^D$ with
hyperbolic spatial section. In particular we shall give general
expressions for free energy and thermodynamic potential and, for the
more important 4-dimensional case $\cal M^4$, we shall also give low
and high temperature expansion of those quantities and we shall
discuss in some detail finite temperature effective potential and
Bose-Einstein condensation. For the sake of completeness we shall also
give free energy and thermodynamic potential for the cases in which
the constant curvature spatial section is $T^N$ and $S^N$.

\subsection{The free energy and thermodynamic potential} \label{S:FETP}

We recall that the complete high temperature expansion for the
thermodynamic potential of a relativistic ideal Bose gas was derived
and discussed in
Refs.~\cite{habe81-46-1497,habe82-23-1852,habe82-25-502}. The
extension of the formula to a Fermi gas was given first in
Ref.~\cite{acto86-256-689}. The expansions mentioned above have been
done with either trivial (that is $\mbox{$I\!\!R$}^{N}$) and toroidal
topology. Generalizations of the high temperature expansion to
particular curved spaces have been given in
Refs.~\cite{dowk89-327-267,camp91-8-529,kirs91-24-3281}. In
Refs.~\cite{dowk89-327-267,camp91-8-529} also spin 1/2 fields have
been discussed. Only recently space-times with hyperbolic spatial
sections have been considered for this purpose
\cite{cogn92-7-3677,byts92-291-26,byts92-7-2669,cogn94-49-5307}.

Here we present the evaluation of the thermodynamic potential for
massive scalar fields in thermal equilibrium at finite temperature
$T=1/\beta$ on an ultrastatic space-time with constant curvature
spatial section of the kind $T^N$, $S^N$ and $H^N/\Gamma$. Of course,
for arbitrary $N$, $\Gamma$ is assumed to contain only hyperbolic
elements. Only for $N=3$ and $N=2$  elliptic elements shall be taken
into account.

At very high temperatures, we expect the dominant term of the
thermodynamic energy to be insensitive to curvature, because particles
at high energy have wave length much smaller than the curvature
radius. We have no physical ground for similar conclusions with
respect to topological effects, if any. It must also be admitted that
there is no clear definition of what a "topological effect" should be,
because it is difficult to disentangle curvature from topology. Here
we leave it understood that by non trivial topology we really mean the
non triviality of the group $\Gamma$, i.e. $\Gamma\neq e$.

Free energy and thermodynamic potential have been  defined in
Sec.~\ref{S:FT}, where general integral representations valid on any
curved ultrastatic manifold have been also given. Here we will
specialize Eqs.~(\ref{logPF-Poisson}-\ref{logPF-Barnes}) and
(\ref{Om-Poisson}-\ref{Om-Barnes}) to torii, spheres and compact
hyperbolic manifolds with elliptic and hyperbolic elements. Moreover,
on these particular manifolds we shall be able to give another
representation of finite temperature quantities in terms of Mc Donald
functions, which is very useful in order to get the low temperature
expansion.

The free energy can be derived from the thermodynamic potential in the
limit $\mu\to0$ and for this reason we directly attach the computation
of thermodynamic potential, deriving the free energy as a particular
case.

\paragraph{The thermodynamic potential on $\mbox{$I\!\!R$}\times T^N$.}
Here we use the notations of Sec.~\ref{S:ZFRR}. There is no
possibility to confuse the volume $\Omega_N$ with the thermodynamic
potential $\Omega_\beta$. Using Eq.~(\ref{KtTN}) in
Eq.~(\ref{Om-Jacobi}), after some calculations we obtain
\begin{eqnarray}
\Omega^T_\beta(\beta,\mu|\L_N)&=&-\frac{2\sqrt2\Omega_N}{\sqrt\pi}
\sum_{n=1}^\infty\sum_{\vec k\neq0} \frac{\cosh
n\beta\mu}{\sqrt{n\beta}}\nonumber\\ &&\qquad\qquad\times\left(\vec
k\cdot {\cal R}_N^{-1}\vec k+\alpha^2\right)^{1/4}\,
K_{1/2}(n\beta[\vec k\cdot {\cal R}_N^{-1}\vec k+\alpha^2]^{1/2})
\:,\label{Om-TN}\end{eqnarray} where $\vec k\in\mbox{$Z\!\!\!Z$}^N$
and for simplicity we have written the formula for untwisted fields.
Replacing $\vec k$ with $\vec k+\vec q$ one obtains the formula for
arbitrary twists. Of course, in the limit $\mu\to0$ this gives the
representation for the free energy. The latter equation is
particularly useful if one is interested in the low temperature
expansion. For high temperature expansion it is convenient to use
another expression, which can be obtained by the Mellin-Barnes
representation, Eq.~(\ref{Om-Barnes}).

\paragraph{The thermodynamic potential on $\mbox{$I\!\!R$}\times S^N$.}

As for the vacuum energy it is sufficient to compute the thermodynamic
potential for $S^1$ and $S^2$ and then apply the recurrence relations
for the $\zeta$-function, Eq.~(\ref{ZFRR}). In fact we have
\begin{eqnarray}
\frac{\Omega_\beta^S(\beta,\mu|\L_{N+2})}{\Omega_{N+2}}
=-\frac{1}{2\pi N\Omega_N}\left[ (\alpha^2+\kappa\varrho_N^2)
\Omega_\beta^S(\beta,\mu|\L_N)-\zeta'(-1|\L_N)\right]
\:.\nonumber\end{eqnarray}

For $S^1$ we have simply to take Eq.~(\ref{Om-TN}) for $N=1$. So we get
\begin{eqnarray}
\Omega_\beta^S(\beta,\mu|\L_1)=-\frac{2\sqrt2\Omega_1}{\sqrt\pi}
\sum_{n=1}^\infty\sum_{k=-\infty}^\infty \frac{\cosh
n\mu\beta}{\sqrt{n\beta}} \left( k^2+\alpha^2\right)^{1/4}\,
K_{1/2}(n\beta\sqrt{k^2+\alpha^2}) \:,\nonumber\end{eqnarray} while
for $S^2$ we use Eq.~(\ref{Kts2}) in Eq.~(\ref{Om-Jacobi}). In  this
way the thermodynamic potential reads \begin{eqnarray}
\Omega_\beta^S(\beta,\mu|\L_2)= -\frac{\Omega_2}{i\sqrt{2\pi}}
\sum_{n=1}^\infty\frac{\cosh n\beta\mu}{{(n\beta)^{3/2}}}
\int_\Gamma\frac{K_{3/2}(n\beta\sqrt{z^2+\alpha^2})\,
(z^2+\alpha^2)^{3/4}}{\cos^2\pi z}\,dz \:.\nonumber\end{eqnarray}

\paragraph{The thermodynamic potential on $\mbox{$I\!\!R$}\times
H^N/\Gamma$.}

In Sec.~\ref{S:FT} three different representations of the temperature
dependent part of the thermodynamic potential have been given by means
of Eqs.~(\ref{Om-Poisson}-\ref{Om-Barnes}). As we shall see in the
following, such representations are useful for low and high
temperature expansion. In order to specialize those equations to an
ultrastatic space-time with a compact spatial section $H^N/\Gamma$, we
separate the contributions coming from different kinds of elements of
the isometry group and first compute the contributions due to the
identity in the two cases $N=3$ and $N=2$.

Using Eqs.~(\ref{Om-Jacobi}), (\ref{K3-I}) and (\ref{K2-I}) we obtain
\begin{eqnarray} \Omega_\beta^I(\beta,\mu|\L_3)=
-\frac{\alpha^4\Omega({\cal F}_3)}{\pi^2}
\sum_{n=1}^{\infty}\frac{\cosh n\beta\mu}{(n\beta\alpha)^2}
K_2(n\beta\alpha) \:,\label{Om3-I} \end{eqnarray} \begin{eqnarray}
\Omega_\beta^I(\beta,\mu|\L_2)= -\frac{2\Omega( {\cal
F}_2)}{\sqrt{2\pi}} \sum_{n=1}^\infty\frac{\cosh
n\beta\mu}{{(n\beta)^{3/2}}}
\int_0^\infty\frac{K_{3/2}(n\beta\sqrt{r^2+\alpha^2})\,
(r^2+\alpha^2)^{3/4}}{\cosh^2\pi r}\,dr \:.\nonumber\end{eqnarray} As
for the spherical case, from the recurrence relations for the
$\zeta$-function, Eq.~(\ref{RRze}), we have the thermodynamic
potential by means of equation \begin{eqnarray}
\frac{\Omega_\beta^I(\beta,\mu|\L_{N+2})}{\Omega( {\cal F}_{N+2})}
=-\frac{1}{2\pi N\Omega( {\cal F}_N)}\left[
(\alpha^2+\kappa\varrho_N^2)
\Omega_\beta^I(\beta,\mu|\L_N)-\zeta'_I(-1|\L_N)\right]
\:.\nonumber\end{eqnarray}

The contribution due to hyperbolic elements can be computed in
arbitrary dimensions using Eq.~(\ref{dzeHN}) or (\ref{KI+KH}) in
Eq.~(\ref{Om-Poisson}) or (\ref{Om-Jacobi}). So we obtain the two
representations \begin{eqnarray} \Omega_\beta^H(\beta,\mu|\L_N)=
\frac{1}{\pi} \sum_{n=1}^\infty\int_{-\infty}^\infty e^{in\beta
t}\,\ln Z\left(\varrho_N+\sqrt{[t+i\mu]^2+\alpha^2}\right)\,dt
\:,\nonumber\end{eqnarray} \begin{eqnarray}
\Omega_\beta^H(\beta,\mu|\L_N)= -\frac{2\alpha^2}{\pi}
\sum_{n=1}^\infty\cosh n\beta\mu \sum_{\{\gamma\}}\sum_{k=1}^{\infty}
\frac{\chi^k(\gamma)l_\gamma}{S_N(k;l_{\gamma})}
\frac{K_1(\alpha\sqrt{(n\beta)^2+(kl_\gamma)^2}}
{\alpha\sqrt{(n\beta)^2+(kl_\gamma)^2}} \:.\label{OmN-H} \end{eqnarray}
As we shall see in a moment, the latter equation is particularly
useful in the low temperature expansion.

Finally, the contributions due to elliptic elements for $N=3$ and
$N=2$ read \begin{eqnarray} \Omega_\beta^E(\beta,\mu|\L_3)=
-\frac{2\alpha^2E}{\pi} \sum_{n=1}^{\infty}\frac{\cosh
n\beta\mu}{n\beta\alpha} K_1(n\beta\alpha) \:,\label{Om3-E}
\end{eqnarray} \begin{eqnarray} \Omega_\beta^E(\beta,\mu|\L_2)=
-\frac1\pi\sqrt{\frac2\pi} \sum_{n=1}^\infty\frac{\cosh
n\beta\mu}{\sqrt{n\beta}} \int_{-\infty}^\infty (r^2+\alpha^2)^{1/4}\,
K_{1/2}(n\beta\sqrt{r^2+\alpha^2})\,E_2(r)\,dr \:.\label{Om2-E}
\end{eqnarray}

\paragraph{The free energy on $\mbox{$I\!\!R$}\times H^N/\Gamma$.}

Putting $\mu=0$ in Eqs.~(\ref{Om3-I})-(\ref{Om2-E}) we obtain the
corresponding formulae for the free energy of a charged scalar field.
In the case of a neutral scalar field, we have to multiply all results
by a factor 1/2. Here we simply write down the formulae of free energy
for $N=3$ and $N=2$. They read \begin{eqnarray}
F_\beta(\beta|\L_3)&=&-\nu\sum_{n=1}^\infty\left\{
\frac{\alpha^4\Omega( {\cal F}_3)}{\pi^2}
\frac{K_2(n\beta\alpha)}{(n\beta\alpha)^2} +\frac{2\alpha^2E}{\pi}
\frac{K_1(n\beta\alpha)}{(n\beta\alpha)}
\right.\nonumber\\&&\qquad\qquad\left. -\frac{2}{\pi}\int_0^\infty
\cos n\beta t\,\ln
Z\left(\varrho_3+\sqrt{t^2+\alpha^2}\right)\,dt\right\} \:,\label{FE3}
\end{eqnarray} \begin{eqnarray}
F_\beta(\beta|\L_2)&=&-\nu\sum_{n=1}^\infty\left\{
\frac{\sqrt{2}\Omega( {\cal F}_2)}{\sqrt{\pi}}
\int_{0}^\infty\frac{K_{3/2}(n\beta\sqrt{r^2+\alpha^2})\,
(r^2+\alpha^2)^{3/4}}{(n\beta)^{3/2}\,\cosh^2\pi r}\,dr
\right.\nonumber\\ &&\qquad\qquad+\frac{2^{3/2}}{\pi^{3/2}}
\int_{0}^\infty\frac{K_{1/2}(n\beta\sqrt{r^2+\alpha^2})\,
(r^2+\alpha^2)^{1/4}\,E_2(r)}{\sqrt{n\beta}}\,dr
\nonumber\\&&\qquad\qquad\qquad\qquad\left. -\frac{2}{\pi}\int_0^\infty
\cos n\beta t\,\ln
Z\left(\varrho_2+\sqrt{t^2+\alpha^2}\right)\,dt\right\}
\:.\nonumber\end{eqnarray}

In the rest of this section we focus our attention on a 4-dimensional
space-time with a hyperbolic spatial part. Then we simplify the
notation using  $ {\cal F}$ in place of $ {\cal F}_3$ and we leave to
drop the argument of the functions. For example, $\Omega_\beta$ has to
be understood as $\Omega_\beta(\beta,\mu|\L_3)$. We also substitute
$M$, the effective mass, in place of $\alpha$.

\paragraph{The low temperature expansion of thermodynamic potential on
$\mbox{$I\!\!R$}\times H^3/\Gamma$.}

Recalling that for real values of $z$ and $\nu$ the asymptotics for Mc
Donald functions reads \begin{equation}
K_\nu(z)\sim\sqrt{\frac{\pi}{2z}}e^{-z}
\sum_{k=0}^{\infty}\frac{\Gamma(\nu+k+1/2)}{\Gamma(k+1)
\Gamma(\nu-k+1/2)}(2z)^{-k} \nonumber\end{equation} and taking
Eqs.~(\ref{Om3-I}), (\ref{OmN-H}) and (\ref{Om3-E}) into account we
get the low temperature expansion \begin{eqnarray} \Omega_\beta&\sim&
-\sum_{n=1}^\infty\sum_{\{\gamma\}}\sum_{j=1}^{\infty}
\frac{\chi^j(\gamma)l_\gamma}{S_3(j;l_{\gamma})}
\frac{M^2e^{-n\beta\left[ M\sqrt{1+(jl_\gamma/n\beta)^2}-|\mu|\right]}}
{(2\pi)^{1/2}\left( M\sqrt{n^2\beta^2+j^2l_\gamma^2}\right)^{3/2}}
\nonumber\\&&\qquad\qquad\qquad\qquad\times\:
\sum_{k=0}^{\infty}\frac{\Gamma(k+3/2)
\left(2M\sqrt{n^2\beta^2+j^2l_\gamma^2}\right)^{-k}}
{\Gamma(k+1)\Gamma(-k+3/2)} \nonumber\\&& -\frac{M^2E}{(2\pi)^{1/2}}
\sum_{n=1}^{\infty}\frac{e^{-n\beta(M-|\mu|)}}{(n\beta M)^{3/2}}
\sum_{k=0}^{\infty} \frac{\Gamma(k+3/2)(2n\beta
M)^{-k}}{\Gamma(k+1)\Gamma(-k+3/2)} \nonumber\\&&
\qquad\qquad-\frac{M^4\Omega({\cal F})}{(2\pi)^{3/2}}
\sum_{n=1}^{\infty}\frac{e^{-n\beta(M-|\mu|)}}{(n\beta M)^{5/2}}
\sum_{k=0}^{\infty}\frac{\Gamma(k+5/2)(2n\beta M)^{-k}}
{\Gamma(k+1)\Gamma(-k+5/2)} \:.\label{Om3-LTE} \end{eqnarray} It has
to be noted that the leading terms come from the topological part
since they dominate the volume part by one power of $1/\beta$. This is
in accord with the intuitive expectation that at very low temperature
the quantum field is probing the full manifold because the most
occupied states are the low energy ones. The last term of the latter
formula, which does not contain topological contributions, gives the
correct non relativistic limit if we set $\mu_{NR}=|\mu|-M$ for the
non-relativistic chemical potential.

\paragraph{The high temperature expansion of thermodynamic potential
on $\mbox{$I\!\!R$}\times H^3/\Gamma$.}

In order to derive the high temperature expansion, it is convenient to
use the Mellin-Barnes representation, Eq.~(\ref{Om-Barnes}) and
integrate it on a closed path enclosing a suitable number of poles. To
carry out the integration we recall that the $\zeta$-function in
3-dimensional compact manifold without boundary has simple poles at
the points $s=3/2-k$ $(k=0,1,\dots)$ and simple zeros at the points
$s=0,-1,-2,\dots$ \cite{seel67-10-172}. On the other hand,
$\zeta_R(s)$ has a simple pole at $s=1$ and simple zeros at all the
negative even numbers. Then the integrand function \begin{equation}
\Gamma(s+2n-1)\zeta_R(s)\zeta(\frac{s+2n-1}{2}|\L_3)\;\beta^{-s}
\:,\nonumber\end{equation} has simple poles at the points $s=1$ and
$s=-2(n+k)$ ($k=0,1,2,\dots$). Moreover, for $n=2$ we have another
simple pole at $s=0$, for $n=1$ at $s=0$ and $s=2$ and finally for
$n=0$ at $s=0, 2, 4$. In the latter case $s=0$ is a double pole. Hence
integrating this function and recalling that in our case the residues
of $\zeta(s|\L_3)$ are given by Eq.~(\ref{ZF3-res}) we obtain the high
temperature expansion \begin{eqnarray}
\Omega_\beta&=&-\frac{\Omega({\cal F})\pi^2}{45}\beta^{-4}
+\left[\frac{\Omega({\cal F})}{12}(M^2-2\mu^2) -\frac{\pi
E}{3}\right]\beta^{-2} \nonumber\\&&-\left\{ \frac{\Omega({\cal
F})(M^2-\mu^2)^{3/2}}{6\pi}-E(m^2-\mu^2)^{1/2}
\right.\nonumber\\&&\left.
\qquad\qquad+\sum_{n=0}^{\infty}\frac{\mu^{2n}}{\sqrt{\pi}n!}
\sum_{\{\gamma\}}\sum_{j=1}^\infty
\frac{\chi^j(\gamma)l_\gamma}{S_3(j;l_\gamma}
\left[\frac{jl_\gamma}{2M}\right]^{n-1/2} K_{n-1/2}(Mjl_\gamma)
\right\}\beta^{-1} \nonumber\\&&
-\left(\ln\frac{M\beta}{4\pi}+\gamma-\frac{3}{4}\right)
\left[\frac{M^4\Omega({\cal F})}{16\pi^2}-\frac{M^2E}{2\pi}\right]
-\mu^2\left[\frac{M^2\Omega({\cal F})}{8\pi^2} -\frac{E}{2\pi}\right]
+\mu^4\frac{\Omega({\cal F})}{24\pi^2} \nonumber\\&& -\frac{M^2E}{4\pi}
-\frac{M^2}{\pi}\sum_{\{\gamma\}}\sum_{j=1}^\infty
\frac{\chi^j(\gamma)l_\gamma}{S_3(j;l_\gamma}
\frac{K_1(Mjl_\gamma)}{Mjl_\gamma} +O(\beta^2) \:,\label{Om3-HTE}
\end{eqnarray} \begin{eqnarray}
O(\beta^2)&\sim&-\sum_{k=1}^\infty\beta^{2k}\zeta'_R(-2k)
\sum_{n=0}^k\frac{\mu^{2n}}{(2n)!(k-n)!}\:\:\times\nonumber\\
&&\qquad\qquad\qquad\qquad \left[\frac{\Omega({\cal
F})M^4}{16\pi^2(k-n+2)!}
-\frac{EM^2}{\pi(k-n+1)!}\right]\left(\frac{M}{2}\right)^{2(k-n)}
\:,\nonumber\end{eqnarray} which is in agreement with results of
Ref.~\cite{habe82-23-1852} in the case of a flat and topological
trivial space.

{}From the high-temperature expansion we see that hyperbolic elements
enter the thermodynamic potential and the free energy with two main
contributions, one linear in $T=1/\beta$ and one independent of $T$.
In sharp contrast are placed elliptic elements, which enter the
formulae in all terms of the expansion, but the leading one. Even with
zero chemical potential, the elliptic number gives a negative
contribution to the coefficient of $T^2=1/\beta^2$. Thus the pattern
of symmetry breaking in the finite temperature effective potential
will be probably modified by the elliptic topology of the manifold.

By looking at Eq.~(\ref{Om3-HTE}), we also note that $\Omega_\beta$ in
the complex $\mu$-plane presents, as expected, two branch points at
$|\mu|=M$ in agreement with flat-space results.

Another form of the high temperature expansion which looks quite
similar to the one which one has on a flat space-time, can derived by
means of Eq.~(\ref{Om-B2}) (see Ref.~\cite{cogn93-47-4575}).

The results here obtained can also be extended to fermions once a
given spin structure has been chosen on the manifold, the different
spin structures being parametrized by the first cohomology group
$H^1(H^3/\Gamma,\mbox{$Z\!\!\!Z$}_2)$. The expansion for fermions can
be obtained using the relation
$\Omega_{f}(\beta,\mu)=2\Omega_{b}(2\beta,\mu)-\Omega_{b}(\beta,\mu)$
\cite{dowk89-327-267}, where $f/b$ stands for fermion/boson degrees of
freedom, by referring to our previous results.

\subsection{The Bose-Einstein condensation on $\mbox{$I\!\!R$}\times
H^3$} \label{S:BEC}

As a first application of the formalism we have developed in this
section we discuss Bose-Einstein condensation for a non relativistic
ideal gas. The physical phenomenon, which in the non relativistic case
has a long story~\cite{eins24-22-261}, is well described in many text
books (see for example Ref.~\cite{huan63b}) and a rigorous
mathematical discussion of it was given by many authors
\cite{arak63-4-637,lewi74-36-1,land79-70-43}. The generalization to a
relativistic ideal Bose gas is non trivial and only recently has been
discussed in a series of papers
\cite{habe81-46-1497,habe82-23-1852,habe82-25-502,kirs91-24-3281}. The
effect of self-interaction has been taken into account, at least in
the one-loop approximation with a proper relativistic treatment
\cite{bens91-44-2480,bern91-66-683} and the interesting interpretation
of Bose-Einstein condensation as a symmetry breaking effect both in
flat as well as in curved space-time has been analyzed
\cite{habe82-25-502,bens91-44-2480,toms92-69-1152}. Related and recent
works on Bose-Einstein condensation in curved space-times have been
presented in Refs. \cite{park91-44-2421,park93-47-416,park95-51-2703}

It is well known that in the thermodynamic limit (infinite volume and
fixed density) there is a phase transition of the first kind in
correspondence of the critical temperature at which the condensation
manifests itself. At that temperature, the first derivative of some
continuous thermodynamic quantities has a jump. If the volume is kept
finite there is no phase transition, nevertheless the phenomenon of
condensation still occurs, but the critical temperature in this case
is not well defined.

As an example here we focus our attention on $\mbox{$I\!\!R$}\times
H^3$. We shall derive the thermodynamic potential for a free, charged
scalar field of mass $m$. As we know, the thermodynamic potential has
two branch points when the chemical potential $\mu$ reaches
$\pm\omega_0$, $\omega_0^2=M^2=m^2+|\kappa|$ being the lower bound of
the spectrum of the operator $\L_3=-\Delta_3+m^2$ and $\kappa$ the
negative constant curvature of $H^3$. The values $\pm\omega_0$ will be
reached by $\mu=\mu(T)$ of course for $T=0$, but also for some
$T=T_c>0$. This is the critical temperature at which the Bose gas
condensates.

The elementary properties of the Laplace-Beltrami operator on $H^N$,
that is the spectrum and the density of states has been derived in
Sec.~\ref{S:LDO}. In particular, for the density of zero angular
momentum radial functions we have found $\Phi_3(r)=r^2/2\pi^2$. The
continuum spectrum of the Laplacian on $H^3$ has a lower bound at
$\lambda=1$ (or $\lambda=|\kappa|$ in standard units) in contrast with
which in general happens on the compact manifold $H^3/\Gamma$. We also
notice that the wave operator propagates the field excitations on the
light cone, hence the gap should not be interpreted as a physical mass.

In the following, we shall need low and high temperature expansions of
thermodynamic potential. Then we shall use Eqs.~(\ref{Om3-LTE}) and
(\ref{Om3-HTE}) disregarding topological contributions. As usual, in
order to avoid divergences we shall consider a large volume $\Omega$
in $H^3$ and the limit $\Omega\to\infty$ shall be understood when
possible.

\paragraph{The Bose-Einstein condensation.} In order to discuss
Bose-Einstein condensation we have to analyze the behaviour of the
charge density \begin{equation}
\varrho=-\frac{\partial\Omega_\beta(\beta,\mu)}{\partial\mu}
=-\frac{\beta z}{\Omega} \frac{\partial\Omega_\beta(\Omega,\beta,z)}
{\partial z}\equiv f(z)-f(1/z) \label{roz} \end{equation} in the
infinite volume limit. Here $z=\exp\beta\mu$ is the activity and
\begin{equation} f(z)=\sum_{j}\frac{z}{\Omega(e^{\beta\omega_{j}}-z)}
\:.\nonumber\end{equation} The $\omega_{j}$ in the sum are meant to be
the Dirichlet eigenvalues for any normal domain $\Omega\subset H^{3}$.
That is, $\Omega$ is a smooth connected submanifold of $H^{3}$ with
non empty piecewise $C^{\infty}$ boundary. By the infinite volume
limit we shall mean that a nested sequence of normal domains
$\Omega_{k}$ has been chosen together with Dirichlet boundary
conditions and such that $\bigcup_k \Omega_{k}\equiv H^{3}$. The
reason for this choice is the following theorem (see for example
\cite{chav84b}): \begin{Theorem}[Mc Kean] if $\omega_0^k$ denotes the
smallest Dirichlet eigenvalue for any sequence of normal domains
$\Omega_{k}$ filling all of $H^{3}$ then $\omega_0^k\geq M$ and
$\lim_{k\to\infty}\omega_0^k=M$. \end{Theorem} Although the above
inequality is also true for Neumann boundary conditions, the existence
of the limit in not assured to the authors knowledge.

Now we can show the convergence of the finite volume activity $z_{k}$
to a limit point $\bar z$ as $k\rightarrow\infty$. To fix ideas, let
us suppose $\varrho\geq0$: then $z_{k}\in(1,\exp\beta\omega_0^k)$.
Since $\varrho(\Omega,\beta,z)$ is an increasing function of $z$ such
that $\varrho(\Omega,\beta,1)=0$ and
$\varrho(\Omega,\beta,\infty)=\infty$, for each fixed $\Omega_{k}$
there is a unique $z_{k}(\bar\varrho,\beta)\in
(1,\beta\exp\omega_0^k)$ such that
$\bar\varrho=\varrho(\Omega_{k},\beta,z_{k})$. By compactness, the
sequence $z_{k}$ must have at least one fixed point $\bar z$ and as
$\omega_0^k\to M^2$ as $k$ goes to infinity, by Mc Kean theorem, $\bar
z\in [1,\exp\beta M]$.

{}From this point on, the mathematical analysis of the infinite volume
limit exactly parallels the one in flat space for non relativistic
systems, as it is done for example in
Refs.~\cite{lewi74-36-1,ziff77-32-169,land79-70-43}. An accurate
analysis for the relativistic ideal gas in Minkowski space can be
found in Ref.~\cite{habe81-46-1497}. The difference between $R\times
H^3$ and $R^4$ is simply due to the fact that the mass $m$ is replaced
by $M$. For this reason, here we skip all details of computation and
refer the reader to the literature.

We recall that there is a critical temperature $T_c$, related to
$\varrho$ by a complicated integral equation, over which there are no
particles in the ground state. The solution of such an equation states
that for $T>T_c$ one always has $|\mu|<M$ and a vanishing charge
density $\varrho_0$ of the particles in the ground state, while for
$T\leq T_c$  $|\mu|$ remains equal to $M$ and the charge density of
the particles in the ground state is non vanishing. That is, below
$T_c$ one has Bose-Einstein condensation.

The critical temperature can be easily obtained in the two cases
$\beta M\gg 1$ and $\beta M\ll 1$ (in the case of massive bosons these
correspond to non relativistic and ultrarelativistic limits
respectively). In fact one has \begin{equation}
T_c=\frac{2\pi}{M}\left(\frac{\varrho}{\zeta_R(3/2)}\right)^{2/3}\:,
\qquad\qquad \beta M\gg 1 \:,\label{Tc1} \end{equation}
\begin{equation} T_c=\left(\frac{3\varrho}{M}\right)^{1/2}\:,
\qquad\qquad\beta M\ll 1 \label{Tc2} \end{equation} and the
corresponding charge densities of particles in the ground state
\begin{equation} \varrho_0=\varrho\left[1-(T/T_c)^{3/2}\right]\:,
\qquad\qquad\beta M\gg 1\:, \nonumber\end{equation} \begin{equation}
\varrho_0=\varrho\left[1-(T/T_c)^2\right]\:, \qquad\qquad \beta M\ll
1\:. \nonumber\end{equation}

It has to bo noted that for massless bosons, the condition $|\mu|\leq
M$ does not require $\mu=0$ like in the flat space, because $M>0$ also
for massless particles. This implies that the critical temperature is
always finite and so, unlike in the flat case, $\varrho$ is always
different from $\varrho_0$. As has been noticed in
Ref.~\cite{habe81-46-1497} that on a flat manifold the net charge of
massless bosons resides in the Bose-Einstein condensed ground state.
This never happens if the spatial manifold is $H^3$. In fact, because
of constant curvature massive and massless bosons have a similar
behaviour.

As it is well known, at the critical temperature, continuous
thermodynamic quantities may have a discontinuous derivative (first
order phase transition). This stems from the fact that the second
derivative with respect to $T$ of the chemical potential is a
discontinuos function for $T=T_c$. This implies that the first
derivative of the specific heat $C_V$ has a jump for $T=T_c$ given by
\begin{equation} \left.\frac{dC_V}{dT}\right|_{T_c^+}
-\left.\frac{dC_V}{dT}\right|_{T_c^-} =\left. \Omega T_c^+\mu''(T_c^)
\frac{\partial\varrho(T,\mu)}{\partial T}\right|_{T=T_c^+}
\label{jumpDTCV} \end{equation} where by the prime we indicate the
total derivative with respect to $T$.

Now we again consider the low and high temperature limits and compute
the discontinuity of the derivative of the specific heat using
Eq.~(\ref{jumpDTCV}). The charge density as given in Eq.~(\ref{roz})
can be obtained by deriving $-\Omega_\beta(\beta,\mu)/\Omega$ with
respect to $\mu$. For $T>T_c$ this gives $\mu$ as an implicit function
of $T$ and $\varrho$.

Deriving Eq.~(\ref{Om3-LTE}) with respect to $\mu$ and taking only
leading terms into account (of course disregarding topological
contributions), we get an expansion for $\varrho$ valid for small $T$.
It reads \begin{equation}
\varrho\simeq\left(\frac{MT}{2\pi}\right)^{3/2}
\sum_{n=1}^{\infty}\frac{e^{-n\beta(M-|\mu|)}}{n^{3/2}}
\:.\label{rolow} \end{equation} Of course, for $|\mu|=M$ this gives
Eq.~(\ref{Tc1}). From Eq.~(\ref{rolow}) we obtain \begin{equation}
\mu''(T_c^+)\sim-\frac{2.44}{T_c}\:,\qquad\qquad
T\frac{\partial\varrho}{\partial T}=\frac{3\varrho}{2}
\nonumber\end{equation} and the standard result \begin{equation}
\left.\frac{dC_V}{dT}\right|_{T_c^+}
-\left.\frac{dC_V}{dT}\right|_{T_c^-} =-\frac{3.66 Q}{T_c}
\end{equation} immediately follows ($Q=\varrho \Omega$ is the total
charge).

In a similar way, deriving Eq.~(\ref{Om3-HTE}) with respect to $\mu$
(always disregarding topological contributions), we get the expansion
for the charge density \begin{eqnarray} \varrho&\simeq&\frac{\mu
T^2}{3}-\frac{\mu T(M^2-\mu^2)^{1/2}}{2\pi}
+\frac{\mu(3M^2-2\mu^2)}{12\pi^2}\nonumber\\ &&+
\sum_{k=1}^\infty\beta^{2k}\zeta'_R(-2k)
\sum_{n=1}^k\frac{\mu^{2n-1}}{(2n-1)!(k-n)!}
\frac{M^4}{16\pi^2(k-n+2)!}\left(\frac{M}{2}\right)^{2(k-n)}
\:,\label{rohigh}\end{eqnarray} valid for $T>Tc$. For $\mu=M$, the
leading term of the latter expression gives again the result
(\ref{Tc2}). From Eq.~(\ref{rohigh}), by a straightforward computation
and taking only leading terms into account we get \begin{equation}
\mu''(T^+_c)\simeq -\frac{16\pi^2}{9M}\:,\qquad\qquad\qquad\qquad
\left. T\frac{\partial\varrho}{\partial T}\right|_{T=T^+_c} \simeq
2\varrho\:, \nonumber\end{equation} \begin{equation}
\left.\frac{dC_V}{dT}\right|_{T_c^+}
-\left.\frac{dC_V}{dT}\right|_{T_c^-} \simeq-\frac{32 Q\pi^2}{9M}
\:,\nonumber\end{equation} in agreement with the result on
$\mbox{$I\!\!R$}^4$ given in Ref.~\cite{habe81-46-1497}.

We have shown that both massless and massive scalar fields on the
Lobachevsky space $H^{3}$ exhibits Bose-Einstein condensation at a
critical temperature depending on the curvature of the space. The
higher is the curvature radius the higher is the critical temperature.
The treatment is not intended to be complete in any sense but we have
not been able to display a curvature effect on thermodynamic
quantities at the most elementary level. In particular, due to
curvature, massless charged bosons have a finite $T_{c}$ in contrast
with the flat space result. The difference can be traced back to the
existence of a gap in the spectrum of the Laplace operator on $H^{3}$.
We also pointed out that the infinite volume limit is under good
control only for Dirichlet boundary conditions, for which the smallest
eigenvalue has its limit value precisely at the gap of the continuous
spectrum.

\subsection{The finite temperature effective potential for a
self-interacting scalar field on $\mbox{$I\!\!R$}\times H^3/\Gamma$}
\label{S:FTEP}

As a last example, we shall consider finite temperature effects
associated with a self-interacting scalar field on ultrastatic
space-time of the form $\mbox{$I\!\!R$}\times H^3/\Gamma$. We shall
give low and high temperature expansions and we shall obtain the
one-loop, finite temperature effective potential as a by-product of
free energy.

To start with, let $\phi$ a neutral scalar field non-minimally coupled
to the gravitational field and with a self-interacting term of the
kind $\lambda\phi^4$. We concentrate on the temperature dependent part
of the one loop effective potential, the zero temperature
contributions, which require renormalization, being computed in
Sec.~\ref{S:OLEP}. We shall also assume the isometry group to contain
only hyperbolic elements.

After the compactification in the imaginary time $\tau$, the classical
action for such a system reads \begin{eqnarray}
S_c[\phi,g]=\int_0^{\beta}\,d\tau\,\int_{H^3/\Gamma}\left[
-\frac12\phi\left(\frac{\partial^2}{\partial\tau^2}+\Delta_3\right)
+V(\phi,R)\right]\,|g|^{\frac12}\,d^3x \nonumber\end{eqnarray} and the
small disturbance operator $A=-\partial^2_\tau+\L_3$, with
$\L_3=-\Delta_3+V''(\phi_c,R)$. As usual $\phi_c$ represents a
classical solution around which we expand the action. The potential is
assumed to be of the form \begin{eqnarray}
V(\phi,R)=\frac{m^2\phi^2}2+\frac{\xi R\phi^2}2
+\frac{\lambda\phi^4}{24} \:,\nonumber\end{eqnarray} which for
$\phi=\phi_c$ becomes the classical potential $V_c(\phi_c,R)$. From
now on we shall leave understood the explicit dependence on $R$.

Assuming a constant background field $\phi_c$ the concept of one-loop,
finite temperature effective potential is well defined
\cite{berk92-46-1551} and given by the one-loop free energy density.
As in Sec.~\ref{S:OLEA} we formally have
$V_{eff}(\phi_c)=V_c(\phi_c)+V^{(1)}(\phi_c)$, with the one-loop
quantum corrections \begin{eqnarray} V^{(1)}(\phi_c)=\frac1{2\beta
\Omega( {\cal F})}\ln\det(A\ell^2)
=V_0^{(1)}(\phi_c)+V_\beta^{(1)}(\phi_c)\:, \nonumber\end{eqnarray}
where we have separated the zero ($V_0^{(1)}(\phi_c)$) and finite
($V_\beta^{(1)}(\phi_c)$) temperature contributions.

Using Eqs.~(\ref{Ombemu}) and (\ref{FE3}) we directly have
\begin{eqnarray} V_\beta^{(1)}(\phi_c)&=&\frac{1}{\beta \Omega( {\cal
F})} \,\mbox{Tr}\,\ln\left(1-e^{-\beta Q}\right) \label{FTEP-log}\\
&=&-\frac{M^4}{2\pi^2} \sum_{n=1}^\infty\frac{K_2(n\beta M)}{(n\beta
M)^2}\nonumber\\ &&+\frac{1}{\pi \Omega( {\cal F})}\sum_{n=1}^{\infty}
\int_{0}^{\infty}\cos n\beta t\, \ln
Z\left(1+\sqrt{[t^2+M^2]/|\kappa|}\right)\,dt \:,\label{FTEP-Z}
\end{eqnarray} where $Q=|L_3|^{1/2}$ and \begin{eqnarray}
M^2=V''(\phi_c)+\kappa\varrho_3^2=m^2+\left(\xi-\frac16R\right)
+\frac{\lambda\phi_c^2}2 \:.\nonumber\end{eqnarray} Of course, other
representations can be derived from the results concerning the
thermodynamic potential in the limit $\mu\to0$. In particular we write
down the low and high temperature expansions, which we need in the
following. They read respectively \begin{eqnarray}
V_\beta^{(1)}(\phi_c)&\sim& -\frac{1}{2\Omega( {\cal
F})}\frac{M^2}{(2\pi)^{1/2}}
\sum_{n=1}^\infty\sum_{\{\gamma\}}\sum_{k=1}^{\infty}
\frac{\chi^k(\gamma)l_\gamma e^{-n\beta
M\sqrt{1+(kl_\gamma/n\beta)^2}}} {S_3(k;l_{\gamma})\left( n\beta
M\sqrt{1+(kl_\gamma/n\beta)^2}\right)^{3/2}} \nonumber\\
&&\qquad\times\left[1 +\frac3{4\left(n\beta
M\sqrt{1+(kl_\gamma/n\beta)^2}\right)} \right]
-\frac12\frac{M^4}{(2\pi)^{3/2}} \sum_{n=1}^\infty\frac{e^{-n\beta
M}}{(n\beta M)^{5/2}} \:.\label{EP-LTE} \end{eqnarray} \begin{eqnarray}
V_\beta^{(1)}(\hat\phi)&\sim&-\frac{\pi^2}{90\beta^4}
+\frac{M^2}{24\beta^2} -\left[\frac{M^3}{12\pi}-\frac{M}{2\Omega(
{\cal F})}\ln Z(1+M) \right]\frac{1}{\beta}\nonumber\\
&&-\frac{M^4}{32\pi^2} \left[\ln\frac{\beta M}{4\pi}+\gamma
-\frac{3}{4}\right]\label{EP-HTE} \\ &&-\frac{M^2}{2\pi \Omega( {\cal
F})|\kappa|^{1/2}}
\int_1^\infty\Xi(1+tM|\kappa|^{-1/2})\,\sqrt{t^2-1}\,dt
+O(\beta^2)\nonumber \:,\end{eqnarray} where only leading terms have
been written. Of course, also in this case the coefficients of the
positive powers of $\beta$ do not depend on the topology. It is
interesting to observe that the topological term independent of
$\beta$ in the latter formula is the same, but the sign, as the
topological contribution to the zero temperature effective potential,
also after renormalization (see Eq.~(\ref{OLEPtop}) in
Sec.~\ref{S:OLEP} and Ref.~\cite{cogn93-48-790}). This means that in
the high temperature expansion of the one-loop effective potential
only the term proportional to $T=1/\beta$ feels the non trivial
topology (hyperbolic elements only).

\paragraph{Phase transitions.}

The relevant quantity for analyzing the phase transitions of the
system is the mass of the field. The quantum corrections to the mass
are defined by means of equation \begin{eqnarray}
V_{eff}(\phi_c)=\Lambda_{eff} +\frac12(m_0^2+m_\beta^2)\phi_c^2
+O(\phi_c^4) \:,\nonumber\end{eqnarray} where $\Lambda_{eff}$ (the
cosmological constant) in general represents a complicated expression
not depending on the background field $\phi_c$ while $m_0$ and
$m_\beta$ represent the zero and finite temperature quantum
corrections to the mass $m$. As has been shown in Sec.~\ref{S:OLEP}
(see also Ref.~\cite{cogn93-47-4575}), $m_0$ has curvature and
topological contributions, which help to break symmetry (for
$\xi<1/6$). On the contrary, here we shall see that $m_\beta$ always
helps to restore the symmetry.

By evaluating the second derivative with respect to $\phi_c$ at the
point $\phi_c=0$ of Eqs.~(\ref{FTEP-log}) and (\ref{FTEP-Z}), we
obtain for $m_\beta$ the two equations \begin{eqnarray}
m_\beta^2&=&\frac{\lambda}{2\Omega( {\cal F})}
\,\mbox{Tr}\,\frac{e^{-\beta Q_0}}{Q_0(1-e^{-\beta Q_0})}
\label{mT0}\\ &=&\frac{\lambda M_0^2}{4\pi^2}
\sum_{n=1}^\infty\frac{K_1(n\beta M_0)}{n\beta M_0}\nonumber\\
&&+\frac{\lambda}{2\pi \Omega( {\cal F})|\kappa|}\sum_{n=1}^\infty
\int_0^\infty\cos n\beta t\, \frac{\Xi(1+\sqrt{[t^2+M^2_0]/|\kappa|})}
{\sqrt{[t^2+M^2_0]/|\kappa|}}\,dt \:,\label{mT} \end{eqnarray} where
$Q_0=Q_{\phi=0}=|-\Delta_3+m^2+\xi R|^{1/2}$ and
$M_0=M_{\phi=0}=|m^2+(\xi-1/6)R|^{1/2}$. From the exact formula
(\ref{mT0}) we see that the finite temperature quantum corrections to
the mass are always positive, their strength mainly depending on the
smallest eigenvalues of the operator $Q_0$. This means that such a
contribution always helps to restore the symmetry. The second
expression (\ref{mT}) gives the mass in terms of geometry and topology
of the manifold. In fact, wew recall that the $\Xi$-function is
strictly related to the isometry group $\Gamma$, which realizes the
non trivial topology of $\cal M$.

To go further, we compute the corrections to the mass in the high
temperature limit. Using Eq.~(\ref{EP-HTE}) we obtain \begin{eqnarray}
m_\beta^2&\sim&\frac{\lambda}{24\beta^2}
-\frac{3M_0\lambda}{8\pi\beta}\nonumber\\
&&+\frac{|\kappa|^{-1/2}}{4M_0^2 \Omega( {\cal F})} \left[\ln
Z(1+M_0|\kappa|^{-1/2})+M_0|\kappa|^{-1/2}\,
\Xi(1+M_0|\kappa|^{-1/2})\right]\,\frac{\lambda M_0}{\beta}
\:,\nonumber\end{eqnarray} from which we see that if temperature is
high enough, quantum corrections always help to restore symmetry. By
using Eq.~(\ref{EP-LTE}) one can also compute the quantum corrections
to the mass in the low temperature limit. Of course one obtains
exponentially damped corrections dominated by the topological part
\cite{cogn94-49-5307}.

\newpage
\setcounter{equation}{0}
\section{Strings at finite temperature}

Since the early days of dual string models, it was been known that an
essential ingredient of string theory at non-zero temperature was the
so-called Hagedorn temperature
\cite{hage65-3-147,fubi69-64-1640,huan70-25-895}. It was soon
recognized that the appearance of the Hagedorn temperature was a
consequence of the fact that the asymptotic form of the state level
density had an exponential dependence on the mass. A naive argument
led to the conclusion that above such a temperature the free energy
was diverging. After these pioneering works, the success of string
theory yielded a lot of investigations on finite temperature effects
for these extended objects
\cite{rhom84-237-553,kikk84-149-357,glei85-164-36,bowi85-54-2485},
\cite{alva85-31-418,saka86-75-692,polc87-104-539,anto87-199-402},
\cite{alva87-36-1175,ahar87-199-366,mcgu88-38-552}. In
Refs.~\cite{witt88-117-353,atic88-310-291} the possible occurrence of
a first order transition above the Hagedorn temperature has been
discussed. Further recent references are
\cite{odin89-8-207,ndeo89-40-2626,anto91-261-369,odin92-15-1}. One of
the main reasons for these investigations is connected with the
thermodynamic of the early universe (see
Refs.~\cite{bran89-316-391,alva86-269-596} and references therein) as
well as with the attempts to use extended objects for the description
of the high temperature limit of the confining phase of large N-SU(N)
Yang-Mills theory
\cite{polc92-68-1267,gree92-282-380}.

We shall try to investigate the finite temperature effects in string
theory within the canonical ensemble approach. The existence of the
Hagedorn spectrum (the exponentially growing density of states) leads
to the breakdown of the correspondence between canonical and
microcanonical ensemble above the Hagedorn temperature. Thus our
considerations are only valid below this critical temperature. We also
mention that the microcanonical ensemble, in some sense more
appropriate for the study of strings at finite temperature, has been
advocated and used in Ref.~\cite{mitc87-294-1138}.

In this section we put the Regge slope parameter $\alpha=1/2$, thus
the string tension is normalized at $T=1/{\pi}$.

\subsection{The Mellin-Barnes representation for one-loop string free
energy}

We start by recalling that there are different representations for the
string free energy. One of these, which is very useful for formal
manipulations, gives a modular invariant expression for the free energy
\cite{saka86-75-692,polc87-104-539,odin90-252-573}. However, this and
all other well known representations (see for example
Ref.~\cite{odin92-15-1}) are integral ones, in which the Hagedorn
temperature appears as the  convergence condition in the ultraviolet
limit of a certain integral. These may be called proper-time
representations. In order to discuss high- or low-temperature limits
in such representations one has to expand the integral in terms of a
corresponding series. Thus, a specific series expansion appears in
string theory at non-zero temperature. Here, making use of the
so-called Mellin-Barnes representation, we shall exhibit a Laurent
representation for the one-loop open (super)string free energy
\cite{byts93-394-423,byts93-8-1131}.

We have shown how to arrive at Mellin-Barnes representation for field
theory free energy at finite temperature in Sec.~\ref{S:FT}. Recalling
that the $\zeta$-function density related to the operator
$L_N=-\Delta_N+m^2=\vec p^2+m^2$ acting on functions in
$\mbox{$I\!\!R$}^N$ reads \begin{eqnarray}
\tilde\zeta(s|L_N)=\frac{\Gamma(s-\frac{N}2)m^{N-2s}}
{(4\pi)^{\frac{N}{2}}\Gamma(s)} \:\nonumber\end{eqnarray} and making
use of Eq.~(\ref{logPF-Barnes}), for the statistical sum of a free
massive field in a $D=N+1$-dimensional, flat space-time one has ($b$
stands for bosons and $f$ for fermions) \begin{eqnarray} \tilde
F_\beta^{b,f}=-\frac{(4\pi)^{-\frac D2}}{4\pi i}
\int_{c-i\infty}^{c+i\infty}\,\left(\frac\beta2\right)^{-s}
\Gamma(\frac{s}{2})\Gamma(\frac{s-D}{2})\zeta_{b,f}(s) m^{D-s}\,ds
\:.\label{bu}\end{eqnarray} Here $\zeta_b(s)=\zeta_R(s)$,
$\zeta_f(s)=(1-2^{1-s})\zeta_R(s)$ and the mass $m$ is a c-number. In
order to generalize the above representation to (super)strings, it is
sufficient to note that in that case one has to deal with a mass
operator $M^2$. As a consequence, for (super)string theory one can
generalize the representation (\ref{bu}) in the form \begin{eqnarray}
{\cal F}_{\mbox{string}}=-\frac{(4\pi)^{-\frac D2}}{4\pi i}
\int_{c-i\infty}^{c+i\infty}\,\left(\frac\beta2\right)^{-s}
\Gamma(\frac{s}{2})\Gamma(\frac{s-D}{2})\zeta_R(s) \,\mbox{Tr}\,
[M^2]^{\frac{D-s}2}\,ds \:,\nonumber\end{eqnarray} \begin{eqnarray}
{\cal F}_{\mbox{superstring}} =-\frac{(4\pi)^{-\frac D2}}{2\pi i}
\int_{c-i\infty}^{c+i\infty}\,\left(\frac\beta2\right)^{-s}
\Gamma(\frac{s}{2})\Gamma(\frac{s-D}{2}) (1-2^{-s})\zeta_R(s)
\,\mbox{Tr}\, [M^2]^{\frac{D-s}2}\,ds \:,\label{Fs1} \end{eqnarray}
where $D=26$ for strings and $D=10$ for superstrings. The symbol
$\,\mbox{Tr}\,$ means trace over boson and fermion fields. The
quantity $\,\mbox{Tr}\,[M^2]^{(D-s)/2}$ which appears in above
equations requires a regularization because a naive definition of it
leads to a formal divergent expression. For this reason we make use of
the Mellin transform \begin{eqnarray}
\,\mbox{Tr}\,[M^2]^{-s}=\frac{1}{\Gamma(s)}
\int_0^{\infty}\,t^{s-1}\,\mbox{Tr}\, e^{-tM^2} \,dt
\nonumber\end{eqnarray} and the heat-kernel expansion of
$\,\mbox{Tr}\, e^{-tM^2}$.

For bosonic strings the mass operator contains both infrared (due to
the presence of the tachyon in the spectrum) and ultraviolet
divergences, while for superstrings it contains only ultraviolet
divergences. Furthermore, for closed (super)strings the constraints
should be introduced via the usual identity (see for example
Ref.~\cite{odin92-15-1}). Hence the consideration of open superstrings
is simpler from a technical point of view and in the following we
shall consider them in some detail.

A simple and standard way to arrive at the heat-kernel expansion is
the following. For open superstrings (without gauge group) the mass
operator is given by (see for example Ref.~\cite{gree87b})
\begin{eqnarray} M^2=2\sum_{i=1}^{D-2}\sum_{n=1}^{\infty}  n \left(
N_{ni}^b + N_{ni}^f \right), \qquad\qquad D=10
\:.\nonumber\end{eqnarray} This leads to \cite{gree87b}
\begin{eqnarray} \,\mbox{Tr}\, e^{-tM^2} = 8\prod_{n=1}^{\infty}
\left( \frac{1-e^{-2tn}}{1+e^{-2tn}}
\right)^{-8}=8\left[\theta_4\left(0|e^{-2t} \right)\right]^{-8},
\label{ht1} \end{eqnarray} where $\theta_4(x,y)$ is the Jacobi
elliptic theta-function (see Appendix \ref{S:UR} for definition and
properties) and the presence of the factor 8 is due to the degeneracy
of the ground states.

Using Eq.~(\ref{theta4-small}) or alternatively Meinardus theorem in
Eq.~(\ref{ht1}) we obtain the asymptotics for small $t$
\begin{eqnarray} \left[\theta_4(0,e^{-2t})\right]^{-8}
\sim\frac{t^4}{(2\pi)^4}\,e^{\pi^2/t} -\frac{t^4}{2\pi^4}+
O\left(e^{-\pi^2/t}\right). \nonumber\end{eqnarray} So we may define
the regularized trace of the complex power $\,\mbox{Tr}\,
(M^2)^{\frac{D-s}2}$ ($D=10$) in the following way \begin{eqnarray}
\,\mbox{Tr}\, (M^2)^{5-\frac{s}2}
&=&\frac8{\Gamma\left(\frac{s}2-5\right)} \int_0^{\infty}
\,t^{\frac{s}2-6} \left[ [\theta_4(0,e^{-2t})]^{-8}
-\frac{t^4}{(2\pi)^4}\left(e^{\pi^2/t}-8\right)\right]\,dt
\nonumber\\&&\qquad\qquad\qquad\qquad
+\frac8{(2\pi)^4\Gamma\left(\frac{s}2-5\right)}
\int_0^{\infty}t^{\frac{s}{2}-2} \left(e^{2\pi^2/t}-8\right)\,dt
\:,\nonumber\end{eqnarray} where the latter integral has to be
understood in the sense of analytical continuation. This means that
$\int_0^{\infty}t^{s/2-2}\,dt=0$, while \begin{eqnarray}
\int_0^{\infty}t^{ \frac{s}{2}-2} e^{\pi^2/t} =(-\pi^2)^{\frac{s}{2}-1}
\Gamma\left(1-\frac{s}{2}\right) \:.\nonumber\end{eqnarray} The final
result then assumes the form \begin{eqnarray}
\mbox{Tr}\,(M^2)^{5-\frac{s}{2}}
=\frac1{2\pi^6\Gamma\left(\frac{s}{2}-5\right)}
\left[(-1)^{\frac{s}{2}-1}\pi^s\Gamma\left(1-\frac{s}{2}\right)
+\pi^{3/2}G(s,\Lambda)\right] \:,\label{Fs3}\end{eqnarray} where we
have set \begin{eqnarray}
G(s,\Lambda)=\sqrt\pi\left(\frac\pi2\right)^{\frac s2-1}
\int_0^{\Lambda} t^{\frac{s}{2}-6} \left\{\left[\frac{1}{2} \theta_4
\left(0,e^{-\pi t}
\right)\right]^{-8}-t^4\left(e^{2\pi/t}-8\right)\right\}\,dt.
\label{fs4} \end{eqnarray} In Eq. (\ref{fs4}) the infrared cutoff
parameter $\Lambda$  has been introduced. On the next stage of our
calculations this regularization will be removed (i.e. we will take
$\Lambda\rightarrow\infty$).

Making use of Eqs. (\ref{Fs1}) and (\ref{Fs3}) we get \begin{eqnarray}
 {\cal F}_{\mbox{superstring}}=-\frac{(2\pi)^{-11}}{2\pi i}
\int_{c-i\infty}^{c+i\infty}\left[\varphi(s)+\psi(s)\right]\,ds
\:,\label{fs5}\end{eqnarray} \begin{eqnarray}
\varphi(s)&=&(-1)^{\frac{s}{2}-1}(1-2^{-s})
\zeta_R(s)\frac{\pi}{\sin\frac{\pi s}{2}}
\left(\frac{\beta}{2\pi}\right)^{-s}, \\
\psi(s)&=&(1-2^{-s})\pi^{\frac32-s}\zeta_R(s) \Gamma(\frac
s2)G(s,\Lambda)\,(\frac{\beta}{2\pi})^{-s}. \nonumber\end{eqnarray}
The meromorphic function $\varphi(s)$ has first order poles at $s=1$
and $s=2k$, $k=1,2,\dots$. The pole of $\varphi$ at $s=1$ has
imaginary residue. The meromorphic function $\psi(s)$ has first order
poles at $s=1$. One can see that the regularization cutoff parameter
can be removed. If $c>1$, closing the contour in the right half-plane,
we obtain \begin{eqnarray}  {\cal
F}_{\mbox{superstring}}&=&-\frac2{(2\pi)^{11}}
\left[\sum_{k=1}^{\infty}(1-2^{-2k}) \zeta_R(2k)x^{2k}-\frac{\pi x}{4}
G(1,\infty) \right]\nonumber\\ &&+I_R(x)+I_R(x,\Lambda)
,\nonumber\end{eqnarray} where $x=\beta_c/\beta$, $\beta_c=2\pi$,
while $I_R(x)$ and $I_R(x,\Lambda)$  are the contributions coming from
the contour integrals of the functions $\phi(s)$ and $\varphi(s)$
respectively along the arc of radius $R$ in the right half-plane. The
series converges when $\beta>\beta_c=2\pi$, $\beta_c$ being the
Hagedorn temperature (see for example Ref.~\cite{alva87-36-1175}). The
sum of the series can be explicitly evaluated and the result is
\begin{eqnarray} \sum_{k=1}^{\infty}(1-2^{-2k})
\zeta_R(2k)\:x^{2k}=\frac{\pi x}{4} \tan\frac{\pi x}{2},\qquad\qquad
|x|<1 \ . \nonumber\end{eqnarray} As a consequence the statistical sum
contribution to the one-loop free energy is given by \begin{eqnarray}
{\cal F}_{\mbox{superstring}}=-\frac{\pi x}{2(2\pi)^{11}}
\left[\tan\frac{\pi x}{2}-G(1,\infty) \right]+I_R(x)+I_R(x,\Lambda)
\:.\label{SS-Laurent} \end{eqnarray} If $|x|<1$ then the value of the
contour integral $I_R(x)$ is vanishing when $R\rightarrow\infty$. With
regard to the contour contribution $I_R(x,\Lambda)$, we observe that
we can remove the cut-off $\Lambda$ and it has the corrected low
temperature limit (see for example Ref.~\cite{mats87-36-289}). For
high temperature it is negligible.

We conclude by observing that for the open bosonic string one can
repeat all the steps and arrive at the series representation
\begin{eqnarray} \sum_{k=1}^{\infty}\zeta_R(2k)\:y^{2k}
=\frac{1}{2}-\frac{\pi y}{2}\cot (\pi y),\qquad\qquad|y|<1
\:.\nonumber\end{eqnarray} As a result, for $y<1$ one can show that
the high temperature expansion assumes the form \begin{eqnarray}
{\cal F}_{\mbox{bosonic string}}\simeq
-\frac{1}{2^{23}\pi^{16}}\left[-y\cot(\pi y)
-\frac{2y}{\beta_c}D(1,\Lambda)+\frac1\pi D(0,\Lambda)\right]
\:.\label{BS-Laurent}\end{eqnarray} Here $y=\beta_c/\beta$, the
related Hagedorn temperature is $\beta_c=\sqrt8\pi$, \begin{eqnarray}
D(s,\Lambda)=2\pi\int_0^{\Lambda} t^{\frac s2-14}
\left[\eta(it)^{-24}-t^{12}e^{2\pi/t}\right]\,dt
\nonumber\end{eqnarray} and finally $\eta(\tau)=e^{i\pi \tau
/12}\prod_{n=1}^{\infty}(1-e^{2\pi in\tau})$ is the Dedekind eta
function. However in this case, the infrared cutoff parameter
$\Lambda$ cannot be removed for the presence of a tachyon in the
spectrum. We observe that a  similar asymptotic behaviour was already
been pointed out in Ref.~\cite{huan70-25-895}, but in a different
context.

The Laurent series have been obtained for $|x|<1$ and $|y|<1$, namely
for $\beta>\beta_c$. The right hand sides of the above formulae,
Eqs.~(\ref{SS-Laurent}) and (\ref{BS-Laurent}), may be understood as
an analytic continuations of those  series for $|x|>1$, $|y|>1$ (i.e.
$\beta<\beta_c$). As a consequence we have exhibited a kind of
periodic structure for the one-loop free energy of (super)strings.

The results we have obtained here are based on the Mellin-Barnes
representation for the one-loop free energy of the critical
(super)strings. Such a novel representation for the lowest order in
string perturbation theory has permitted to obtain  explicit
thermodynamic expressions in term of a Laurent series. The critical
temperature arises in this formalism as the convergence condition
(namely the radius of convergence) of these series. Furthermore, the
explicit analytic continuation of the free energy for temperatures
beyond the critical Hagedorn one ($\beta\leq\beta_c$) has been found.
As a result, there might be the possibility to analyze the breakdown
of the canonical ensemble and possible new string phases.

It is somewhat surprising that exists such a finite temperature
periodic structure in the behaviour of (super)string thermodynamic
quantities. The typical widths of the periodic sectors depends on the
Regge slope parameter $\alpha$. The widths of the sectors grow
together with the parameter $\alpha$ and in the limit $\alpha
\rightarrow 0$ (string tension goes to $\infty$), the thermodynamic
system can be associated with ideal massless gas of quantum fields
present in the normal modes of the string (see for example
Ref.~\cite{rhom84-237-553}). In addition, from Eq. (\ref{fs5}) it
follows that $ {\cal F}_{\mbox{superstring}}\sim\beta^{-10}$ and such
behaviour is consistent with the ordinary statistical mechanics
results
(see for example Ref.~\cite{mats87-36-289}).

\subsection{High genus contributions to string free energy}

The physical meaning of the Hagedorn temperature as the critical one
corresponding to the behaviour of thermodynamic ensembles, may also be
grasped by investigating the interplay between free strings and their
interactions (i.e. higher loops). In this subsection we would like  to
generalize to arbitrary genus-$g$ strings the Mellin-Barnes
representation previously obtained. Such a generalization will allow
us to identify the critical temperature at arbitrary loop order. Early
attempts to study the critical temperature for multi-loop strings has
been presented in
Refs.~\cite{mccl87-111-539,murp89-233-322,alva90-241-215}. Here we
closely follow Ref. \cite{byts93-311-87}.

It is  well-known that the genus-$g$ temperature contribution to the
free energy for the bosonic string can be written as
\cite{mccl87-111-539} \begin{eqnarray}
F_g(\beta)={\sum^\infty_{m_i,n_j=-\infty}}' \int (d\tau)_{WP}\, (\det
P^{\dag} P)^{1/2}(\det\Delta_g)^{-13} e^{-\Delta S(\beta,\vec m,\vec
n)}\,, \label{r1} \end{eqnarray} where $(d\tau)_{WP}$ is the
Weil-Petersson measure on the Teichm\"{u}ller space. This measure as
well as the factors $\det(P^{\dag}P)$ and $\det\Delta_g$ are each
individually modular invariant \cite{mccl87-111-539}. In addition
\begin{eqnarray} I_g(\tau)=(\det P^{\dag}P)^{1/2}(\det
\Delta_g)^{-13}=e^{C(2g-2)}Z'(1)^{-13}Z(2)\,, \label{r2} \end{eqnarray}
where $Z(s)$ is the Selberg zeta function and $C$  an universal
constant \cite{dhok88-60-917}. Furthermore, the winding-number factor
has the form of a metric over the space of windings, namely
\begin{eqnarray} \Delta S(\beta,\vec m,\vec n)=\frac{T\beta^2 }{2}
[m_l\Omega_{li}-n_i][(\,\mbox{Im}\, \Omega)^{-1}]_{ij}
[\bar{\Omega}_{jk}m_k-n_j] =g^{\mu\nu}(\Omega)N_{\mu}N_{\nu},
\nonumber\end{eqnarray} $T$ being the string tension (which here is
explicitly written) and $\mu,\nu=1,2,\dots,2g$,
$\{N_1,...,N_{2g}\}\equiv \{m_1,n_1,...,m_g,n_g\}$. The periodic
matrix $\Omega$ corresponding to the string world-sheet of genus $g$
is a holomorphic function of the moduli, $\Omega_{ij}=\Omega_{ji}$ and
$\,\mbox{Im}\, \Omega >0$. The matrix $\Omega$ admits a decomposition
into real symmetric $g \times g$ matrices, that is $\Omega
=\Omega_1+i\Omega_2$. As a result \begin{eqnarray}
g(\Omega_1+i\Omega_2)=\left( \begin{array}{cc}
\Omega_1\Omega_2^{-1}\Omega_1+\Omega_2&-\Omega_1\Omega_2^{-1}\\
-\Omega_2^{-1}\Omega_1 & \Omega_2^{-1} \end{array}
\right).\nonumber\end{eqnarray} Besides,
$g(\Omega)=\hat{\Lambda}^tg(\Lambda(\Omega))\hat{\Lambda}$
\cite{dhok88-60-917}, where $\Lambda$ is an element of the symplectic
modular group $Sp(2g,\mbox{$Z\!\!\!Z$})$ and the associated
transformation of the periodic matrix reads $\Omega \mapsto
\Omega'=\Lambda(\Omega)=(A\Omega+B)(C\Omega+D)^{-1}$. As a
consequence, the winding factor \begin{eqnarray} \sum_{\vec m,\vec
n}{}' \exp[-\Delta S(\beta,\vec m,\vec n)]\nonumber \end{eqnarray} is
also modular invariant.

It can be shown that the $2g$ summations present in the expression for
$F_g(\beta)$ can be replaced by a single summation together with a
change in the region of integration from the fundamental domain to the
analogue of the strip $S_{a_1}$ related to the cycle $a_1$, whose
choice is entirely arbitrary \cite{mccl87-111-539}. Then, one has
\begin{eqnarray} F_{g}(\beta)=\sum^\infty_{r=1} \int (d\tau)_{WP}\,
I_g(\tau) \exp\left[-\frac{T\beta^2r^2}{2}\Omega_{1i} [(\,\mbox{Im}\,
\Omega)^{-1}]_{ij} \bar{\Omega}_{j1}\right]. \label{r5} \end{eqnarray}
Let us now consider the Mellin-Barnes representation for the genus-$g$
free energy. A simple way to arrive at it is to make use of the Mellin
transform of the exponential factor, i.e. \begin{eqnarray}
e^{-v}=\frac{1}{2\pi i}\int_{c-i\infty}^{c+i\infty}\Gamma(s)v^{-s}
\,ds , \nonumber\end{eqnarray} with $\,\mbox{Re}\, v>0$ and $c>0$.
Therefore one gets \begin{eqnarray} \sum_{\vec m,\vec n}{}'
\exp[-\Delta S(\beta,\vec m,\vec n)] &=&\sum _{\vec m,\vec n}{}'
\frac1{2\pi i}\int_{c-i\infty}^{c+i\infty} \Gamma(s)(\Delta
S(\beta,\vec m,\vec n))^{-s}\,ds \nonumber\\&=& \frac{1}{2\pi
i}\int_{c-i\infty}^{c+i\infty} \Gamma(s)\left(\frac{\beta^2}{2
\pi}\right)^{-s}G_g(s;\Omega)\,ds, \label{r7} \end{eqnarray} where
\begin{eqnarray} G_g(s;\Omega)\equiv \sum_{\vec
N\in\mbox{$Z\!\!\!Z$}^{2g}/\{0\}} (\vec N^t \Omega \vec N)^{-s}
\nonumber\:,\end{eqnarray} \begin{eqnarray} \sum^\infty_{r=1}
\exp\left[-\frac{\beta^2r^2}{2\pi}\, \Omega_{1i}[(\,\mbox{Im}\,
\Omega)^{-1}]_{ij} \bar{\Omega}_{j1}\right] &=&\nonumber\\
&&\hspace{-5cm} \frac{1}{2\pi i}
\int_{c-i\infty}^{c+i\infty}ds\,\Gamma(s)\zeta_R(2s)
\left(\frac{\beta^2}{2\pi}\right)^{-s}
\left\{\Omega_{1i}[(\,\mbox{Im}\, \Omega)^{-1}]_{ij}
\bar{\Omega}_{j1}\right\}^{-s}. \label{r9} \end{eqnarray} Finally,
using the formulae (\ref{r7}) and (\ref{r9}) in Eqs.~(\ref{r1}) and
(\ref{r5}) respectively, one obtains \begin{eqnarray}
F_g(\beta)=\frac{1}{2\pi i}\int_{c-i\infty}^{c+i\infty}
ds\,\Gamma(s)(\frac{\beta^2}{2\pi})^{-s} \left\{\int d(\tau)_{WP}\,
I_g(\tau) G_g(s;\Omega)\right\} _{(Reg)} \:,\nonumber\end{eqnarray}
\begin{eqnarray} F_g(\beta)&=&\frac{1}{2\pi
i}\int_{c-i\infty}^{c+i\infty}ds
\Gamma(s)\zeta_R(2s)\left(\frac{\beta^2}{2\pi}\right)^{-s}\nonumber\\
&&\qquad\qquad\times\left\{\int(d\tau)_{WP}\, I_g(\tau)
\left[\Omega_{1i}[(\,\mbox{Im}\, \Omega)^{-1}]_{ij}
\bar{\Omega}_{j1}\right]^{-s}\right\} _{(Reg)}. \nonumber\end{eqnarray}
These are the main formulae which will be used for the evaluation of
the genus-$g$ string contribution. In order to deal with such
expressions, the integrals on a suitable variable in $(d\tau)_{WP}$
should be understood as the regularized ones. In this way the order of
integration may be interchanged.

For the sake of simplicity let us reconsider briefly the $g=1$ case.
It is well known that \cite{dhok88-60-917} \begin{eqnarray}
(d\tau)_{WP}=\frac{d\tau_1d\tau_2}{2\tau_2^2}\:,\qquad\qquad
I_1(\tau)=-\frac{\mbox{Vol}(\mbox{$I\!\!R$}^{26})}{(2\pi)^{13}}
[\tau_2|\eta(\tau)|^4]^{-12} \,,\nonumber\end{eqnarray} where
$\eta(\tau)$ is Dedekind eta function defined above and
$\mbox{Vol}(\mbox{$I\!\!R$}^{26})$ is the volume of a large region in
$\mbox{$I\!\!R$}^{26}$. In the case of an open bosonic string we have
$\Omega_1=0$, $\Omega_2=\tau_2$ and
$\Omega=diag\,(\tau_2,\tau_2^{-1})$. In the limit $\tau_2\rightarrow0$
we get \begin{eqnarray} \exp\left[-\frac{\beta^2}{2\pi} (\vec N^t
\Omega \vec N)^{-s}\right]\mapsto
\exp\left[-\frac{\beta^2}{2\pi}n^{-2s}\tau_2^{-1}\right]
\nonumber\:,\end{eqnarray} \begin{eqnarray} G_1(s;\Omega) \mapsto
\sum^\infty_{n=1}(n^2\tau_2^{-1}) ^{-s}=\tau_2^s\zeta_R(2s) .
\label{15} \end{eqnarray} The corresponding contribution to free
energy is given by \begin{eqnarray} \frac{1}{2\pi
i}\int_{c-i\infty}^{c+i\infty}ds\,
\Gamma(s)\left(\frac{\beta^2}{2\pi}\right)^{-s}\zeta_R(2s) \left\{
\int_0^\infty d\tau_2 \tau_2^{s-14} \eta(i \tau_2)^{-24}
\right\}_{(Reg)}. \nonumber\end{eqnarray} After having regularized the
ultraviolet region ($\tau_2\rightarrow0$) one has again Eq.
(\ref{BS-Laurent}) of previous Subsection.

By analogy with the above one-loop evaluation, now we shall consider
the open string genus-$g$ contribution to the free energy. The matrix
$\Omega$ may be chosen as $\Omega=diag\,(\Omega_2,\Omega_2^{-1})$. In
the limit $\Omega_2 \rightarrow 0$ one has \begin{eqnarray}
\exp\left[-\frac{\beta^2}{2\pi} (\vec N^t \Omega \vec N)^{-s}\right]
\mapsto \exp\left[-\frac{\beta^2}{2\pi}\Omega_2^{-1}\vec N^t \vec
N\right] \:,\nonumber\end{eqnarray} \begin{eqnarray} G_g(s;\Omega)
\mapsto \Omega_2^s \sum_{\vec N\in\mbox{$Z\!\!\!Z$}^g/\{0\}}(\vec N^t
\vec N) ^{-s} =\Omega_2^sZ_{I_g}(\frac{2s}g;0,0) \, ,
\nonumber\end{eqnarray} where $Z$ is the Epstein zeta function (see
Appendix \ref{S:UR} for definition and properties) and $I$ the
$g\times g$ identity matrix. In this way the contribution to free
energy reads \begin{eqnarray} \frac{1}{2\pi
i}\int_{c-i\infty}^{c+i\infty}ds\,
\Gamma(s)\left(\frac{\beta^2}{2\pi}\right)^{-s}
Z_{I_g}(\frac{2s}g;0,0)\left\{\int d\tau_{WP} \Omega_2^{s}
I_g(\tau)\right\} _{(Reg)}. \nonumber\end{eqnarray}

Since a tachyon is present in the spectrum, the total free energy will
be divergent for any $g$. The infrared divergence may be regularized
by means of a suitable cutoff parameter. Such a kind of behaviour can
be associated with the procedure of pinching a cycle non homologous at
zero (see for example Ref.~\cite{alva90-241-215}). It is well known
that the behaviour of the factor $(d\tau)_{WP}\,I_g(\tau)$ is given by
the Belavin-Knizhnik double-pole result and it has a universal
character for any $g$. It should also be noticed that this divergence
is $\beta$-independent and the meromorphic structure is similar to the
genus-one case. As a consequence, the contribution to the free energy
relative to high temperature, may be obtained again in terms of a
Laurent-like series and the whole genus dependence of the critical
temperature is encoded in the analogue of Riemann zeta-function,
namely Epstein zeta function. For this reason we have to determine the
asymptotic properties of $Z$. To this aim we make use of the following
general result: \begin{eqnarray} C_g\equiv\lim _{\,\mbox{Re}\,
s\rightarrow\infty} \frac{Z_{I_g}(\frac{2(s+1)}g;\vec
b,0)}{Z_{I_g}(\frac{2s}g;\vec b,0)}=
[(\hat{b}_1-\eta_1)^2+\cdots+(\hat{b}_g-\eta_g)^2]^{-1 } \:,\label{r27}
\end{eqnarray} where at least one of the $b_i$ is noninteger. By
$\hat{b}_i=b_i-[b_i]<1$ we indicate the decimal part of $b_i$, while
$\eta_i$ is equal to 0 or 1 according to whether $\hat b_i\leq1/2$ or
$\hat b_i\geq1/2$. Furthermore, if $\vec b=(0,0,...,0)$ then $C_g=1$.
This is just our case. Then we arrive at the conclusion that the
interactions of bosonic strings do not modify the critical Hagedorn
temperature in full agreement with other computations
\cite{mccl87-111-539,alva90-241-215}.

One can consider also different linear real bundles over compact
Riemann surfaces and spinorial structures on them. The procedure of
evaluation of the free energy in terms of the path integral over the
metrics $g_{\mu\nu}$ does not depend on whatever type of real scalars
are considered. This fact leads to new contributions to the genus-$g$
integrals (\ref{r1}) and Eq. (\ref{r2}). On the other hand, one can
investigate the role of these contributions for the torus
compactification \cite{dhok88-60-917,lerc89-177-1}. In this case, the
sum in Eq.~(\ref{r1}) should be taken over the vectors on the lattice
on which some space dimensions are compactified. The half-lattice
vectors can be labelled by the multiplets $(b_1,..,b_p)$, with
$b_i=1/2$. The critical temperature related to the multiplet $\vec
b=(b_1,...b_p,0,...,0)$ can be easily evaluated by means of
Eq.~(\ref{r27}), which gives $C_p=4p^{-1}$. As a result
\begin{eqnarray} T_{c,p}=\frac{\sqrt p}{2}\,T_c.
\nonumber\end{eqnarray} We note that the three particular multiplets
in which only one or two or three components of $\vec b$ are equal to
$1/2$ and all the others are vanishing are associated with "minimal"
critical temperatures given by $T_{c,1}=T_c/2$,  $T_{c,2}=T_c/{\sqrt
2}$ and  $T_{c,3}=T_c/{\sqrt 3}$ respectively.

 \par

\section*{Acknowledgments}{We would like to thank Alfred A.~Actor,
Roberto Camporesi, Emilio Elizalde, Klaus Kirsten and Sergei
D.~Odintsov for useful discussions and suggestions. A.A.~Bytsenko
thanks the Istituto Nazionale di Fisica Nucleare and the Dipartimento
di Fisica dell'Universit\`a di Trento for financial support and
Prof.~Marco Toller for the kind hospitality at the Theoretical Group
of the Department of Physics of the Trento University.}

\newpage
\section*{Appendices} \addcontentsline{toc}{section}{Appendices}
\appendix
\renewcommand{\theequation}{\mbox{\Alph{section}.\arabic{equation}}}
\setcounter{equation}{0}
\section{Admissible regularization functions for the determinant}
\label{S:RF}

For the sake of completeness we give in this appendix a list of
admissible regularization functions for the logarithm of the
determinant discussed in Sec.~\ref{S:RT}, which are often used in the
literature (see also Ref.~\cite{ball89-182-1}). We limit our analysis
to the physically interesting case $N=4$. The $N$-dimensional case can
be treated along the same line.

\paragraph{1.} Let us start with \begin{equation}
\varrho_1(\varepsilon,t)=\frac{d}{d\varepsilon}
\frac{t^\varepsilon}{\Gamma(\varepsilon)}
=\frac{t^{\varepsilon}}{\Gamma(\varepsilon)}\{\ln t-\psi(\varepsilon)\}
\:,\nonumber\end{equation} where $\psi(\varepsilon)$ is the
logarithmic derivative of $\Gamma(\varepsilon)$. This is the
zeta-function regularization introduced in Refs. \cite{ray71-7-145}
and popularized in the physical literature in the seminal paper
\cite{hawk77-55-133}. All requirements are satisfied. The related
$B_e(\varepsilon,y)$ and $V(\varepsilon,f)$, say $B_1(\varepsilon,y)$
and $V_1(\varepsilon,f)$ read \begin{equation} B_1(\varepsilon,y)=
y^{-\varepsilon}\ln y=\ln y + O(\varepsilon) \nonumber,\end{equation}
\begin{eqnarray} V_1(\varepsilon)&=&-\frac{\mu^4y^2}{32\pi^2}
\frac{d}{d\varepsilon}\frac{\Gamma(\varepsilon-2)
y^{-\varepsilon}}{\Gamma(\varepsilon)}
=\frac{\mu^4y^{2-\varepsilon}}{64\pi^2}
\frac{\Gamma(\varepsilon-2)}{\Gamma(\varepsilon)} \left[\ln
y-\psi(\varepsilon-2)+\psi(\varepsilon)\right]\nonumber\\
&=&\frac{M^4}{64 \pi^2}\left[\ln\frac{M^2}{\mu^2}-\frac{3}{2}\right]
+O(\varepsilon) \nonumber.\end{eqnarray} We see that the effective
potential is finite, the divergent terms being removed by the
particular structure of $\varrho_1$. This is the regularization we
have been used throughout the paper.

\paragraph{2.} The next regularization we shall consider is closely
related to the above one, and in some sense is associated with the
familiar dimensional regularization in momentum space. It is defined
by \cite{dowk76-13-3224} \begin{equation}
\varrho_2(\varepsilon,t)=t^{\varepsilon} \nonumber.\end{equation} From
this, \begin{equation} B_2(\varepsilon,y)= -y^{-\varepsilon}
\Gamma(\varepsilon)= \ln y+\gamma-\frac{1}{\varepsilon} +
O(\varepsilon) \nonumber,\end{equation} \begin{eqnarray}
V_2(\varepsilon)&=&-\frac{\mu^4y^{2-\varepsilon}}{32\pi^2}
\Gamma(\varepsilon-2)\nonumber\\
&=&\frac{M^4}{64\pi^2}\left[\ln\frac{M^2}{\mu^2}-\frac{3}{2}+\gamma\right]
-\frac{M^4}{64\pi^2\varepsilon}+O(\varepsilon) \nonumber,\end{eqnarray}
easily follow, again in agreement with the general result. Within this
regularization, only one divergent term is present.

\paragraph{3.} Another often used regularization is the ultraviolet
cut-off regularization, defined by \begin{equation}
\varrho_3(\varepsilon,t)=\theta(t-\varepsilon) \nonumber.\end{equation}
In this case we have \begin{equation}
B_3(\varepsilon,y)=-\Gamma(0,y\varepsilon) =\ln
y+\gamma+\ln\varepsilon + O(\varepsilon) \nonumber,\end{equation}
\begin{eqnarray}
V_3(\varepsilon)&=&-\frac{\mu^4y^2\Gamma(-2,\varepsilon
y)}{32\pi^2}\nonumber\\ &=&\frac{M^4}{64\pi^2}
\left[\ln\frac{M^2}{\mu^2}-\frac{3}{2}+\gamma\right]
-\frac{\mu^4}{64\pi^2\varepsilon^2}+\frac{\mu^2M^2}{32\pi^2\varepsilon}
+\frac{M^4}{64\pi^2}\ln\varepsilon+O(\varepsilon)
\nonumber.\end{eqnarray} Here we have three divergent terms. From
these examples, it is obvious that they depend on the regularization
function.

\paragraph{4.} The fourth regularization reads \begin{equation}
\varrho_4(\varepsilon,t)=e^{-\varepsilon/4t} \nonumber.\end{equation}
We have \begin{equation} B_4(\varepsilon,y)= -2K_0(\sqrt{\varepsilon
y}) =\ln y+2\gamma-2\ln 2+\ln\varepsilon + O(\varepsilon)
\nonumber,\end{equation} \begin{eqnarray}
V_4(\varepsilon)&=&-\frac{\mu^4y}{4\varepsilon\pi^2}
K_2(\sqrt{\varepsilon y})\nonumber\\ &=&\frac{M^4}{64\pi^2}
\left[\ln\frac{M^2}{\mu^2}-\frac{3}{2}+2\gamma-2\ln 2\right]
-\frac{\mu^4}{2\pi^2\varepsilon^2}+\frac{\mu^2M^2}{8\pi^2\varepsilon}+
\frac{M^4}{64\pi^2}\ln\varepsilon+O(\varepsilon)
\nonumber,\end{eqnarray} $K_{\nu}$ being the Mc Donald functions.

\paragraph{5.} The last regularization we would like to consider is
presented as an example of the freedom one has. It is similar to a
Pauli-Villars type and it is defined by ($\alpha$ being an arbitrary
positive constant) \begin{equation}
\varrho_5(\varepsilon,t)=(1-e^{-\alpha t/\varepsilon})^3
\nonumber,\end{equation} the power 3 being related to the fact that we
are working in four dimensions. This is a general feature of the
Pauli-Villars regularization. We have \begin{equation}
B_5(\varepsilon,y)=\ln\frac{\varepsilon y}{\varepsilon y+3\alpha}
+3\ln\frac{\varepsilon y+2\alpha}{\varepsilon y+\alpha} =\ln
y+\ln\frac{8}{3\alpha}+\ln\varepsilon+O(\varepsilon)
\nonumber,\end{equation} \begin{eqnarray}
V_5(\varepsilon)&=&\frac{\mu^4y^2}{64\pi^2}\left\{
\left[\ln\frac{\varepsilon y}{\alpha}-\frac{3}{2}\right] -3\left(
1+\frac{\alpha}{\varepsilon y}\right)^2\left[ \ln\left(
1+\frac{\varepsilon y}{\alpha}\right)-\frac{3}{2}\right]
\right.\nonumber\\&&\phantom{\frac{\mu^4y^2}{64\pi^2}} +\left. 3\left(
1+\frac{2\alpha}{\varepsilon y}\right)^2\left[ \ln\left(
2+\frac{\varepsilon y}{\alpha}\right)-\frac{3}{2}\right] -\left(
1+\frac{3\alpha}{\varepsilon y}\right)^2\left[ \ln\left(
3+\frac{\varepsilon y}{\alpha}\right)-\frac{3}{2}\right]
\right\}\nonumber\\ &=&\frac{M^4}{64\pi^2}
\left[\ln\frac{M^2}{\mu^2}-\frac{3}{2}+\ln\frac{8\alpha}{3}\right]
+\frac{3\alpha^2\ln(16/27)\mu^4}{64\pi^2\varepsilon^2}\\
&&\phantom{\frac{M^4}{64\pi^2}
\left[\ln\frac{M^2}{\mu^2}-\frac{3}{2}+\ln\frac{8\alpha}{3}\right]}
+\frac{3\alpha\ln(16/9)\mu^2M^2}{64\pi^2\varepsilon}
+\frac{M^4}{64\pi^2}\ln\varepsilon +O(\varepsilon)\nonumber
.\end{eqnarray} With this example we conclude the list of possible
regularization functions.

\setcounter{equation}{0}
\section{The heat kernel on a Riemannian manifold without boundary}

\subsection{Spectral coefficients for a Laplace-like operator}
\label{S:SC}

In the paper we need some coefficients of the heat kernel expansion
for a second order elliptic differential operator $A=-\Delta+V(x)$
acting on neutral scalar fields on Riemannian manifold without
boundary. In this case some coefficients have been computed by many
authors \cite{dewi65b,gilk75-10-601}. They are given by
\begin{eqnarray} a_0(x)&=&1\:,\qquad\qquad\qquad\qquad
a_1(x)=-V(x)+\frac{1}{6} R(x)\:, \label{a1}\\
a_2(x)&=&\frac12(a_1)^2+\frac16\Delta a_1(x) +\frac1{180}\left(\Delta
R+R^{ijrs}R_{ijrs}-R^{ij}R_{ij}\right) \:,\label{a2}\\
a_3(x)&=&\frac16(a_1)^3+a_1\left[ a_2-\frac12(a_1)^2\right]\nonumber\\
&&+\frac1{12}V^{;i}V_{;i}-\frac1{60}\Delta^2V +\frac1{90}R^{ij}V_{;ij}
-\frac1{30}R^{;i}V_{;i}\nonumber\\
&&+\frac{1}{7!}\left\{\phantom{\frac11}
18\Delta^2R+17R_{;i}R^{;i}-2R_{ij;r}R^{ij;r} -4R_{ij;r}R^{ir;j}
\right.\nonumber\\ &&\qquad\qquad+9R_{ijrs;q}R^{ijrs;q} -8R_{ij}\Delta
R^{ij} +24R_{ij}R^{ir;j}{}_{r} +12R_{ijrs}\Delta R^{ijrs} \nonumber\\
&&\qquad\qquad-\frac{208}9R_{ij}R^{ir}R^j_r
+\frac{64}3R_{ij}R_{rs}R^{irjs} -\frac{16}3R_{ij}R^i{}_{rsq}R^{jrsq}
\nonumber\\&&\left.\qquad\qquad
+\frac{44}9R_{ijrs}R^{ijmn}R^{rs}{}_{mn} +\frac{80}9R_{ijrs}R^{imrn}
R^j{}_m{}^s{}_n \right\}\:.\nonumber\end{eqnarray} We refer the
interested reader to Ref.~\cite{bran90-15-245} for the spectral
coefficients on a Riemannian manifold with boundary and to
Ref.~\cite{cogn88-214-70} for the spectral coefficients on a
Riemann-Cartan manifold.

In some physical problems it may be convenient to factorize the
exponential $\exp(ta_1)$ and consider an expansion (as introduced in
Refs.~\cite{park85-31-953,park85-31-2424}) very closely related to
Eq.~(\ref{HKEnB}), that is \begin{equation}
K_t^0(x,x)\sim\frac{e^{ta_1}}{(4\pi t)^{N/2}}
\sum_{n=0}^{\infty}b_n(x)t^n \:,\nonumber\end{equation} with $b_0=1$,
$b_1=0$ and more generally \begin{eqnarray} b_n(x)=\sum_{l=0}^n
\frac{(-1)^l a_{n-l}a_1^l}{l!}\:, \qquad\qquad a_n(x)=\sum_{l=0}^n
\frac{b_{n-l}a_1^l}{l!} \:.\nonumber\end{eqnarray} In this way, as
proved in Ref.~\cite{jack85-31-2439}, all coefficients $b_n$ do not
depend explicitly on $V(x)$, but eventually on its derivatives. The
coefficients $b_n$ up to $b_3$ have been computed in
Ref.~\cite{jack85-31-2439} and read \begin{eqnarray}
b_2(x)&=&\frac1{36}\Delta R-\frac16\Delta V +\frac1{180}\left(\Delta
R+R_{ijrs}R^{ijrs} -R_{ij}R^{ij}\right)\:,\label{b2}\\
b_3(x)&=&\frac1{12}V^{;i}V_{;i} -\frac1{60}\Delta^2V
+\frac1{90}R^{ij}V_{;ij} -\frac1{30}R^{;i}V_{;i}\nonumber\\
&&+\frac{1}{7!}\left\{\phantom{\frac11}
18\Delta^2R+17R_{;i}R^{;i}-2R_{ij;r}R^{ij;r} -4R_{ij;r}R^{ir;j}
\right.\nonumber\\ &&\qquad\qquad+9R_{ijrs;q}R^{ijrs;q} -8R_{ij}\Delta
R^{ij} +24R_{ij}R^{ir;j}{}_{r} +12R_{ijrs}\Delta R^{ijrs} \nonumber\\
&&\qquad\qquad-\frac{208}9R_{ij}R^{ir}R^j_r
+\frac{64}3R_{ij}R_{rs}R^{irjs} -\frac{16}3R_{ij}R^i{}_{rsq}R^{jrsq}
\nonumber\\&&\left.\qquad\qquad
+\frac{44}9R_{ijrs}R^{ijmn}R^{rs}{}_{mn} +\frac{80}9R_{ijrs}R^{imrn}
R^j{}_m{}^s{}_n \right\}\:.\nonumber\end{eqnarray}

\subsection{Heat kernel exact solutions on constant curvature
manifolds} \label{S:HKES}

Here we derive recurrence relations for the heat kernel related to the
Laplacian of functions on spheres and hyperbolic manifolds. Then we
just write down some known results concerning $S^N$, $H^N$ and $T^N$.
Other exact solutions on homogeneous spaces and bibliography on this
subject can be found for example in Ref.~\cite{camp90-196-1}.

\paragraph{The recurrence relation.} As simple examples of manifolds
in which the heat kernel related to the Laplacian $\Delta_N$ on
functions can be exactly computed and given in a closed form, here we
consider Riemannian manifolds with constant curvature. We use normal
coordinates $y^k(x,x')$ about the point $x$ and indicate by $\sigma$
the geodesic distance between $x$ and $x'$. Because of homogeneity, we
expect the heat kernel $K_t(x,x'|-\Delta_N)$ to depend only upon the
geodesic distance between $x$ and $x'$. Then, choosing a scalar
density $\tilde f(\sigma)$ of weight $-1/2$ depending only upon
$\sigma$, a  direct computation shows (note that $\tilde
f(\sigma)=[g_N(\sigma)]^{1/4}f(\sigma)$, $f(\sigma)$ being a true
scalar and $g_N(\sigma)$ the determinant of the metric tensor)
\begin{equation} \Delta_N\tilde f(\sigma)= \tilde f''(\sigma)
+\frac{N-1}{\sigma}\tilde f'(\sigma)
-\left(\beta'_N+\beta^2N+\frac{N-1}{\sigma}\beta_N\right)\tilde
f(\sigma) \:,\nonumber\end{equation} where by the prime we indicate
the derivative with respect to $\sigma$ and
$\beta_N(\sigma)=\partial_\sigma\ln[g_N(\sigma)]^{1/4}$. For
$g_N(\sigma)$ one has \begin{eqnarray} g_N(\sigma)&=&\left(\frac{\sin
\sigma}{\sigma}\right)^{2(N-1)}, \qquad\qquad\mbox{ for }S^N, \\
g_N(\sigma)&=&\left(\frac{\sinh \sigma}{\sigma}\right)^{2(N-1)},
\qquad\qquad\mbox{ for }H^N. \nonumber\end{eqnarray} For convenience
and without loss of generality, we have normalized the constant
curvature $\kappa$ to 1 for the sphere $S^{N}$ and to $-1$ for the
hyperbolic space $H^{N}$. In this way $\sigma$ is dimensionless. The
curvature can be restored in all the formulae by simple dimensional
arguments.

It is known (unpublished result of Millson reported in
Ref.~\cite{chav84b}, proved in Ref.~\cite{camp90-196-1}), that the
heat kernel $K_t(\sigma|-\Delta_{N+2})$ on a $N+2$ dimensional space
of constant curvature can be obtained from $K_t(\sigma|-\Delta_N)$ by
applying what is called "intertwining operator" in
Ref.~\cite{camp90-196-1}. In our notation and taking into account that
we are working with scalar densities, such an operator is given by
\begin{equation} D_N=-\frac{1}{2\pi}g_N(\sigma)^{1/4}
\frac{\partial}{\sigma\partial \sigma}g_N(\sigma)^{-1/4}
=-\frac{1}{2\pi \sigma}(\partial_\sigma-\beta_N(\sigma))
\:.\nonumber\end{equation} The following properties of $D_N$ can be
directly proved: \begin{equation} D_N[\Delta_N+\kappa N]\tilde
f(\sigma)=\Delta_{N+2}D_N\tilde f(\sigma) \:,\label{inter}
\end{equation} \begin{equation} D_N
[g_N(\sigma)]^{1/4}\delta_N=[g_{N+2}(\sigma)]^{1/4}\delta_{N+2}
\:,\nonumber\end{equation} where $\delta_N$ is the Dirac delta
function on $S^N$ or $H^N$. From Eq.~(\ref{inter}) we then obtain
\begin{equation} (\partial_t-\Delta_{N+2}-\kappa
N)D_NK_t(\sigma|-\Delta_N)=0 \:,\nonumber\end{equation}
\begin{equation} \lim_{t\to 0_+}D_NK_t(x,x'|-\Delta_N)
=[g_{N+2}(x)]^{1/4}\delta_{N+2}(x,x')[g_{N+2}(x')]^{1/4}
\:,\nonumber\end{equation} from which directly follows \begin{equation}
K_t(\sigma|-\Delta_{N+2})=e^{t\kappa N}D_NK_t(\sigma|-\Delta_N)
\label{KN} \:,\end{equation} where $N\geq 1$ for $S^N$ ($\kappa=1$)
and $N\geq 2$ for $H^N$ ($\kappa=-1$). The integral version of such
recurrence relation, Eq.~(\ref{KN}), was obtained in
Secs.~\ref{S:ZFRR} and \ref{S:HKZFHM}. Iterating this equation we
obtain the heat kernel for the Laplacian on $S^N$ by knowing the
kernels $K^{S^1}_t(\sigma)$ or $K^{S^2}_t(\sigma)$ according to
whether $N$ is odd or even. In a similar way we get the kernel on
$H^N$ starting from $K^{H^2}_t(\sigma)$ or $K^{H^3}_t(\sigma)$.

It has been shown in Ref.~\cite{camp90-196-1} that
$K_t(\sigma|-\Delta_N)$ can be related to $K_t(\sigma|-\Delta_{N-1})$
by introducing fractional derivatives of semi-integer order. In this
manner, the knowledge of $K^{S^1}_t(\sigma)$ or $K^{H^2}_t(\sigma)$ is
sufficient in order to get the heat kernel in any dimensional smooth
Riemannian space of constant curvature.

\paragraph{The torus case.} Because of flatness, the solution of the
heat equation for the Laplacian on $S^1=T^1$ can be easily derived
from the solution on $\mbox{$I\!\!R$}$, making use of the method of
images. The heat kernel on $\mbox{$I\!\!R$}^N$ is well known to be
\begin{equation} K^{\mbox{$I\!\!R$}^N}_t(x,x')=\frac{1}{(4\pi
t)^{N/2}}e^{-(x-x')^2/4t} \:.\label{HKRN} \end{equation} Putting $N=1$
and replacing $(x-x')^2$ with $(\sigma+2\pi nr)^2$ ($r>0$ being the
radius of curvature of $S^1$), we obtain an expression which of course
satisfy the heat equation on $S^1$. By summing over all $n$, we have
\begin{equation} K^{S^1}_t(\sigma)=\frac{1}{\sqrt{4\pi
t}}\sum_{n=-\infty}^{\infty} e^{-(\sigma+2\pi nr)^2/4t} \:.\label{HKS1}
\end{equation}

The heat kernel for the Laplacian on a $N$-dimensional torus can be
directly obtained by observing that $T^N=S^1\times\dots\times S^1$ is
the direct product of $N$ circles. Then using Eqs.~(\ref{HKS1}) and
(\ref{HKfact}) one gets \begin{equation}
K^{T^N}_t(\vec{\sigma})=\frac{1}{(4\pi t)^{N/2}}\sum_{\vec k}
e^{-(\vec{\sigma}+2\pi \vec{k})\cdot {\cal R}(\vec{\sigma}+2\pi
\vec{k})/4t} \label{HKTN} \:,\end{equation} where $\vec
k,\vec\sigma\in\mbox{$Z\!\!\!Z$}^N$ and $ {\cal R}$ is the diagonal
matrix $ {\cal R}=diag(r_1^2,\dots,r_N^2)$, $r_i$ being the radii of
the $N$ circles $S^1$.

\paragraph{The sphere case.} Looking at Eq.~(\ref{KN}), it follows
that the solution on the odd dimensional  spheres looks quite similar
to Eq.~(\ref{HKS1}). For example, for $N=3$ we immediately get
\begin{equation} K^{S^3}_t(\sigma)=\frac{e^{ta^2}}{(4\pi t)^{3/2}}
\sum_{n=-\infty}^{\infty} \left(1+\frac{2\pi
n/a}{\sigma}\right)e^{-(\sigma+2\pi n/a)^2/4t} \:,\label{HKS3}
\end{equation} where $a^2=\kappa$ has been put. Note that this
expression looks different from the one given, for example, in
Ref.~\cite{camp90-196-1}, because we are working with scalar
densities.

The solution on $S^2$ looks very differently and can be expressed for
example in terms of Legendre polynomials. It has also the integral
representation \cite{camp90-196-1} \begin{equation}
K^{S^2}_t(\sigma)=\frac{\sqrt{2}e^{ta^2/4}}{(4\pi t)^{3/2}}
\left(\frac{\sin a\sigma}{a\sigma}\right)^{1/2}
\sum_{n=-\infty}^{\infty}(-1)^n\int_{\sigma}^{\pi/a}
\frac{(a\sigma'+2\pi n)e^{-(\sigma'+2\pi n/a)^2/4t}} {(\cos
a\sigma-\cos a\sigma')^{1/2}}\,d\sigma' \:.\label{HKS2} \end{equation}

\paragraph{The hyperbolic case.} The solution on $H^N$ can be derived
from the corresponding solution on $S^N$, noting that the passage from
$S^N$ to $H^N$ is formally given by the replacement
$|\kappa|\to-|\kappa|$, that is $a\to ia$ ($a=\sqrt{|\kappa|}$). Of
course, due to the non compactness of $H^N$, one has to take into
account only the ``direct path'' $n=0$. Then, from Eqs.~(\ref{HKS3})
and (\ref{HKS2}) for $H^3$ and $H^2$ we get respectively
\begin{equation}
K^{H^3}_t(\sigma)=\frac{e^{-ta^2}e^{-\sigma^2/4t}}{(4\pi t)^{3/2}}
\:,\nonumber\end{equation} \begin{equation}
K^{H^2}_t(\sigma)=\frac{\sqrt{2}e^{-a^2t/4}}{(4\pi t)^{3/2}}
\left(\frac{\sinh a\sigma}{a\sigma}\right)^{1/2} \int_{\sigma}^{\infty}
\frac{a\sigma'e^{-\sigma'^2/4t}} {(\cosh a\sigma'-\cosh
a\sigma)^{1/2}}\,d\sigma' \:.\nonumber\end{equation}

\setcounter{equation}{0}
\section{The explicit computation of $\zeta$-function on compact
manifolds without boundary}
\label{S:ZF-ExComp}

Here we collect the explicit representations for $\zeta$-function on
several manifolds which are used in the text. The operator is assumed
to be of the form $\L_N=-\Delta_N+\alpha^2+\kappa\varrho_N^2$, with
$\varrho_N=(N-1)/2$ and $\alpha$ an arbitrary constant.

\paragraph{Example: $S^1$.} Using Eq.~(\ref{KtSN}) and Poisson
summation formula (\ref{PSF}) (or alternatively Eq.~(\ref{HKS1})) in
Eq.~(\ref{ZFdef}) we get \begin{eqnarray}
\zeta(s|\L_1)-\frac{\Omega_1\Gamma(s-1/2)\alpha^{1-2s}}{\sqrt{4\pi}\Gamma(s)}
&=&\frac{2\Omega_1\alpha^{1-2s}}{\sqrt{\pi}\Gamma(s)}
\sum_{n=1}^{\infty}\frac{K_{1/2-s}(2\pi n\alpha)}{(\pi
n\alpha)^{1/2-s}} \label{ZFS1-K}
\\&=&\frac{2\Omega_1\alpha^{1-2s}\sin\pi s}{\pi}
\int_{1}^{\infty}\frac{(u^2-1)^{-s}}{e^{2\pi\alpha u}-1}
\,du\:,\label{ZFS1-I}\end{eqnarray} from which we directly read off
the residues of the poles of $\zeta$-function at the points $s=1/2-k$
($k=0,1,\dots$). As for the torus, the representation of $\zeta$ in
terms of Mc Donald functions is valid for any $s$, while the last
integral is convergent only for $\,\mbox{Re}\, s<1$, but this is
sufficient for our aims. We incidentally note that the last term in
Eq.~(\ref{ZFS1-I}) looks quite similar to the contribution of
hyperbolic elements on a compact hyperbolic manifold (see
Eq.~(\ref{ZF-H-Z})). The right hand sides of the above formula, in the
limit $\alpha\to0$, give the $\zeta$-function in the massless case.

\paragraph{Example: $S^3$.} Using Eqs.~(\ref{ZFRR}) (\ref{ZFS1-K}) and
(\ref{ZFS1-I}) we obtain \begin{eqnarray}
\zeta(s|\L_3)-\frac{\Omega_3\Gamma(s-3/2)\alpha^{3-2s}}{(4\pi)^{3/2}\Gamma(s)}
&&\nonumber\\
&&\hspace{-3cm}=\frac{\Omega_3\alpha^{3-2s}}{(\pi)^{3/2}\Gamma(s)}
\sum_{n=1}^{\infty}\left[\frac{(s-1)K_{3/2-s}(2\pi n\alpha)} {(\pi
n\alpha)^{3/2-s}} -\frac{K_{1/2-s}(2\pi n\alpha)}{(\pi
n\alpha)^{1/2-s}}\right] \label{ZFS3}\\
&&=-\frac{\Omega_3\alpha^{3-2s}\sin\pi s}{\pi^2}
\int_{1}^{\infty}\frac{u^2(u^2-1)^{-s}}{e^{2\pi\alpha u}-1}
\,du\:.\nonumber\end{eqnarray} A similar representation in terms of Mc
Donald functions has been derived in
Refs.~\cite{mina49-13-41,mina52-4-26}.

\paragraph{Example: $S^2$.} Choosing the path of integration
$z=re^{\pm\pi/2}$ in Eq.~(\ref{ZFs2}) we get \begin{eqnarray}
\zeta(s|\L_2)=\frac{\Omega_2} {4(s-1)}\left[\int_{0}^{\alpha}
\frac{(\alpha^2-r^2)^{-(s-1)}}{\cosh^2\pi r}\,dr -\cos\pi
s\int_\alpha^\infty \frac{(r^2-\alpha^2)^{-(s-1)}}{\cosh^2\pi
r}\,dr\right] \:,\label{ZFS2} \end{eqnarray} which is valid for
$\,\mbox{Re}\, s<2$ and $\alpha>0$.

\paragraph{Example: $H^3/\Gamma$.} Here we only report identity and
elliptic contributions to heat kernel and $\zeta$-function, hyperbolic
contributions being expressed in terms of Selberg $Z$-function for any
$N$ (see Sec.~\ref{S:HKZFHM}). We have \begin{eqnarray}
K_I(t|\L_3)=\frac{\Omega( {\cal F}_3)e^{-t\alpha^2}}{(4\pi t)^{3/2}}
\:,\label{K3-I} \end{eqnarray} \begin{eqnarray}
\zeta_I(s|\L_3)=\frac{\Omega( {\cal F}_3)\Gamma(s-\frac32)}
{(4\pi)^{\frac32}\Gamma(s)}\alpha^{3-2s} \:,\label{ze3} \end{eqnarray}
\begin{eqnarray} \zeta'_I(0|\L_3) =\frac{\Omega( {\cal
F}_3)\alpha}{6\pi} \:,\nonumber\end{eqnarray} \begin{eqnarray}
K_E(t|\L_3)=\frac{Ee^{-t\alpha^2}}{\sqrt{4\pi t}} \:,\label{K3-E}
\end{eqnarray} \begin{eqnarray}
\zeta_E(s|\L_3)=\frac{E\Gamma(s-\frac12)}
{\sqrt{4\pi}\Gamma(s)}\alpha^{1-2s} \:,\label{ZF3-E} \end{eqnarray}
\begin{eqnarray} \zeta'_E(0|\L_3) =-E\alpha\:.\nonumber\end{eqnarray}

\paragraph{Example: $H^2/\Gamma$.} Again, identity and elliptic
contributions read \begin{eqnarray} K_I(t|\L_2)&=&\frac{\Omega( {\cal
F}_2)e^{-t\alpha^2}}{4\pi t} \int_{0}^{\infty}\frac{\pi
e^{-tr^2}}{\cosh^2\pi r}\;dr\nonumber\\ &=&\frac{\Omega( {\cal
F}_2)e^{-t\alpha^2}}{4\pi t} \sum_{n=0}^\infty\frac{B_{2n}}{n!}\left(
2^{1-2n}-1\right)\,t^n \:,\label{K2-I} \end{eqnarray} \begin{eqnarray}
\zeta_I(s|\L_2)=\Omega( {\cal F}_2)\left[
\frac{\alpha^{2-2s}}{4\pi(s-1)} -\frac1\pi\int_{0}^{\infty}
\frac{r\,(r^2+\alpha^2)^{-s}}{1+e^{2\pi r}}\,dr\right] \:,\label{ze2}
\end{eqnarray} \begin{eqnarray} \zeta'_I(0|\L_2)=\Omega( {\cal
F}_2)\left[ \frac{\alpha^2}{4\pi}(\ln\alpha^2-1)
+\frac1\pi\int_{0}^{\infty} \frac{r\,\ln(r^2+\alpha^2)}{1+e^{2\pi
r}}\,dr\right] \:,\nonumber\end{eqnarray} \begin{eqnarray}
K_E(t|\L_2)=\frac{e^{-t\alpha^2}}{2\pi}
\int_{-\infty}^{\infty}e^{-tr^2}\,E_2(r)\;dr \:,\nonumber\end{eqnarray}
\begin{eqnarray} \zeta_E(s|\L_2)=\int_{-\infty}^{\infty}
(r^2+\alpha^2)^{-s}E_2(r)\,dr \:,\label{ZF2-E} \end{eqnarray}
\begin{eqnarray} \zeta'_E(0|\L_2)= -\int_{-\infty}^{\infty}
\ln(r^2+\alpha^2)\,E_2(r)\,dr \:,\nonumber\end{eqnarray} where for
convenience we have introduced the function \begin{eqnarray}
E_2(r)=\sum_{\{\alpha\}}\sum_{j=1}^{m_\alpha-1}
\left[2m_\alpha\sin\frac{\pi j}{m_\alpha}\right]^{-1}
\frac{\exp\left(-\frac{2\pi rj}{m_\alpha}\right)} {1+\exp(-2\pi r)}
\nonumber\:.\end{eqnarray} In Eq.~(\ref{K2-I}), $B_n$ are the
Bernoulli numbers and the series is convergent for $0<t<2\pi$.

Looking at the above equations we see that elliptic elements of
isometry group modify the heat coefficients. This is due to the fact
that $H^N/\Gamma$ is not a smooth manifold when elliptic groups are
taken into consideration. In particular for $N=3$, from
Eqs.~(\ref{K3-I}) and (\ref{K3-E}) we easily obtain \begin{eqnarray}
K_{2n}(\L_3)=\frac{(-\alpha^2)^n}{n!} \left[\frac{\Omega({\cal
F})}{(4\pi)^{\frac32}}\;-\; \frac{4\pi
nE}{(4\pi)^{\frac12}\alpha^2}\right] \label{K2nL3}\end{eqnarray} and
of course $K_{2n+1}(\L_3)=0$ since $H^3/\Gamma$ is a manifold without
boundary. The residues of $\zeta(s|\L_3)$ at the poles $s=\frac32-n$
($n\geq0$) immediately follow from Eqs.~(\ref{ZFpoles}) and
(\ref{K2nL3}). We have \begin{eqnarray}
\,\mbox{Res}\,(\zeta(\frac32-n|\L_3))=
\frac{(-\alpha^2)^n}{\Gamma(n+1)\Gamma(\frac32-n)}
\left[\frac{\Omega({\cal F})}{(4\pi)^{\frac32}} \;-\;
\frac{nE}{(4\pi)^{\frac12}\alpha^2}\right]
\:.\label{ZF3-res}\end{eqnarray}

\setcounter{equation}{0}
\section{Useful relations}
\label{S:UR}

For reader convenience we collect some definitions and properties of
special functions which we used throughout the article.

\paragraph{The Poisson summation formula.} This is one of the most
useful summation formulae. It can be regarded as a non-abelian version
of the Selberg trace formula. There exist several versions. To begin
with, in the sense of distributions we have \begin{eqnarray}
\sum_{n=-\infty}^{\infty}\delta(x-2\pi n)
=\frac{1}{2\pi}\sum_{n=-\infty}^{\infty}e^{inx}\:. \label{PSF}
\end{eqnarray} In general, for any suitable function $f(\vec x)$,
$\vec x\in\mbox{$I\!\!R$}^N$ and any $\vec q\in\mbox{$I\!\!R$}^N$, the
Poisson summation formula reads \begin{eqnarray} \sum_{\vec
k\in\mbox{$Z\!\!\!Z$}^N}f(\vec k+\vec q) =\sum_{\vec
k\in\mbox{$Z\!\!\!Z$}^N} \int_{\mbox{$I\!\!R$}^N}f(\vec x) e^{2\pi
i\,\vec k\cdot(\vec x-\vec q)}\,d^Nx \:.\nonumber\end{eqnarray}

\paragraph{The Mellin transform.} Let $x^{z-1} f(x)$ belong to
$L(0,\infty)$ and let $f(x)$ have bounded variation on every finite
interval. Then the Mellin transform is defined by \begin{eqnarray}
\hat f(z)=\int_0^\infty x^{z-1} f(x) \,dx\,. \nonumber\end{eqnarray}
In the case when $f(x)$ is continuos, the Mellin inversion formula is
\begin{eqnarray} f(x)=\frac1{2\pi i} \int_{\,\mbox{Re}\,
z=c}x^{-z}\hat f(z)\,dz \nonumber\end{eqnarray} $c$ being a real
number in the strip in which $\hat f(z)$ is analytic. For any pair of
functions $f$, $g$ with Mellin transforms $\hat f$, $\hat g$, we have
the  useful Mellin-Parseval identity \begin{eqnarray} \int_0^\infty
f(x) g(x)\,dx=\frac1{2\pi i} \int_{\,\mbox{Re}\, z=c}\hat f(z)\hat
g(1-z) \,dz \label{M-P} \end{eqnarray} where $c$ is in the common
strip in which $\hat f(z)$ and $\hat g(1-z)$ are analytic.

\paragraph{Mc Donald $K_\nu(z)$-functions.} The following integral
representations \cite{grad80b} are frequently used in the paper:
\begin{eqnarray} K_\nu(z)&=&\frac{\left(\frac
z2\right)^\nu\Gamma\left(\frac12\right)}
{\Gamma\left(\nu+\frac12\right)} \int_1^\infty
e^{-zt}(t^2-1)^{\nu-\frac12}\,dt\nonumber\\
&&\:\:\:\left[\,\mbox{Re}\,\left(\nu+\frac12\right)>0         \mbox{
and }|\arg z|<\frac\pi2         \:\:\:\mbox{ or }\,\mbox{Re}\,
z=0\mbox{ and }\nu=0\right] \:,\\ K_\nu(z)&=&\frac12\left(\frac
z2\right)^\nu \int_0^\infty\frac{e^{-t-\frac{z^2}{4t}}}
{t^{\nu+1}}\,dt         \qquad\qquad\left[|\arg z|<\frac\pi2
\mbox{ and } \,\mbox{Re}\, z^2>0\right] \:.\nonumber\end{eqnarray}

\paragraph{Riemann-Hurwitz functions.} The prototype of the
zeta-functions is the celebrated Riemann-Hurwitz $\zeta_H$-function.
For $\,\mbox{Re}\, s>0$, it may defined by means of equation (see for
example \cite{grad80b}) \begin{eqnarray}
\zeta_H(s;q)=\sum_{n=0}^\infty(n+q)^{-s} \:,\label{RHzf}\end{eqnarray}
the sum being extended to all non-negative $n$ such that $n+q\neq0$.
Here $q$ is an arbitrary real number. We have the simple relation
$\zeta_H(s;1)=\zeta_R(s)$, where $\zeta_R(s)$ is the usual Riemann
function. It satisfies the useful functional equation \begin{eqnarray}
\pi^{-\frac{s}{2}}\Gamma(\frac{s}{2})\zeta_R(s)
=\pi^{-\frac{1-s}{2}}\Gamma(\frac{1-s}{2})\zeta_R(1-s)
\,.\nonumber\end{eqnarray}

\paragraph{The Epstein $Z$-function.} The Epstein $Z$-function can be
considered as a generalization of the one of Riemann-Hurwitz. We
consider three $N$ dimensional vectors $\vec q$ and $\vec h$ in
$\mbox{$I\!\!R$}^N$ and $\vec k\in\mbox{$Z\!\!\!Z$}^N$ and an
invertible $N\times N$ matrix $ {\cal R}$. For $\,\mbox{Re}\, s>1$ the
Epstein Z-function is defined by \cite{erde55b} \begin{eqnarray} Z_{
{\cal R}}(s;\vec q,\vec h) =\sum_{\vec k}e^{2\pi i \vec k\cdot\vec h}
[(\vec k+\vec q) {\cal R}(\vec k+\vec q)]^{-Ns/2}
\:,\label{Epstein}\end{eqnarray} where the sum run over $\vec
k\in\mbox{$Z\!\!\!Z$}^N$ for which $\vec k+\vec q\neq0$. If all
components of $\vec h$ are not integer, then Eq.~(\ref{Epstein}) can
be analytically continued to an entire function in the complex plane,
otherwise it has a simple pole at $s=1$ with residue $\pi^{Ns/2}\det
{\cal R}^{-1/2}\Gamma(1+N/2)$. $Z_{ {\cal R}}(s;\vec q,\vec h)$
satisfies the functional equation \begin{eqnarray}
\pi^{-\frac{Ns}2}\Gamma\left(\frac{Ns}2\right) Z_{ {\cal R}}(s;\vec
q,\vec h)&=&\nonumber\\ &&\hspace{-3cm} \det {\cal
R}^{-\frac12}\pi^{-\frac{N(1-s)}2} e^{-2\pi i\vec q\cdot\vec h}
\Gamma\left(\frac{N(1-s)}2\right) Z_{ {\cal R}^{-1}}(1-s;\vec h,-\vec
q) \:.\label{FR-Epstein}\end{eqnarray}

\paragraph{The Jacobi elliptic $\theta_4$-function.} In the paper we
shall use just $\theta_4(u,q)$, which is defined by (we follow
Ref.~\cite{grad80b}) \begin{eqnarray}
\theta_4(u,q)=\sum_{n=-\infty}^{\infty}(-1)^nq^{n^2}e^{2\pi i u}
\label{theta4}\end{eqnarray} and can be represented also as an
infinite product. In particular one has \begin{eqnarray}
\theta_4(0,q)=\prod_{n=1}^{\infty} \frac{1-q^n}{1+q^n}
\:.\nonumber\end{eqnarray} Setting $q=e^{-t}$ and using the dual
property \begin{eqnarray} \theta_4(0,e^{-t})= \sqrt{\frac{\pi}{t}}
\sum_{n=-\infty}^{\infty} e^{-\pi^2(n-1/2)^2/t}
\:,\nonumber\end{eqnarray} we obtain the asymptotic expansion (for
small $t$) \begin{eqnarray} \theta_4(0,e^{-t})\sim 2\sqrt{\frac\pi t}\,
\left[e^{-\frac{\pi^2}{4t}}
+e^{-\frac{(3\pi)^2}{4t}}+e^{-\frac{(5\pi)^2}{4t}}+ \dots\right]
\:.\label{theta4-small}\end{eqnarray}

\end{document}